\documentclass[12pt,twoside,english]{extarticle}
\usepackage{lmodern}
\usepackage{helvet}
\usepackage[T1]{fontenc}
\usepackage[latin9]{inputenc}
\usepackage{geometry}
\usepackage{color}
\usepackage{babel}
\usepackage{float}
\usepackage{mathrsfs}
\usepackage{amsmath}
\usepackage{amsthm}
\usepackage{amssymb}
\usepackage{setspace}
\usepackage[unicode=true,pdfusetitle,
 bookmarks=true,bookmarksnumbered=false,bookmarksopen=false,
 breaklinks=true,pdfborder={0 0 0},pdfborderstyle={},backref=false,colorlinks=true]
 {hyperref}

\makeatletter

\providecommand{\tabularnewline}{\\}

\numberwithin{equation}{section}
\numberwithin{table}{section}
\numberwithin{figure}{section}
\usepackage[natbibapa]{apacite}
\theoremstyle{plain}
\newtheorem{assumption}{\protect\assumptionname}
\theoremstyle{plain}
\newtheorem{thm}{\protect\theoremname}[section]
\theoremstyle{plain}
\newtheorem{prop}{\protect\propositionname}[section]
\theoremstyle{definition}
\newtheorem{defn}{\protect\definitionname}[section]
\theoremstyle{plain}
\newtheorem{lem}{\protect\lemmaname}[section]

\usepackage{graphicx}
\usepackage{amsmath}
\usepackage{dcolumn}
\usepackage{listings}
\usepackage{color}
\usepackage{setspace}
\usepackage[normalem]{ulem}  
\definecolor{hellgelb}{rgb}{1,1,0.8}
\definecolor{colKeys}{rgb}{0,0,1}
\definecolor{colIdentifier}{rgb}{0,0,0}
\definecolor{colComments}{rgb}{1,0,0}
\definecolor{colString}{rgb}{0,0.5,0}

\usepackage{lmodern}
\usepackage[T1]{fontenc}
\usepackage{geometry}
\usepackage{fancyhdr}
\usepackage{amsthm}
\usepackage{thmtools}
\usepackage{amsmath}
\usepackage{amssymb}
\usepackage{esint}
\usepackage{multirow}
\usepackage{mathrsfs} 
\usepackage{textgreek}
\usepackage{amsfonts}
\usepackage{amssymb}
\usepackage{mathrsfs} 

\usepackage[USenglish]{isodate}
\usepackage{chngcntr}

\numberwithin{equation}{section}
\numberwithin{table}{section}
\numberwithin{assumption}{section}
\counterwithout{figure}{section}
\counterwithout{table}{section}

\makeatother

  \providecommand{\assumptionname}{Assumption}

  \providecommand{\definitionname}{Definition}

  \providecommand{\lemmaname}{Lemma}

  \providecommand{\propositionname}{Proposition}

  \providecommand{\theoremname}{Theorem}
 
 \providecommand{\theoremname}{Theorem}
 
\usepackage{thmtools}

\newtheoremstyle{MyTheoremstyle}
  {\topsep} 
  {\topsep} 
  {} 
  {} 
  {\bfseries} 
  {.} 
  {.90em} 
  {} 
\theoremstyle{MyTheoremstyle} 
\theoremstyle{MyTheoremstyle} 
\theoremstyle{MyTheoremstyle} 
\theoremstyle{MyTheoremstyle} 
\theoremstyle{MyTheoremstyle}

\declaretheoremstyle[
    headfont=\bfseries,
    notefont=\normalfont,
    bodyfont=\itshape,
    headpunct=\newline,
    headformat={%
        \makebox{\NAME\ \NUMBER\ }{\NOTE}%
    },
]{theorem}

\newlength{\spacelength}
\declaretheoremstyle[
    headfont=\bfseries,
    notefont=\normalfont,
    bodyfont=\itshape,
    headpunct=\newline,
    headformat={%
        \makebox[0pt][l]{\NAME\ \NUMBER\ }\hskip-\spacelength{\NOTE}%
    },
]{theore}

 \usepackage{fancyhdr}
\usepackage{booktabs}
\usepackage{float}
\usepackage{graphicx}
\usepackage{epstopdf}
\usepackage{morefloats}
\usepackage[referable]{threeparttablex}
\usepackage{footnote}

\usepackage{setspace} 
\usepackage{graphicx,color}

\usepackage{listings}
\RequirePackage[T1]{fontenc}
\RequirePackage{ae,fancyvrb}
\DefineVerbatimEnvironment{Sinput}{Verbatim}{fontshape=sl}
\DefineVerbatimEnvironment{Soutput}{Verbatim}{}
\DefineVerbatimEnvironment{Scode}{Verbatim}{fontshape=sl}

\title{\bf Prewhitened Double Kernel HAC Estimation}
\author{
\textsc{\textcolor{MyBlue}{Alessandro Casini}}\thanks{Corresponding author at: Department of Economics and Finance, University of Rome Tor Vergata, Via Columbia 2, Rome 00133, IT. 
Email: 
\texttt{\textcolor{MyBlue}{{alessandro.casini@uniroma2.it}}}.} 
\\
\small{University of Rome Tor Vergata}
\and
\textsc{\textcolor{MyBlue}{Pierre Perron}}\thanks{Department of Economics, Boston University, 270 Bay State Road, Boston, MA 02215, US. 
Email: 
\texttt{\textcolor{MyBlue}{\mbox{perron@bu.edu}}}.} 
\\
\small{Boston University}
}

\date{\small{\today}}

 \makeatletter

\numberwithin{equation}{section}
\makeatother

\usepackage{babel}
\addto\captionsenglish{%
}

\usepackage{color}
\usepackage{xcolor}

\renewcommand*{\thesection}{\arabic{section}}

\definecolor{MyRed}{rgb}{0.8,0,0}
\definecolor{MyBlue}{rgb}{0,0,0.7}
\definecolor{Green}{rgb}{0,0.5,0}
\definecolor{hellgelb}{rgb}{1,1,0.8}
\definecolor{colKeys}{rgb}{0,0,1}
\definecolor{colIdentifier}{rgb}{0,0,0}
\definecolor{colComments}{rgb}{1,0,0}
\definecolor{colString}{rgb}{0,0.5,0}

\definecolor{MyLightRed}{rgb}{2.2,0.2,0.4} 
\definecolor{MyLightRed2}{rgb}{0.6,0.2,0.3} 
\definecolor{MyLightRed2temp}{rgb}{0.6,0.2,0.3}
\definecolor{MyLightRed3}{rgb}{0.8,0.1,0.1} 
\definecolor{MyRed}{rgb}{0.7,0.0,0}

\definecolor{MyLigthBlue13}{rgb}{0,0.2,0.7}
 \definecolor{MyLigthBlack}{rgb}{0.2,0.25,0.3} 

\hypersetup{%
  bookmarks=true
  pdftitle = {Title Here},
  pdfsubject = {},
  pdfkeywords = {Keywords},
  pdfauthor = {Alessandro Casini},}
\hypersetup{ 
  colorlinks = {true}, 
  citecolor = {MyBlue},
  linkcolor = {MyRed},
  filecolor= {Green},	
  urlcolor = {MyBlue},
  hyperindex = {true},
  linktocpage = {true},
  linkbordercolor ={1 0 0},
 citebordercolor ={0 1 1},
 urlbordercolor ={0 1 1}
}

\usepackage{bookmark}

\usepackage[bottom]{footmisc}
\usepackage{multibib}
\newcites{ReferencesSupp}{References}

\makeatother

\providecommand{\assumptionname}{Assumption}
\providecommand{\definitionname}{Definition}
\providecommand{\lemmaname}{Lemma}
\providecommand{\propositionname}{Proposition}
\providecommand{\theoremname}{Theorem}

\begin{document}
\pagebreak{}

\setcounter{page}{0}

\raggedbottom
\title{\textbf{\Large{}}\textbf{Prewhitened Long-Run Variance Estimation
Robust to Nonstationarity}\textbf{\large{}}\textbf{}\thanks{This paper previously circulated with the title ``Minimax MSE Bounds
and Nonlinear VAR Prewhitening for Long-Run Variance Estimation Under
Nonstationarity''. We thank two anonymous referees for helpful suggestions.
We thank Oliver Linton, Whitney Newey and Tim Vogelsang for helpful
discussions. We also thank seminar participants at University College
London, University of Cambridge and University of Connecticut for
comments. }}
\maketitle
\begin{abstract}
We introduce a nonparametric nonlinear VAR prewhitened long-run variance
(LRV) estimator for the construction of standard errors robust to
autocorrelation and heteroskedasticity that can be used for hypothesis
testing in a variety of contexts including the linear regression model.
Existing methods either are theoretically valid only under stationarity
and have poor finite-sample properties under nonstationarity (i.e.,
fixed-$b$ methods), or are theoretically valid under the null hypothesis
but lead to tests that are not consistent under nonstationary alternative
hypothesis (i.e., both fixed-$b$ and traditional HAC estimators).
The proposed estimator accounts explicitly for nonstationarity, unlike
previous prewhitened procedures which are known to be unreliable,
and leads to tests with accurate null rejection rates and good monotonic
power. We also establish MSE bounds for LRV estimation that are sharper
than previously established and use them to determine the data-dependent
bandwidths.
\end{abstract}

\onehalfspacing
\thispagestyle{empty}

\pagebreak{}

\section{Introduction}

Heteroskedasticity and autocorrelation robust (HAR) inference requires
estimation of the relevant asymptotic variance or simply the long-run
variance (LRV). A large literature has considered this problem.
In econometrics, \citet{andrews:91} and \citeauthor{newey/west:87}
(\citeyear{newey/west:87}; \citeyear{newey/west:94}) extended the
scope of kernel-based autocorrelation and heteroskedastic consistent
(HAC) estimators of the LRV {[}see also \citet{dejong/davidson:00}
and \citet{hansen:92ecma}{]}. Test statistics normalized by HAC estimators
follow standard asymptotic distributions under the null hypothesis
under mild conditions.

It was early noted that classical HAC estimators lead to test statistics
that do not correctly control the rejection rates under the null hypothesis
when there is strong serial dependence in the data. A vast literature
has considered this issue.  \citet{Kiefer/vogelsang/bunzel:00} and
\citeauthor{Kiefer/vogelsang:02} (\citeyear{Kiefer/vogelsang:02};
\citeyear{kiefer/vogelsang:05}) introduced the fixed-$b$ LRV estimators
for stationary sequences which are characterized by using a fixed
bandwidth {[}e.g., the Newey-West/Bartlett estimator including all
lags{]}. The crucial difference relative to HAC estimators is that
the LRV estimator is not consistent under fixed-$b$ asymptotics and
inference is nonstandard. Test statistics  under the null hypotheses
asymptotically follow nonstandard distributions whose critical values
are obtained numerically. This has limited the use of fixed-$b$ in
practice. The advantage of the fixed-$b$ framework is that it yields
test statistics with more accurate null rejection rates when there
is strong  dependence.\footnote{See \citet{jansson:04} and \citet{sun/phillips/jin:08} for theoretical
results based on asymptotic expansions. } 

There is widespread evidence that the processes governing economic
data  are nonstationary. By nonstationary we mean non-constant moments.
As in the literature, we consider processes whose sum of absolute
autocovariances is finite. That is, we rule out processes with unbounded
second moments (e.g., unit root). The latter can be handled by taking
first-differences or applying some de-trending technique. Nonstationarity
can occur for several reasons: changes in the moments induced by changes
in the model parameters that govern the data (e.g., the Great Moderation
with the decline in variance for many macroeconomic variables or the
effects of the COVID-19 pandemic); smooth changes in the distributions
of the processes that arise from transitory dynamics; and so on. HAR
inference requires the estimation of the LRV of some relevant process,
$V_{t}$ say.\footnote{For example, in the linear regression model $V_{t}=x_{t}e_{t}$ where
$x_{t}$ is a vector of regressors and $e_{t}$ is a disturbance. } We first analyze the case with $\mathbb{E}(V_{t})=0$ for all $t$,
since it is the leading case that applies under the null hypothesis.
This will allow us to derive useful properties to construct bandwidths
(and so on) to have tests with the correct null rejection rates. Thus,
under the null hypothesis, nonstationary occurs through time-varying
autocovariances $\mathbb{E}(V_{t}V_{t-k})$. We recognize that in
some cases, the null hypothesis may involve a non-constant mean (e.g.,
when the model is misspecified). As in the literature, we do not address
this case since the results can only be obtained on a case by case
basis. Under the alternative hypothesis, $\mathbb{E}(V_{t})\neq0$,
and $\mathbb{E}(V_{t})$ as well as $\mathbb{E}(V_{t}V_{t-k})$ can
be time-varying. In most HAR inference problems the leading case is
with a non-zero mean. The literature has so far not properly addressed
this leading case. Our aim is to devise a method for this leading
case that delivers useful estimates such that the tests have good
power. Hence, we shall also consider the properties of our estimator
when the mean of $V_{t}$ is non-zero and show that it leads to test
having good monotonic power, unlike what is available in the literature.

The objective of this paper is to propose an estimator of the LRV
that has the following properties: (i) it can be used for any hypothesis
testing problem both within and outside the linear regression model
and is valid under both stationarity and nonstationarity; (ii) it
can be used  without the need to develop further asymptotic analyses
to determine the null limiting distribution of the test statistics;
(iii) it leads to tests that have accurate null rejection rates even
with strong dependence; such tests are consistent in any hypothesis
testing problem, and in particular, in testing problems characterized
by a nonstationary alternative hypothesis.\footnote{By nonstationary alternative hypothesis we mean alternative hypothesis
such that $\mathbb{E}(V_{t})$ is time-varying.} None of the existing procedures satisfies all three properties. Fixed-$b$
methods rely on nonstandard limit theory and require one to derive
the null limiting distribution on a case-by-case basis.\footnote{\citet{lazarus/lewis/stock:17} pointed out the usefulness for empirical
work of having test statistics that follow asymptotically standard
distributions rather than nonstandard distributions whose critical
value has to be obtained by simulations.} \citet{casini_fixed_b_erp} showed that the original fixed-$b$ methods
are not theoretically valid under nonstationarity since the null limiting
distribution of the test statistics is then not pivotal.  More recently,
a variant of the fixed-$b$ approach {[}see, e.g., \citet{sun:14}
and \citet{lazarus/lewis/stock/watson:18}{]} considered the use of
small-$b$ asymptotics (i.e., small-bandwidths) in conjunction with
fixed-$b$ critical values. In general, the latter methods do not
satisfy (i)-(ii) since they use fixed-$b$ critical values, and we
show below that they may not lead to consistent tests under nonstationary
alternative hypothesis. Traditional HAC estimators satisfy (i)-(ii)
since they are consistent for the LRV so that a test statistic studentized
by an HAC estimator follows asymptotically a standard distribution.
A long-lasting problem with HAC estimators is that they lead to HAR
tests that can be oversized when there is strong dependence. To address
this issue, \citet{andrews/monahan:92} proposed the prewhitened HAC
estimators which substantially reduce the oversized problem under
stationarity with HAR tests having null rejection rates similar to
those of recent methods based on fixed-$b$ {[}e.g., the EWP and EWC
methods of \citet{lazarus/lewis/stock:17} and \textcolor{MyBlue}{Lazarus et al.}
\citeyearpar{lazarus/lewis/stock/watson:18}, respectively{]}. However,
we show theoretically that existing prewhitened and non-prewhitened
LRV estimators lead to HAR tests that are not consistent in contexts
characterized by nonstationary alternative hypotheses. This has been
a recurrent problem in the time series econometrics literature.\footnote{Simulation evidence of serious (e.g., non-monotonic) power problems
was documented by \citet{altissimo/corradi:2003}, \citet{casini_CR_Test_Inst_Forecast},
\citeauthor{casini/perron_Lap_CR_Single_Inf} (\citeyear{casini/perron_Oxford_Survey},
\citeyear{casini/perron_Lap_CR_Single_Inf}, \citeyear{casini/perron_SC_BP_Lap}),
\citet{chan:2020}, \citet{chang/perron:18}, \citet{crainiceanu/vogelsang:07},
\citet{demetrescu/salish:2020}, \citet{deng/perron:06}, \citet{juhl/xiao:09},
\citet{kim/perron:09}, \citet{martins/perron:16}, \citet{otto/breitung:2021},
\citet{perron:1991}, \citet{perron/yamamoto:18}, \citet{shao/zhang:2010},
\citet{vogeslang:99} and \citet{zhang/lavitas:2018} among others.} It occurs, for instance, when using tests involving structural breaks
based on estimating the model under the null hypothesis; e.g., tests
for forecast evaluation {[}e.g., \citet{diebold/mariano:95}, \citet{giacomini/white:06}
and \citet{west:96}{]}, tests for forecast instability {[}cf. \citet{casini_CR_Test_Inst_Forecast},
\citeauthor{giacomini/rossi:09} (\citeyear{giacomini/rossi:09},
\citeyear{giacomini/rossi:10}) and \citet{perron/yamamoto:18}{]},
CUSUM tests for structural change {[}see, e.g., \citet{brown/durbin/evans:1975}
and \citet{ploberger/kramer:1992}{]} tests and inference in time-varying
parameters models {[}e.g., \citet{cai:07} and \citet{chen/hong:12}{]},
tests and inference for regime switching models {[}e.g., \citet{hamilton:89}
and \citet{qu/zhuo:2020}{]}. 

To improve the power properties of HAR tests based on HAC estimators,
\citet{casini_hac} proposed to modify the HAC estimators by adding
a second kernel which applies smoothing over time. Such double kernel
HAC estimators (DK-HAC) satisfy properties (i)-(iii) except that they
can be oversized when there is high serial correlation. We introduce
a novel nonparametric nonlinear VAR prewhitening procedure to apply
prior to constructing the DK-HAC estimators. The key property is that
our prewhitening procedure is applied locally in time through nonparametric
time smoothing. This allows us to account flexibly for the time-varying
second-order properties of the data and to reduce the asymptotic bias
arising from nonparametric estimation. Our prewhitening is robust
to nonstationarity unlike previous prewhitened procedures {[}e.g.,
\citet{andrews/monahan:92}, \citet{preinerstorfer:17}, \citet{rho/shao:13}
and \citet{xiao/linton:02}{]}. The latter are sensitive to estimation
errors in the whitening step when there is nonstationarity in the
autoregressive dynamics. For example, with AR(1) prewhitening the
resulting LRV estimator is given by $\widehat{J}_{\mathrm{HAC},\mathrm{pw}}=\widehat{J}_{\mathrm{HAC},V^{*}}/(1-\widehat{a}_{1})^{2}$
where $\widehat{a}_{1}$ is the estimated parameter in the regression
$V_{t}=a_{1}V_{t-1}+V_{t}^{*}$ involving the process of interest
$\{V_{t}\}$ and $\widehat{J}_{\mathrm{HAC},V^{*}}$ is a classical
HAC estimator applied to the prewhitened residuals $\{V_{t}^{*}\}$.
Under nonstationarity in $\{V_{t}\}$, $\widehat{a}_{1}$ is biased
toward one, {[}cf. \citet{perron:89}{]}. This makes the recoloring
step unstable as $(1-\widehat{a}_{1})^{2}$ approaches zero and more
so as the nonstationarity increases. Hence, $\widehat{J}_{\mathrm{HAC},\mathrm{pw}}$
is inflated and test statistics lose power.

The consistency, rate of convergence and MSE of the new prewhitening
procedure are established under nonstationarity using the segmented
locally stationary framework. We then establish the consistency, rate
of convergence and minimax MSE bounds for the DK-HAC estimator under
general nonstationarity (i.e., unconditionally heteroskedastic processes)
and discuss how these results can be used to show that the prewhitened
DK-HAC estimators are valid under general nonstationarity. The new
minimax MSE bounds generalizes the MSE bounds in \citet{andrews:91}
as follows. \citet{andrews:91} expressed the bounds in terms of the
distributions of two different second-order stationary processes.
The two distributions provide upper and lower bounds, respectively,
to the autocovariances of the nonstationary processes in some class.
We show that this class can be enlarged substantially if the two distributions
are taken to be those of some nonstationary processes that satisfy
segmented locally stationarity. This allows for more variability of
$\mathbb{E}(V_{t}V_{t-k})$ and serial dependence of $\left\{ V_{t}\right\} $.
Thus, our bounds apply to a richer class of processes. The new bounds
also provide information on how nonstationarity influences the estimation
bias. 

The paper makes several theoretical contributions to the HAR inference
literature. First, it establishes the consistency and MSE-optimality
under nonstationarity of a local prewhitening procedure applied to
the double-kernel HAC estimator. Most of the existing literature focused
on the stationary case, (i.e., $\mathbb{E}\left(V_{t}V_{t-k}\right)$
depends on $k$ but not on $t$), and considered a typical LRV estimator
that applies smoothing only over lagged sample autocovariances. We
allow $\mathbb{E}\left(V_{t}V_{t-k}\right)$ to depend on $t$ as
well as $k$ and consider a prewhitened LRV estimator that applies
non-parametric smoothing over lagged sample autocovariances and time.
We establish the theoretical properties of the prewhitened LRV estimator
using data-dependent bandwidths that flexibly account for nonstationarity
unlike previously proposed data-dependent bandwidths. Second, we show
that imposing restrictions on nonstationarity allows one to obtain
superior minimax MSE bounds relative to those obtained under stationarity.
The usefulness of these bounds is twofold. On the one hand, they allow
the construction of data-dependent bandwidths that lead to a more
efficient LRV estimator. On the other hand, they are used to show
the validity of the (prewhitened) LRV estimator under general forms
of nonstationarity (e.g., more general than segmented locally stationarity).

The prewhitened DK-HAC estimators lead to HAR tests with null rejection
rates close to the nominal even with strong dependence. Furthermore,
we show theoretically that the prewhitened DK-HAC estimators lead
to HAR tests that are consistent even under nonstationary alternative
hypotheses whereas existing HAC-based and fixed-$b$-based HAR tests
are not consistent with their power converging to zero as nonstationarity
increases. The simulations demonstrate that these theoretical results
provide accurate predictions about the finite-sample behavior of the
tests.

The paper is organized as follows. Section \ref{Section: Statistical Enviromnent}
introduces the  nonlinear VAR prewhitening procedure and its asymptotic
results are established in Section \ref{Section Prewhitended DKHAC- Theory}.
Section \ref{Section: Extension-to-Unrestrcited Nonstationary} establishes
the theoretical validity of the DK-HAC estimators under general nonstationarity
and presents new minimax MSE bounds. Section \ref{Section Power}
presents some theoretical results about the power of HAR tests under
nonstationary alternative hypotheses. Section \ref{Section Monte Carlo}
presents the simulation results. Section \ref{Section Conclusions}
concludes. The supplemental materials {[}cf. \citet{casini/perron_PrewhitedHAC_Supp}{]}
contain all mathematical proofs. 

\section{The Statistical Environment\label{Section: Statistical Enviromnent}}

Suppose $\left\{ V_{t}\right\} _{t=1}^{T}$ is defined on an abstract
probability space $\left(\Omega,\,\mathscr{F},\,\mathbb{P}\right)$,
where $\Omega$ is the sample space, $\mathscr{F}$ is the $\sigma$-algebra
and $\mathbb{P}$ is a probability measure. HAR inference requires
the estimation of asymptotic variances of the form $J\triangleq\mathrm{lim}_{T\rightarrow\infty}J_{T}$
where 
\begin{align*}
J_{T} & =T^{-1}\sum_{s=1}^{T}\sum_{t=1}^{T}\mathbb{E}(V_{s}(\beta_{0})V_{t}(\beta_{0})'),
\end{align*}
 with $V_{t}(\beta)$ a random $p$-vector for each $\beta\in\Theta\subset\mathbb{R}^{p_{\beta}}$
and $\mathbb{E}(V_{t}(\beta_{0}))=0$ for all $t$ under the null
hypothesis provided that the underlying model is correctly specified.
We allow for $\mathbb{E}(V_{t})\neq0$ in Section \ref{Section Power}
when we analyze the theoretical properties of the power of the tests.
For the linear regression model $y_{t}=x'_{t}\beta_{0}+e_{t}$, we
have $V_{t}(\beta_{0})=x_{t}e_{t}.$ More generally, in nonlinear
dynamic models, we have under mild conditions,
\[
(B_{T}J_{T}B_{T})^{-1/2}\sqrt{T}(\widehat{\beta}-\beta_{0})\overset{d}{\rightarrow}\mathscr{N}(0,\,I_{p_{\beta}}),
\]
where $B_{T}$ is a nonrandom $p_{\beta}\times p$ matrix. Often it
is easy to construct estimators $\widehat{B}_{T}$ such that $\widehat{B}_{T}-B_{T}\overset{\mathbb{P}}{\rightarrow}0$.
Thus, one needs a consistent estimator of $J=\lim_{T\rightarrow\infty}J_{T}$
to construct a consistent estimator of $\lim_{T\rightarrow\infty}B_{T}J_{T}B'_{T}.$
 Our goal is to consider the estimation of $J$ under nonstationarity. 

Under nonstationarity the autocovariance of $V_{t}$ depends on the
calendar time at which it is computed in addition to the lag. That
is, $\Gamma_{u}\left(k\right)\triangleq\mathbb{E}(V_{Tu}V'_{Tu-k})$
where $u=t/T$ for some lag $k\in\mathbb{Z}$. The rescaled time index
$u\in\left[0,\,1\right]$ is introduced because under nonstationarity
we use the infill asymptotics. We now define the local spectral density
of $V_{t}$ at time $u$ and frequency $\omega,$ $f\left(u,\,\omega\right)$.
It is an important quantity because it summarizes the second-order
properties of $V_{t}$. It is defined as the squared modulus of the
transfer function $A\left(u,\,\omega\right)$ where the latter appears
in the spectral representation of $V_{t}$ {[}see eq. \eqref{Eq. Spectral Rep of SLS}
in the supplement{]}. That is, $f\left(u,\,\omega\right)=|A\left(u,\,\omega\right)|^{2}$.
The local spectral density can also be defined implicitly from the
definition of $c\left(u,\,k\right)$ which is the approximation to
the local autocovariance $\Gamma_{u}\left(k\right)$ where 
\begin{equation}
c\left(u,\,k\right)\triangleq\int_{-\pi}^{\pi}e^{i\omega k}f\left(u,\,\omega\right)d\omega,\label{Eq. c(u,k) =00003D f(u,w)}
\end{equation}
and $i=\sqrt{-1}.$ In fact, Lemma S.A.1 in \citet{casini_hac} showed
that, under the assumptions we introduce below, $\Gamma_{u}\left(k\right)=c\left(u,\,k\right)+O\left(T^{-1}\right)$
where $O\left(T^{-1}\right)$ is the error due to the infill asymptotic
approximation. Eq. \eqref{Eq. c(u,k) =00003D f(u,w)} relates the
local autocovariance of $\{V_{t}\}$ at rescaled time $u$ and lag
$k$ to its local spectral density at $u$. Thus, the nonstationary
properties of $\{V_{t}\}$, which are reflected in the time-varying
behavior of the autocovariance function $\Gamma_{u}\left(k\right)$,
depend on the smoothness properties of $f\left(u,\,\omega\right)$
in $u$. For example, if $\{V_{t}\}$ is stationary, then $\Gamma_{u}\left(k\right)=\Gamma\left(k\right)$
for all $u$, $c\left(u,\,k\right)$ is constant in $u$, $f\left(u,\,\omega\right)=f\left(\omega\right)$
and \eqref{Eq. c(u,k) =00003D f(u,w)} reduces to $\Gamma\left(k\right)=\int_{-\pi}^{\pi}e^{i\omega k}f\left(\omega\right)d\omega$.
These coincide with textbook definitions used under stationarity {[}see,
e.g., \citet{brillinger:75}{]}. If $f\left(u,\,\omega\right)$ is
continuous in $u$ then $\{V_{t}\}$ is locally stationary {[}cf.
\citet{dahlhaus:96}{]}.\footnote{In econometrics, locally stationary processes are often referred to
as time-varying parameter processes.} For example, consider a time-varying AR(1) $V_{t}=a\left(t/T\right)V_{t-1}+u_{t}$
where $u_{t}$ is a zero-mean i.i.d. process with unit variance and
$a\left(\cdot\right)$ is continuous with $a\left(t/T\right)\in(-1,\,1)$
for all $t.$ Then $V_{t}$ is a locally stationary AR(1) with a local
spectral density $f\left(u,\,\omega\right)$ that is continuous in
$u$. We impose restrictions on the smoothness of $f\left(u,\,\omega\right)$
in $u$ which allow for considerable forms of nonstationarity in $\left\{ V_{t}\right\} $
including most of the nonstationary models used in econometrics.\footnote{A function $g\left(\cdot\right):\,\left[0,\,1\right]\mapsto\mathbb{R}$
is said to be piecewise (Lipschitz) continuous if there exists a finite
subdivision $\left\{ x_{0},\,x_{1},\ldots,\,x_{n}\right\} $ of $\left[0,\,1\right]$
where $x_{0}=0$ and $x_{n}=1$, such that for all $i\in\left\{ 1,\,2,\ldots,\,n\right\} $
$g$ is (Lipschitz) continuous on $\left(x_{i-1},\,x_{i}\right)$. }
\begin{assumption}
\label{Assumption Smothness f(u,w)}(i) $\{V_{t}\}$ is zero-mean
with local spectral density $f\left(u,\,\omega\right)$ that is piecewise
Lipschitz continuous with $m_{0}$ discontinuity points; (ii) $f\left(u,\,\omega\right)$
is \textcolor{red}{ }twice continuously differentiable in $u$ at
all continuity points with  bounded derivatives $\left(\partial/\partial u\right)f\left(u,\,\cdot\right)$
and $\left(\partial^{2}/\partial u^{2}\right)f\left(u,\,\cdot\right)$,
and Lipschitz continuous in the second component; (iii) $\left(\partial^{2}/\partial u^{2}\right)f\left(u,\,\cdot\right)$
is Lipschitz continuous at all continuity points; (iv) $f\left(u,\,\omega\right)$
is twice left-differentiable at all discontinuity points with  bounded
derivatives $\left(\partial/\partial_{-}u\right)f\left(u,\,\cdot\right)$
and $\left(\partial^{2}/\partial_{-}u^{2}\right)f\left(u,\,\cdot\right)$
and has piecewise Lipschitz continuous derivative $\left(\partial^{2}/\partial_{-}u^{2}\right)f\left(u,\,\cdot\right)$.
\end{assumption}
Assumption \ref{Assumption Smothness f(u,w)} implies that $\left\{ V_{t}\right\} $
is segmented locally stationary (SLS) (see Definition \ref{Definition Segmented-Locally-Stationary}
in the supplement). It is similar to Assumption 3.1 in \citet{casini_hac}
where the latter imposes smoothness conditions on the transfer function
$A\left(u,\,\omega\right)$ whereas here we directly make assumptions
on the local spectral density $f\left(u,\,\omega\right)$. The class
of SLS processes allows for relevant features such as structural change,
regime switching-type and threshold model and includes general time-varying
parameter processes, locally stationary processes and stationary processes.\footnote{For general time-varying parameter processes we mean linear and nonlinear
processes whose parameters can change smoothly as well as abruptly.
See Example 2.1 in \citet{casini_hac} for some examples.} Assumption \ref{Assumption Smothness f(u,w)} requires $f\left(u,\,\cdot\right)$
to be twice differentiable at the continuity points and left-differentiable
at the discontinuity points. The zero-mean assumption holds under
the null hypothesis. To focus on the main intuition, we first consider
the case of SLS processes and then extend the results to general nonstationarity
processes in Section \ref{Section: Extension-to-Unrestrcited Nonstationary}.\footnote{For general nonstationarity we mean a process with a time-varying
spectral density that does not satisfy piecewise Lipschitz continuity.} The latter require more technical notations and assumptions. In
Section \ref{Subsection Prewhitening-DK-HAC-Estimator} we present
the prewhitening DK-HAC estimator while in Section \ref{Subsection Data-depedent Bandwdiths}
we discuss its data-dependent bandwidths. 

\subsection{\label{Subsection Prewhitening-DK-HAC-Estimator}Prewhitening DK-HAC
Estimator}

Under Assumption \ref{Assumption Smothness f(u,w)}, the argument
at the beginning of Section 2.1 in \citet{casini_hac} suggests that
$J=2\pi\int_{0}^{1}f\left(u,\,0\right)du$. The right-hand side can
be seen as a function, say $\widetilde{f}\left(\omega\right)$, evaluated
at the zero frequency $\omega=0.$ The intuition behind prewhitening
is simple, though the mechanics under nonstationarity are quite different.
Suppose one is estimating $\widetilde{f}\left(0\right)$ nonparametrically
by averaging asymptotically unbiased estimators of $\widetilde{f}\left(\omega\right)$
at a number of points $\omega$ in a neighborhood of $0$. The flatter
is the function $\widetilde{f}\left(\omega\right)$ around 0, the
smaller the estimation bias. The idea is to transform the data such
that the function of the transformed data, say $\widetilde{f}^{*}(\omega)$,
is flatter in the neighborhood of $\omega=0$. Then, using the transformed
data one can estimate $\widetilde{f}^{*}(0)$ by averaging asymptotically
unbiased estimators of $\widetilde{f}^{*}(\omega)$ at points $\omega$
in the neighborhood of 0. The resulting bias should be less than that
incurred by estimating $\widetilde{f}\left(0\right)$ since $\widetilde{f}^{*}(\omega)$
is flatter than $\widetilde{f}\left(\omega\right)$. Finally, one
can apply the inverse of the transformation from $\widetilde{f}\left(\omega\right)$
to $\widetilde{f}^{*}(\omega)$ to obtain an estimator of $\widetilde{f}\left(\omega\right)$
from the estimator of $\widetilde{f}^{*}(\omega)$. This is how it
works under stationarity. However, under nonstationarity one applies
both the transformation and the inverse transformation locally in
time, otherwise the prewhitening procedure may be unreliable as nonstationarity
induces an additional source of bias in both the transformation and
its inverse.  

The proposed prewhitening procedure is based on the following three
steps:

\textbf{Step 1 (whitening step):} Divide the sample in $\left\lfloor T/n_{T}\right\rfloor $
blocks, each with $n_{T}$ observations. Let $\widehat{V}_{t}=V_{t}(\widehat{\beta})$,
where $\widehat{\beta}$ is a $\sqrt{T}$-consistent estimator of
$\beta_{0}$. For each block $r=0,\ldots,\,\left\lfloor T/n_{T}\right\rfloor $,
run the following VAR$(p_{A})$, 
\begin{align}
\widehat{V}_{t} & =\sum_{j=1}^{p_{A}}\widehat{A}_{r,j}\widehat{V}_{t-j}+\widehat{V}_{t}^{*}\quad\mathrm{for\quad}t=rn_{T}+1,\ldots,\,\left(r+1\right)n_{T},\label{Eq. (2.3) Andrews 90}
\end{align}
where $\widehat{A}_{r,j}$ for $j=1,\ldots,\,p_{A}$ are $p\times p$
least-squares estimators and $\widehat{V}_{t}^{*}=V_{t}^{*}(\widehat{\beta})$
are the prewhitened residuals. The VAR in \eqref{Eq. (2.3) Andrews 90}
is used to ``soak up'' some of the serial dependence in $\widehat{V}_{t}$
and to leave one with residuals $\{\widehat{V}_{t}^{*}\}$ that are
closer to white noise.\footnote{Since the residuals $\{\widehat{V}_{t}^{*}\}$ are closer to a white
noise process, they have a flatter spectral density at $\omega=0$
than $\{\widehat{V}_{t}\}$ because a white noise process has a flat
spectral density.} That is why it is called ``whitening step''.

\textbf{Step 2 (recoloring step): }Take the prewhitened residuals
$\widehat{V}_{t}^{*}$, transform them by applying an inverse transformation
$\widehat{V}_{t}^{*}\mapsto\widehat{V}_{D,t}^{*}=\widehat{D}_{t}\widehat{V}_{t}^{*}$
where $\widehat{D}_{t}=(I_{p}-\sum_{j=1}^{p_{A}}\widehat{A}_{D,t,j})^{-1}$
with $\widehat{A}_{D,t,j}=\widehat{A}_{r,j}$ for $t=rn_{T}+1,\ldots,\,\left(r+1\right)n_{T}$.
This implies that the transformed residuals $\widehat{V}_{D,t}^{*}$
have been ``recolored'' (i.e., the dependence has been added back).
Note that the matrix $\widehat{D}_{t}$ is the same for all $t$ in
a given block. In this way the appropriate amount of dependence is
added back, i.e., no contamination from possibly different strengths
of dependence occurring in other blocks.

\textbf{Step 3 (prewhitened DK-HAC estimation): }Construct the prewhitened
DK-HAC estimator $\widehat{J}_{\mathrm{pw},T}$ using $\widehat{V}_{D,t}^{*}$:
\begin{align}
\widehat{J}_{\mathrm{pw},T}\left(\widehat{b}_{1,T}^{*},\,\widehat{b}_{2,T}^{*}\right) & =\frac{T}{T-p}\sum_{k=-T+1}^{T-1}K_{1}\left(\widehat{b}_{1,T}^{*}k\right)\widehat{\Gamma}_{D}^{*}\left(k\right),\label{Eq. (J_hat_PW)}\\
\mathrm{where}\quad\widehat{\Gamma}_{D}^{*}\left(k\right) & \triangleq\frac{n_{T}}{T-n_{T}}\sum_{r=0}^{\left\lfloor \left(T-n_{T}\right)/n_{T}\right\rfloor }\widehat{c}_{T,D}^{*}\left(rn_{T}/T,\,k\right),\nonumber 
\end{align}
 with $K_{1}\left(\cdot\right)$ a real-valued kernel in the class
$\boldsymbol{K}_{3}$ defined below, $\widehat{b}_{1,T}^{*}$ is a
data-dependent bandwidth sequence to be discussed below, $n_{T}\rightarrow\infty$,
and 
\begin{align*}
\widehat{c}_{D,T}^{*}\left(rn_{T}/T,\,k\right) & \triangleq\begin{cases}
\left(T\widehat{b}_{2,T}^{*}\right)^{-1}\sum_{s=k+1}^{T}K_{2}^{*}\left(\frac{\left(\left(r+1\right)n_{T}-\left(s-k/2\right)\right)/T}{\widehat{b}_{2,T}^{*}}\right)\widehat{V}_{D,s}^{*}\widehat{V}_{D,s-k}^{*\prime}, & k\geq0\\
\left(T\widehat{b}_{2,T}^{*}\right)^{-1}\sum_{s=-k+1}^{T}K_{2}^{*}\left(\frac{\left(\left(r+1\right)n_{T}-\left(s+k/2\right)\right)/T}{\widehat{b}_{2,T}^{*}}\right)\widehat{V}_{D,s+k}^{*}\widehat{V}_{D,s}^{*\prime}, & k<0
\end{cases},
\end{align*}
$K_{2}^{*}$ being a kernel, $\widehat{b}_{2,T}^{*}$ a data-dependent
bandwidth sequence to be defined below.   

In order to guarantee positive semi-definiteness, one needs to use
a data taper or, e.g., for $k\geq0$,
\begin{align*}
K_{2}^{*} & \left(\frac{\left(r+1\right)n_{T}-\left(s-k/2\right)}{T\widehat{b}_{2,T}^{*}}\right)=\left(K_{2}\left(\frac{\left(r+1\right)n_{T}-s}{T\widehat{b}_{2,T}^{*}}\right)K_{2}\left(\frac{\left(r+1\right)n_{T}-\left(s-k\right)}{T\widehat{b}_{2,T}^{*}}\right)\right)^{1/2},
\end{align*}
see \citet{casini_hac}.

In Step 1 the last block is $t=\left\lfloor T/n_{T}\right\rfloor n_{T}+1,\ldots,\,T$.
The order of the VAR, $p_{A},$ can potentially change across blocks
but, for notational ease, we assume it is the same for each $r$.
 The choices of  $n_{T}$ and how to optimally split the sample
depend on the property of the spectrum of $\{\widehat{V}_{t}\}$.
A test for breaks versus smooth changes in the spectrum of $\{\widehat{V}_{t}\}$
is introduced in \citet{casini:change-point-spectra}. The latter
could be employed here to efficiently determine the sample-splitting.
This would result in the sample being split in blocks with the property
that within each block $\{\widehat{V}_{t}\}$ is locally stationary.
However, this is not required for the theoretical validity. The least-squares
estimation within blocks yields consistent estimators $\widehat{A}_{r,j}$
for some $A_{r,j}$   even when the fitted VAR is not the true model.
The fitted VAR is used only to yield residuals $\{\widehat{V}_{t}^{*}\}$
that are closer to white noise so that their spectral density at zero
is flatter, implying less asymptotic bias when estimating it nonparametrically. 

Below we assume that $\widehat{A}_{r,j}\overset{\mathbb{P}}{\rightarrow}A_{r,j}$
for some $A_{r,j}\in\mathbb{R}^{p\times p}$  for all $r$ and $j$
which follow from standard arguments. For $K_{1}$ we suggest using
the Quadratic Spectral (QS) kernel 
\[
K_{\mathrm{1}}^{\mathrm{QS}}\left(x\right)=\left(25/\left(12\pi^{2}x^{2}\right)\right)\left[\frac{\sin\left(6\pi x/5\right)}{6\pi x/5}-\cos\left(6\pi x/5\right)\right],
\]
and for $K_{2}$ a quadratic-type kernel {[}cf. \citet{epanechnikov:69}{]}
given by $K_{2}\left(x\right)=6x\left(1-x\right),\,0\leq x\leq1$.
These kernels are optimal under an MSE criterion {[}see \citet{casini_hac}{]}. 

There has been some recent works on LRV estimation in statistics that
relate to ours. \citet{kawka:2020} studied the asymptotic properties
of classical spectral estimators for a linear time-varying AR process
where the AR coefficients can have a finite number of discontinuities.
Since classical spectral estimators do not involve any local smoothing
over time, and since he focused on linear processes and did not consider
data-dependent bandwidths, his framework required simpler assumptions.
He also considered an estimate of the spectrum profile which is defined
similarly to the variance profile of \citet{cavaliere/taylor:2007}.
That is based on a recursive estimate of the spectral density which
is, however, different from applying local smoothing. The local smoothing
is important to better account for nonstationarity as shown in \citet{casini/perron_Low_Frequency_Contam_Nonstat:2020}.
 \citet{potiron/mykland:2020} showed that in the context of estimation
of higher powers of volatility for high frequency data the local smoothing
can lead to substantial efficiency gains. Although our setting is
complicated by serial dependence and the fact that the class of estimators
has a slower rate of convergence than the parametric $\sqrt{T}$-rate,
the theoretical results on the power of the HAR tests below suggest
that the local smoothing yields more powerful tests. In addition,
\citet{casini/perron_Low_Frequency_Contam_Nonstat:2020} showed that
under nonstationarity the sample autocovariance can be upward biased
asymptotically relative to the integrated local sample autocovariance,
both for fixed lag $k$ and for $k\rightarrow\infty.$ An alternative
way to deal with a time-varying mean has been considered by \citeauthor{chan:2020}
(\citeyear{chan:2022}, \citeyear{chan:2020}) who proposed a LRV
estimator which uses difference-based statistics that combine local
smoothing and lagged differences of the series. His results confirmed
that the local smoothing is important to enhance efficiency. However,
he required covariance stationarity and did not study the theoretical
properties of HAR tests normalized by the proposed LRV estimator. 

\subsection{\label{Subsection Data-depedent Bandwdiths}Data-Dependent Bandwidths }

For data-dependent bandwidths, we use plug-in estimates of the optimal
value that minimizes some MSE criterion, see Section \ref{Section: Extension-to-Unrestrcited Nonstationary}
and \citet{casini_hac}. Let $\Gamma_{D,u}\left(k\right)=\mathrm{Cov}(V_{D,Tu}^{*},\,V_{D,Tu-k}^{*})$
and $C_{pp}=\sum_{j=1}^{p}\sum_{l=1}^{p}\iota_{j}\iota_{l}'\otimes\iota_{l}\iota_{j}'$,
where $\iota_{i}$ is the $i$-th elementary $p$-vector. The notation
$W$ and $\widetilde{W}$ are used for some $p^{2}\times p^{2}$ weight
matrices. Let $F(K_{2})\triangleq\int_{0}^{1}K_{2}^{2}\left(x\right)dx,$
$H\left(K_{2}\right)\triangleq(\int_{0}^{1}x^{2}K_{2}\left(x\right)dx)^{2}$,
\begin{align*}
D_{1,D}\left(u\right) & \triangleq\mathrm{vec}\left(\partial^{2}c_{D}^{*}\left(u,\,k\right)/\partial u^{2}\right)'\widetilde{W}\mathrm{\,vec}\left(\partial^{2}c_{D}^{*}\left(u,\,k\right)/\partial u^{2}\right),\\
D_{2,D}\left(u\right) & \triangleq\mathrm{tr}[\widetilde{W}(I_{p^{2}}+C_{pp})\sum_{l=-\infty}^{\infty}c_{D}^{*}\left(u,\,l\right)\otimes[2c_{D}^{*}\left(u,\,l\right)]],
\end{align*}
where $c_{D}^{*}\left(u,\,l\right)=\mathrm{Cov}(V_{D,Tu}^{*},\,V_{D,Tu-l}^{*})$,
$V_{D,t}^{*}=D_{t}V_{t}^{*}$,
\begin{align*}
V_{t}^{*} & =V_{t}-\sum_{j=1}^{p_{A}}A_{r,j}V_{t-j}\qquad\qquad\mathrm{for}\,t=rn_{T}+1,\ldots,\,\left(r+1\right)n_{T},\\
D_{t} & =(I_{p}-\sum_{j=1}^{p_{A}}A_{D,t,j})^{-1},\,\qquad A_{D,t,j}=A_{r,j}\qquad\mathrm{for}\,t=rn_{T}+1,\ldots,\,\left(r+1\right)n_{T}.
\end{align*}
 The optimal $b_{2,T}$ is given by {[}see \citet{casini_hac}{]}
\[
b_{2,T}^{\mathrm{opt,*}}\left(u\right)=[H\left(K_{2}^{\mathrm{}}\right)D_{1,D}\left(u\right)]^{-1/5}\left(F\left(K_{2}\right)\left(D_{2,D}\left(u\right)\right)\right)^{1/5}T^{-1/5}.
\]
 Let 
\begin{align}
K_{1,q} & \triangleq\lim_{x\downarrow0}\left(1-K_{1}\left(x\right)\right)/\left|x\right|^{q}\qquad\mathrm{for}\qquad q\in[0,\,\infty);\label{Eq. Definition of K_1,q}
\end{align}
$K_{1,q}<\infty$ if and only if $K_{1}\left(x\right)$ is $q$ times
differentiable at zero.  Let $f_{D}^{*}\left(u,\,\omega\right)=\sum_{k=-\infty}^{\infty}c_{D}^{*}\left(u,\,k\right)e^{-i\omega k}$
and define the index of smoothness of $f_{D}^{*}\left(u,\,\omega\right)$
at $\omega=0$ by $f_{D}^{*\left(q\right)}\left(u,\,0\right)\triangleq\left(2\pi\right)^{-1}\sum_{k=-\infty}^{\infty}\left|k\right|^{q}c_{D}^{*}\left(u,\,k\right)$.
Let 
\begin{align}
\phi_{D}\left(q\right) & =\frac{\mathrm{vec}\left(\int_{0}^{1}f_{D}^{*\left(q\right)}\left(u,\,0\right)du\right)'W\mathrm{vec}\left(\int_{0}^{1}f_{D}^{*\left(q\right)}\left(u,\,0\right)du\right)}{\mathrm{tr}\left[W\left(I_{p^{2}}+C_{pp}\right)\left(\int_{0}^{1}f_{D}^{*}\left(u,\,0\right)du\right)\otimes\left(\int_{0}^{1}f_{D}^{*}\left(v,\,0\right)dv\right)\right]}.\label{Eq. (phi_D(q))}
\end{align}
The optimal $b_{1,T}$ given the optimal value $b_{2,T}^{\mathrm{opt,*}}$
is given by {[}see \citet{casini_hac}{]}, 
\[
b_{1,T}^{\mathrm{opt,*}}=(2qK_{1,q}^{2}\phi_{D}\left(q\right)T\overline{b}_{2,T}^{\mathrm{opt}}/\left(\smallint K_{1}^{2}\left(y\right)dy\smallint K_{2}^{2}\left(x\right)dx\right))^{-1/\left(2q+1\right)},
\]
 with $\overline{b}_{2,T}^{\mathrm{opt,*}}=\int_{0}^{1}b_{2,T}^{\mathrm{opt},*}\left(u\right)du$.
 For the QS kernel, $q=2$, $K_{1,2}=1.421223$, and $\int K_{1}^{2}\left(x\right)dx=1$.
For the optimal $K_{2}^{\mathrm{}}$ we have $H(K_{2}^{\mathrm{opt}})=0.09$
and $F(K_{2}^{\mathrm{opt}})=1.2$. 

The bandwidths $(b_{1,T}^{\mathrm{opt,*}},\,\overline{b}_{2,T}^{\mathrm{opt,*}})$
are optimal under a sequential MSE criterion that determines the optimal
$b_{1}$ as a function of the optimal $b_{2}\left(u\right)$. Thus,
the latter influences the former but not vice-versa. However, this
has the advantage that the optimal $b_{2}\left(u\right)$ is allowed
to change over time. \textcolor{MyBlue}{Belotti et al.} \citeyearpar{belotti/casini/catania/grassi/perron_HAC_Sim_Bandws}
proposed an alternative criterion that determines the optimal $b_{1}$
and $b_{2}$ that jointly minimize the global MSE. The latter yields
an optimal $b_{2}$ that does not depend on $u$ and so it does not
perform as well as the sequential method when the data is far from
stationary. 

In order to construct a data-dependent bandwidth for $b_{2,T}\left(u\right)$,
we need consistent estimators of $D_{1,D}\left(u\right)$ and $D_{2,D}\left(u\right)$.
We set $\widetilde{W}^{\left(r,r\right)}=p^{-1}$ for all $r$ which
corresponds to the normalization used below for $W$. In order to
replace $D_{1,D}\left(u\right)$ we make a parametric assumption and
estimate $D_{1,D}\left(u\right)$ under this assumption. Following
\citet{casini_hac}, the approximating parametric assumption is that
$V_{D,t}^{*}$ belongs to the class of class of locally stationary
first-order autoregressive (AR(l)) models with certain restrictions
on the smoothness of the parameters. Under this approximating parametric
assumption, the estimator of $D_{1,D}\left(u\right)$  is
\begin{align*}
\widehat{D}_{1,D}\left(u\right) & \triangleq\left[S_{\omega}\right]^{-1}\sum_{s\in S_{\omega}}\left[\left(3/\pi\right)\left(1+0.8\left(\cos1.5+\cos4\pi u\right)\exp\left(-i\omega_{s}\right)\right)^{-4}\left(0.8\left(-4\pi\sin\left(4\pi u\right)\right)\right)\exp\left(-i\omega_{s}\right)\right.\\
 & \quad\left.-\pi^{-1}\left|1+0.8\left(\cos1.5+\cos4\pi u\right)\exp\left(-i\omega_{s}\right)\right|^{-3}\left(0.8\left(-16\pi^{2}\cos\left(4\pi u\right)\right)\right)\exp\left(-i\omega_{s}\right)\right],
\end{align*}
 where $\left[S_{\omega}\right]$ is the cardinality of $S_{\omega}$
and $\omega_{s+1}>\omega_{s}$ with $\omega_{1}=-\pi,\,\omega_{\left[S_{\omega}\right]}=\pi.$
We set $S_{\omega}=\{-\pi,\,-3,\,-2,\,-1,\,0,\,1,\,2,\,3,\,\pi\}$.
  The estimator of $D_{2,D}\left(u\right)$ is given by 
\[
\widehat{D}_{2,D}\left(u_{0}\right)\triangleq2p^{-1}\sum_{r=1}^{p}\sum_{l=-\left\lfloor T^{4/25}\right\rfloor }^{\left\lfloor T^{4/25}\right\rfloor }\left(\widehat{c}_{D,T}^{*,\left(r,r\right)}\left(u_{0},\,l\right)\right)^{2},
\]
where the number of summands  grows at the same rate as the inverse
of the optimal bandwidth $b_{1,T}^{\mathrm{opt,*}}$. Hence, the estimator
of the optimal bandwidth $b_{2,T}^{\mathrm{opt},*}$ is given by 
\begin{align}
\widehat{\overline{b}}_{2,T}^{*} & =\left(n_{T}/T\right)\sum_{r=1}^{\left\lfloor T/n_{T}\right\rfloor -1}\widehat{b}_{2,T}^{*}\left(u_{r}\right),\quad\label{Eq. (b2_bar)}\\
\mathrm{where}\quad\widehat{b}_{2,T}^{*}\left(u_{r}\right) & =1.7781(\widehat{D}_{1,D}\left(u_{r}\right)){}^{-1/5}(\widehat{D}_{2,D}\left(u_{r}\right))^{1/5}T^{-1/5},\quad u_{r}=rn_{T}/T.\label{Eq. (b2)}
\end{align}

The data-dependent bandwidth parameter $\widehat{b}_{1,T}^{*}$ is
then defined as follows. First, one specifies $p$ univariate approximating
parametric models given by $\{V_{D,t}^{*\left(r\right)}\}$ for $r=1,\ldots,\,p$.
Second, one estimates the parameters of the approximating parametric
model by least-squares. Third, one substitutes these estimates into
$\phi_{D}\left(q\right)$ with the estimate denoted by $\widehat{\phi}_{D}\left(q\right)$.
This yields the data-dependent bandwidth parameter 
\begin{equation}
\widehat{b}_{1,T}^{*}=\left[2qK_{1,q}^{2}\widehat{\phi}_{D}\left(q\right)T\widehat{\overline{b}}_{2,T}^{*}/\left(\int K_{1}^{2}\left(x\right)dx\smallint K_{2}^{2}\left(x\right)dx\right)\right]{}^{-1/\left(2q+1\right)}.\label{Eq. (b1)}
\end{equation}
 For the QS kernel, we have $\widehat{b}_{1,T}^{*}=0.6828(\widehat{\phi}_{D}\left(2\right)T\widehat{\overline{b}}_{2,T}^{*})^{-1/5}$.
As mentioned above, the suggested approximating parametric models
are the locally stationary AR(l) models given by $V_{D,t}^{*\left(r\right)}=a_{1}^{\left(r\right)}\left(t/T\right)$
$V_{D,t-1}^{*\left(r\right)}+u_{t}^{\left(r\right)}$, $r=1,\ldots,\,p$.
Let $\widehat{a}_{1}^{\left(r\right)}\left(u\right)$ and $(\widehat{\sigma}^{\left(r\right)}\left(u\right))^{2}$
be the least-squares estimators of the autoregressive and innovation
variance parameters computed using data close to $u=t/T$: 
\begin{align*}
\widehat{a}_{1}^{\left(r\right)}\left(u\right) & =\frac{\sum_{j=t-n_{2,T}+1}^{t}\widehat{V}_{D,j}^{*,\left(r\right)}\widehat{V}_{D,j-1}^{*,\left(r\right)}}{\sum_{j=t-n_{2,T}+1}^{t}\left(\widehat{V}_{D,j-1}^{*,\left(r\right)}\right)^{2}},\qquad\widehat{\sigma}^{\left(r\right)}\left(u\right)=\left(\sum_{j=t-n_{2,T}+1}^{t}\left(\widehat{V}_{D,j}^{*,\left(r\right)}-\widehat{a}_{1}^{\left(r\right)}\left(u\right)\widehat{V}_{D,j-1}^{*,\left(r\right)}\right)^{2}\right)^{1/2},
\end{align*}
where $n_{2,T}\rightarrow\infty$.\footnote{See, for example, \citet{Dahlhaus/Giraitis:98} for a discussion about
nonparametric local parameter estimates in the context of locally
stationary time series.} These are simply least-squares estimators based on rolling windows.
Then, for $q=2$, we have 
\begin{align*}
\widehat{\phi}_{D}\left(2\right) & =\sum_{r=1}^{p}W^{\left(r,r\right)}\left(18\left(\frac{n_{3,T}}{T}\sum_{j=0}^{\left\lfloor T/n_{3,T}\right\rfloor -1}\frac{\left(\widehat{\sigma}^{\left(r\right)}\left(\left(jn_{3,T}+1\right)/T\right)\widehat{a}_{1}^{\left(r\right)}\left(\left(jn_{3,T}+1\right)/T\right)\right)^{2}}{\left(1-\widehat{a}_{1}^{\left(r\right)}\left(\left(jn_{3,T}+1\right)/T\right)\right)^{4}}\right)^{2}\right)/\\
 & \quad\sum_{r=1}^{p}W^{\left(r,r\right)}\left(\frac{n_{3,T}}{T}\sum_{j=0}^{\left\lfloor T/n_{3,T}\right\rfloor -1}\frac{\left(\widehat{\sigma}^{\left(r\right)}\left(\left(jn_{3,T}+1\right)/T\right)\right)^{2}}{\left(1-\widehat{a}_{1}^{\left(r\right)}\left(\left(jn_{3,T}+1\right)/T\right)\right)^{2}}\right)^{2},
\end{align*}
where $W^{\left(r,r\right)},\,r=1,\ldots,\,p$ are pre-specified weights
and $n_{3,T}\rightarrow\infty$. The usual choice for $W^{\left(r,r\right)}$
is one for all $r$ except that which corresponds to an intercept
in which case it is zero. Let $\widehat{\theta}=(\int_{0}^{1}\widehat{a}_{1}^{\left(1\right)}\left(u\right)du,\,\int_{0}^{1}(\widehat{\sigma}^{\left(1\right)}\left(u\right))^{2}du,\ldots,\,\int_{0}^{1}\widehat{a}_{1}^{\left(p\right)}\left(u\right)du,\,\int_{0}^{1}(\widehat{\sigma}^{\left(p\right)}\left(u\right))^{2}du)'$
and let $\theta^{*}$ denote the probability limit of $\widehat{\theta}$.
If the locally stationary AR(1) parametric model is not correctly
specified for $V_{D,t}^{*}$, then the probability limit of $\widehat{\phi}_{D}\left(q\right)$
need not be equal to $\phi_{D}\left(q\right)$. Let $\phi_{\theta^{*}}\in\mathbb{R}$
be the probability limit of $\widehat{\phi}_{D}\left(q\right)$ (i.e.,
$\widehat{\phi}_{D}\left(q\right)-\phi_{\theta^{*}}=o_{\mathbb{P}}\left(1\right))$.
When the locally stationary AR(1) parametric model is correctly specified
we have $\phi_{\theta^{*}}=\phi_{D}\left(q\right)$. 

\section{\label{Section Prewhitended DKHAC- Theory}Large-Sample Results
When $\mathbb{E}\left(V_{t}\right)=0$}

In this section, we analyze the asymptotic properties of  $\widehat{J}_{\mathrm{pw},T}$
for the case with $\mathbb{E}\left(V_{t}\right)=0$ for all $t$,
which is relevant under the null hypothesis provided that the model
is correctly specified. Let $K$ denote a generic kernel and $K^{(q)}$
be defined as $K_{1,q}$ in \eqref{Eq. Definition of K_1,q} with
$K_{1}$ replaced by $K$. Let 
\begin{align*}
\boldsymbol{K}_{3} & =\biggl\{ K\left(\cdot\right):\,\mathbb{R}\rightarrow\left[-1,\,1\right],\,(i)\,K\left(0\right)=1,\,K\left(x\right)=\left(-x\right),\,\int_{-\infty}^{\infty}\left|K\left(x\right)\right|dx\,\int_{-\infty}^{\infty}K^{2}\left(x\right)dx<\infty\\
 & \quad\left(ii\right)\,\left|K\left(x\right)\right|\leq C_{1}\left|x\right|^{-b}\,\mathrm{with\,}b>\max\left(1+1/q,\,4\right)\,\mathrm{for}\,\left|x\right|\in\left[\overline{x}_{L},\,D_{T}h_{T}\overline{x}_{U}\right],\\
 & \quad T^{-1/2}h_{T}\rightarrow\infty,\,D_{T}>0,\,\overline{x}_{L},\,\overline{x}_{U}\in\mathbb{R},\,1\leq\overline{x}_{L}<\overline{x}_{U},\,\mathrm{and}\,\mathrm{with\,}b>1+1/q\,\\
 & \quad\mathrm{for}\,\left|x\right|\notin\left[\overline{x}_{L},\,D_{T}h_{T}\overline{x}_{U}\right]\,\mathrm{and\,some\,}C_{1}<\infty,\,\mathrm{where}\,q\in\left(0,\,\infty\right)\,\mathrm{is\,such\,that\,}K^{(q)}\in\left(0,\,\infty\right),\\
 & \quad\,\left(iii\right)\,\left|K\left(x\right)-K\left(y\right)\right|\leq C_{2}\left|x-y\right|\,\forall x,\,y\in\mathbb{R}\,\mathrm{for\,some\,costant\,}C_{2}<\infty,\,\mathrm{and}\,(iv)\,q<17/2\biggr\}.
\end{align*}
Note that $\boldsymbol{K}_{3}$ depends on $T$, though we omit this
dependence. $\boldsymbol{K}_{3}$ contains commonly used kernels,
e.g., QS, Bartlett, Parzen, and Tukey-Hanning, with the exception
of the truncated kernel. For the QS, Parzen, and Tukey-Hanning kernels,
$q=2$. For the Bartlett kernel, $q=1$. The condition $q<17/2$ in
part (iv) is a technical condition needed to control the deviation
$|\widehat{b}_{1,T}^{*}-b_{\theta_{1},T}|$, where $b_{\theta_{1},T}$
is defined as $\widehat{b}_{1,T}^{*}$ {[}cf. \eqref{Eq. (b1,T_theta)}
below{]} but with $\widehat{\phi}_{D}\left(q\right)$ replaced by
$\phi_{\theta^{*}}$.  

For $K_{2}$ we consider the same class of kernels $\boldsymbol{K}_{2}$
as considered by \citet{casini_hac}: 
\begin{align*}
\boldsymbol{K}_{2} & =\biggl\{ K\left(\cdot\right):\,\mathbb{R}\rightarrow\left[0,\,\infty\right]:\,K\left(x\right)=K\left(1-x\right),\,\int K\left(x\right)dx=1,\,\\
 & \qquad\int_{0}^{1}K^{2}\left(x\right)dx<\infty,\,K\left(x\right)=0,\,\mathrm{for\,}\,x\notin\left[0,\,1\right]\\
 & \qquad\left|K\left(x\right)-K\left(y\right)\right|\leq C_{4}\left|x-y\right|\,\mathrm{for\,all\,\mathit{x,\,y\in\mathbb{R}\,}and\,some\,constant\,C_{4}<\infty}\biggr\}.
\end{align*}
We define 
\[
\mathrm{MSE}(Tb_{1,T}b_{2,T},\,\widetilde{J}_{T},\,J_{T},\,W)=Tb_{1,T}b_{2,T}\mathbb{E}[\mathrm{vec}(\widetilde{J}_{T}-J_{T})'W\mathrm{vec}(\widetilde{J}_{T}-J_{T})].
\]
We need to impose conditions on the temporal dependence of $\{V_{t}\}$.
Let 
\begin{align*}
\kappa_{V,t}^{\left(a_{1},a_{2},a_{3},a_{4}\right)}\left(u,\,v,\,w\right) & \triangleq\kappa^{\left(a_{1},a_{2},a_{3},a_{4}\right)}\left(t,\,t+u,\,t+v,\,t+w\right)-\kappa_{\mathscr{N}}^{\left(a_{1},a_{2},a_{3},a_{4}\right)}\left(t,\,t+u,\,t+v,\,t+w\right)\\
 & \triangleq\mathbb{E}(V_{t}^{\left(a_{1}\right)}V_{t+u}^{\left(a_{2}\right)}V_{t+v}^{\left(a_{3}\right)}V_{t+w}^{\left(a_{4}\right)})-\mathbb{E}V_{\mathscr{N},t}^{\left(a_{1}\right)}V_{\mathscr{N},t+u}^{\left(a_{2}\right)}V_{\mathscr{N},t+v}^{\left(a_{3}\right)}V_{\mathscr{N},t+w}^{\left(a_{4}\right)},
\end{align*}
where $\{V_{\mathscr{N},t}\}$ is a Gaussian sequence with the same
mean and covariance structure as $\left\{ V_{t}\right\} $. $\kappa_{V,t}^{\left(a_{1},a_{2},a_{3},a_{4}\right)}\left(u,\,v,\,w\right)$
is the time-$t$ fourth-order cumulant of $(V_{t}^{\left(a_{1}\right)},\,V_{t+u}^{\left(a_{2}\right)},\,V_{t+v}^{\left(a_{3}\right)},$
$\,V_{t+w}^{\left(a_{4}\right)})$ while $\kappa_{\mathscr{N}}^{\left(a_{1},a_{2},a_{3},a_{4}\right)}$
$(t,\,t+u,\,t+v,\,t+w)$ is the time-$t$ centered fourth moment of
$V_{t}$ if $V_{t}$ were Gaussian. Let $\lambda_{\max}\left(A\right)$
denote the largest eigenvalue of the matrix $A$.
\begin{assumption}
\label{Assumption A - Dependence}(i) $\sum_{k=-\infty}^{\infty}\sup_{u\in\left[0,\,1\right]}$
$\left\Vert c\left(u,\,k\right)\right\Vert <\infty$ and $\sum_{k=-\infty}^{\infty}\sum_{j=-\infty}^{\infty}\sum_{l=-\infty}^{\infty}\sup_{u}|\kappa_{V,\left\lfloor Tu\right\rfloor }^{\left(a_{1},a_{2},a_{3},a_{4}\right)}$
$\left(k,\,j,\,l\right)|<\infty$ for all $a_{1},a_{2},a_{3},a_{4}\leq p$.
(ii) For all $a_{1},a_{2},a_{3},a_{4}\leq p$ there exists a function
$\widetilde{\kappa}_{a_{1},a_{2},a_{3},a_{4}}:\,\left[0,\,1\right]\times\mathbb{Z}\times\mathbb{Z}\times\mathbb{Z}\rightarrow\mathbb{R}$
that is piecewise continuous in the first argument such that $\sup_{u\in\left[0,\,1\right]}|\kappa_{V,\left\lfloor Tu\right\rfloor }^{\left(a_{1},a_{2},a_{3},a_{4}\right)}\left(k,\,s,\,l\right)-\widetilde{\kappa}_{a_{1},a_{2},a_{3},a_{4}}$
$\left(u,\,k,\,s,\,l\right)|\leq CT^{-1}$ for some $C<\infty$; $\widetilde{\kappa}_{a_{1},a_{2},a_{3},a_{4}}(u,\,k,$
$\,s,\,l)$ is twice differentiable in $u$ at all continuity points
with  bounded derivatives $\left(\partial/\partial u\right)$ $\widetilde{\kappa}_{a_{1},a_{2},a_{3},a_{4}}$
$\left(u,\cdot,\cdot,\cdot\right)$ and $\left(\partial^{2}/\partial u^{2}\right)\widetilde{\kappa}_{a_{1},a_{2},a_{3},a_{4}}\left(u,\cdot,\cdot,\cdot\right)$,
and twice left-differentiable in $u$ at all discontinuity points
with  bounded derivatives $\left(\partial/\partial_{-}u\right)\widetilde{\kappa}_{a_{1},a_{2},a_{3},a_{4}}\left(u,\cdot,\cdot,\cdot\right)$
and $\left(\partial^{2}/\partial_{-}u^{2}\right)$$\widetilde{\kappa}_{a_{1},a_{2},a_{3},a_{4}}$
$\left(u,\cdot,\cdot,\cdot\right)$, and piecewise Lipschitz continuous
derivative $\left(\partial^{2}/\partial_{-}u^{2}\right)\widetilde{\kappa}_{a_{1},a_{2},a_{3},a_{4}}\left(u,\cdot,\cdot,\cdot\right)$.
\end{assumption}
If $\left\{ V_{t,T}\right\} $ is stationary then the cumulant condition
of Assumption \ref{Assumption A - Dependence}-(i) reduces to the
standard one used in the time series literature {[}see, e.g., Assumption
A in \citet{andrews:91}{]}. We do not require fourth-order stationarity
but only that the time-$t=Tu$ fourth order cumulant is locally constant
in a neighborhood of a continuity point $u$. As explained in \citet{casini_hac},
using an argument similar to that used in Lemma 1 in \citet{andrews:91},
one can show that $\alpha$-mixing and moment conditions imply that
the cumulant condition of Assumption \ref{Assumption A - Dependence}-(i)
holds. Part (ii) essentially requires that the approximating cumulant
function $\widetilde{\kappa}_{a_{1},a_{2},a_{3},a_{4}}(u,\,k,$ $\,s,\,l)$
satisfies similar smoothness restrictions as $f\left(u,\,\cdot\right)$
(i.e., twice differentiability at the continuity points and twice
left-differentiable at the discontinuity points). 
\begin{assumption}
\label{Assumption B}(i) $\sqrt{T}(\widehat{\beta}-\beta_{0})=O_{\mathbb{P}}\left(1\right)$;
(ii) $\sup_{u\in\left[0,\,1\right]}\mathbb{E}||V_{\left\lfloor Tu\right\rfloor }||^{2}<\infty$;
(iii) $\sup_{u\in\left[0,\,1\right]}\mathbb{E}\sup_{\beta\in\Theta}$
$||\left(\partial/\partial\beta'\right)V_{\left\lfloor Tu\right\rfloor }\left(\beta\right)||^{2}<\infty$.
\end{assumption}
Assumption \ref{Assumption B}-(i,iii) is an extension of Assumption
B in \citet{andrews:91} to a nonstationary setting. Part (i) follows
from asymptotic normality of $\sqrt{T}(\widehat{\beta}-\beta_{0})$.
Part (ii)-(iii) are common conditions used to obtain the asymptotic
normality of $\sqrt{T}(\widehat{\beta}-\beta_{0})$ under nonstationarity.
In order to obtain  rate of convergence results we shall replace
Assumption \ref{Assumption A - Dependence} with the following assumption.

\begin{assumption}
\label{Assumption C Andrews 91}(i) Assumption \ref{Assumption A - Dependence}-(i)
holds with $V_{t}$ replaced by 
\begin{align*}
\left(V'_{\left\lfloor Tu\right\rfloor },\,\mathrm{vec}\left(\left(\frac{\partial}{\partial\beta'}V_{\left\lfloor Tu\right\rfloor }\left(\beta_{0}\right)\right)-\mathbb{E}\left(\frac{\partial}{\partial\beta'}V_{\left\lfloor Tu\right\rfloor }\left(\beta_{0}\right)\right)\right)'\right)' & .
\end{align*}
(ii) $\sup_{u\in\left[0,\,1\right]}\mathbb{E}(\sup_{\beta\in\Theta}||\left(\partial^{2}/\partial\beta\partial\beta'\right)V_{\left\lfloor Tu\right\rfloor }^{\left(r\right)}\left(\beta\right)||)^{2}<\infty$
for all $r=1,\ldots,\,p$.
\end{assumption}
\begin{assumption}
\label{Assumption W_T and unbounded kernel and Cumulant 8}Let $W_{T}$
denote a $p^{2}\times p^{2}$ weight matrix such that $W_{T}\overset{\mathbb{P}}{\rightarrow}W$.
\end{assumption}
\begin{assumption}
\label{Assumption E-F-G}(i) $\widehat{\phi}_{D}\left(q\right)=O\mathbb{_{P}}\left(1\right)$
and $1/\widehat{\phi}_{D}\left(q\right)=O\mathbb{_{P}}\left(1\right)$;
(ii) $\min\{T/n_{3,T},\,\sqrt{n_{2,T}}\}(\widehat{\phi}_{D}\left(q\right)-\phi_{\theta^{*}})=O_{\mathbb{P}}\left(1\right)$
for some $\phi_{\theta^{*}}\in\left(0,\,\infty\right)$ where $n_{2,T}/T+n_{3,T}/T\rightarrow0,$
$n_{2,T}^{5/4}/T\rightarrow[c_{2},\,\infty),$ $n_{3,T}^{10/6}/T\rightarrow[c_{3},\,\infty)$
with $0<c_{2},\,c_{3}<\infty$; (iii) $\sup_{u\in\left[0,\,1\right]}\lambda_{\max}(\Gamma_{D,u}^{*}\left(k\right))\leq C_{3}k^{-l}$
for all $k\geq0$ for some $C_{3}<\infty$ and some $l>\max\{2,\,\left(4q+2\right)/\left(2+q\right)\}$,
where $q$ is as in $\boldsymbol{K}_{3}$; (iv) uniformly in $u\in\left[0,\,1\right]$,
$\widehat{D}_{1,D}\left(u\right),\,\widehat{D}_{2,D}\left(u\right)$,
$1/\widehat{D}_{1,D}\left(u\right)$ and $\,1/\widehat{D}_{2,D}\left(u\right)$
are $O\mathbb{_{P}}\left(1\right)$; (v) $\omega_{s+1}-\omega_{s}\rightarrow0,\,\left[S_{\omega}\right]^{-1}\rightarrow\infty$
at rate $O\left(T^{-1}\right)$ and $O\left(T\right),$ respectively;
(vi) $\sqrt{Tb_{2,T}\left(u\right)}(\widehat{D}_{2,D}\left(u\right)$
$-D_{2,D}\left(u\right))=O_{\mathbb{P}}\left(1\right)$ for all $u\in\left[0,\,1\right]$.
\end{assumption}
Assumption \ref{Assumption C Andrews 91} is needed to show that the
effect of using $\widehat{\beta}$ rather than $\beta_{0}$ when constructing
$\widehat{J}_{\mathrm{pw},T}$ is at most $o_{\mathbb{P}}\left(1\right)$;
it is an extension of Assumption C in \citet{andrews:91}. Parts (i)-(ii)
of Assumption \ref{Assumption E-F-G} are the nonparametric analogue
to Assumption E-F in \citet{andrews:91}. Part (iii) is satisfied
if $\left\{ V_{t}\right\} $ is strong mixing with mixing numbers
that are less stringent than those sufficient for the cumulant condition
in Assumption \ref{Assumption A - Dependence}-(i). Part (iv) and
(vi) extend (i)-(ii) to $\widehat{D}_{1}$ and $\widehat{D}_{2}$.
Part (v) is needed to apply the convergence of Riemann sums. Under
Assumption \ref{Assumption E-F-G} the effect of using the bandwidths
$\widehat{b}_{1,T}^{*}$ and $\widehat{b}_{2,T}^{*}$ rather than
$b_{\theta_{1},T}$ and $\overline{b}_{\theta_{2},T}$ (defined below
in \eqref{Eq. (b1,T_theta)}) when constructing $\widehat{J}_{\mathrm{pw},T}$
is at most $o_{\mathbb{P}}\left(1\right)$.
\begin{assumption}
\label{Assumption H Andrews 1990} $\sqrt{n_{T}}(\widehat{A}_{r,j}-A_{r,j})=O_{\mathbb{P}}\left(1\right)$
for some $A_{r,j}\in\mathbb{R}^{p\times p}$ for all $j=1,\ldots,\,p_{A}$
and all $r=0,\ldots,\,\left\lfloor T/n_{T}\right\rfloor $. 
\end{assumption}
Given the restrictions below on $n_{T}$, Assumption \ref{Assumption H Andrews 1990}
is satisfied by standard nonparametric estimators. For the consistency
of $\widehat{J}_{T,\mathrm{pw}},$ Assumption \ref{Assumption Smothness f(u,w)},
\ref{Assumption A - Dependence}-\ref{Assumption B}, \ref{Assumption E-F-G}-(i,iv)
and \ref{Assumption H Andrews 1990} are sufficient. For the rate
of convergence and asymptotic MSE results additional conditions are
needed. Let 
\begin{align}
b_{\theta_{1},T}=\left(2qK_{1,q}^{2}\phi_{\theta^{*}}T\overline{b}_{\theta_{2},T}/\left(\int K_{1}^{2}\left(y\right)dy\int_{0}^{1}K_{2}^{2}\left(x\right)dx\right)\right)^{-1/\left(2q+1\right)} & ,\label{Eq. (b1,T_theta)}
\end{align}
where $\overline{b}_{\theta_{2},T}\triangleq\int_{0}^{1}[H\left(K_{2}\right)$
$D_{1,D}\left(u\right)]^{-1/5}\left(F\left(K_{2}\right)D_{2,D}\left(u\right)\right)^{1/5}T^{-1/5}du$.
  Recall that the bandwidths $\widehat{\overline{b}}_{2,T}^{*},$
$\widehat{b}_{2,T}^{*}$ and $\widehat{b}_{1,T}^{*}$ are defined
by \eqref{Eq. (b2_bar)}, \eqref{Eq. (b2)} and \eqref{Eq. (b1)},
respectively. 
\begin{thm}
\label{Theorem 1 MSE Prew - Andrews 90}Suppose $K_{1}\left(\cdot\right)\in\boldsymbol{K}_{3}$,
$q$ is as in $\boldsymbol{K}_{3}$, $K_{2}\left(\cdot\right)\in\boldsymbol{K}_{2}$,
$||\int_{0}^{1}f_{D}^{*\left(q\right)}\left(u,\,0\right)||<\infty$.
Then, we have:

(i) If Assumption \ref{Assumption Smothness f(u,w)}, \ref{Assumption A - Dependence}-\ref{Assumption B},
\ref{Assumption E-F-G}-(i,iv) and \ref{Assumption H Andrews 1990}
hold, $\sqrt{n_{T}}\widehat{b}_{1,T}^{*}\rightarrow\infty,$  and
$q>1/2$, then $\widehat{J}_{\mathrm{pw},T}(\widehat{b}_{1,T}^{*},\,\widehat{b}_{2,T}^{*})-J_{T}\overset{\mathbb{P}}{\rightarrow}0$.

(ii) If Assumption \ref{Assumption Smothness f(u,w)}, \ref{Assumption A - Dependence}-(ii),
\ref{Assumption B}-\ref{Assumption C Andrews 91}, \ref{Assumption E-F-G}-(ii,iii,v,vi)
and \ref{Assumption H Andrews 1990} hold, and $n_{T}/(T\widehat{b}_{1,T}^{*})\rightarrow0$,
$n_{T}/(T(\widehat{b}_{1,T}^{*})^{q})\rightarrow0$, $T\widehat{\overline{b}}_{2,T}^{*}/(n_{T}^{2}\widehat{b}_{1,T}^{*})\rightarrow0,$
$T\widehat{\overline{b}}_{2,T}^{*}\widehat{b}_{1,T}^{*}/n_{T}\rightarrow0$,
then $\sqrt{Tb_{\theta_{1},T}b_{\theta_{2},T}}(\widehat{J}_{\mathrm{pw},T}(\widehat{b}_{1,T}^{*},\,\widehat{\overline{b}}_{2,T}^{*})-J_{T})=O_{\mathbb{P}}\left(1\right)$.

(iii) Let $\gamma_{K,q}=2qK_{1,q}^{2}\phi_{\theta^{*}}/(\int K_{1}^{2}\left(y\right)dy\int_{0}^{1}K_{2}^{2}\left(x\right)dx).$
If Assumption \ref{Assumption B}-\ref{Assumption W_T and unbounded kernel and Cumulant 8}
and \ref{Assumption E-F-G}-(ii,iii,v,vi) hold, then
\begin{align*}
\lim_{T\rightarrow\infty} & \mathrm{MSE}(T^{8q/5\left(2q+1\right)},\,\widehat{J}_{\mathrm{pw},T}(\widehat{b}_{1,T}^{*},\,\widehat{\overline{b}}_{2,T}^{*}),\,J_{T},\,W_{T})\\
 & =4\pi^{2}\left[\gamma_{K,q}K_{1,q}^{2}\mathrm{vec}\left(\int_{0}^{1}f_{D}^{*\left(q\right)}\left(u,\,0\right)du\right)'W\mathrm{vec}\left(\int_{0}^{1}f_{D}^{*\left(q\right)}\left(u,\,0\right)du\right)\right]\\
 & \quad+\int K_{1}^{2}\left(y\right)dy\int K_{2}^{2}\left(x\right)dx\,\mathrm{tr}\left[W\left(I_{p^{2}}-C_{pp}\right)\left(\int_{0}^{1}f_{D}^{*}\left(u,\,0\right)du\right)\otimes\left(\int_{0}^{1}f_{D}^{*}\left(u,\,0\right)du\right)\right].
\end{align*}
\end{thm}
 A result corresponding to Theorem \ref{Theorem 1 MSE Prew - Andrews 90}
for non-prewhitened DK-HAC estimators is established in Theorem 5.1
in \citet{casini_hac} under the same assumptions with the exception
of Assumption \ref{Assumption H Andrews 1990}. Note that for $u$
a continuity point, $f_{D}^{*}\left(u,\,\omega\right)=D\left(u,\,\omega\right)f^{*}\left(u,\,\omega\right)D\left(u,\,\omega\right)',$
where $D\left(u,\,\omega\right)=(I_{p}-\sum_{j=1}^{p_{A}}A_{D,j}\left(u\right)e^{-ij\omega})^{-1}$
with $A_{D,j}\left(u\right)=A_{D,Tu,j}+O\left(T^{-1}\right)$ and
$f^{*}\left(u,\,\omega\right)$ is the local spectral density  of
$\left\{ V_{t}^{*}\right\} $. Since $D\left(u-k/T,\,\omega\right)=D\left(u,\,\omega\right)+O\left(T^{-1}\right)$
by local stationarity, we have
\begin{align*}
f_{D}^{*\left(q\right)}\left(u,\,0\right) & =\left(-1\right)^{q/2}\frac{d^{q}}{d\omega^{q}}\left[D\left(u,\,\omega\right)^{-1}f\left(u,\,\omega\right)\left(D\left(u,\,\omega\right)'\right)^{-1}\right]|_{\omega=0}+O\left(T^{-1}\right),\quad q\,\,\mathrm{even}.
\end{align*}
 A meaningful comparison between prewhitened and non-prewhitened DK-HAC
estimators $\widehat{J}_{T}$ can be made only if reasonable choices
of the bandwdiths $b_{1,T}$ and $b_{2,T}$ are made. When the optimal
bandwidths for $\widehat{J}_{\mathrm{pw},T}$ and $\widehat{J}_{T}$
are used we find that $\widehat{J}_{\mathrm{pw},T}$ has smaller asymptotic
MSE than $\widehat{J}_{T}$ if and only if (assuming $p=1$, i.e.,
the scalar case, with $w_{1,1}=1$)
\begin{align}
\underset{\mathrm{squared\,bias}}{\underbrace{\int_{0}^{1}f_{D}^{*\left(q\right)}\left(u,\,0\right)du}}\underset{\mathrm{variance}}{\underbrace{\left(\int_{0}^{1}f_{D}^{*}\left(u,\,0\right)du\right)^{2q}}} & <\underset{\mathrm{squared\,bias}}{\underbrace{\int_{0}^{1}f^{\left(q\right)}\left(u,\,0\right)du}}\underset{\mathrm{variance}}{\underbrace{\left(\int_{0}^{1}f\left(u,\,0\right)du\right)^{2q}}}.\label{Eq. (MSE comparison Non-Pre and Pre)}
\end{align}
 A numerical comparison would be tedious since the condition depends
on the true data-generating process of $\{V_{t}\}$ and the VAR approximation
for $\widehat{V}_{t}=V_{t}(\widehat{\beta})$. Under stationarity,
\citet{grenander/rosenblatt:57} and \citet{andrews/monahan:92} considered
a few examples. We can make a few observations on the difference
between the condition \eqref{Eq. (MSE comparison Non-Pre and Pre)}
and an analogous condition for the case with $\{V_{t}\}$ second-order
stationary and $D_{s}=D=(1-\sum_{j=1}^{p_{A}}A_{j})^{-1}$ for all
$s$ {[}cf. \citet{andrews/monahan:92}{]}. The condition in \citet{andrews/monahan:92}
is then
\begin{align}
|f^{*\left(q\right)}\left(0\right)|D^{2} & <|f^{q}\left(0\right)|,\label{Eq. Condition Andrews Monahan 92}
\end{align}
 where the quantities $f^{q}\left(0\right)$ and $f^{*\left(q\right)}\left(0\right)$
do not depend on $u$ by stationarity. The main difference between
the two conditions \eqref{Eq. (MSE comparison Non-Pre and Pre)}-\eqref{Eq. Condition Andrews Monahan 92}
is that the part involving the asymptotic variance is missing in
\eqref{Eq. Condition Andrews Monahan 92}. The quantities $|f^{*\left(q\right)}\left(0\right)|D^{2}$
and $|f^{q}\left(0\right)|$ are from the asymptotic squared bias.
This is a consequence of the fact that prewhitened and non-prewhitened
HAC estimators have the same asymptotic variance under stationarity
when the optimal bandwidths are used. This property does not hold
when $\{V_{t}\}$ is nonstationary. The condition \eqref{Eq. (MSE comparison Non-Pre and Pre)}
suggests instead that, in general, both the asymptotic squared bias
and asymptotic variance of prewhitened and non-prewhitened HAC estimators
can be different. Simulations in \citet{andrews/monahan:92} showed
that this is indeed the case even under stationarity: the variance
of the prewhitened HAC estimators is larger than that of the non-prewhitened
HAC estimators\textemdash this feature is consistent with our theoretical
results but not with theirs. 

Both the smoothing over lagged autocovariances and over time influence
the bias of $\widehat{J}_{\mathrm{pw},T}$. The contribution to the
bias due to smoothing over lagged autocovariances is $O(b_{1,T}^{q})$
while the contribution due to smoothing over time is $O(b_{2,T}^{2})$.
Note that the continuity points and the discontinuity points here
induce a bias of the same order $b_{2,T}^{2}$. For the continuity
points, $O(b_{2,T}^{2})$ follows from the usual argument. In the
neighborhood of a discontinuity point $[\lambda_{j}^{0}-b_{2,T},\,\lambda_{j}^{0}+b_{2,T}]$,
the bias of the local smoothing is $O(b_{2,T})$. However, when averaging
over blocks or equivalently integrating over $u\in\left[0,\,1\right]$,
this bias becomes $O(b_{2,T}^{2})$ since there are only a finite
number of discontinuity points and so each discontinuity point contributes
$O(b_{2,T}^{2})$ to the integrated bias. For $\widehat{J}_{\mathrm{pw},T}$
we have $(\widehat{\overline{b}}_{2,T}^{*})^{2}/(\widehat{b}_{1,T}^{*})^{q}\rightarrow0$
since $q=2.$ Thus, the bias due to smoothing over lagged autocovariances
dominates the bias due to smoothing over time.

\section{\label{Section: Extension-to-Unrestrcited Nonstationary}Extension
to General Nonstationary  Random Variables}

In this section we discuss the case where $\left\{ V_{t}\right\} $
is unconditionally heteroskedastic and establish new MSE bounds which
we compare to existing ones. To focus on the main intuition and for
comparison purposes, we consider the non-prewhitened DK-HAC estimator
\[
\widehat{J}_{T}(b_{1,T},\,b_{2,T})=\sum_{k=-T+1}^{T-1}K_{1}(b_{1,T}k)\widehat{\Gamma}\left(k\right),
\]
 where $\widehat{\Gamma}\left(k\right)$ is defined analogously to
$\widehat{\Gamma}_{D}^{*}\left(k\right)$ but with $\widehat{V}_{t}$
in place of $\widehat{V}_{D,t}^{*}$. We use the new MSE bounds to
show that the data-dependent bandwidths for the DK-HAC estimator
are minimax MSE-optimal also under general nonstationarity. Corresponding
results for the prewhitened estimator $\widehat{J}_{\mathrm{pw},T}$
can be obtained by using the results of Section \ref{Section Prewhitended DKHAC- Theory},
though the proofs are more lengthy with no special gain in intuition.

We provide theoretical results under the assumption that $\{V_{t}\}$
is generated by some distribution $\mathscr{P}$ and so defined on
the probability space $(\Omega,\,\mathscr{F},\,\widetilde{\mathbb{P}})$
where $\mathscr{P}=\widetilde{\mathbb{P}}\circ V^{-1}$, $\widetilde{\mathbb{P}}$
is different from $\mathbb{P}$ used in Section \ref{Section: Statistical Enviromnent}-\ref{Section Prewhitended DKHAC- Theory}
and $V$ is a random variable that is a measurable function $V:\,\Omega\mapsto\mathbb{R}$.
$\mathbb{E}_{\mathscr{P}}$ denotes the expectation taken under $\mathscr{P}$.
We establish lower and upper bounds on the MSE under $\mathscr{P}$
and use a minimax MSE criterion for optimality. Define the sample
size dependent spectral density of $\{V_{t}\}$ as 
\[
f_{\mathscr{P},T}\left(\omega\right)\triangleq\left(2\pi\right)^{-1}\sum_{k=-T+1}^{T-1}\Gamma_{\mathscr{P},T}\left(k\right)\exp\left(-i\omega k\right),\,\,\,\mathrm{for\,\,\,}\omega\in\left[-\pi,\,\pi\right],
\]
where 
\begin{align*}
\Gamma_{\mathscr{P},T}\left(k\right) & =\begin{cases}
T^{-1}\sum_{t=k+1}^{T}\mathbb{E}_{\mathscr{P}}(V_{t}V'_{t-k}), & \mathrm{for\,}k\geq0\\
T^{-1}\sum_{t=-k+1}^{T}\mathbb{E}_{\mathscr{P}}(V_{t+k}V'_{t}), & \mathrm{for\,}k<0
\end{cases}.
\end{align*}
 The estimand is then given by 
\begin{align}
J_{\mathscr{P},T} & \triangleq\sum_{k=-T+1}^{T-1}\Gamma_{\mathscr{P},T}\left(k\right).\label{Eq. (Definition J_P,T)}
\end{align}

The theoretical bounds  are derived in terms of two distributions
$\mathscr{P}_{w}$, $w=L,\,U$, under which $\{V_{t}\}$ is zero-mean
SLS with $m_{0}+1$ regimes and satisfies Assumption \ref{Assumption Smothness f(u,w)}
and \ref{Assumption A - Dependence} with autocovariance function
$\{\Gamma_{\mathscr{P}_{w},t/T}\left(k\right)\}$. Then, $\{a'V_{t}\}$
has spectral density $f_{\mathscr{P}_{w},a}\left(\omega\right)\triangleq\int_{0}^{1}f_{\mathscr{P}_{w},a}\left(u,\,\omega\right)du,$
where 
\begin{align*}
f_{\mathscr{P}_{w},a}\left(u,\,\omega\right)\triangleq\left(2\pi\right)^{-1}\sum_{k=\infty}^{\infty}a'\Gamma_{\mathscr{P}_{w},u}\left(k\right)a\exp\left(-i\omega k\right), & \mathrm{\,\,\,for\,all\,}a\in\mathbb{R}^{p}.
\end{align*}
 Let $\kappa_{\mathscr{P},aV,t}\left(k,\,j,\,m\right)$ denote the
time-$t$ fourth-order cumulant of $(a'V_{t},\,a'V_{t+k},\,a'V_{t+j},\,a'V_{t+m})$
under $\mathscr{P}$. For two matrices $A$ and $B$, $A\leq B$ if
and only if $A_{ij}\leq B_{ij}$ for all $i$ and $j$. Define 
\begin{align*}
\boldsymbol{P}_{U} & \triangleq\biggl\{\mathscr{P}:\,-\Gamma_{\mathscr{P}_{U},t/T}\left(k\right)\leq\Gamma_{\mathscr{P},t/T}\left(k\right)\leq\Gamma_{\mathscr{P}_{U},t/T}\left(k\right),\,\mathrm{and}\,\left|\kappa_{\mathscr{P},aV,t}\left(k,\,j,\,m\right)\right|\leq\,\left|\kappa_{t}^{*}\left(k,\,j,\,m\right)\right|\\
 & \qquad\forall t\geq1,\,k,\,j,\,m\geq-t+1,\,a\in\mathbb{R}^{p}\,\mathrm{that\,satisfies}\sum_{k=-\infty}^{\infty}\sum_{j=-\infty}^{\infty}\sum_{m=-\infty}^{\infty}\sup_{t}\kappa_{t}^{*}\left(k,\,j,\,m\right)<\infty\biggr\},
\end{align*}
\begin{align*}
\mathrm{and}\qquad\boldsymbol{P}_{L} & \triangleq\biggl\{\mathscr{P}:\,0\leq\Gamma_{\mathscr{P}_{L},t/T}\left(k\right)\leq\Gamma_{\mathscr{P},t/T}\left(k\right),\,\forall t\geq1,\,k\geq-t+1\,\mathrm{and}\,\kappa_{\mathscr{P},aV,t}\left(k,\,j,\,m\right)\\
 & \qquad\mathrm{satisfies\,the\,same\,condition\,as\,in\,}\boldsymbol{P}_{U}\biggr\}.
\end{align*}
To derive the MSE bounds for a given class of general nonstationary
processes one needs to impose restrictions on the autocovariance function
of the processes in the class relative to the autocovariance function
of some process whose second-order properties are known. This approach
was also used by \citet{andrews:91} who, however, relied on stationarity.
$\boldsymbol{P}_{U}$ includes all distributions such that the autocovariances
of $\left\{ V_{t}\right\} $ are bounded above by those of some SLS
process with distribution $\mathscr{P}_{U}$, thereby allowing considerable
variability of $\Gamma_{\mathscr{P},t/T}\left(k\right)$ for given
$t$ and $k.$  The set $\boldsymbol{P}_{L}$ requires the autocovariances
of $\left\{ V_{t}\right\} $ to be bounded below by positive semidefinite
autocovariances of some SLS process with distribution $\mathscr{P}_{L}$.
Let $c_{\mathscr{P}_{w}}\left(u,\,k\right)=\int e^{i\omega k}\Gamma_{\mathscr{P}_{w},u}\left(k\right)d\omega$
denote the local autocovariance associated to the distribution $\mathscr{P}_{w},$
$w=L,\,U.$ Let 
\begin{align*}
\boldsymbol{K}_{1} & =\biggl\{ K\left(\cdot\right):\,\mathbb{R}\rightarrow\left[-1,\,1\right]:\,K\left(0\right)=1,\,K\left(x\right)=K\left(-x\right),\,\forall x\in\mathbb{R}\\
 & \quad\int_{-\infty}^{\infty}K^{2}\left(x\right)dx<\infty,\,K\left(\cdot\right)\,\mathrm{is\,continuous\,at\,0\,and\,at\,all\,but\,finite\,number\,of\,points}\biggr\}.
\end{align*}
 Note that $\boldsymbol{K}_{3}\subset\boldsymbol{K}_{1}$. In particular,
$\boldsymbol{K}_{1}$ includes also the truncated kernel. 

\subsection{Consistency, Rate of Convergence and MSE Bounds}

Consider the following generalization of Assumption \ref{Assumption Smothness f(u,w)}
and \ref{Assumption A - Dependence}:
\begin{assumption}
\label{Assumption A*}$\left\{ V_{t}\right\} $ is a mean-zero sequence
and satisfies $\sum_{k=0}^{\infty}\sup_{t\geq1}||\mathbb{E}_{\mathscr{P}}(V_{t}V'_{t+k})||<\infty$
and for all $a_{1},\,a_{2},\,a_{3},\,a_{4}\leq p,$ $\sum_{k=1}^{\infty}\sum_{j=1}^{\infty}$
$\sum_{m=1}^{\infty}\sup_{t\geq1}|\kappa_{\mathscr{P},V,t}^{\left(a_{1},\,a_{2},\,a_{3},\,a_{4}\right)}\left(k,\,j,\,m\right)|<\infty$. 
\end{assumption}
Let $\mathrm{MSE}_{\mathscr{P}}\left(\cdot\right)$ denote the MSE
of $\cdot$ under $\mathscr{P}$ and let $\boldsymbol{K}_{1,+}=\left\{ K_{1}\left(\cdot\right)\in\boldsymbol{K}_{1}:\,K_{1}\left(x\right)\geq0\,\forall x\right\} $.
 $\boldsymbol{K}_{1,+}$ is a subset of $\boldsymbol{K}_{1}$ that
contains all kernels that are non-negative and is used for some results
below. The QS kernel is not in $\boldsymbol{K}_{1,+}$. The smoothness
of $f_{\mathscr{P}_{w},a}\left(u,\,\omega\right)$ at $\omega=0$
is indexed by 
\[
f_{\mathscr{P}_{w},a}^{\left(q\right)}\left(u,\,0\right)=\left(2\pi\right)^{-1}\sum_{k=-\infty}^{\infty}\left|k\right|^{q}a'\Gamma_{\mathscr{P}_{w},u}\left(k\right)a,\,\,\,\,\mathrm{for}\,q\in[0,\,\infty),\,w=L,\,U.
\]
 We first consider the MSE bounds for  $\widetilde{J}_{T}$ which
is constructed using $V_{t}(\beta_{0})$ rather than $\widehat{V}_{t}$.
\begin{thm}
\label{Theorem MSE J DKHAC Nonstationary}Suppose Assumption \ref{Assumption A*}
holds, $K_{2}\left(\cdot\right)\in\boldsymbol{K}_{2}$, $b_{1,T},\,b_{2,T}\rightarrow0$,
$n_{T}\rightarrow\infty,\,n_{T}/T\rightarrow0$ and $1/(Tb_{1,T}b_{2,T})\rightarrow0$.
If $n_{T}/(Tb_{1,T}^{q})\rightarrow0$, $b_{2,T}^{2}/b_{1,T}^{q}\rightarrow0$
and $Tb_{1,T}^{2q+1}b_{2,T}\rightarrow\gamma\in\left(0,\,\infty\right)$
for some $q\in[0,\,\infty)$ for which $K_{1,q},\,|\int_{0}^{1}f_{\mathscr{P}_{w},a}^{\left(q\right)}\left(u,\,0\right)du|\in[0,\,\infty)$,
$w=L,\,U$, $a\in\mathbb{R}^{p}$, then we have:

(i) for all $K_{1}\left(\cdot\right)\in\boldsymbol{K}_{1}$, 
\begin{align*}
\lim_{T\rightarrow\infty} & Tb_{1,T}b_{2,T}\underset{\mathscr{P}\in\boldsymbol{P}_{U}}{\sup}\mathrm{MSE}_{\mathscr{P}}\left(a'\widetilde{J}_{T}a\right)=4\pi^{2}\left[\gamma K_{1,q}^{2}\left(\int_{0}^{1}f_{\mathscr{P}_{U},a}^{\left(q\right)}\left(u,\,0\right)du\right)^{2}\right.\\
 & \quad\left.+2\int K_{1}^{2}\left(y\right)dy\int_{0}^{1}K_{2}^{2}\left(x\right)dx\,\left(\int_{0}^{1}f_{\mathscr{P}_{U},a}\left(u,\,0\right)du\right)^{2}\right].
\end{align*}

(ii) for all $K_{1}\left(\cdot\right)\in\boldsymbol{K}_{1,+}$, 
\begin{align*}
\lim_{T\rightarrow\infty} & Tb_{1,T}b_{2,T}\underset{\mathscr{P}\in\boldsymbol{P}_{L}}{\inf}\mathrm{MSE}_{\mathscr{P}}\left(a'\widetilde{J}_{T}a\right)=4\pi^{2}\left[\gamma K_{1,q}^{2}\left(\int_{0}^{1}f_{\mathscr{P}_{L},a}^{\left(q\right)}\left(u,\,0\right)du\right)^{2}\right.\\
 & \quad\left.+2\int K_{1}^{2}\left(y\right)dy\int_{0}^{1}K_{2}^{2}\left(x\right)dx\,\left(\int_{0}^{1}f_{\mathscr{P}_{L},a}\left(u,\,0\right)du\right)^{2}\right].
\end{align*}
\end{thm}
The theoretical  bounds in Theorem \ref{Theorem MSE J DKHAC Nonstationary}
are sharper than the ones in \textcolor{MyBlue}{Andrews (1988; 1991)}
which are based on stationarity (i.e., the autocovariances that dominate
the autocovariances of any $\mathscr{P}\in\boldsymbol{P}_{U}$ are
assumed by \citet{andrews:91} to be those of a  stationary process).\footnote{There are a couple of technical issues in Section 8 in \citet{andrews:91}.
In particular, the MSE bound is not correct. See \citet{casini_comment_andrews91}
for details.} Given that stationarity is a special case of SLS, our bounds apply
to a wider class of processes. Furthermore, they are more informative
because they change with the specific type of nonstationarity unlike
\textcolor{MyBlue}{Andrews'} \citeyearpar{andrews:91} bounds that
depend on the spectral density of a stationary process. 

The theorem is derived under $b_{2,T}^{2}/b_{1,T}^{q}\rightarrow0$
(i.e., the bias due to smoothing over time is of smaller order than
that due to smoothing over lagged autocovariances). When instead $b_{2,T}^{2}/b_{1,T}^{q}\rightarrow\nu\in\left(0,\,\infty\right)$,
there is an additional term in the bound. For example, in part (i)
this term is  
\[
\left(\pi\nu\int_{0}^{1}x^{2}K_{2}\left(x\right)dx\int_{\mathbf{\mathbf{\widetilde{\mathbf{C}}}_{\mathscr{P}_{U}}}}\left(\partial^{2}/\partial u^{2}\right)f_{\mathscr{P}_{U},a}\left(u,\,0\right)du+2\pi\nu\Delta_{f_{\mathscr{P}_{U},a}}\left(0\right)\right)^{2}+\Xi,
\]
 where  $\mathbf{\widetilde{\mathbf{C}}}_{\mathscr{P}_{U}}$ is the
set of continuity points under $\mathscr{P}_{U}$,
\begin{align*}
\Delta_{f_{\mathscr{P}_{U},a}}\left(\omega\right) & =\sum_{j=1}^{m_{0}}\int_{0}^{1}\left(\frac{\partial}{\partial u_{-}}f_{\mathscr{P}_{U},a}\left(\lambda_{j}^{0},\,\omega\right)\int_{0}^{1-s}xK_{2}\left(x\right)dx+\frac{\partial}{\partial u_{+}}f_{\mathscr{P}_{U},a}\left(\lambda_{j}^{0},\,\omega\right)\int_{1-s}^{1}xK_{2}\left(x\right)dx\right)ds,
\end{align*}
with $\left\{ \lambda_{j}^{0}\right\} _{j=1}^{m_{0}}$ being the discontinuity
points, $m_{0}$ being a finite integer, 
\begin{align*}
\frac{\partial}{\partial u_{-}}f_{\mathscr{P}_{U},a}\left(\lambda_{j}^{0},\,\omega\right) & =\underset{h\uparrow0}{\lim}\frac{f_{\mathscr{P}_{U},a}\left(\lambda_{j}^{0}+h,\,\omega\right)-f_{\mathscr{P}_{U},a}\left(\lambda_{j}^{0},\,\omega\right)}{h},\\
\frac{\partial}{\partial u_{+}}f_{\mathscr{P}_{U},a}\left(\lambda_{j}^{0},\,\omega\right) & =\underset{h\downarrow0}{\lim}\frac{f_{\mathscr{P}_{U},a}\left(\lambda_{j}^{0}+h,\,\omega\right)-f_{\mathscr{P}_{U},a}\left(\lambda_{j}^{0},\,\omega\right)}{h},
\end{align*}
 and $\Xi$ depends on the cross-products of the bias terms due to
smoothing over time and lagged autocovariances. Some of the results
of this paper are extended to the case $b_{2,T}^{2}/b_{1,T}^{q}\rightarrow\nu\in\left(0,\,\infty\right)$
in \textcolor{MyBlue}{Belotti et al. (2021)}. \nocite{belotti/casini/catania/grassi/perron_HAC_Sim_Bandws}
Thus, our bounds show how nonstationarity influences the bias-variance
trade-off. They also highlight how it is affected by the smoothing
over the time direction versus the autocovariance lags direction.
These are important elements in order to understand the properties
of HAR tests normalized by LRV estimators.  

We now extend the results in Theorem \ref{Theorem MSE J DKHAC Nonstationary}
to the estimator $\widehat{J}_{T}$ that uses $V_{t}(\widehat{\beta})$.
The following assumptions extend Assumption \ref{Assumption B}-\ref{Assumption C Andrews 91}
to the distribution $\mathscr{P}.$
\begin{assumption}
\label{Assumption B EP}Assumption \ref{Assumption B} holds with
$\mathbb{E}$ replaced by $\mathbb{E}_{\mathscr{P}}$.
\end{assumption}
\begin{assumption}
\label{Assumption C*}(i) Assumption \ref{Assumption A*} holds with
$V_{t}$ replaced by $(V'_{\left\lfloor Tu\right\rfloor },\,\mathrm{vec}(((\partial/\partial\beta')V_{\left\lfloor Tu\right\rfloor }(\beta_{0}))-$
$\mathbb{E}_{\mathscr{P}}($ $\left(\partial/\partial\beta'\right)V_{\left\lfloor Tu\right\rfloor }(\beta_{0}))')'$;
(ii) $\sup_{u\in\left[0,\,1\right]}\mathbb{E}_{\mathscr{P}}(\sup_{\beta\in\Theta}||(\partial^{2}/\partial\beta\partial\beta')V_{\left\lfloor Tu\right\rfloor }^{\left(a_{r}\right)}\left(\beta\right)||^{2})<\infty$
$(r=1,\ldots,\,p)$. 
\end{assumption}
To show the asymptotic equivalence of the MSE of $a'\widehat{J}_{T}a$
to that of $a'\widetilde{J}_{T}a$ we need an additional assumption
which was also used by \citet{andrews:91}. Let $|A|$ denote the
vector or matrix of absolute values of the elements of $A.$ Define
\begin{align*}
H_{1,T} & \triangleq b_{1,T}\sum_{k=-T+1}^{T-1}\biggl|K_{1}\left(b_{1,T}k\right)\biggr|\\
 & \quad\times\biggl|\frac{n_{T}}{T}\sum_{r=0}^{\left\lfloor T/n_{T}\right\rfloor }\left(Tb_{2,T}\right)^{-1/2}\sum_{s=k+1}^{T}K_{2}^{*}\left(\frac{\left(\left(r+1\right)n_{T}-\left(s-k/2\right)\right)/T}{b_{2,T}}\right)\frac{\partial}{\partial\beta}a'V_{s}\left(\beta_{0}\right)a'V{}_{s-k}\left(\beta_{0}\right)\biggr|,\\
H_{2,T} & \triangleq b_{1,T}\sum_{k=-T+1}^{T-1}\biggl|K_{1}\left(b_{1,T}k\right)\biggr|\underset{\beta\in\Theta}{\sup}\biggl|\frac{n_{T}}{T}\sum_{r=0}^{\left\lfloor T/n_{T}\right\rfloor }\left(Tb_{2,T}\right)^{-1}\\
 & \quad\times\sum_{s=k+1}^{T}K_{2}^{*}\left(\frac{\left(\left(r+1\right)n_{T}-\left(s-k/2\right)\right)/T}{b_{2,T}}\right)\frac{\partial^{2}}{\partial\beta\partial\beta'}a'V_{s}\left(\beta\right)a'V{}_{s-k}\left(\beta\right)\biggr|.
\end{align*}
Let $H_{1,T}^{\left(r\right)}$, $\widehat{\beta}^{\left(r\right)}$
and $\beta_{0}^{\left(r\right)}$ denote the $r$-th elements of $H_{1,T}$,
$\widehat{\beta}$ and $\beta_{0}$, respectively, for $r=1,\ldots,\,p$. 

\begin{assumption}
\label{Assumption B Andrews 88}For all $r=1,\ldots,\,p$, $\limsup_{T\rightarrow\infty}\sup_{\mathscr{P}\in\mathscr{\boldsymbol{P}}_{U}}\mathbb{E}_{\mathscr{P}}(H_{1,T}^{\left(r\right)}\sqrt{T}(\widehat{\beta}^{\left(r\right)}-\beta_{0}^{\left(r\right)}))^{2}<\infty$
and $\limsup_{T\rightarrow\infty}\sup_{\mathscr{P}\in\boldsymbol{P}_{U}}\mathbb{E}_{\mathscr{P}}(\sqrt{T}(\widehat{\beta}-\beta_{0})'H_{2,T}\sqrt{T}(\widehat{\beta}-\beta_{0}))^{2}<\infty.$
\end{assumption}
\begin{thm}
\label{Theorem 1 Andrews 91 - Consistency and Rate - Nonstationary}Suppose
$K_{1}\left(\cdot\right)\in\boldsymbol{K}_{1}$, $K_{2}\left(\cdot\right)\in\boldsymbol{K}_{2}$,
$b_{1,T},\,b_{2,T}\rightarrow0$,\textbf{ }$n_{T}\rightarrow\infty,\,n_{T}/T\rightarrow0$
and $1/Tb_{1,T}b_{2,T}\rightarrow0$. We have: 

(i) If Assumption \ref{Assumption A*}-\ref{Assumption B EP} hold,
$\sqrt{T}b_{1,T}\rightarrow\infty$, then $\widehat{J}_{T}-J_{\mathscr{P},T}\overset{\mathbb{P}}{\rightarrow}0$
and $\widehat{J}_{T}-\widetilde{J}_{T}\overset{\mathbb{P}}{\rightarrow}0$
where $J_{\mathscr{P},T}$ is defined in \eqref{Eq. (Definition J_P,T)}.

(ii) If Assumption \ref{Assumption A*}-\ref{Assumption C*} hold,
$n_{T}/(Tb_{1,T})\rightarrow0$, $n_{T}/(Tb_{1,T}^{q})\rightarrow0$
and $Tb_{1,T}^{2q+1}b_{2,T}\rightarrow\gamma\in\left(0,\,\infty\right)$
for some $q\in[0,\,\infty)$ for which $K_{1,q},\,|\int_{0}^{1}f_{\mathscr{P}_{w},a}^{\left(q\right)}\left(u,\,0\right)du|\in[0,\,\infty)$,
$w=U,\,L$, $a\in\mathbb{R}^{p}$, then $\sqrt{Tb_{1,T}b_{2,T}}(\widehat{J}_{T}-J_{\mathscr{P},T})=O_{\mathbb{\mathscr{P}}}\left(1\right)$
and $\sqrt{Tb_{1,T}}(\widehat{J}_{T}-\widetilde{J}_{T})=o_{\mathscr{P}}\left(1\right).$ 

(iii) Under the assumptions of part (ii) and Assumption \ref{Assumption B Andrews 88},
\[
\lim_{T\rightarrow\infty}\sup_{\mathscr{P}\in\boldsymbol{P}_{U}}Tb_{1,T}b_{2,T}|\mathrm{MSE}_{\mathscr{P}}(a'\widehat{J}_{T}a)-\mathrm{MSE}_{\mathscr{P}}(a'\widetilde{J}_{T}a)|=0
\]
 for all $a\in\mathbb{R}^{p}$ such that $|\int_{0}^{1}f_{\mathscr{P}_{U},a}^{\left(q\right)}\left(u,\,0\right)du|<\infty$. 
\end{thm}
Theorem \ref{Theorem 1 Andrews 91 - Consistency and Rate - Nonstationary}
extends the consistency, rate of convergence, MSE results of Theorem
3.2 in \citet{casini_hac}. The asymptotic equivalence of the MSE
implies that the bounds in Theorem \ref{Theorem MSE J DKHAC Nonstationary}
apply to $\widehat{J}_{T}$ as well. The MSE equivalence is used to
show that the optimal kernels and bandwidths results below apply to
$\widehat{J}_{T}$ as well as to $\widetilde{J}_{T}$. Similar results
can be shown for the prewhitened estimator $\widehat{J}_{T,\mathrm{pw}}$.
For this case, the sets $\boldsymbol{P}_{U}$ and $\boldsymbol{P}_{L}$
would need to be defined in terms of the autocovariance function of
$V_{D,t}^{*}=D_{t}V_{t}^{*}.$ The distributions $\mathscr{P}_{U}$
and $\mathscr{P}_{L}$ that form an envelope for the autocovariances
of $V_{D,t}^{*}$ may  depend on  different prewhitening models. 

\subsection{\label{subsec: Optimal-Bandwidths-and Kernels - Nonstationarity}Optimal
Bandwidths and Kernels}

We use the sequential MSE procedure  that first determines the optimal
$b_{2,T}\left(u\right)$ and then determines the optimal $b_{1,T}$
as function of the integrated optimal $\overline{b}_{2,T}$, see \citet{casini_hac}.
The results for the global MSE criterion can easily be extended using
similar arguments as those used in this section. 

We consider distributions $\mathscr{P}\in\boldsymbol{P}_{U,2}$ where
$\boldsymbol{P}_{U,2}\subseteq\boldsymbol{P}_{U}$ is defined below.
We need to restrict attention to a subset $\boldsymbol{P}_{U,2}$
of $\boldsymbol{P}_{U}$ for technical reasons related to the derivation
of the optimal bandwidth $b_{2,T}^{\mathrm{opt}}\left(u\right)$.
The distributions in $\boldsymbol{P}_{U,2}$ restrict the degree of
nonstationarity  by requiring some smoothness of the local autocovariance.
This is intuitive since the optimality of $b_{2,T}^{\mathrm{opt}}\left(u\right)$
is justified under smoothness locally in time. We remark, however,
that the optimality of $b_{1,T}$ and $K_{1}$ determined below holds
over all distributions $\mathscr{P}\in\boldsymbol{P}_{U}.$ We show
that the resulting optimal kernels are $K_{1}^{\mathrm{opt}}\left(\cdot\right)$
and $K_{2}^{\mathrm{opt}}\left(\cdot\right)$ from Section \ref{Section Prewhitended DKHAC- Theory}. 

Let $\mathbf{\widetilde{C}}_{\mathscr{\mathscr{P}}_{U}}$ denote the
set of continuity points $u\in\left(0,\,1\right)$ under $\mathscr{\mathscr{P}}_{U}$.
For any $a\in\mathbb{R}^{p}$ and $u_{0}\in\mathbf{\widetilde{C}}_{\mathscr{\mathscr{P}}_{U}}$
consider the following inequality,
\begin{align}
\left|a'\left(\frac{\partial^{2}}{\partial^{2}u}c_{\mathscr{P}}\left(u_{0},\,k\right)\right)a\right| & \leq\Biggl|a'\left(\frac{\partial^{2}}{\partial^{2}u}c_{\mathscr{\mathscr{P}}_{U}}\left(u_{0},\,k\right)\right)a\Biggr|,\label{Eq. Inequality for Pu2}
\end{align}
which essentially requires that the distribution $\mathscr{\mathscr{P}}_{U}$
has locally a larger degree of nonstationarity than that of the distribution
$\mathscr{P}$. We consider the following class of distributions,
\begin{align*}
\boldsymbol{P}_{U,2} & \triangleq\left\{ \mathscr{P}:\,\mathscr{P}\in\boldsymbol{P}_{U},\,m_{0}=0,\,\mathrm{and\,}\eqref{Eq. Inequality for Pu2}\,\mathrm{holds}\,\forall k\in\mathbb{R}\,\mathrm{and}\,\forall u_{0}\in\left(0,\,1\right)\right\} .
\end{align*}
Let 
\begin{align*}
D_{1,U,a}(u_{0}) & \triangleq\left(a'\left(\frac{\partial^{2}c_{\mathscr{\mathscr{P}}_{U}}(u_{0},\,k)}{\partial u^{2}}\right)a\right)^{2},\\
D_{2,U,a}(u_{0}) & \triangleq\sum_{l=-\infty}^{\infty}a'(c_{\mathscr{\mathscr{P}}_{U}}(u_{0},\,l)[c_{\mathscr{\mathscr{P}}_{U}}(u_{0},\,l)+c_{\mathscr{\mathscr{P}}_{U}}(u_{0},\,l+2k)]')a.
\end{align*}

\begin{prop}
\label{Proposition: Optimal Local Covariance - Nonstationary}Suppose
Assumption \ref{Assumption A*}-\ref{Assumption B Andrews 88} hold
and $u_{0}\in\mathbf{\widetilde{C}}_{\mathscr{\mathscr{P}}_{U}}$.
For any sequence of bandwidth parameters $\left\{ b_{2,T}\right\} $
such that $b_{2,T}\rightarrow0,$ we have 
\begin{align}
\sup_{\mathscr{P}\in\boldsymbol{P}_{U,2}} & \mathrm{MSE}_{\mathscr{P}}\left(a'\widehat{c}_{T}\left(u_{0},\,k\right)a\right)=\sup_{\mathscr{P}\in\boldsymbol{P}_{U,2}}\mathbb{E}_{\mathscr{P}}\left(a'\widehat{c}_{T}\left(u_{0},\,k\right)a-a'c_{\mathscr{P}}\left(u_{0},\,k\right)a\right)^{2}\label{Eq. MSE Local Autocov in Prop - Nonstationary}\\
 & \leq\frac{1}{4}b_{2,T}^{4}\left(\int_{0}^{1}xK_{2}\left(x\right)dx\right)^{2}\left(\frac{\partial^{2}}{\partial^{2}u}a'c_{\mathscr{\mathscr{P}}_{U}}\left(u_{0},\,k\right)a\right)^{2}\nonumber \\
 & \quad+\frac{1}{Tb_{2,T}}\int_{0}^{1}K_{2}^{2}\left(x\right)dx\sum_{l=-\infty}^{\infty}a'\left(c_{\mathscr{\mathscr{P}}_{U}}\left(u_{0},\,l\right)\left[c_{\mathscr{\mathscr{P}}_{U}}\left(u_{0},\,l\right)+c_{\mathscr{\mathscr{P}}_{U}}\left(u_{0},\,l+2k\right)\right]'\right)a\nonumber \\
 & \quad+\frac{1}{Tb_{2,T}}\int_{0}^{1}K_{2}^{2}\left(x\right)dx\sum_{h_{1}=-\infty}^{\infty}\sum_{h_{2}=-\infty}^{\infty}\kappa_{\mathscr{\mathscr{P}}_{U},aV,Tu_{0}}\left(h_{1},\,0,\,h_{2}\right)+o\left(b_{2,T}^{4}\right)+O\left(1/\left(b_{2,T}T\right)\right),\nonumber 
\end{align}
 which is minimized for 
\[
b_{2,T}^{\mathrm{opt}}\left(u_{0}\right)=[H\left(K_{2}^{\mathrm{opt}}\right)D_{1,U,a}\left(u_{0}\right)]^{-1/5}\left(F\left(K_{2}^{\mathrm{opt}}\right)\left(D_{2,U,a}\left(u_{0}\right)+D_{3,U}\left(u_{0}\right)\right)\right)^{1/5}T^{-1/5},
\]
 where 
\[
D_{3,U}\left(u_{0}\right)=\sum_{h_{1}=-\infty}^{\infty}\sum_{h_{2}=-\infty}^{\infty}\kappa_{\mathscr{\mathscr{P}}_{U},aV,Tu_{0}}\left(h_{1},\,0,\,h_{2}\right),
\]
 and $K_{2}^{\mathrm{opt}}\left(x\right)=6x\left(1-x\right),\,0\leq x\leq1$.
In addition, if $\{V_{t}\}$ is Gaussian, then $D_{3,U}\left(u_{0}\right)=0$
for all $u_{0}\in\left(0,\,1\right)$. 
\end{prop}
We now obtain the optimal $K_{1}\left(\cdot\right)$ and $b_{1,T}$
as a function of $\overline{b}_{2,T}^{\mathrm{opt}}=\int_{0}^{1}b_{2,T}^{\mathrm{opt}}\left(u\right)du$
and $K_{2}^{\mathrm{opt}}\left(\cdot\right)$. For some results
below, we consider a subset of $\boldsymbol{K}_{1}$ defined by $\boldsymbol{\widetilde{K}}_{1}=\{K_{1}\left(\cdot\right)\in\boldsymbol{K}_{1}|\,\widetilde{K}\left(\omega\right)\geq0\,\forall\,\omega\in\mathbb{R}\}$
where $\widetilde{K}\left(\omega\right)=\left(2\pi\right)^{-1}\int_{-\infty}^{\infty}K_{1}\left(x\right)e^{-ix\omega}dx.$
The function $\widetilde{K}\left(\omega\right)$ is referred to as
the spectral window generator corresponding to the kernel $K_{1}\left(\cdot\right)$.
The set $\widetilde{\boldsymbol{K}}_{1}$ contains all kernels $K_{1}$
that generate positive semidefinite estimators in finite samples.
$\boldsymbol{\widetilde{K}}_{1}$ contains the Bartlett, Parzen, and
QS kernels, but not the truncated or Tukey-Hanning kernels. 

We adopt the notation $\widehat{J}_{T}(b_{1,T})=\widehat{J}_{T}(b_{1,T},\,b_{2,T},\,K_{2,0})$
for the estimator $\widehat{J}_{T}$ that uses $K_{2,0}\left(\cdot\right)\in\boldsymbol{K}_{2}$,
$b_{1,T}$ and $b_{2,T}=\overline{b}_{2,T}^{\mathrm{opt}}+o(T^{-1/5})$
where $\overline{b}_{2,T}^{\mathrm{opt}}=\int_{0}^{1}b_{2,T}^{\mathrm{opt}}\left(u\right)du$.
Let $\widehat{J}_{T}^{\mathrm{QS}}(b_{1,T})$ denote the estimator
based on the QS kernel $K_{1}^{\mathrm{QS}}\left(\cdot\right)$. 
We then compare two kernels $K_{1}$ using comparable\textcolor{red}{{}
}bandwidths $b_{1,T}$ which are defined as follows. Given $K_{1}\left(\cdot\right)\in\widetilde{\boldsymbol{K}}_{1}$,
the QS kernel $K_{1}^{\mathrm{QS}}\left(\cdot\right)$, and a bandwidth
 $\left\{ b_{1,T}\right\} $ to be used with the QS kernel, define
a comparable bandwidth  $\{b_{1,T,K_{1}}\}$ for use with $K_{1}\left(\cdot\right)$
such that both kernel/bandwidth combinations have the same maximum
asymptotic variance over $\mathscr{P}\in\boldsymbol{P}_{U}$ when
scaled by the same factor $Tb_{1,T}b_{2,T}$. This means that $b_{1,T,K_{1}}$
is such that 
\begin{align*}
\lim_{T\rightarrow\infty}\sup_{\mathscr{P}\in\boldsymbol{P}_{U}}Tb_{1,T}b_{2,T}\mathrm{MSE}_{\mathscr{P}} & (a'(\widehat{J}_{T}^{\mathrm{QS}}(b_{1,T})-\mathbb{E}(\widetilde{J}_{T}^{\mathrm{QS}}(b_{1,T}))+J_{T})a)\\
=\lim_{T\rightarrow\infty}\sup_{\mathscr{P}\in\boldsymbol{P}_{U}}Tb_{1,T}b_{2,T}\mathrm{MSE}_{\mathscr{P}} & (a'(\widehat{J}_{T}^{\mathrm{}}(b_{1,T,K_{1}})-\mathbb{E}(\widetilde{J}_{T}(b_{1,T,K_{1}}))+J_{T})a).
\end{align*}
This definition yields $b_{1,T,K_{1}}=b_{1,T}/(\int K_{1}^{2}\left(x\right)dx).$
Note that for the QS kernel, $K_{1}^{\mathrm{QS}}\left(x\right)$,
we have $b_{1,T,\mathrm{QS}}=b_{1,T}$ since $\int(K_{1}^{\mathrm{QS}}(x))^{2}dx=1$.
\begin{thm}
\label{Theorem Optimal Kernels Nonstationarity}Suppose Assumption
\ref{Assumption A*}-\ref{Assumption B Andrews 88} hold, $\int_{0}^{1}|f_{U,a}^{\left(2\right)}\left(u,\,0\right)|du<\infty$,
and $b_{2,T}\rightarrow0$, $b_{2,T}^{5}T\rightarrow\eta\in\left(0,\,\infty\right)$.
For any bandwidth sequence $\{b_{1,T}\}$ such that $b_{2,T}/b_{1,T}\rightarrow0$,
$n_{T}/Tb_{1,T}^{2}\rightarrow0$ and $Tb_{1,T}^{5}b_{2,T}\rightarrow\gamma\in\left(0,\,\infty\right)$,
and for any kernel $K_{1}\left(\cdot\right)\in\boldsymbol{\widetilde{K}}_{1}$
used to construct $\widehat{J}_{T}^{\mathrm{}}$, the QS kernel is
preferred to $K_{1}\left(\cdot\right)$ in the sense that for all
$a\in\mathbb{R}^{p}$, 
\begin{align*}
\liminf_{T\rightarrow\infty} & Tb_{1,T}b_{2,T}\left(\sup_{\mathscr{P}\in\boldsymbol{P}_{U}}\mathrm{MSE}_{\mathscr{P}}\left(a'\widehat{J}_{T}^{\mathrm{}}\left(b_{1,T,K_{1}}\right)a\right)-\sup_{\mathscr{P}\in\boldsymbol{P}_{U}}\mathrm{MSE}_{\mathscr{P}}\left(a'\widehat{J}_{T}^{\mathrm{QS}}\left(b_{1,T}\right)a\right)\right)\\
 & =4\gamma\pi^{2}\left(\int_{0}^{1}f_{U,a}^{\left(2\right)}\left(u,\,0\right)du\right)^{2}\int_{0}^{1}\left(K_{2,0}\left(x\right)\right)^{2}dx\times\left[K_{1,2}^{2}\left(\int K_{1}^{2}\left(y\right)dy\right)^{4}-\left(K_{1,2}^{\mathrm{QS}}\right)^{2}\right]\geq0.
\end{align*}
 The inequality is strict if $K_{1}\left(x\right)\neq K_{1}^{\mathrm{QS}}\left(x\right)$
with positive Lebesgue measure.
\end{thm}
We now consider the asymptotically optimal choice of $b_{1,T}$ for
a given kernel $K_{1}\left(\cdot\right)$ for which $K_{1,q}\in\left(0,\,\infty\right)$
for some $q$, and given $K_{2}^{\mathrm{opt}}$ and $\overline{b}_{2,T}^{\mathrm{opt}}$.
We continue to use a minimax optimality criterion. However, unlike
the results of Proposition \ref{Proposition: Optimal Local Covariance - Nonstationary}
and Theorem \ref{Theorem Optimal Kernels Nonstationarity}, in which
an optimal kernel was found that was the same for any dominating distribution
$\boldsymbol{P}_{U,2}$ and $\boldsymbol{P}_{U}$, respectively, the
optimal bandwidth $b_{1,T}$ depends on a scalar parameter $\phi\left(q\right)$
that is a function of $\mathscr{P}_{U}$ and $q$. 

Let $w_{r},\,\,r=1,\ldots,\,p$, be a set of non-negative weights
summing to one. We consider a weighted squared error loss function
\[
\mathrm{L}(\widehat{J}_{T},\,J_{\mathscr{P},T})=\sum_{r=1}^{p}w_{r}(\widehat{J}_{T}^{\left(r,r\right)}(b_{1,T})-J_{\mathscr{P},T}^{\left(r,r\right)})^{2}.
\]
 A common choice is $w_{r}=1/p$ for $r=1,\ldots,\,p$. For a given
dominating distribution $\mathscr{P}_{U}$, define
\begin{align}
\phi\left(q\right) & =\sum_{r=1}^{p}w_{r}\left(\int_{0}^{1}f_{U,a^{\left(r\right)}}^{\left(q\right)}\left(u,\,0\right)du\right)^{2}/\sum_{r=1}^{p}w_{r}\left(\int_{0}^{1}f_{U,a^{\left(r\right)}}\left(u,\,0\right)du\right)^{2},\label{Eq (5.2) Andrews 88 (phi(q)) Nonstationary}
\end{align}
 where $a^{\left(r\right)}$ is a $p$-vector with the $r$-th element
one and all other elements zero. For any given $\phi\left(q\right)\in\left(0,\,\infty\right)$,
let $\mathscr{\boldsymbol{P}}_{U}\left(\phi\right)$ denote some set
$\mathscr{\boldsymbol{P}}_{U}$ whose dominating distribution $\mathscr{P}_{U}$
satisfies \eqref{Eq (5.2) Andrews 88 (phi(q)) Nonstationary}. 
\begin{thm}
\label{Theorem 3 Andrews 88}Suppose Assumption \ref{Assumption A*}-\ref{Assumption B Andrews 88}
hold. For any given $K_{1}\left(\cdot\right)\in\boldsymbol{K}_{1}$
such that $0<K_{1,q}<\infty$ for some $q\in\left(0,\,\infty\right)$,
and any given sequence $\{b_{1,T}\}$ such that $b_{2,T}/b_{1,T}\rightarrow0$,
$Tb_{1,T}^{2q+1}b_{2,T}\rightarrow\gamma\in\left(0,\,\infty\right)$,
the bandwidth defined by 
\[
b_{1,T}^{\mathrm{opt}}=\left(2qK_{1,q}^{2}\phi\left(q\right)T\overline{b}_{2,T}^{\mathrm{\mathrm{opt}}}/\left(\int K_{1}^{2}\left(y\right)dy\int_{0}^{1}K_{2}^{2}\left(x\right)dx\right)\right){}^{-1/\left(2q+1\right)},
\]
is optimal in the sense that, 
\begin{align*}
\liminf_{T\rightarrow\infty} & T^{8q/5\left(2q+1\right)}\left(\sup_{\mathscr{P}\in\boldsymbol{P}_{U}\left(\phi\right)}\mathbb{E}_{\mathscr{P}}\mathrm{L}\left(\widehat{J}_{T}^{\mathrm{}}\left(b_{1,T}\right),\,J_{\mathscr{P},T}\right)-\sup_{\mathscr{P}\in\boldsymbol{P}_{U}\left(\phi\right)}\mathbb{E}_{\mathscr{P}}\mathrm{L}\left(\widehat{J}_{T}^{\mathrm{}}\left(b_{1,T}^{\mathrm{opt}}\right),\,J_{\mathscr{P},T}\right)\right)\geq0,
\end{align*}
 provided $f_{U,a^{\left(r\right)}}>0$ and $f_{U,a^{\left(r\right)}}^{\left(q\right)}>0$
for some $r$ for which $w_{r}>0.$ The inequality is strict unless
$b_{1,T}=b_{1,T}^{\mathrm{opt}}+o(T^{-4/5\left(2q+1\right)})$. 
\end{thm}

\subsection{\label{Subsec: Data-dependent-DKHAC-Estimation}Data-dependent DK-HAC
Estimation}

We now show that the DK-HAC estimators based on data-dependent bandwidths
with similar form as $\widehat{b}_{1,T}^{*}$ and $\widehat{\overline{b}}_{2,T}^{*}$
(cf. Section \ref{Section: Statistical Enviromnent}) have the same
asymptotic MSE properties as the estimators based on optimal fixed
bandwidth sequences $b_{1,T}^{\mathrm{opt}}$ and $\overline{b}_{2,T}^{\mathrm{opt}}$
that depend on the unknown distribution $\mathscr{P}$. 

We consider the data-dependent bandwidths $\widehat{b}_{1,T}$ and
$\widehat{\overline{b}}_{2,T}$ from \citet{casini_hac} which are
defined as $\widehat{b}_{1,T}^{*}$ and $\widehat{\overline{b}}_{2,T}^{*}$,
repetitively, with $\widehat{V}_{t}$ in place of $\widehat{V}_{D,t}^{*}$.
We choose a parametric model for $\{a^{\left(r\right)\prime}V_{t}\}$,
$r=1,\,\ldots,\,p$, where $a^{\left(r\right)}$ is a $p$-vector
with the $r$-th element one and all other elements zero. We use the
same locally stationary AR(1) models as in Section \ref{Section Prewhitended DKHAC- Theory},
i.e., 
\[
V_{t}^{\left(r\right)}=a_{1}^{\left(r\right)}\left(t/T\right)V_{t-1}^{\left(r\right)}+u_{t}^{\left(r\right)},
\]
 with estimated parameters $\widehat{a}_{1}^{\left(r\right)}\left(\cdot\right)$
and $\widehat{\sigma}^{\left(r\right)}\left(\cdot\right).$ Let 
\[
\widehat{\theta}=\left(\int_{0}^{1}\widehat{a}_{1}^{\left(1\right)}\left(u\right)du,\,\int_{0}^{1}\left(\widehat{\sigma}^{\left(1\right)}\left(u\right)\right){}^{2}du,\ldots,\,\int_{0}^{1}\widehat{a}_{1}^{\left(p\right)}\left(u\right)du,\,\int_{0}^{1}\left(\widehat{\sigma}^{\left(p\right)}\left(u\right)\right){}^{2}du\right)',
\]
and $\theta_{\mathscr{P}}^{*}$ denote the probability limit of $\widehat{\theta}$.
We only consider distributions $\mathscr{P}$ for which $\theta_{\mathscr{P}}^{*}$
exists. We construct $\widehat{\phi}\left(q\right)=\widehat{\phi}_{D}\left(q\right)$
as in Section \ref{Section: Statistical Enviromnent} but using the
estimate $\widehat{\theta}$. The probability limit of $\widehat{\phi}\left(q\right)$
is denoted by $\phi_{\theta^{*}}\left(q\right)$. Let $\phi_{\mathscr{P}}\left(\cdot\right)$
be the value of $\phi\left(\cdot\right)$ from \eqref{Eq (5.2) Andrews 88 (phi(q)) Nonstationary}
obtained when $\mathscr{P}_{U}$ is given by the approximating  distribution
with parameter $\theta_{\mathscr{P}}^{*}$. For some $\underline{\phi}$,
$\overline{\phi}$ such that $0<\underline{\phi}\leq\overline{\phi}<\infty$,
define   
\begin{align*}
\boldsymbol{P}_{U,3} & \triangleq\biggl\{\mathscr{P}\in\boldsymbol{P}_{U}:\,(i)\,\widehat{\theta}\overset{\mathscr{P}}{\rightarrow}\theta_{\mathscr{P}}^{*}\,\mathrm{for\,some\,}\theta_{\mathscr{P}}^{*}\in\overline{\Theta}\,\mathrm{such}\,\mathrm{that\,}\phi_{\mathscr{P}}\left(q\right)\in[\underline{\phi},\,\overline{\phi}]\mathrm{\,for\,any\,}q,\\
 & \qquad(ii)\,\sup_{u\in\left[0,\,1\right]}|a'\Gamma_{\mathscr{P}_{U},u}\left(k\right)a|\leq C_{3}\left|k\right|^{-l}\,\mathrm{for}\,k=0,\,\pm1,\ldots,\,\mathrm{for\,some\,}C_{3}<\infty,\\
 & \qquad\,\mathrm{for\,some\,}l>\max\{2,\,\left(4q+2\right)/\left(2+q\right)\},\,\mathrm{for\,all\,}a\in\mathbb{R}^{p}\,\mathrm{with}\,||a||=1,\\
 & \qquad\,\mathrm{where}\,q\,\mathrm{is\,as\,in\,}\boldsymbol{K}_{3}\mathrm{\,and\,satisfying\,}8/q-20q<6,\,\mathrm{and}\,q<11/2,\\
 & \qquad(iii)\,\sup_{k\geq1}\mathrm{\mathrm{Var}}_{\mathscr{P}_{U}}(a'\widehat{\Gamma}\left(k\right)a)=O(1/Tb_{2,T}^{\mathrm{opt}}),\,\mathrm{and}\,\\
 & \qquad(iv)\,\limsup_{T\rightarrow\infty}\mathbb{E}_{\mathscr{P}}\left(\frac{1}{S_{\mathscr{P},T}}\sum_{k=1}^{S_{\mathscr{P},T}}\sqrt{Tb_{2,T}^{\mathrm{opt}}}\left|a'\widehat{\Gamma}\left(k\right)a-a'\Gamma_{\mathscr{P},T}\left(k\right)a\right|\right)^{4}\leq C_{4}\\
 & \qquad\,\mathrm{for}\,\mathrm{some\,}C_{4}<\infty\,\mathrm{with\,\mathit{S}_{\mathscr{P},\mathit{T}}=\bigl\lfloor(\mathit{b}_{1,\mathit{T}}^{\mathrm{opt}})^{-r}\bigr\rfloor\,some\,}r\in\boldsymbol{S}\left(q,\,b,\,l\right)\biggr\},
\end{align*}
where 
\begin{align*}
\boldsymbol{S}\left(q,\,b,\,l\right) & =(\max\{(b-3/4-q/2)/\left(b-1\right),\,q/\left(l-1\right),\\
 & \quad\min\left\{ \left(6+4q\right)/8,\,15/16+3q/8\right\} ),
\end{align*}
with $b>1+1/q$. The class of distributions $\boldsymbol{P}_{U,3}$
corresponds to the class $P_{1,1}$ used by \citet{andrews:88hac}.
The lower bound $0<\underline{\phi}\leq\phi_{\mathscr{P}}\left(q\right)$
in part (i) eliminates any distribution for which $\phi_{\mathscr{P}}\left(\cdot\right)=0$.
For example, white noise sequences do not belong to $\boldsymbol{P}_{U,3}$
since then $\phi\left(q\right)=0$.  We discuss these cases at the
end of the section. Part (ii) imposes a condition on the temporal
dependence of the distribution $\mathscr{P}_{U}$ and is similar to
Assumption \ref{Assumption E-F-G}-(iii). Part (iii) is satisfied
by a wide class of SLS processes as shown by \citet{casini_hac}.
Part (iv) was also used by \citet{andrews:88hac}, though the interval
$\boldsymbol{S}\left(q,\,b,\,l\right)$ is tighter as it takes into
account of the time smoothing. 

Let 
\begin{align*}
b_{1,\theta_{\mathscr{P}},T}=\left(2qK_{1,q}^{2}\phi_{\theta_{\mathscr{P}}^{*}}\left(q\right)T\overline{b}_{2,T}^{\mathrm{opt}}/\left(\int K_{1}^{2}\left(y\right)dy\int_{0}^{1}K_{2}^{2}\left(x\right)dx\right)\right){}^{-1/\left(2q+1\right)} & ,
\end{align*}
 denote the optimal bandwidth for the case in which $\mathscr{P}_{U}$
equals the approximating parametric model with parameter $\theta_{\mathscr{P}}^{*}$.
Let
\[
\widehat{D}_{2,a}\left(u\right)\triangleq2\sum_{l=-\left\lfloor T^{4/25}\right\rfloor }^{\left\lfloor T^{4/25}\right\rfloor }a'\widehat{c}_{T}\left(u_{0},\,l\right)\widehat{c}_{T}\left(u_{0},\,l\right)'a,
\]
 where $\widehat{c}_{T}$ is defined as $\widehat{c}_{D,T}^{*}$ with
$\widehat{V}_{t}$ and $\widehat{b}_{2,T}$ in place of $\widehat{V}_{D,t}^{*}$
and $\widehat{b}_{2,T}^{*}$, respectively, 
\begin{align*}
\widehat{c}_{T}\left(rn_{T}/T,\,k\right) & \triangleq\begin{cases}
\left(T\widehat{b}_{2,T}\right)^{-1}\sum_{s=k+1}^{T}K_{2}^{*}\left(\frac{\left(\left(r+1\right)n_{T}-\left(s-k/2\right)\right)/T}{\widehat{b}_{2,T}}\right)\widehat{V}_{s}\widehat{V}_{s-k}^{\prime}, & k\geq0\\
\left(T\widehat{b}_{2,T}\right)^{-1}\sum_{s=-k+1}^{T}K_{2}^{*}\left(\frac{\left(\left(r+1\right)n_{T}-\left(s+k/2\right)\right)/T}{\widehat{b}_{2,T}}\right)\widehat{V}_{s+k}\widehat{V}_{s}^{\prime}, & k<0
\end{cases}.
\end{align*}

\begin{assumption}
\label{Assumption C Andrews 88}(i) We have $\underset{\mathscr{P}\in\boldsymbol{P}_{U,3}}{\sup}\mathbb{E}_{\mathscr{P}}\left[\frac{\min\left\{ T/n_{3,T},\,\sqrt{n_{2,T}}\right\} \left(\widehat{\phi}\left(q\right)^{1/\left(2q+1\right)}-\phi_{\theta_{\mathscr{P}}^{*}}^{1/\left(2q+1\right)}\right)}{\widehat{\phi}\left(q\right)^{1/\left(2q+1\right)}}\right]^{4}=O\left(1\right)$
as $T\rightarrow\infty$, where $q$ is as defined in $\boldsymbol{K}_{3}$,
$\widehat{\phi}\left(q\right)\leq\overline{\phi}<\infty$, and $n_{2,T}/T+n_{3,T}/T\rightarrow0,$
$n_{2,T}^{10/6}/T\rightarrow[c_{2},\,\infty),$ $n_{3,T}^{10/6}/T\rightarrow[c_{3},\,\infty)$
with $0<c_{2},\,c_{3}<\infty$; (ii)  $\sqrt{Tb_{2,T}\left(u\right)}(\widehat{D}_{2,a}\left(u\right)-D_{2,U,a}\left(u\right))=O_{\mathscr{P}}\left(1\right)$
for all $u\in\left[0,\,1\right]$; (iii) Assumption \ref{Assumption E-F-G}-(v)
hold.
\end{assumption}
Any estimator $\widehat{\phi}$ based on kernel nonparametric estimators
of $\widehat{a}_{1}^{\left(r\right)}\left(\cdot\right)$ and $\widehat{\sigma}^{\left(r\right)}\left(\cdot\right)$
satisfies Assumption \ref{Assumption C Andrews 88}-(i). Assumption
\ref{Assumption C Andrews 88}-(ii) extends Assumption \ref{Assumption E-F-G}-(vi)
to the distribution $\mathscr{P}$ and is are useful to show that
the effect of using $\widehat{b}_{1,T}$ and $\widehat{\overline{b}}_{2,T}$
rather than $b_{1,\theta_{\mathscr{P}},T}$ and $\overline{b}_{2,T}^{\mathrm{opt}}$
when constructing $\widehat{J}_{T}$ is at most $o_{\mathbb{P}}\left(1\right)$.
The following result shows that $\widehat{J}_{T}(\widehat{b}_{1,T},\,\widehat{\overline{b}}_{2,T})$
has the same asymptotic MSE properties under $\mathscr{P}$ as the
estimator $\widehat{J}_{T}(b_{1,\theta_{\mathscr{P}},T},\,\overline{b}_{2,T}^{\mathrm{opt}})$.
Since the asymptotic MSE properties of the estimators with fixed bandwidth
parameters have been determined in Section \ref{subsec: Optimal-Bandwidths-and Kernels - Nonstationarity},
from this result follows the consistency of $\widehat{J}_{T}(\widehat{b}_{1,T},\,\widehat{\overline{b}}_{2,T})$
and its asymptotic optimality properties. 
\begin{thm}
\label{Theorem 6 Andrews 88}Consider any kernel $K_{1}\left(\cdot\right)\in\boldsymbol{K}_{3},$
$q$ as in $\boldsymbol{K}_{3}$ and any $K_{2}\left(\cdot\right)\in\boldsymbol{K}_{2}$.
Suppose Assumption \ref{Assumption A*}-\ref{Assumption C Andrews 88}
hold. Then, for all $a\in\mathbb{R}^{p}$, 
\begin{align*}
T^{8q/5\left(2q+1\right)}\sup_{\mathscr{P}\in\boldsymbol{P}_{U,3}}\left|\mathrm{MSE}_{\mathscr{P}}(a'\widehat{J}_{T}(\widehat{b}_{1,T},\,\widehat{\overline{b}}_{2,T})a)-\mathrm{MSE}_{\mathscr{P}}(a'\widehat{J}_{T}(b_{1,\theta_{\mathscr{P}},T},\,\overline{b}_{2,T}^{\mathrm{opt}})a)\right| & \rightarrow0.
\end{align*}
\end{thm}
Theorem \ref{Theorem 6 Andrews 88} combined with Theorem \ref{Theorem MSE J DKHAC Nonstationary}
and Theorem \ref{Theorem 1 Andrews 91 - Consistency and Rate - Nonstationary}-(iii)
establish upper and lower bounds on the asymptotic MSE.  Results
on asymptotic minimax optimality for data-dependent bandwidths parameters
can be obtained using Theorem \ref{Theorem MSE J DKHAC Nonstationary},
Theorem \ref{Theorem 1 Andrews 91 - Consistency and Rate - Nonstationary}-(iii)
and Theorem \ref{Theorem 3 Andrews 88}-\ref{Theorem 6 Andrews 88}. 

It remains to consider the case $\phi_{\mathscr{P}}\left(\cdot\right)=0$.
When this occurs, $\widehat{\phi}^{-1}\left(\cdot\right)$ is $O_{\mathscr{P}}((T/n_{3,T})^{2}+n_{2,T})$.
Under the additional condition $((T/n_{3,T})^{2}+n_{2,T})/T^{4/5}\rightarrow c\in[0,\,\infty)$
in Assumption \ref{Assumption C Andrews 88}-(i) we have $\widehat{b}_{1,T}=O_{\mathscr{P}}\left(1\right)$.
Thus, $\widehat{J}_{T}(\widehat{b}_{1,T},\,\widehat{\overline{b}}_{2,T})-J_{\mathscr{P},T}\overset{\mathscr{P}}{\rightarrow}0$
also when the series is white noise. This is important in applied
work because often researchers use robust standard errors even when
they are not aware of whether any dependence is present at all. 

\section{\label{Section Power}Theoretical Results About the Power of HAR
Tests Under General $\mathbb{E}\left(V_{t}\right)$}

A long-lasting problem in time series econometrics is the low/non-monotonic
power of HAR inference tests under nonstationary alternative hypotheses.
 The problem involves HAR tests outside the regression model that
can be characterized by an alternative hypothesis involving $\mathbb{E}\left(V_{t}\right)=\mu_{t}$
with $\mu_{t}\neq0$ for at least one $t$. The process $\mu_{t}$
can be any piecewise continuous function of $t$. For example, tests
for structural breaks, tests for regime switching and tests for time-varying
parameters can be framed in this way. To see this, consider a linear
regression model,
\begin{align}
y_{t} & =x'_{t}\beta_{t}+e_{t},\qquad t=1,\ldots,\,T.\label{Eq. (Linear Model SC)}
\end{align}
The null hypothesis of no break in the regression coefficient of $x_{t}$
is written as $H_{\beta,0}:\,\beta_{t}=\beta_{0}$ for all $t$ for
some $\beta_{0}\in\mathbb{R}^{p}$ {[}see, e.g., \citet{andrews:93}{]}.
The alternative hypothesis may be of several forms. Let $H_{\beta,1}:\,\beta_{t}=\beta\left(t/T\right)$
$\left(t=1,\ldots,\,T\right)$ for some piecewise continuous function
$\beta\left(\cdot\right)$. Estimating \eqref{Eq. (Linear Model SC)}
by least-squares yields $y_{t}=x'_{t}\widehat{\beta}+\widehat{e}_{t}$
for all $t$ where $\widehat{\beta}$ is the  least-squares estimate
and $\left\{ \widehat{e}_{t}\right\} $ are the least-squares residuals.
Letting $V_{t}=x_{t}\widehat{e}_{t}$, the null hypothesis $H_{\beta,0}$
can be rewritten as $H_{0}:\,\mathbb{E}\left(V_{t}\right)=0$ for
all $t$ while the alternative hypothesis $H_{\beta,1}$ can be rewritten
as $H_{1}:\,\mathbb{E}\left(V_{t}\right)=\mu_{t}$ where $\mu_{t}\neq0$
for at least one $t.$ Structural break tests are based on an estimate
of the LRV of $V_{t}=x_{t}\widehat{e}_{t}$. While under $H_{0}$
$V_{t}$ is zero-mean, under $H_{1}$ the mean of $V_{t}$ is time-varying.
Under the alternative hypothesis, it is sufficient for consistency
of the test that the LRV estimator converges to some positive semidefinite
matrix since the numerator of the test statistics diverges to infinity.
However, time-variation in the mean of $V_{t}$ severely biases upward
traditional LRV estimators which then lead to tests with non-monotonic
power. \citet{casini/perron_Low_Frequency_Contam_Nonstat:2020} established
analytical results for this phenomenon, which they referred to as
low frequency contamination. We show that the proposed nonlinear prewhitened
DK-HAC estimator accounts for nonstationarity also under the alternative
hypothesis and leads to consistent tests with good monotonic power.
 Although the main theoretical result of this section is presented
for a particular HAR test and a particular form of $H_{1}$, this
result is general enough to provide guidance for most cases discussed
in the literature.

We present theoretical results about the power of a popular forecast
evaluation test, namely the Diebold-Mariano test {[}cf. \citet{diebold/mariano:95}{]},
which can be also framed as above. We focus on the Diebold-Mariano
test for ease of the exposition. Similar results hold for the other
HAR inference tests that can be framed as above, though the proofs
change slightly depending on the specific test statistic. Suppose
the goal is to forecast some variable $y_{t}$. Two forecast models
are used: $y_{t}=\beta^{\left(1\right)}+\beta^{\left(2\right)}x_{t-1}^{(i)}+e_{t}$
where $x_{t-1}^{(i)}$ is some predictor and $i=1,\,2$. That is,
each forecast model uses an intercept and a predictor. The parameters
$\beta^{\left(1\right)}$ and $\beta^{\left(2\right)}$ are estimated
using least-squares in the in-sample $t=1,\ldots,\,T_{m}$ with a
fixed forecasting scheme. Each forecast model generates a sequence
of $\tau\left(=1\right)$-step ahead out-of-sample losses $L_{t}^{(i)}$
$\left(i=1,\,2\right)$ for $t=T_{m}+1,\ldots,\,T-\tau.$ Then $d_{t}\triangleq L_{t}^{(2)}-L_{t}^{(1)}$
denotes the loss differential at time $t$. Let $\overline{d}_{L}$
denote the average of the loss differentials. The Diebold-Mariano
test statistic is defined as $t_{\mathrm{DM}}\triangleq T_{n}^{1/2}\overline{d}_{L}/\widehat{J}_{d_{L},T}^{1/2}$,
where $\widehat{J}_{d_{L},T}$ is an estimate of the LRV of the loss
differentials and $T_{n}$ is the number of observations in the out-of-sample.
Throughout, we use the quadratic loss. The true model is $y_{t}=\beta_{0}^{\left(1\right)}+\beta_{0}^{\left(2\right)}x_{t-1}^{(0)}+e_{t}$
where $x_{t-1}^{(0)}$ is a predictor and $e_{t}$ is a zero-mean
error. We assume that the conditions for consistency and asymptotic
normality of the least-squares estimates of $\beta_{0}^{\left(1\right)}$
and $\beta_{0}^{\left(2\right)}$ are satisfied. 

In this setting, $\widehat{V}_{t}=d_{t}.$ The hypothesis testing
problem is given by 
\begin{align}
H_{0} & :\,\mathbb{E}\left(\widehat{V}_{t}\right)=0,\quad\mathrm{for\,all\,}t,\label{Eq. Hypothesis Testing Problem DM Test}\\
H_{1} & :\,\mathbb{E}\left(\widehat{V}_{t}\right)=\mu_{t},\quad\mathrm{with}\,\mu_{t}\neq0\,\mathrm{for\,at\,least\,one\,}t.\nonumber 
\end{align}
 $H_{0}$ corresponds to equal predictive ability between the two
forecast models while $H_{1}$ corresponds to the two forecast models
performing differently. 

Since we want to study the power of $t_{\mathrm{DM}}$, we need to
work under the alternative hypothesis. The two competing forecast
models are as follows: the first model uses the actual true predictor
(i.e., $x_{t-1}^{(1)}=x_{t-1}^{(0)}$ for all $t$) while the second
model differs in that in place of $x_{t-1}^{(0)}$ it uses $x_{t-1}^{(2)}=x_{t-1}^{(0)}+u_{X_{2},t}$
for $t\leq T_{b}$ and $x_{t-1}^{(2)}=\delta+x_{t-1}^{(0)}+u_{X_{2},t}$
for $t>T_{b}$ with $T_{b}>T_{m}$, and $u_{X_{2},t}$ is a zero-mean
error term. Evidently, the null hypotheses of equal predictive ability
should be rejected whenever $\delta>0$. We consider $t_{\mathrm{DM}}$
normalized by different LRV estimators. The HAC estimator is defined
as
\begin{align*}
\widehat{J}_{d_{L},\mathrm{HAC,}T}\triangleq\sum_{k=-T+1}^{T-1}K_{1}\left(b_{T}k\right)\widehat{\Gamma}\left(k\right), & \qquad\widehat{\Gamma}\left(k\right)=T^{-1}\sum_{t=|k|+1}^{T}\widehat{V}_{t}\widehat{V}_{t-|k|},
\end{align*}
 where $K_{1}\left(\cdot\right)$ is a kernel (e.g., the Bartlett
and QS) and $b_{T}$ a bandwidth. \citet{Kiefer/vogelsang/bunzel:00}
proposed to use a LRV estimator that keeps $b_{T}$ at a fixed fraction
of $T$, i.e., $\widehat{J}_{\mathrm{\mathrm{KVB},}T}\triangleq T^{-1}\sum_{t=1}^{T}\sum_{s=1}^{T}$
$\left(1-\left|t-s\right|/T\right)\widehat{V}_{t}\widehat{V}_{s}$
which is equivalent to the Newey-West estimator with $b_{T}=T^{-1}$.

We present theoretical results about the power of $t_{\mathrm{DM}}$.
Let $t_{\mathrm{DM},i}=T_{n}^{1/2}\overline{d}_{L}/\sqrt{\widehat{J}_{d_{L},i,T}}$
denote the DM test statistic where $i=\mathrm{DK},\,\mathrm{pwDK},\,\mathrm{KVB},\,\mathrm{EWC},$
$\mathrm{A91},\,\mathrm{pwA}91$, $\mathrm{NW87}$ and $\mathrm{pwNW87}$.
$\widehat{J}_{d_{L},\mathrm{A91},T}$ and $\widehat{J}_{d_{L},\mathrm{NW87},T}$
are $\widehat{J}_{d_{L},\mathrm{HAC,}T}$ where $K_{1}\left(\cdot\right)$
is the Bartlett and QS kernel, respectively.\footnote{Since $\{\widehat{V}_{t}\}$ is only observed in the out-of-sample,
the LRV estimators use a sample of $T_{n}$ observations.} $\widehat{J}_{d_{L},\mathrm{pwA91},T}$ and $\widehat{J}_{d_{L},\mathrm{pwNW87},T}$
are the prewhitened HAC estimators using the QS and Bartlett kernel,
respectively, and the prewhitening procedure of \citet{andrews/monahan:92}.
``DK'' refers to the DK-HAC estimator from \citet{casini_hac} with
the MSE-optimal kernels and bandwidths whereas ``pwDK'' refers to
the prewhitened DK-HAC estimator $\widehat{J}_{\mathrm{pw},T}$ in
\eqref{Eq. (J_hat_PW)}. Define the power of $t_{\mathrm{DM},i}$
as $\mathbb{P}_{\delta}(|t_{\mathrm{DM},i}|>z_{\alpha})$ where $z_{\alpha}$
is the two-sided standard normal critical value and $\alpha\in\left(0,\,1\right)$
is the significance level. To avoid repetitions we present the results
only for $i=\mathrm{DK},\,\mathrm{pwDK},\,\mathrm{KVB},\,\mathrm{NW87}$
and $\mathrm{pwNW87}$. The results concerning the EWC estimator are
the same as those for the KVB's fixed-$b$ estimator. The results
pertaining to \citeauthor{andrews:91}' \citeyearpar{andrews:91}
HAC estimator (with and without prewhitening) are the same as those
corresponding to \citeauthor{newey/west:87}'s \citeyearpar{newey/west:87}
estimator (with and without prewhitening, respectively). For the HAC
and DK-HAC estimators we report the results for the MSE-optimal bandwidth
{[}see \citet{andrews:91}, \citet{casini_hac} and \citet{whilelm:2015}{]}.\footnote{For the HAC estimators we also report the result for any bandwidth
choice $b_{T}\rightarrow0$ such that $Tb_{T}\rightarrow\infty$,
which is sufficient for the consistency of the estimator. } We set $n_{T}=n_{2,T}=n_{3,T}=T^{2/3}$ which satisfy the growth
rate bounds {[}see \citet{casini_hac} for details{]}. Let $n_{\delta}=T-T_{b}-2$
denote the length of the regime in which $x_{t}^{(2)}$ exhibits a
shift $\delta$ in the mean. The alternative hypothesis depends on
the shift magnitude $\delta$ and on how long the shift lasts for.
Here the latter is $n_{\delta}$. More generally, this is the set
of time points such that $\mathbb{E}(\widehat{V}_{t})=\mu_{t}\neq0$
holds. 
\begin{thm}
\label{Theorem Power DM HAR Tests}Let $\left\{ d_{t}-\mathbb{E}(d_{t})\right\} $
be a SLS process satisfying Assumption \ref{Assumption Smothness f(u,w)}
and \ref{Assumption A - Dependence}, and $n_{\delta}=O(T_{n}^{1/2+\zeta})$
where $\zeta\in\left(0,\,1/2\right)$ such that $T_{n}^{\zeta}b_{T}^{1/2}\rightarrow0$
and $T_{n}^{\zeta}(\widehat{b}_{1,T}^{*})^{1/2}\rightarrow0$. Then,
we have:

(i) If $b_{T}\rightarrow0$, then $\mathbb{P}_{\delta}(|t_{\mathrm{DM},\mathrm{NW87}}|>z_{\alpha})\rightarrow0$$.$
If $b_{T}=O(T^{-1/3})$, then $|t_{\mathrm{DM},\mathrm{NW87}}|=O_{\mathbb{P}}(T_{n}^{\zeta-1/6})$
and $\mathbb{P}_{\delta}(|t_{\mathrm{DM},\mathrm{NW87}}|>z_{\alpha})$$\rightarrow0.$ 

(ii) If $b_{T}\rightarrow0$, then $\mathbb{P}_{\delta}(|t_{\mathrm{DM},\mathrm{pwNW87}}|>z_{\alpha})\rightarrow0$$.$
If $b_{T}=O(T^{-1/3})$, then $|t_{\mathrm{DM},\mathrm{pwNW87}}|=O_{\mathbb{P}}(T_{n}^{\zeta-1/6})$
and $\mathbb{P}_{\delta}(|t_{\mathrm{DM},\mathrm{pwNW87}}|>z_{\alpha})$$\rightarrow0.$ 

(iii) If $b_{T}=T^{-1}$, then $|t_{\mathrm{DM},\mathrm{KVB}}|=O_{\mathbb{P}}(T_{n}^{\zeta-1/2})$
and $\mathbb{P}_{\delta}(|t_{\mathrm{DM},\mathrm{KVB}}|>z_{\alpha})$$\rightarrow0.$

(iv) Under Assumption  \ref{Assumption B}-(i-iii), $|t_{\mathrm{DM},\mathrm{DK}}|=\delta^{2}O_{\mathbb{P}}(T_{n}^{\zeta})$
and $\mathbb{P}_{\delta}(|t_{\mathrm{DM},\mathrm{DK}}|>z_{\alpha})$$\rightarrow1$.

(v) Under Assumption \ref{Assumption B}-(i-iii), \ref{Assumption E-F-G}-(i,iv)
and \ref{Assumption H Andrews 1990}, $|t_{\mathrm{DM},\mathrm{pwDK}}|=\delta^{2}O_{\mathbb{P}}(T_{n}^{\zeta})$
and $\mathbb{P}_{\delta}(|t_{\mathrm{DM},\mathrm{pwDK}}|>z_{\alpha})$$\rightarrow1$.
 
\end{thm}
 Note that $b_{T}=O(T^{-1/3})$ in parts (i)-(ii) refers to the MSE-optimal
bandwidth for the \citeauthor{newey/west:87}'s \citeyearpar{newey/west:87}
estimator. The conditions $T_{n}^{\zeta}b_{T}^{1/2}\rightarrow0$
and $T_{n}^{\zeta}(\widehat{b}_{1,T}^{*})^{1/2}\rightarrow0$ mean
that the length of the regime in which $x_{t}^{(2)}$ exhibits a shift
$\delta$ in the mean increases to infinity at a slower rate than
$T$. Theorem \ref{Theorem Power DM HAR Tests} implies that when
the (prewhitened or non-prewhitened) HAC estimators or the fixed-$b$
LRV estimators are used, the DM test is not consistent and its power
converges to zero. The theorem suggests that prewhitened and non-prewhitened
HAC estimators suffer from this problem in a similar way. The theorem
also implies that the power functions corresponding to tests based
on HAC estimators lie above the power functions corresponding to those
based on fixed-$b$/EWC LRV estimators. An additional feature is
that $|t_{\mathrm{DM},\mathrm{NW87}}|$, $|t_{\mathrm{DM},\mathrm{pwNW87}}|$
and $|t_{\mathrm{DM},\mathrm{KVB}}|$ do not increase in magnitude
with $\delta$ because $\delta$ appears in both the numerator and
denominator. The results concerning the DK-HAC estimator and the
prewhitened DK-HAC estimator $\widehat{J}_{\mathrm{pw},T}$ show that
these issues do not occur when these estimators are used. In fact,
the test is consistent and its power increases with $\delta$ and
with the sample size. We provide finite-sample evidence in support
of these theoretical results in Section \ref{Section Monte Carlo}.

\section{\label{Section Monte Carlo}Small-Sample Evaluations}

We now show that the prewhitened DK-HAC estimators lead to HAR inference
tests that have accurate null rejection rates when there is strong
dependence and have superior power properties relative to those based
on traditional LRV estimators.  We consider HAR tests in the linear
regression model as well as applied to the forecast evaluation literature,
namely the Diebold-Mariano test and the forecast breakdown test of
\citet{giacomini/rossi:09}. 

The linear regression models have an intercept and a stochastic regressor.
We focus on the $t$-statistics $t_{r}=\sqrt{T}(\widehat{\beta}^{\left(r\right)}-\beta_{0}^{\left(r\right)})/\sqrt{\widehat{J}_{X,T}^{\left(r,r\right)}}$
where $\widehat{J}_{X,T}$ is a consistent estimator of the limit
of $\mathrm{Var}(\sqrt{T}(\widehat{\beta}-\beta_{0}))$ and $r=1,\,2$.
$t_{1}$ is the $t$-statistic for the parameter associated to the
intercept while $t_{2}$ is associated to the stochastic regressor.
Two regression models are considered. We run a $t$-test on the intercept
in model M1 whereas a $t$-test on the coefficient of the stochastic
regressor is run in model M2. The models are,
\begin{align}
y_{t} & =\beta_{0}^{\left(1\right)}+\delta+\beta_{0}^{\left(2\right)}x_{t}+e_{t},\qquad\qquad t=1,\ldots,\,T,\label{eq: Model P1}
\end{align}
for the $t$-test on the intercept and
\begin{align}
y_{t} & =\beta_{0}^{\left(1\right)}+(\beta_{0}^{\left(2\right)}+\delta)x_{t}+e_{t},\qquad\qquad t=1,\ldots,\,T,\label{eq Model P1 beta2}
\end{align}
for the $t$-test on $\beta_{0}^{\left(2\right)}$ where $\delta=0$
under the null hypotheses. In model M1 we set $\beta_{0}^{\left(1\right)}=0$,
$\beta_{0}^{\left(2\right)}=1$, $x_{t}\sim\mathrm{i.i.d}.\,\mathscr{N}\left(1,\,1\right)$
and $e_{t}=\rho e_{t-1}+u_{t},\,\rho=0.4,\,0.9,\,u_{t}\sim\mathrm{i.i.d.}\,\mathscr{N}\left(0,\,0.7\right)$.
Model M2 involves segmented locally stationary errors: $\beta_{0}^{\left(1\right)}=\beta_{0}^{\left(2\right)}=0$,
$x_{t}=0.6+0.8x_{t-1}+u_{x,t},\,u_{x,t}\sim\mathrm{i.i.d.\,}\mathscr{N}\left(0,\,1\right)$
and $e_{t}=\rho_{t}e_{t-1}+u_{t},\,\rho_{t}=\max\left\{ 0,\,0.8\left(\cos\left(1.5-\cos\left(5t/T\right)\right)\right)\right\} $
for $t<4T/5$ and $e_{t}=0.5e_{t-1}+u_{t},\,u_{t}\sim\mathrm{\mathrm{i.i.d.\,}}\mathscr{N}\left(0,\,1\right)$
for $t\geq4T/5$. Note that $\rho_{t}$ varies smoothly between 0
and 0.7021. Then, $\widehat{J}_{X,T}=(X'X/T)^{-1}\widehat{J}_{T}(X'X/T)^{-1}$
where $X=\left[X_{1},\ldots,\,X_{T}\right]'$ and $X_{t}=[1,\,x_{t}]'$.

Next, we move to the forecast evaluation tests. The Diebold-Mariano
test statistic is defined as in Section \ref{Section Power}, $t_{\mathrm{DM}}\triangleq T_{n}^{1/2}\overline{d}_{L}/\widehat{J}_{d_{L},T}^{1/2}$.
 In model M3 we consider an out-of-sample forecasting exercise with
a fixed  scheme where, given a sample of $T$ observations, $0.5T$
observations are used for the in-sample and the remaining half is
used for prediction. To evaluate the empirical size, we specify the
following data-generating process and the two forecasting models that
have equal predictive ability. The true model for $y_{t}$ is given
by $y_{t}=\beta_{0}^{\left(1\right)}+\beta_{0}^{\left(2\right)}x_{t-1}^{(0)}+e_{t}$
where $x_{t-1}^{(0)}\sim\mathrm{i.i.d.}\,\mathscr{N}\left(1,\,1\right)$,
$e_{t}=0.8e_{t-1}+u_{t}$ with $u_{t}\sim\mathrm{i.i.d.\,}\mathscr{N}\left(0,\,1\right)$
and we set $\beta_{0}^{\left(1\right)}=0,\,\beta_{0}^{\left(2\right)}=1.$
The two competing models differ on the predictor used in place of
$x_{t}^{(0)}$. The first forecast model uses $x_{t}^{(1)}$ while
the second uses $x_{t}^{(2)}$ where $x_{t}^{(1)}$ and $x_{t}^{(2)}$
are $\mathrm{i.i.d.}\,\mathscr{N}\left(1,\,1\right)$ sequences, both
independent from $x_{t}^{(0)}$. Each forecast model generates a sequence
of $\tau\left(=1\right)$-step ahead out-of-sample losses $L_{t}^{(i)}$
$\left(i=1,\,2\right)$ for $t=T/2+1,\ldots,\,T-\tau.$ Then $d_{t}\triangleq L_{t}^{(2)}-L_{t}^{(1)}$
denotes the loss differential at time $t$. The  test rejects the
null of equal predictive ability when (after normalization) $\overline{d}_{L}$
is sufficiently far from zero. 

Next, we specify the alternative hypotheses for the Diebold-Mariano
test. The two competing forecast models are as follows: the first
model uses the actual true data-generating process while the second
model differs in that in place of $x_{t-1}^{(0)}$ it uses $x_{t-1}^{(2)}=x_{t-1}^{(0)}+u_{X_{2},t}$
for $t\leq3T/4$ and $x_{t-1}^{(2)}=\delta+x_{t-1}^{(0)}+u_{X_{2},t}$
for $t>3T/4$, with $u_{X_{2},t}\sim\mathrm{i.i.d.\,}\mathscr{N}\left(0,\,1\right)$.
The null hypotheses of equal predictive ability should be rejected
whenever $\delta>0$.

Finally, we consider model M4 which we use for investigating the performance
the $t$-test for forecast breakdown of \citet{giacomini/rossi:09}.
Suppose we want to forecast a variable $y_{t}$ which follows $y_{t}=\beta_{0}^{\left(1\right)}+\beta_{0}^{\left(2\right)}x_{t-1}+\delta x_{t-1}\mathbf{1}\{t>T_{1}^{0}\}+e_{t}$
where $x_{t}\sim\mathrm{i.i.d.\,}\mathscr{N}\left(1.5,\,1.5\right)$
and $e_{t}=0.3e_{t-1}+u_{t}$ with $u_{t}\sim\mathrm{i.i.d.\,}\mathscr{N}\left(0,\,0.7\right)$,
$\beta_{0}^{\left(1\right)}=\beta_{0}^{\left(2\right)}=1$ and $T_{1}^{0}=T\lambda_{1}^{0}$
with $\lambda_{1}^{0}=0.85$. The test detects a forecast breakdown
when the average of the out-of-sample losses differs significantly
from the average of the in-sample losses. The in-sample is used to
obtain estimates of $\beta_{0}^{\left(1\right)}$ and $\beta_{0}^{\left(2\right)}$
which are in turn used to construct out-of-sample forecasts $\widehat{y}_{t}=\widehat{\beta}_{0}^{\left(1\right)}+\widehat{\beta}_{0}^{\left(2\right)}x_{t-1}$.
The test is defined as $t^{\mathrm{GR}}\triangleq T_{n}^{1/2}\overline{SL}/\widehat{J}_{SL}^{1/2}$
where $\overline{SL}\triangleq T_{n}^{-1}\sum_{t=T_{m}+1}^{T-\tau}SL_{t+\tau}$,
$SL_{t+\tau}$ is the surprise loss at time $t+\tau$, i.e., the difference
between the time $t+\tau$ out-of-sample loss and in-sample-average
loss, $SL_{t+\tau}=L_{t+\tau}-\overline{L}_{t+\tau}$. Here $T_{n}$
is the sample size in the out-of-sample, $T_{m}$ is the sample size
in the in-sample and $\widehat{J}_{SL}$ is a LRV estimator. We consider
a fixed forecasting scheme and $\tau=1.$ 

We consider the following DK-HAC estimators: $\widehat{J}_{T,\mathrm{pw},\mathrm{SLS}}=\widehat{J}_{T,\mathrm{pw}}$
as discussed in Section \ref{Section: Statistical Enviromnent}, $\widehat{J}_{T,\mathrm{pw},1}$
which uses prewhitening with a single block {[}$n_{T}=T$ in \eqref{Eq. (2.3) Andrews 90}{]}
(i.e., stationary prewhitening), $\widehat{J}_{T,\mathrm{pw},\mathrm{SLS,}\mu}$
which uses prewhitening involving a VAR(1) with time-varying intercept
{[}i.e., with $\widehat{\mu}_{t}$ in \eqref{Eq. (2.3) Andrews 90}{]}.
The asymptotic properties of $\widehat{J}_{T,\mathrm{pw},\mathrm{SLS,}\mu}$
are the same as those of $\widehat{J}_{T,\mathrm{pw,\mathrm{SLS}}}$
since $\widehat{\mu}_{t}$ plays no role in the theory given the zero-mean
assumption on $\left\{ V_{t}\right\} $. However, it leads to power
enhancement under nonstationary alternative hypotheses. The asymptotic
properties of $\widehat{J}_{T,\mathrm{pw},1}$ follow as a special
case from the properties of $\widehat{J}_{T,\mathrm{pw,\mathrm{SLS}}}$.
We set $n_{T}=n_{2,T}=n_{3,T}=T^{2/3}.$ For the test of \citet{giacomini/rossi:09}
we do not report the results for $\widehat{J}_{T,\mathrm{pw},1}$
because the stationarity assumption is clearly violated under the
alternative. We compare tests using these estimators to those using
the following estimates: \citeauthor{andrews:91}' \citeyearpar{andrews:91}
HAC estimator with automatic bandwidth; \citeauthor{andrews:91}'
\citeyearpar{andrews:91} HAC estimator with automatic bandwidth and
the prewhitening procedure of \citet{andrews/monahan:92}; \citeauthor{newey/west:87}'s
\citeyearpar{newey/west:87} HAC estimator with the automatic bandwidth
as proposed in \citet{newey/west:94}; \citeauthor{newey/west:87}'s
\citeyearpar{newey/west:87} HAC estimator with the automatic bandwidth
as proposed in \citet{newey/west:94} and the prewhitening procedure;
Newey-West with the fixed-$b$ method of \citet{Kiefer/vogelsang/bunzel:00};
the Empirical Weighted Cosine (EWC) of \citet{lazarus/lewis/stock/watson:18}.
We consider the following sample sizes: $T=200,\,400$ for M1-M2 and
$T=400,\,800$ for model M3-M4. We set $T_{m}=200,\,400$ for M3 and
$T_{m}=240,\,480$ for M4. The nominal size is $\alpha=0.05$ throughout.

Table \ref{Table Size M1-M2}-\ref{Table Size M3-M4} report the rejection
rates under the null hypothesis for model M1-M4. We begin with model
M1 with medium dependence ($\rho=0.4$). The prewhitened DK-HAC estimators
lead to tests with accurate rejection rates that are slightly better
than those obtained with Newey-West with fixed-$b$ and to EWC. In
contrast, the classical HAC estimators of \citet{andrews:91} and
\citet{newey/west:87} are less accurate with rejection rates higher
than the nominal level. The prewhitening of \citet{andrews/monahan:92}
helps to reduce the size distortions but they still persist for the
Newey-West estimator even for $T=400.$ For higher dependence (i.e.,
$\rho=0.9$), using EWC and $\widehat{J}_{T,\mathrm{pw},\mathrm{SLS,}\mu}$
yield oversized tests, though by a small margin. The best size control
is achieved using the Newey-West with fixed-$b$ (KVB), $\widehat{J}_{T,\mathrm{pw},1}$
and $\widehat{J}_{T,\mathrm{pw,\mathrm{SLS}}}$.

For model M2, Newey-West with fixed-$b$ and the prewhitened DK-HAC
$(\widehat{J}_{T,\mathrm{pw},1},\,\widehat{J}_{T,\mathrm{pw},\mathrm{SLS}},$
$\widehat{J}_{T,\mathrm{pw},\mu})$ allow accurate rejection rates.
In some cases, tests based on the prewhitened DK-HAC are superior
to those based on fixed-$b$ (KVB). The tests with EWC are slightly
oversized when $T=200$ but close to the nominal level when $T=400$.
The classical HAC of \citet{andrews:91} and \citet{newey/west:87},
either prewhitened or not, imply oversized tests with $T=200.$ 

Turning to the HAR tests for forecast evaluation, Table \ref{Table Size M3-M4}
reports some striking results. First, tests based on the Newey-West
with fixed-$b$ (KVB) have size essentially equal to zero, while those
based on the EWC and prewhitened or non-prewhitened classical HAC
estimators are oversized. The prewhitened DK-HAC allows more accurate
tests. For model M4, many of the tests have size equal to or close
to zero. This occurs using the classical HAC, either prewhitened or
not and EWC. The prewhitened DK-HAC estimators and Newey-West with
fixed-$b$ (KVB) allow controlling the size reasonably well. Overall,
Table \ref{Table Size M1-M2}-\ref{Table Size M3-M4} in part confirm
previous evidence and in part suggest new facts. Newey-West with
fixed-$b$ (KVB) leads to better size control than using the classical
HAC estimators of \citet{andrews:91} and \citet{newey/west:87} even
when the latter are used in conjunction with the prewhitening device
of \citet{andrews/monahan:92}. The new result is that several of
the LRV estimators proposed in the literature can lead to tests having
null rejection rates equal to or close to zero. This occurs because
the null hypotheses involves nonstationary data generating mechanisms.
These LRV estimators are inflated and the associated test statistics
are undersized. This is expected to have negative consequences for
the power of the tests, as we will see below. The estimators proposed
in this paper perform well in leading to tests that control the null
rejection rates for all cases. They are in general competitive with
using the Newey-West with fixed-$b$ (KVB) when the latter does not
fail and in some cases can also outperform it. 

Table \ref{Table Power M1-M2}-\ref{Table Power M3-M4} report the
empirical power of the tests for model M1-M4.  For model M1 with
$\rho=0.9$ and M2 we see that all tests have good and monotonic power.
It is fair to compare tests based on the DK-HAC estimators relative
to using Newey-West with fixed-$b$ (KVB) since they have similar
well-controlled null rejection rates. Tests based on the Newey-West
with fixed-$b$ (KVB) sacrifice power more than using the DK-HAC estimators
and the difference is substantial. The classical HAC estimators have
higher power but it is unfair to compare them since they are often
oversized. A similar argument applies to using the EWC. 

We now move to the forecast evaluation tests. For both models M3 and
M4 we observe several features of interests. Essentially all tests
proposed previously experience severe power issues. The power is either
non-monotonic, very low or equal zero. This holds when using the classical
HAC estimators of \citet{andrews:91} as well as \citet{newey/west:87}
irrespective of whether prewhitening is used, with the EWC and the
Newey-West with fixed-$b$ (KVB). The only exceptions are tests based
on the \citeauthor{newey/west:87}'s \citeyearpar{newey/west:87}
and \citeauthor{andrews:91}' \citeyearpar{andrews:91} HAC estimator
with prewhitening in model M4 that display some power but much lower
compared to using the prewhitened DK-HAC estimators. The latter have
excellent power. The reason for the severe power problems for the
previous LRV-based tests is that  models M3 and M4 involve nonstationary
alternative hypotheses. The sample autocovariances become inflated
and overestimate the true autocovariances. The theoretical results
about the power in Theorem \ref{Theorem Power DM HAR Tests} suggest
that this issue becomes more severe as $\delta$ increases, which
explains the non-monotonic power for some of the tests, with tests
based on fixed-$b$ methods that include many lags suffering most.
The double smoothing in the DK-HAC estimators allows to avoid this
problem because it flexibly accounts for nonstationarity. The key
idea is not to mix observations belonging to different regimes. Simulation
results for additional data-generating processes involving ARMA, ARCH
and heteroskedastic errors are not discussed here because the results
are qualitatively equivalent.

\section{\label{Section Conclusions}Conclusions}

We introduce a nonparametric nonlinear VAR prewhitened long-run variance
(LRV) estimator for the construction of standard errors robust to
autocorrelation and heteroskedasticity that can be used for hypothesis
testing both within and outside the linear regression model. HAR tests
normalized by the proposed estimator exhibit accurate null rejection
rates even when there is strong dependence. We show theoretically
that existing estimators lead to HAR tests that have low/non-monotonic
power under nonstationary alternative hypotheses while the proposed
estimator has good monotonic power thereby addressing a long-lasting
problem in time series econometrics. The proposed method is theoretically
valid under general nonstationary random variables. We also establish
mean-squared error bounds for LRV estimation that are sharper than
previously established and use them to determine the data-dependent
bandwidths.

\subsection*{Supplemental Materials}

The supplement for online publication {[}cf. \citet{casini/perron_PrewhitedHAC_Supp}{]}
presents the proofs of the results in the paper. 

\newpage{}

\bibliographystyle{elsarticle-harv}
\bibliography{References_JoE}

\begin{thebibliography}{67}
\expandafter\ifx\csname natexlab\endcsname\relax\def\natexlab#1{#1}\fi
\providecommand{\url}[1]{\texttt{#1}}
\providecommand{\href}[2]{#2}
\providecommand{\path}[1]{#1}
\providecommand{\DOIprefix}{doi:}
\providecommand{\ArXivprefix}{arXiv:}
\providecommand{\URLprefix}{URL: }
\providecommand{\Pubmedprefix}{pmid:}
\providecommand{\doi}[1]{\href{http://dx.doi.org/#1}{\path{#1}}}
\providecommand{\Pubmed}[1]{\href{pmid:#1}{\path{#1}}}
\providecommand{\bibinfo}[2]{#2}
\ifx\xfnm\relax \def\xfnm[#1]{\unskip,\space#1}\fi
\bibitem[{Altissimo and Corradi(2003)}]{altissimo/corradi:2003}
\bibinfo{author}{Altissimo, F.}, \bibinfo{author}{Corradi, V.},
  \bibinfo{year}{2003}.
\newblock \bibinfo{title}{Strong rules for detecting the number of breaks in a
  time series}.
\newblock \bibinfo{journal}{Journal of Econometrics} \bibinfo{volume}{117},
  \bibinfo{pages}{207--244}.
\bibitem[{Andrews(1988)}]{andrews:88hac}
\bibinfo{author}{Andrews, {\relax D.W.K}.}, \bibinfo{year}{1988}.
\newblock \bibinfo{title}{{H}eteroskedasticity and autocorrelation consistent
  covariance matrix estimation}.
\newblock \bibinfo{journal}{{Cowles Foundation Discussion Paper No. 877, Yale
  University}} .
\bibitem[{Andrews(1991)}]{andrews:91}
\bibinfo{author}{Andrews, {\relax D.W.K}.}, \bibinfo{year}{1991}.
\newblock \bibinfo{title}{{H}eteroskedasticity and autocorrelation consistent
  covariance matrix estimation}.
\newblock \bibinfo{journal}{Econometrica} \bibinfo{volume}{59},
  \bibinfo{pages}{817--858}.
\bibitem[{Andrews(1993)}]{andrews:93}
\bibinfo{author}{Andrews, {\relax D.W.K}.}, \bibinfo{year}{1993}.
\newblock \bibinfo{title}{Tests for parameter instability and structural change
  with unknown change-point}.
\newblock \bibinfo{journal}{Econometrica} \bibinfo{volume}{61},
  \bibinfo{pages}{821--56}.
\bibitem[{Andrews and Monahan(1992)}]{andrews/monahan:92}
\bibinfo{author}{Andrews, {\relax D.W.K}.}, \bibinfo{author}{Monahan, {\relax
  J.C}.}, \bibinfo{year}{1992}.
\newblock \bibinfo{title}{An improved heteroskedasticity and autocorrelation
  consistent covariance matrix estimator}.
\newblock \bibinfo{journal}{Econometrica} \bibinfo{volume}{60},
  \bibinfo{pages}{953--966}.
\bibitem[{Belotti et~al.(2023)Belotti, Casini, Catania, Grassi and
  Perron}]{belotti/casini/catania/grassi/perron_HAC_Sim_Bandws}
\bibinfo{author}{Belotti, F.}, \bibinfo{author}{Casini, A.},
  \bibinfo{author}{Catania, L.}, \bibinfo{author}{Grassi, S.},
  \bibinfo{author}{Perron, P.}, \bibinfo{year}{2023}.
\newblock \bibinfo{title}{Simultaneous bandwidths determination for
  double-kernel {HAC} estimators and long-run variance estimation in
  nonparametric settings}.
\newblock \bibinfo{journal}{Econometric Reviews} \bibinfo{volume}{42},
  \bibinfo{pages}{281--306.}
\bibitem[{Brillinger(1975)}]{brillinger:75}
\bibinfo{author}{Brillinger, D.}, \bibinfo{year}{1975}.
\newblock \bibinfo{title}{Time {S}eries {D}ata {A}nalysis and {T}heory}.
\newblock \bibinfo{publisher}{New York: Holt, Rinehart and Winston}.
\bibitem[{Brown et~al.(1975)Brown, Durbin and Evans}]{brown/durbin/evans:1975}
\bibinfo{author}{Brown, {\relax R.L}.}, \bibinfo{author}{Durbin, J.},
  \bibinfo{author}{Evans, {\relax J.M}.}, \bibinfo{year}{1975}.
\newblock \bibinfo{title}{Techniques for testing the constancy of regression
  relationships over time}.
\newblock \bibinfo{journal}{Journal of the Royal Statistical Society. Series B
  (Methodological)} \bibinfo{volume}{37}, \bibinfo{pages}{149--192}.
\bibitem[{Cai(2007)}]{cai:07}
\bibinfo{author}{Cai, Z.}, \bibinfo{year}{2007}.
\newblock \bibinfo{title}{Trending time-varying coefficient time series models
  with serially correlated errors}.
\newblock \bibinfo{journal}{Journal of Econometrics} \bibinfo{volume}{136},
  \bibinfo{pages}{163--188}.
\bibitem[{Casini(2018)}]{casini_CR_Test_Inst_Forecast}
\bibinfo{author}{Casini, A.}, \bibinfo{year}{2018}.
\newblock \bibinfo{title}{Tests for forecast instability and forecast failure
  under a continuous record asymptotic framework}.
\newblock \bibinfo{journal}{arXiv preprint} \bibinfo{volume}{arXiv:1803.10883}.
\bibitem[{Casini(2022)}]{casini_comment_andrews91}
\bibinfo{author}{Casini, A.}, \bibinfo{year}{2022}.
\newblock \bibinfo{title}{Comment on {A}ndrews (1991) "{H}eteroskedasticity and
  autocorrelation consistent covariance matrix estimation"}.
\newblock \bibinfo{journal}{Econometrica} \bibinfo{volume}{90},
  \bibinfo{pages}{1--2}.
\bibitem[{Casini(2023)}]{casini_hac}
\bibinfo{author}{Casini, A.}, \bibinfo{year}{2023}.
\newblock \bibinfo{title}{Theory of evolutionary spectra for heteroskedasticity
  and autocorrelation robust inference in possibly misspecified and
  nonstationary models}.
\newblock \bibinfo{journal}{Journal of Econometrics,} \bibinfo{volume}{235},
  \bibinfo{pages}{372--392}.
\bibitem[{Casini(2024)}]{casini_fixed_b_erp}
\bibinfo{author}{Casini, A.}, \bibinfo{year}{2024}.
\newblock \bibinfo{title}{The {F}ixed-b limiting distribution and the {ERP} of
  {HAR} tests under nonstationarity}.
\newblock \bibinfo{journal}{Journal of Econometrics} \bibinfo{volume}{238},
  \bibinfo{pages}{105625}.
\bibitem[{Casini et~al.(2024)Casini, Deng and
  Perron}]{casini/perron_Low_Frequency_Contam_Nonstat:2020}
\bibinfo{author}{Casini, A.}, \bibinfo{author}{Deng, T.},
  \bibinfo{author}{Perron, P.}, \bibinfo{year}{2024}.
\newblock \bibinfo{title}{Theory of low frequency contamination from
  nonstationarity and misspecification: consequences for {HAR} inference}.
\newblock \bibinfo{journal}{arXiv preprint} \bibinfo{volume}{arXiv:2103.01604}.
\bibitem[{Casini and Perron(2019)}]{casini/perron_Oxford_Survey}
\bibinfo{author}{Casini, A.}, \bibinfo{author}{Perron, P.},
  \bibinfo{year}{2019}.
\newblock \bibinfo{title}{Structural breaks in time series}.
\newblock \bibinfo{journal}{Oxford Research Encyclopedia of Economics and
  Finance,} \bibinfo{volume}{Oxford University Press}.
\bibitem[{Casini and Perron(2020)}]{casini/perron_SC_BP_Lap}
\bibinfo{author}{Casini, A.}, \bibinfo{author}{Perron, P.},
  \bibinfo{year}{2020}.
\newblock \bibinfo{title}{Generalized {L}aplace inference in multiple
  change-points models}.
\newblock \bibinfo{journal}{Econometric Theory} \bibinfo{volume}{38},
  \bibinfo{pages}{35--65}.
\bibitem[{Casini and Perron(2021)}]{casini/perron_Lap_CR_Single_Inf}
\bibinfo{author}{Casini, A.}, \bibinfo{author}{Perron, P.},
  \bibinfo{year}{2021}.
\newblock \bibinfo{title}{Continuous record {L}aplace-based inference about the
  break date in structural change models}.
\newblock \bibinfo{journal}{Juornal of Econometrics} \bibinfo{volume}{224},
  \bibinfo{pages}{3--21}.
\bibitem[{Casini and Perron(2023a)}]{casini:change-point-spectra}
\bibinfo{author}{Casini, A.}, \bibinfo{author}{Perron, P.},
  \bibinfo{year}{2023}a.
\newblock \bibinfo{title}{{C}hange-point analysis of time series with
  evolutionary spectra}.
\newblock \bibinfo{journal}{arXiv preprint} \bibinfo{volume}{arXiv 2106.02031}.
\bibitem[{Casini and Perron(2023b)}]{casini/perron_PrewhitedHAC_Supp}
\bibinfo{author}{Casini, A.}, \bibinfo{author}{Perron, P.},
  \bibinfo{year}{2023}b.
\newblock \bibinfo{title}{Supplement to ``prewhitened long-run variance
  estimation robust to nonstationarity"}.
\newblock \bibinfo{journal}{arXiv preprint} \bibinfo{volume}{arXiv:2103.02235}.
\bibitem[{Cavaliere and Taylor(2007)}]{cavaliere/taylor:2007}
\bibinfo{author}{Cavaliere, G.}, \bibinfo{author}{Taylor, A.M.R.},
  \bibinfo{year}{2007}.
\newblock \bibinfo{title}{Testing for unit roots in time series models with
  non-stationary volatility}.
\newblock \bibinfo{journal}{Journal of Econometrics} \bibinfo{volume}{140},
  \bibinfo{pages}{919--947}.
\bibitem[{Chan(2022a)}]{chan:2022}
\bibinfo{author}{Chan, K.W.}, \bibinfo{year}{2022}a.
\newblock \bibinfo{title}{Optimal difference-based variance estimators in time
  series: a general framework}.
\newblock \bibinfo{journal}{Annals of Statistics} \bibinfo{volume}{50},
  \bibinfo{pages}{1376--1400}.
\bibitem[{Chan(2022b)}]{chan:2020}
\bibinfo{author}{Chan, {\relax K.W}.}, \bibinfo{year}{2022}b.
\newblock \bibinfo{title}{Mean-structure and autocorrelation consistent
  covariance matrix estimation}.
\newblock \bibinfo{journal}{Journal of Business and Economic Statistics,}
  \bibinfo{volume}{40}, \bibinfo{pages}{201--215}.
\bibitem[{Chang and Perron(2018)}]{chang/perron:18}
\bibinfo{author}{Chang, {\relax S.Y}.}, \bibinfo{author}{Perron, P.},
  \bibinfo{year}{2018}.
\newblock \bibinfo{title}{A comparison of alternative methods to construct
  confidence intervals for the estimate of a break date in linear regression
  models}.
\newblock \bibinfo{journal}{Econometric Reviews} \bibinfo{volume}{37},
  \bibinfo{pages}{577--601}.
\bibitem[{Chen and Hong(2012)}]{chen/hong:12}
\bibinfo{author}{Chen, B.}, \bibinfo{author}{Hong, Y.}, \bibinfo{year}{2012}.
\newblock \bibinfo{title}{Testing for smooth structural changes in time series
  models via nonparametric regression}.
\newblock \bibinfo{journal}{Econometrica} \bibinfo{volume}{80},
  \bibinfo{pages}{1157--1183}.
\bibitem[{Crainiceanu and Vogelsang(2007)}]{crainiceanu/vogelsang:07}
\bibinfo{author}{Crainiceanu, {\relax C.M}.}, \bibinfo{author}{Vogelsang,
  {\relax T.J}.}, \bibinfo{year}{2007}.
\newblock \bibinfo{title}{Nonmonotonic power for tests of a mean shift in a
  time series}.
\newblock \bibinfo{journal}{Journal of {S}tatistical {C}omputation and
  {S}imulation} \bibinfo{volume}{77}, \bibinfo{pages}{457--476}.
\bibitem[{Dahlhaus(1997)}]{dahlhaus:96}
\bibinfo{author}{Dahlhaus, R.}, \bibinfo{year}{1997}.
\newblock \bibinfo{title}{Fitting time series models to nonstationary
  processes}.
\newblock \bibinfo{journal}{Annals of Statistics} \bibinfo{volume}{25},
  \bibinfo{pages}{1--37}.
\bibitem[{Dahlhaus and Giraitis(1998)}]{Dahlhaus/Giraitis:98}
\bibinfo{author}{Dahlhaus, R.}, \bibinfo{author}{Giraitis, L.},
  \bibinfo{year}{1998}.
\newblock \bibinfo{title}{On the optimal segment length for parameter estimates
  for locally stationary time series}.
\newblock \bibinfo{journal}{Journal of Time Series Analysis}
  \bibinfo{volume}{19}, \bibinfo{pages}{629--655}.
\bibitem[{Demetrescu and Salish(2020)}]{demetrescu/salish:2020}
\bibinfo{author}{Demetrescu, M.}, \bibinfo{author}{Salish, N.},
  \bibinfo{year}{2020}.
\newblock \bibinfo{title}{({S}tructural) {VAR} models with ignored changes in
  mean and volatility}.
\newblock \bibinfo{journal}{Unpublished Manuscript,} \bibinfo{volume}{SSRN
  https://ssrn.com/abstract=3544676}.
\bibitem[{Deng and Perron(2006)}]{deng/perron:06}
\bibinfo{author}{Deng, A.}, \bibinfo{author}{Perron, P.}, \bibinfo{year}{2006}.
\newblock \bibinfo{title}{A comparison of alternative asymptotic frameworks to
  analyse a structural change in a linear time trend}.
\newblock \bibinfo{journal}{Econometrics Journal} \bibinfo{volume}{9},
  \bibinfo{pages}{423--447}.
\bibitem[{Diebold and Mariano(1995)}]{diebold/mariano:95}
\bibinfo{author}{Diebold, {\relax F.X}.}, \bibinfo{author}{Mariano, {\relax
  R.S}.}, \bibinfo{year}{1995}.
\newblock \bibinfo{title}{Comparing predictive accuracy}.
\newblock \bibinfo{journal}{Journal of Business and Economic Statistics}
  \bibinfo{volume}{13}, \bibinfo{pages}{253--63}.
\bibitem[{Epanechnikov(1969)}]{epanechnikov:69}
\bibinfo{author}{Epanechnikov, V.}, \bibinfo{year}{1969}.
\newblock \bibinfo{title}{Non-parametric estimation of a multivariate
  probability density}.
\newblock \bibinfo{journal}{Theory of Probability and its Applications}
  \bibinfo{volume}{14}, \bibinfo{pages}{153--158}.
\bibitem[{Giacomini and Rossi(2009)}]{giacomini/rossi:09}
\bibinfo{author}{Giacomini, R.}, \bibinfo{author}{Rossi, B.},
  \bibinfo{year}{2009}.
\newblock \bibinfo{title}{Detecting and predicting forecast breakdowns}.
\newblock \bibinfo{journal}{Review of Economic Studies} \bibinfo{volume}{76},
  \bibinfo{pages}{669--705}.
\bibitem[{Giacomini and Rossi(2010)}]{giacomini/rossi:10}
\bibinfo{author}{Giacomini, R.}, \bibinfo{author}{Rossi, B.},
  \bibinfo{year}{2010}.
\newblock \bibinfo{title}{Forecast comparisons in unstable environments}.
\newblock \bibinfo{journal}{Journal of Applied Econometrics}
  \bibinfo{volume}{25}, \bibinfo{pages}{595--620}.
\bibitem[{Giacomini and White(2006)}]{giacomini/white:06}
\bibinfo{author}{Giacomini, R.}, \bibinfo{author}{White, H.},
  \bibinfo{year}{2006}.
\newblock \bibinfo{title}{Tests of conditional predictive ability}.
\newblock \bibinfo{journal}{Econometrica} \bibinfo{volume}{74},
  \bibinfo{pages}{1545--1578}.
\bibitem[{Grenander and Rosenblatt(1957)}]{grenander/rosenblatt:57}
\bibinfo{author}{Grenander, U.}, \bibinfo{author}{Rosenblatt, M.},
  \bibinfo{year}{1957}.
\newblock \bibinfo{title}{Statistical {A}nalysis of {S}tationary {T}ime
  {S}eries}.
\newblock \bibinfo{publisher}{New York: Wiley}.
\bibitem[{Hamilton(1989)}]{hamilton:89}
\bibinfo{author}{Hamilton, {\relax J.D}.}, \bibinfo{year}{1989}.
\newblock \bibinfo{title}{A new approach to the economic analysis of
  nonstationary time series and the business cycle}.
\newblock \bibinfo{journal}{Econometrica} \bibinfo{volume}{57},
  \bibinfo{pages}{357--384}.
\bibitem[{Hansen(1992)}]{hansen:92ecma}
\bibinfo{author}{Hansen, B.}, \bibinfo{year}{1992}.
\newblock \bibinfo{title}{Consistent covariance matrix estimation for dependent
  heterogeneous processes}.
\newblock \bibinfo{journal}{Econometrica} \bibinfo{volume}{60},
  \bibinfo{pages}{967--972}.
\bibitem[{Jansson(2004)}]{jansson:04}
\bibinfo{author}{Jansson, M.}, \bibinfo{year}{2004}.
\newblock \bibinfo{title}{The error in rejection probability of simple
  autocorrelation robust tests}.
\newblock \bibinfo{journal}{Econometrica} \bibinfo{volume}{72},
  \bibinfo{pages}{937--946}.
\bibitem[{de~Jong and Davidson(2000)}]{dejong/davidson:00}
\bibinfo{author}{de~Jong, {\relax R.M}.}, \bibinfo{author}{Davidson, J.},
  \bibinfo{year}{2000}.
\newblock \bibinfo{title}{Consistency of kernel estimators of heteroskedastic
  and autocorrelated covariance matrices}.
\newblock \bibinfo{journal}{Econometrica} \bibinfo{volume}{68},
  \bibinfo{pages}{407--423}.
\bibitem[{Juhl and Xiao(2009)}]{juhl/xiao:09}
\bibinfo{author}{Juhl, T.}, \bibinfo{author}{Xiao, Z.}, \bibinfo{year}{2009}.
\newblock \bibinfo{title}{Testing for changing mean with monotonic power}.
\newblock \bibinfo{journal}{Journal of Econometrics} \bibinfo{volume}{148},
  \bibinfo{pages}{14--24}.
\bibitem[{Kawka(2020)}]{kawka:2020}
\bibinfo{author}{Kawka, R.}, \bibinfo{year}{2020}.
\newblock \bibinfo{title}{Convergence of spectral density estimators in the
  locally stationary framework}.
\newblock \bibinfo{journal}{Econometrics and Statistics,}
  \bibinfo{volume}{forthcoming}.
\bibitem[{Kiefer and Vogelsang(2002)}]{Kiefer/vogelsang:02}
\bibinfo{author}{Kiefer, {\relax N.M}.}, \bibinfo{author}{Vogelsang, {\relax
  T.J}.}, \bibinfo{year}{2002}.
\newblock \bibinfo{title}{Heteroskedasticity-autocorrelation robust standard
  errors using the {B}artlett kernel without truncation}.
\newblock \bibinfo{journal}{Econometrica} \bibinfo{volume}{70},
  \bibinfo{pages}{2093--2095}.
\bibitem[{Kiefer and Vogelsang(2005)}]{kiefer/vogelsang:05}
\bibinfo{author}{Kiefer, {\relax N.M}.}, \bibinfo{author}{Vogelsang, {\relax
  T.J}.}, \bibinfo{year}{2005}.
\newblock \bibinfo{title}{A new asymptotic theory for
  heteroskedasticity-autocorrelation robust tests}.
\newblock \bibinfo{journal}{Econometric Theory} \bibinfo{volume}{21},
  \bibinfo{pages}{1130--1164}.
\bibitem[{Kiefer et~al.(2000)Kiefer, Vogelsang and
  Bunzel}]{Kiefer/vogelsang/bunzel:00}
\bibinfo{author}{Kiefer, {\relax N.M}.}, \bibinfo{author}{Vogelsang, {\relax
  T.J}.}, \bibinfo{author}{Bunzel, H.}, \bibinfo{year}{2000}.
\newblock \bibinfo{title}{Simple robust testing of regression hypotheses}.
\newblock \bibinfo{journal}{Econometrica} \bibinfo{volume}{69},
  \bibinfo{pages}{695--714}.
\bibitem[{Kim and Perron(2009)}]{kim/perron:09}
\bibinfo{author}{Kim, D.}, \bibinfo{author}{Perron, P.}, \bibinfo{year}{2009}.
\newblock \bibinfo{title}{Assessing the relative power of structural break
  tests using a framework based on the approximate {B}ahadur slope}.
\newblock \bibinfo{journal}{Journal of Econometrics} \bibinfo{volume}{149},
  \bibinfo{pages}{26--51}.
\bibitem[{Lazarus et~al.(2021)Lazarus, Lewis and
  Stock}]{lazarus/lewis/stock:17}
\bibinfo{author}{Lazarus, E.}, \bibinfo{author}{Lewis, {\relax D.J}.},
  \bibinfo{author}{Stock, {\relax J.H}.}, \bibinfo{year}{2021}.
\newblock \bibinfo{title}{The size-power tradeoff in {HAR} inference}.
\newblock \bibinfo{journal}{Econometrica} \bibinfo{volume}{89},
  \bibinfo{pages}{2497--2516}.
\bibitem[{Lazarus et~al.(2018)Lazarus, Lewis, Stock and
  Watson}]{lazarus/lewis/stock/watson:18}
\bibinfo{author}{Lazarus, E.}, \bibinfo{author}{Lewis, {\relax D.J}.},
  \bibinfo{author}{Stock, {\relax J.H}.}, \bibinfo{author}{Watson, {\relax
  M.W}.}, \bibinfo{year}{2018}.
\newblock \bibinfo{title}{{HAR} inference: recommendations for practice}.
\newblock \bibinfo{journal}{Journal of Business and Economic Statistics}
  \bibinfo{volume}{36}, \bibinfo{pages}{541--559}.
\bibitem[{Martins and Perron(2016)}]{martins/perron:16}
\bibinfo{author}{Martins, L.}, \bibinfo{author}{Perron, P.},
  \bibinfo{year}{2016}.
\newblock \bibinfo{title}{Improved tests for forecast comparisons in the
  presence of instabilities}.
\newblock \bibinfo{journal}{Journal of Time Series Analysis}
  \bibinfo{volume}{37}, \bibinfo{pages}{650--659}.
\bibitem[{Newey and West(1987)}]{newey/west:87}
\bibinfo{author}{Newey, {\relax W.K}.}, \bibinfo{author}{West, {\relax K.D}.},
  \bibinfo{year}{1987}.
\newblock \bibinfo{title}{A simple positive semidefinite, heteroskedastic and
  autocorrelation consistent covariance matrix}.
\newblock \bibinfo{journal}{Econometrica} \bibinfo{volume}{55},
  \bibinfo{pages}{703--708}.
\bibitem[{Newey and West(1994)}]{newey/west:94}
\bibinfo{author}{Newey, {\relax W.K}.}, \bibinfo{author}{West, {\relax K.D}.},
  \bibinfo{year}{1994}.
\newblock \bibinfo{title}{Automatic lag selection in covariance matrix
  estimation}.
\newblock \bibinfo{journal}{Review of Economic Studies} \bibinfo{volume}{61},
  \bibinfo{pages}{631--653}.
\bibitem[{Otto and Breitung(2021)}]{otto/breitung:2021}
\bibinfo{author}{Otto, S.}, \bibinfo{author}{Breitung, J.},
  \bibinfo{year}{2021}.
\newblock \bibinfo{title}{Backward {CUSUM} for testing and monitoring
  structural change}.
\newblock \bibinfo{journal}{Econometric Theory,} \bibinfo{volume}{forthcoming}.
\bibitem[{Perron(1989)}]{perron:89}
\bibinfo{author}{Perron, P.}, \bibinfo{year}{1989}.
\newblock \bibinfo{title}{The great crash, the oil price shock and the unit
  root hypothesis}.
\newblock \bibinfo{journal}{Econometrica} \bibinfo{volume}{57},
  \bibinfo{pages}{1361--1401}.
\bibitem[{Perron(1991)}]{perron:1991}
\bibinfo{author}{Perron, P.}, \bibinfo{year}{1991}.
\newblock \bibinfo{title}{A test for changes in a polynomial trend function for
  a dynamic time series}.
\newblock \bibinfo{journal}{Research Memorandum No. 363, Econometrics Research
  Program, Princeton University} .
\bibitem[{Perron and Yamamoto(2021)}]{perron/yamamoto:18}
\bibinfo{author}{Perron, P.}, \bibinfo{author}{Yamamoto, Y.},
  \bibinfo{year}{2021}.
\newblock \bibinfo{title}{Testing for changes in forecast performance}.
\newblock \bibinfo{journal}{Journal of Business and Economic Statistics}
  \bibinfo{volume}{39}, \bibinfo{pages}{148--165}.
\bibitem[{Ploberger and Kr\"{a}mer(1992)}]{ploberger/kramer:1992}
\bibinfo{author}{Ploberger, W.}, \bibinfo{author}{Kr\"{a}mer, W.},
  \bibinfo{year}{1992}.
\newblock \bibinfo{title}{The {CUSUM} test with {OLS} residuals}.
\newblock \bibinfo{journal}{Econometrica} \bibinfo{volume}{60},
  \bibinfo{pages}{271--285}.
\bibitem[{Potiron and Mykland(2020)}]{potiron/mykland:2020}
\bibinfo{author}{Potiron, Y.}, \bibinfo{author}{Mykland, P.},
  \bibinfo{year}{2020}.
\newblock \bibinfo{title}{Local parametric estimation in high frequency data}.
\newblock \bibinfo{journal}{Journal of Business and Economic Statistics}
  \bibinfo{volume}{38}, \bibinfo{pages}{679--692}.
\bibitem[{Preinerstorfer(2017)}]{preinerstorfer:17}
\bibinfo{author}{Preinerstorfer, D.}, \bibinfo{year}{2017}.
\newblock \bibinfo{title}{Finite sample properties of tests based on
  prewhitened nonparametric covariance estimators}.
\newblock \bibinfo{journal}{Electronic Journal of Statistics}
  \bibinfo{volume}{11}, \bibinfo{pages}{2097--2167}.
\bibitem[{Qu and Zhuo(2020)}]{qu/zhuo:2020}
\bibinfo{author}{Qu, Z.}, \bibinfo{author}{Zhuo, F.}, \bibinfo{year}{2020}.
\newblock \bibinfo{title}{Likelihood ratio based tests for {M}arkov regime
  switching}.
\newblock \bibinfo{journal}{Review of Economic Studies} \bibinfo{volume}{88},
  \bibinfo{pages}{937--968}.
\bibitem[{Rho and Shao(2013)}]{rho/shao:13}
\bibinfo{author}{Rho, Y.}, \bibinfo{author}{Shao, X.}, \bibinfo{year}{2013}.
\newblock \bibinfo{title}{Improving the bandwidth-free inference methods by
  prewhitening}.
\newblock \bibinfo{journal}{Journal of Statistical Planning and Inference}
  \bibinfo{volume}{143}, \bibinfo{pages}{1912--1922}.
\bibitem[{Shao and Zhang(2010)}]{shao/zhang:2010}
\bibinfo{author}{Shao, X.}, \bibinfo{author}{Zhang, X.}, \bibinfo{year}{2010}.
\newblock \bibinfo{title}{Testing for change points in time series}.
\newblock \bibinfo{journal}{Journal of the American Statistical Association}
  \bibinfo{volume}{105}, \bibinfo{pages}{122--1240}.
\bibitem[{Sun(2014)}]{sun:14}
\bibinfo{author}{Sun, Y.}, \bibinfo{year}{2014}.
\newblock \bibinfo{title}{Let's fix it: fixed-b asymptotics versus small-b
  asymptotics in heteroskedasticity and autocorrelation robust inference}.
\newblock \bibinfo{journal}{Journal of Econometrics} \bibinfo{volume}{178},
  \bibinfo{pages}{659--677}.
\bibitem[{Sun et~al.(2008)Sun, Phillips and Jin}]{sun/phillips/jin:08}
\bibinfo{author}{Sun, Y.}, \bibinfo{author}{Phillips, {\relax P.C.B}.},
  \bibinfo{author}{Jin, S.}, \bibinfo{year}{2008}.
\newblock \bibinfo{title}{Optimal bandwidth selection in
  heteroskedasticity-autocorrelation robust testing}.
\newblock \bibinfo{journal}{Econometrica} \bibinfo{volume}{76},
  \bibinfo{pages}{175--194}.
\bibitem[{Vogelsang(1999)}]{vogeslang:99}
\bibinfo{author}{Vogelsang, {\relax T.J}.}, \bibinfo{year}{1999}.
\newblock \bibinfo{title}{Sources of nonmonotonic power when testing for a
  shift in mean of a dynamic time series}.
\newblock \bibinfo{journal}{Journal of Econometrics} \bibinfo{volume}{88},
  \bibinfo{pages}{283--299}.
\bibitem[{West(1996)}]{west:96}
\bibinfo{author}{West, K.D.}, \bibinfo{year}{1996}.
\newblock \bibinfo{title}{Asymptotic inference about predictive ability}.
\newblock \bibinfo{journal}{Econometrica} \bibinfo{volume}{64},
  \bibinfo{pages}{1067--1084}.
\bibitem[{Whilelm(2015)}]{whilelm:2015}
\bibinfo{author}{Whilelm, D.}, \bibinfo{year}{2015}.
\newblock \bibinfo{title}{Optimal bandwidth selection for robust generalized
  methods of moments estimation}.
\newblock \bibinfo{journal}{Econometric Theory} \bibinfo{volume}{31},
  \bibinfo{pages}{1054--1077}.
\bibitem[{Xiao and Linton(2002)}]{xiao/linton:02}
\bibinfo{author}{Xiao, Z.}, \bibinfo{author}{Linton, O.}, \bibinfo{year}{2002}.
\newblock \bibinfo{title}{A nonparametric prewhitened covariance estimator}.
\newblock \bibinfo{journal}{Journal of Time Series Analysis}
  \bibinfo{volume}{23}, \bibinfo{pages}{215--250}.
\bibitem[{Zhang and Lavitas(2018)}]{zhang/lavitas:2018}
\bibinfo{author}{Zhang, T.}, \bibinfo{author}{Lavitas, L.},
  \bibinfo{year}{2018}.
\newblock \bibinfo{title}{Unsupervised self-normalized change-point testing for
  time series}.
\newblock \bibinfo{journal}{Journal of the American Statistical Association}
  \bibinfo{volume}{113}, \bibinfo{pages}{637--648}.

\end{thebibliography}


\begin{thebibliography}{5}
\expandafter\ifx\csname natexlab\endcsname\relax\def\natexlab#1{#1}\fi
\providecommand{\url}[1]{\texttt{#1}}
\providecommand{\href}[2]{#2}
\providecommand{\path}[1]{#1}
\providecommand{\DOIprefix}{doi:}
\providecommand{\ArXivprefix}{arXiv:}
\providecommand{\URLprefix}{URL: }
\providecommand{\Pubmedprefix}{pmid:}
\providecommand{\doi}[1]{\href{http://dx.doi.org/#1}{\path{#1}}}
\providecommand{\Pubmed}[1]{\href{pmid:#1}{\path{#1}}}
\providecommand{\bibinfo}[2]{#2}
\ifx\xfnm\relax \def\xfnm[#1]{\unskip,\space#1}\fi
\bibitem[{Andrews(1991)}]{andrews:91}
\bibinfo{author}{Andrews, D.W.K.}, \bibinfo{year}{1991}.
\newblock \bibinfo{title}{{H}eteroskedasticity and {A}utocorrelation
  {C}onsistent {C}ovariance {M}atrix {E}stimation}.
\newblock \bibinfo{journal}{Econometrica} \bibinfo{volume}{59},
  \bibinfo{pages}{817--858}.
\bibitem[{Andrews and Monahan(1992)}]{andrews/monahan:92}
\bibinfo{author}{Andrews, D.W.K.}, \bibinfo{author}{Monahan, J.C.},
  \bibinfo{year}{1992}.
\newblock \bibinfo{title}{An {I}mproved {H}eteroskedasticity and
  {A}utocorrelation {C}onsistent {C}ovariance {M}atrix {E}stimator}.
\newblock \bibinfo{journal}{Econometrica} \bibinfo{volume}{60},
  \bibinfo{pages}{953--966}.
\bibitem[{Brillinger(1975)}]{brillinger:75}
\bibinfo{author}{Brillinger, D.R.}, \bibinfo{year}{1975}.
\newblock \bibinfo{title}{Time {S}eries {D}ata {A}nalysis and {T}heory}.
\newblock \bibinfo{publisher}{New York: Holt, Rinehart and Winston}.
\bibitem[{Casini(2023)}]{casini_hac}
\bibinfo{author}{Casini, A.}, \bibinfo{year}{2023}.
\newblock \bibinfo{title}{Theory of {E}volutionary {S}pectra for
  {H}eteroskedasticity and {A}utocorrelation {R}obust {I}nference in {P}ossibly
  {M}isspecified and {N}onstationary {M}odels}.
\newblock \bibinfo{journal}{Journal of Econometrics} \bibinfo{volume}{235},
  \bibinfo{pages}{372--392}.
\bibitem[{Casini et~al.(2024)Casini, Deng and
  Perron}]{casini/perron_Low_Frequency_Contam_Nonstat:2020}
\bibinfo{author}{Casini, A.}, \bibinfo{author}{Deng, T.},
  \bibinfo{author}{Perron, P.}, \bibinfo{year}{2024}.
\newblock \bibinfo{title}{Theory of {L}ow {F}requency {C}ontamination from
  {N}onstationarity and {M}isspecification: {C}onsequences for {HAR}
  {I}nference}.
\newblock \bibinfo{journal}{arXiv preprint} \bibinfo{volume}{arXiv:2103.01604}.

\end{thebibliography}
\addcontentsline{toc}{section}{References}

\newpage{}

\newpage{}

\clearpage 
\pagenumbering{arabic}
\renewcommand*{\thepage}{A-\arabic{page}}
\appendix

\section{Appendix}

\subsection{Tables}

\begin{table}[H]
\caption{\label{Table Size M1-M2}Empirical small-sample size of $t$-test
for model M1-M2}

\begin{centering}
\begin{tabular}{lcccccc}
 & \multicolumn{2}{c}{{\footnotesize{}M1, $\rho=0.4$}} & \multicolumn{2}{c}{{\footnotesize{}M1, $\rho=0.9$}} & \multicolumn{2}{c}{{\footnotesize{}M2}}\tabularnewline
{\footnotesize{}$\alpha=0.05$} & {\footnotesize{}$T=200$} & {\footnotesize{}$T=400$} & {\footnotesize{}$T=200$} & {\footnotesize{}$T=400$} & {\footnotesize{}$T=200$} & {\footnotesize{}$T=400$}\tabularnewline
\hline 
\hline 
{\footnotesize{}$\widehat{J}_{T}$, QS, prew} & {\footnotesize{}0.054} & {\footnotesize{}0.045} & {\footnotesize{}0.085} & {\footnotesize{}0.065} & {\footnotesize{}0.061} & {\footnotesize{}0.053}\tabularnewline
{\footnotesize{}$\widehat{J}_{T}$, QS, prew, SLS} & {\footnotesize{}0.052} & {\footnotesize{}0.043} & {\footnotesize{}0.086} & {\footnotesize{}0.051} & {\footnotesize{}0.065} & {\footnotesize{}0.054}\tabularnewline
{\footnotesize{}$\widehat{J}_{T}$, QS, prew, SLS, $\mu$} & {\footnotesize{}0.049} & {\footnotesize{}0.048} & {\footnotesize{}0.103} & {\footnotesize{}0.092} & {\footnotesize{}0.063} & {\footnotesize{}0.054}\tabularnewline
{\footnotesize{}Andrews} & {\footnotesize{}0.082} & {\footnotesize{}0.065} & {\footnotesize{}0.162} & {\footnotesize{}0.118} & {\footnotesize{}0.095} & {\footnotesize{}0.050}\tabularnewline
{\footnotesize{}Andrews, prew} & {\footnotesize{}0.063} & {\footnotesize{}0.057} & {\footnotesize{}0.104} & {\footnotesize{}0.083} & {\footnotesize{}0.077} & {\footnotesize{}0.048}\tabularnewline
{\footnotesize{}Newey-West } & {\footnotesize{}0.114} & {\footnotesize{}0.090} & {\footnotesize{}0.351} & {\footnotesize{}0.272} & {\footnotesize{}0.138} & {\footnotesize{}0.057}\tabularnewline
{\footnotesize{}Newey-West, prew} & {\footnotesize{}0.075} & {\footnotesize{}0.064} & {\footnotesize{}0.110} & {\footnotesize{}0.077} & {\footnotesize{}0.090} & {\footnotesize{}0.059}\tabularnewline
{\footnotesize{}Newey-West, fixed-$b$ (KVB)} & {\footnotesize{}0.058} & {\footnotesize{}0.056} & {\footnotesize{}0.091} & {\footnotesize{}0.066} & {\footnotesize{}0.069} & {\footnotesize{}0.052}\tabularnewline
{\footnotesize{}EWC} & {\footnotesize{}0.058} & {\footnotesize{}0.055} & {\footnotesize{}0.149} & {\footnotesize{}0.113} & {\footnotesize{}0.071} & {\footnotesize{}0.048}\tabularnewline
\hline 
\end{tabular}
\par\end{centering}
\end{table}

\begin{table}[H]
\caption{\label{Table Size M3-M4}Empirical small-sample size for model M3-M4}

\begin{centering}
\begin{tabular}{lcccc}
 & \multicolumn{2}{c}{{\footnotesize{}M3}} & \multicolumn{2}{c}{{\footnotesize{}M4}}\tabularnewline
{\footnotesize{}$\alpha=0.05$} & {\footnotesize{}$T=400$} & {\footnotesize{}$T=800$} & {\footnotesize{}$T=400$} & {\footnotesize{}$T=800$}\tabularnewline
\hline 
\hline 
{\footnotesize{}$\widehat{J}_{T}$, QS, prew, SLS} & {\footnotesize{}0.065} & {\footnotesize{}0.060} & {\footnotesize{}0.071} & {\footnotesize{}0.066}\tabularnewline
{\footnotesize{}$\widehat{J}_{T}$, QS, prew, SLS, $\mu$} & {\footnotesize{}0.065} & {\footnotesize{}0.061} & {\footnotesize{}0.077} & {\footnotesize{}0.067}\tabularnewline
{\footnotesize{}Andrews} & {\footnotesize{}0.082} & {\footnotesize{}0.073} & {\footnotesize{}0.000} & {\footnotesize{}0.000}\tabularnewline
{\footnotesize{}Andrews, prew} & {\footnotesize{}0.080} & {\footnotesize{}0.074} & {\footnotesize{}0.005} & {\footnotesize{}0.000}\tabularnewline
{\footnotesize{}Newey-West } & {\footnotesize{}0.080} & {\footnotesize{}0.074} & {\footnotesize{}0.000} & {\footnotesize{}0.000}\tabularnewline
{\footnotesize{}Newey-West, prew} & {\footnotesize{}0.078} & {\footnotesize{}0.073} & {\footnotesize{}0.000} & {\footnotesize{}0.000}\tabularnewline
{\footnotesize{}Newey-West, fixed-$b$ (KVB)} & {\footnotesize{}0.002} & {\footnotesize{}0.002} & {\footnotesize{}0.074} & {\footnotesize{}0.061}\tabularnewline
{\footnotesize{}EWC} & {\footnotesize{}0.080} & {\footnotesize{}0.074} & {\footnotesize{}0.018} & {\footnotesize{}0.022}\tabularnewline
\hline 
\end{tabular}
\par\end{centering}
\end{table}

\begin{table}[H]
\caption{\label{Table Power M1-M2}Empirical small-sample power of $t$-test
for model M1-M2}

\begin{centering}
\begin{tabular}{lcccccc}
 & \multicolumn{3}{c}{{\footnotesize{}M1}} & \multicolumn{3}{c}{{\footnotesize{}M2}}\tabularnewline
{\footnotesize{}$\alpha=0.05$, $T=400$} & {\footnotesize{}$\delta=0.5$} & {\footnotesize{}$\delta=1$} & {\footnotesize{}$\delta=2$} & {\footnotesize{}$\delta=0.1$} & {\footnotesize{}$\delta=0.2$} & {\footnotesize{}$\delta=0.4$}\tabularnewline
\hline 
\hline 
{\footnotesize{}$\widehat{J}_{T}$, QS, prew} & {\footnotesize{}0.344} & {\footnotesize{}0.807} & {\footnotesize{}1.000} & {\footnotesize{}0.387} & {\footnotesize{}0.889} & {\footnotesize{}1.000}\tabularnewline
{\footnotesize{}$\widehat{J}_{T}$, QS, prew, SLS} & {\footnotesize{}0.378} & {\footnotesize{}0.787} & {\footnotesize{}1.000} & {\footnotesize{}0.330} & {\footnotesize{}0.813} & {\footnotesize{}1.000}\tabularnewline
{\footnotesize{}$\widehat{J}_{T}$, QS, prew, SLS, $\mu$} & {\footnotesize{}0.463} & {\footnotesize{}0.849} & {\footnotesize{}1.000} & {\footnotesize{}0.347} & {\footnotesize{}0.833} & {\footnotesize{}1.000}\tabularnewline
{\footnotesize{}Andrews} & {\footnotesize{}0.430} & {\footnotesize{}0.864} & {\footnotesize{}1.000} & {\footnotesize{}0.450} & {\footnotesize{}0.922} & {\footnotesize{}1.000}\tabularnewline
{\footnotesize{}Andrews, prew} & {\footnotesize{}0.360} & {\footnotesize{}0.812} & {\footnotesize{}1.000} & {\footnotesize{}0.433} & {\footnotesize{}0.911} & {\footnotesize{}1.000}\tabularnewline
{\footnotesize{}Newey-West} & {\footnotesize{}0.630} & {\footnotesize{}0.958} & {\footnotesize{}1.000} & {\footnotesize{}0.511} & {\footnotesize{}0.938} & {\footnotesize{}1.000}\tabularnewline
{\footnotesize{}Newey-West, prew} & {\footnotesize{}0.363} & {\footnotesize{}0.811} & {\footnotesize{}1.000} & {\footnotesize{}0.443} & {\footnotesize{}0.911} & {\footnotesize{}1.000}\tabularnewline
{\footnotesize{}Newey-West, fixed-$b$ (KVB)} & {\footnotesize{}0.274} & {\footnotesize{}0.655} & {\footnotesize{}0.980} & {\footnotesize{}0.329} & {\footnotesize{}0.758} & {\footnotesize{}0.990}\tabularnewline
{\footnotesize{}EWC} & {\footnotesize{}0.436} & {\footnotesize{}0.886} & {\footnotesize{}1.000} & {\footnotesize{}0.392} & {\footnotesize{}0.890} & {\footnotesize{}1.000}\tabularnewline
\hline 
\end{tabular}
\par\end{centering}
\end{table}

\begin{table}[H]
\caption{\label{Table Power M3-M4}Empirical small-sample power for model M3-M4}

\begin{centering}
\begin{tabular}{lcccccc}
 & \multicolumn{3}{c}{{\footnotesize{}M3}} & \multicolumn{3}{c}{{\footnotesize{}M4}}\tabularnewline
{\footnotesize{}$\alpha=0.05$, $T=400$} & {\footnotesize{}$\delta=0.5$} & {\footnotesize{}$\delta=2$} & {\footnotesize{}$\delta=6$} & {\footnotesize{}$\delta=0.5$} & {\footnotesize{}$\delta=1$} & {\footnotesize{}$\delta=2$}\tabularnewline
\hline 
\hline 
{\footnotesize{}$\widehat{J}_{T}$, QS, prew, SLS} & {\footnotesize{}0.495} & {\footnotesize{}0.920} & {\footnotesize{}1.000} & {\footnotesize{}0.613} & {\footnotesize{}0.923} & {\footnotesize{}1.000}\tabularnewline
{\footnotesize{}$\widehat{J}_{T}$, QS, prew, SLS, $\mu$} & {\footnotesize{}0.498} & {\footnotesize{}0.940} & {\footnotesize{}1.000} & {\footnotesize{}0.663} & {\footnotesize{}0.957} & {\footnotesize{}1.000}\tabularnewline
{\footnotesize{}Andrews} & {\footnotesize{}0.158} & {\footnotesize{}0.014} & {\footnotesize{}0.000} & {\footnotesize{}0.000} & {\footnotesize{}0.043} & {\footnotesize{}0.073}\tabularnewline
{\footnotesize{}Andrews, prew} & {\footnotesize{}0.224} & {\footnotesize{}0.056} & {\footnotesize{}0.000} & {\footnotesize{}0.351} & {\footnotesize{}0.942} & {\footnotesize{}0.952}\tabularnewline
{\footnotesize{}Newey-West} & {\footnotesize{}0.179} & {\footnotesize{}0.302} & {\footnotesize{}0.587} & {\footnotesize{}0.019} & {\footnotesize{}0.821} & {\footnotesize{}1.000}\tabularnewline
{\footnotesize{}Newey-West, prew} & {\footnotesize{}0.137} & {\footnotesize{}0.014} & {\footnotesize{}0.000} & {\footnotesize{}0.003} & {\footnotesize{}0.278} & {\footnotesize{}0.722}\tabularnewline
{\footnotesize{}Newey-West, fixed-$b$ (KVB)} & {\footnotesize{}0.059} & {\footnotesize{}0.008} & {\footnotesize{}0.000} & {\footnotesize{}0.000} & {\footnotesize{}0.000} & {\footnotesize{}0.000}\tabularnewline
{\footnotesize{}EWC} & {\footnotesize{}0.087} & {\footnotesize{}0.018} & {\footnotesize{}0.000} & {\footnotesize{}0.062} & {\footnotesize{}0.000} & {\footnotesize{}0.000}\tabularnewline
\hline 
\end{tabular}
\par\end{centering}
\end{table}

\clearpage{}

\newpage{}

\clearpage 
\pagenumbering{arabic}
\renewcommand*{\thepage}{A-\arabic{page}}
\appendix

\clearpage{}

\newpage{}

\pagebreak{}

\section*{}
\addcontentsline{toc}{part}{Supplemental Material}
\begin{center}
\Large{{Supplemental Material} to} 
\end{center}

\begin{center}
\title{\textbf{\Large{Prewhitened Long-Run Variance Estimation Robust to Nonstationarity}}} 
\maketitle
\end{center}
\medskip{} 
\medskip{} 
\medskip{} 
\thispagestyle{empty}

\begin{center}
$\qquad$ \textsc{\textcolor{MyBlue}{Alessandro Casini}} $\qquad$ \textsc{\textcolor{MyBlue}{Pierre Perron}}\\
\small{University of Rome Tor Vergata} $\quad$ \small{Boston University} 
\\
\medskip{}
\medskip{} 
\medskip{} 

\medskip{} 
\date{\small{\today}} 
\medskip{} 
\medskip{} 
\medskip{} 
\end{center}
\begin{abstract}
{\footnotesize{}This supplemental material is for online publication
and is structured as follows. Section \ref{Section Preliminaries}
presents some preliminary notions. Section \ref{Section: Proofs-of-the Section Prewhitened DK-HAC},
\ref{Section Proofs of the Results General Nonstationarity} and \ref{Section Proofs of Section Power}
present the proofs of the results of Section \ref{Section Prewhitended DKHAC- Theory},
\ref{Section: Extension-to-Unrestrcited Nonstationary} and \ref{Section Power},
respectively. }{\footnotesize\par}
\end{abstract}
\setcounter{page}{0}
\setcounter{section}{0}
\renewcommand*{\theHsection}{\the\value{section}}

\newpage{}

\begin{singlespace} 
\noindent 
\small

\allowdisplaybreaks


\renewcommand{\thepage}{S-\arabic{page}}   
\renewcommand{\thesection}{S.\Alph{section}}   
\renewcommand{\theequation}{\thesection.\arabic{equation}}




\section{\label{Section Preliminaries}Preliminaries}

In this section we present a formal definition of SLS processes which
is implied by Assumption \ref{Assumption Smothness f(u,w)} on $f\left(u,\,\omega\right).$
Let $0=\lambda_{0}^ {}<\lambda_{1}<\ldots<\lambda_{m_{0}}<\lambda_{m_{0}+1}=1$
where $m_{0}$ may be fixed or grow to infinity. A function $G\left(u,\,\cdot\right):\,\left[0,\,1\right]\times\mathbb{R}\rightarrow\mathbb{C}$
is said to be piecewise (Lipschitz) continuous in $u$ with $m_{0}+1$
segments if  for each segment $j=1,\ldots,\,m_{0}+1$ it satisfies
$\sup_{u\neq v}|G\left(u,\,\omega\right)-G\left(v,\,\omega\right)|\leq K|u-v|$
for any $\omega\in\mathbb{R}$ with $\lambda_{j-1}<u,\,v\leq\lambda_{j}$
for some $K<\infty.$ We define $G_{j}\left(u,\,\omega\right)=G\left(u,\,\omega\right)$
for $\lambda_{j-1}<u\leq\lambda_{j}$. A function $G\left(\cdot,\,\cdot\right):\,\left[0,\,1\right]\times\mathbb{R}\rightarrow\mathbb{C}$
is said to be left-differentiable at $u_{0}$ if $\partial G\left(u_{0},\omega\right)/\partial_{-}u\triangleq\lim_{u\rightarrow u_{0}^{-}}\left(G\left(u_{0},\,\omega\right)-G\left(u,\,\omega\right)\right)/\left(u_{0}-u\right)$
exists for any $\omega\in\mathbb{R}$.
\begin{defn}
\label{Definition Segmented-Locally-Stationary}A sequence of stochastic
processes $V_{t,T}$ $\left(t=1,\ldots,\,T\right)$ is called segmented
locally stationary (SLS) with $m_{0}+1$ regimes, transfer function
$A^{0}$  and trend $\mu_{\cdot}$ if there exists a representation,
\begin{align}
V_{t,T} & =\mu_{j}\left(t/T\right)+\int_{-\pi}^{\pi}\exp\left(i\omega t\right)A_{j,t,T}^{0}\left(\omega\right)d\xi\left(\omega\right),\qquad\qquad\left(t=T_{j-1}^{0}+1,\ldots,\,T_{j}^{0}\right),\label{Eq. Spectral Rep of SLS}
\end{align}
for $j=1,\ldots,\,m_{0}+1$, where by convention $T_{0}^{0}=0$ and
$T_{m_{0}+1}^{0}=T$ and the following holds: 

(i) $\xi\left(\lambda\right)$ is a stochastic process on $\left[-\pi,\,\pi\right]$
with $\overline{\xi\left(\omega\right)}=\xi\left(-\omega\right)$
and 
\begin{align*}
\mathrm{cum}\left\{ d\xi\left(\omega_{1}\right),\ldots,\,d\xi\left(\omega_{r}\right)\right\}  & =\varphi\left(\sum_{j=1}^{r}\omega_{j}\right)g_{r}\left(\omega_{1},\ldots,\,\omega_{r-1}\right)d\omega_{1}\ldots d\omega_{r},
\end{align*}
 where $\mathrm{cum}\left\{ \cdot\right\} $ denotes the cumulant
spectra of $r$-th order, $g_{1}=0,\,g_{2}\left(\omega\right)=1$,
$\left|g_{r}\left(\omega_{1},\ldots,\,\omega_{r-1}\right)\right|\leq M_{r}$
for all $r$ with $M_{r}$ being a constant that may depend on $r$,
and $\varphi\left(\omega\right)=\sum_{j=-\infty}^{\infty}\delta\left(\omega+2\pi j\right)$
is the period $2\pi$ extension of the Dirac delta function $\delta\left(\cdot\right)$.

(ii) There exists a constant $K$  and a piecewise continuous function
$A:\,\left[0,\,1\right]\times\mathbb{R}\rightarrow\mathbb{C}$ such
that, for each $j=1,\ldots,\,m_{0}+1$, there exists a $2\pi$-periodic
function $A_{j}:\,(\lambda_{j-1}^{0},\,\lambda_{j}^{0}]\times\mathbb{R}\rightarrow\mathbb{C}$
with $A_{j}\left(u,\,-\omega\right)=\overline{A_{j}\left(u,\,\omega\right)}$,
$\lambda_{j}^{0}\triangleq T_{j}^{0}/T$ and for all $T,$
\begin{align}
A\left(u,\,\omega\right) & =A_{j}\left(u,\,\omega\right)\,\mathrm{\,for\,}\,\lambda_{j-1}^{0}<u\leq\lambda_{j}^{0},\label{Eq A(u) =00003D Ai}\\
\sup_{1\leq j\leq m_{0}+1} & \sup_{T_{j-1}^{0}<t\leq T_{j}^{0},\,\omega}\left|A_{j,t,T}^{0}\left(\omega\right)-A_{j}\left(t/T,\,\omega\right)\right|\leq KT^{-1}.\label{Eq. 2.4 Smothenss Assumption on A}
\end{align}

(iii) $\mu_{j}\left(t/T\right)$ is piecewise continuous. 
\end{defn}
In the context of HAR inference $\mathbb{E}\left(V_{t}\right)=0$
and so $\mu\left(t/T\right)=0$ for all $t$ in Definition \ref{Definition Segmented-Locally-Stationary}.
In view of Definition \ref{Definition Segmented-Locally-Stationary},
Assumption \ref{Assumption Smothness f(u,w)} also holds with $f\left(u,\,\omega\right)$
replaced by $A\left(u,\,\omega\right)$ and this property is used
in some parts of the proofs. In Assumption \ref{Assumption A - Dependence}-(ii),
the continuity points are those $u\in\left[0,\,1\right]$ such that
$u\neq\lambda_{j}^{0}$ ($j=1,\ldots,\,m_{0}+1$) whereas the discontinuity
points are those $u\in\left[0,\,1\right]$ such that $u=\lambda_{j}^{0}$
($j=1,\ldots,\,m_{0}+1$).

\section{\label{Section: Proofs-of-the Section Prewhitened DK-HAC}Proofs
of the Results in Section \ref{Section Prewhitended DKHAC- Theory}}

In some of the proofs below $\overline{\beta}$ is understood to be
on the line segment joining $\widehat{\beta}$ and $\beta_{0}$. We
discard the degrees of freedom adjustment $T/\left(T-p\right)$ from
the derivations since asymptotically it does not play any role. Similarly,
we use $T/n_{T}$ in place of $\left(T-n_{T}\right)/n_{T}$ in the
expression for $\widehat{\Gamma}_{D}^{*}\left(k\right)$ and $\widehat{\Gamma}\left(k\right)$.
 We collect the break dates in $\mathcal{T}\triangleq\{T_{1}^{0},\,\ldots,\,T_{m_{0}}^{0}\}$.

\subsection{Proof of Theorem \ref{Theorem 1 MSE Prew - Andrews 90}}

Let
\begin{align*}
\widehat{J}_{T}^{*}=\widehat{J}_{T}^{*}\left(b_{\theta_{1},T},\,b_{\theta_{2},T}\right)\triangleq\sum_{k=-T+1}^{T-1}K_{1}\left(b_{\theta_{1},T}k\right)\widehat{\Gamma}^{*}\left(k\right) & ,
\end{align*}
 where $\widehat{\Gamma}^{*}\left(k\right)\triangleq(n_{T}/T)\sum_{r=0}^{\left\lfloor T/n_{T}\right\rfloor }\widehat{c}_{T}^{*}\left(rn_{T}/T,\,k\right)$
and
\begin{align}
\widehat{c}_{T}^{*}\left(rn_{T}/T,\,k\right) & \triangleq\begin{cases}
\left(Tb_{2,T}\right)^{-1}\sum_{s=k+1}^{T}K_{2}^{*}\left(\frac{\left(\left(r+1\right)n_{T}-\left(s-k/2\right)\right)/T}{b_{\theta_{2},T}}\right)\widehat{V}_{s}^{*}\widehat{V}_{s-k}^{*\prime}, & k\geq0\\
\left(Tb_{2,T}\right)^{-1}\sum_{s=-k+1}^{T}K_{2}^{*}\left(\frac{\left(\left(r+1\right)n_{T}-\left(s+k/2\right)\right)/T}{b_{\theta_{2},T}}\right)\widehat{V}_{s+k}^{*}\widehat{V}_{s}^{*\prime}, & k<0
\end{cases},\label{eq: Def. chat}
\end{align}
with $\widehat{V}_{s}^{*}=V_{s}^{*}(\widehat{\beta})$ where $\widehat{\beta}$
is elongated to include $\widehat{A}_{\cdot,j}$ $(j=1,\ldots,\,p_{A})$.
Define $\widetilde{J}_{T}^{*}$ as equal to $\widehat{J}_{T}^{*}$
but with $V_{t}^{*}=V_{t}-\sum_{j=1}^{p_{A}}A_{r,j}V_{t-j}$ in place
of $\widehat{V}_{s}^{*}$ and define $J_{T}^{*}$ as equal to $J_{T}$
but with $V_{t}^{*}$ in place of $V_{t}\left(\beta_{0}\right)$.
The proof uses the following decomposition,
\begin{align}
\widehat{J}_{\mathrm{pw},T}-J_{T}=\left(\widehat{J}_{\mathrm{pw},T}-J_{T,\widehat{D}}^{*}\right)+\left(J_{T,\widehat{D}}^{*}-J_{T,D}^{*}\right)+\left(J_{T,D}^{*}-J_{T}\right) & ,\label{Eq. (decomposition Theorem 1)}
\end{align}
 where $J_{T,D}^{*}=T^{-1}\sum_{s=p_{A}+1}^{T}\sum_{t=p_{A}+1}^{T}D_{s}\mathbb{E}(V_{s}^{*}V_{t}^{*\prime})D_{t}^{\prime},$
and $J_{T,\widehat{D}}^{*}$ is equal to $J_{T,D}^{*}$ but with $\widehat{D}_{s}$
in place of $D_{s}$. 

Given the decomposition \eqref{Eq. (decomposition Theorem 1)} there
are two main steps in the proof of Theorem \ref{Theorem 1 MSE Prew - Andrews 90}.
For part (i) and (ii) of the theorem, it is important to analyze the
behavior of $\widehat{J}_{\mathrm{pw},T}-J_{T,\widehat{D}}^{*}$ and
$J_{T,D}^{*}-J_{T}$. Unlike the proofs involving the non-prewhitened
LRV estimators, the factor $\widehat{J}_{\mathrm{pw},T}$ is a function
of $\{\widehat{V}_{t}^{*}\}$ which depends on the whitening step
(step 1) and on the recoloring step (step 2), and so it needs to be
handled using conditions that are not invoked in the proofs involving
non-prewhitened LRV estimators. The factor $J_{T,D}^{*}-J_{T}$ is
only present in the proofs involving the prewhitened LRV estimator.
However, the proof is more complex than the one in \citeReferencesSupp{andrews/monahan:92}
because our local prewhitening procedure involves the time-smoothing
which appears in $\widehat{\Gamma}^{*}\left(k\right)$ through the
kernel $K_{2}$ and in the whitening step through the estimation of
the VAR based on the time windows of length $n_{T}$. 
\begin{lem}
\label{Lemma J*(b1*, b2*) - J* =00003D o(1)}Under the assumptions
of Theorem \ref{Theorem 1 MSE Prew - Andrews 90}-(i), we have
\begin{align}
\widehat{J}_{T}^{*}\left(b_{\theta_{1},T},\,b_{\theta_{2},T}\right)-J_{T}^{*} & =o_{\mathbb{P}}\left(1\right).\label{Eq. (J_hat* - J*) - 2nd term =00003D op(1)}
\end{align}
\end{lem}
\noindent\textit{Proof.} Under Assumption \ref{Assumption A - Dependence},
$||\int_{0}^{1}f^{*\left(0\right)}\left(u,\,0\right)||<\infty$,
where $f^{*}$ is defined analogously to $f_{D}^{*}$ but with $D_{s}=1$
for all $s$. In view of $K_{1,0}=0$, Theorem 3.1-(i,ii) in \citeReferencesSupp{casini_hac}
{[}with $q=0$ in part (ii){]} implies $\widetilde{J}_{T}^{*}-J_{T}^{*}=o_{\mathbb{P}}\left(1\right)$.
Note that the assumptions of the aforementioned theorem are satisfied
by $\{V_{t}^{*}\}$ since they correspond to Assumption \ref{Assumption Smothness f(u,w)}
and \ref{Assumption A - Dependence} here. Note that $\widehat{J}_{T}^{*}-\widetilde{J}_{T}^{*}=o_{\mathbb{P}}\left(1\right)$
if and only if $a'\widehat{J}_{T}^{*}a-a'\widetilde{J}_{T}^{*}a=o_{\mathbb{P}}\left(1\right)$
for arbitrary $a\in\mathbb{R}^{p}$. We shall provide the proof only
for the scalar case. We show that $\sqrt{n_{T}}b_{\theta_{1},T}(\widehat{J}_{T}^{*}-\widetilde{J}_{T}^{*})=O_{\mathbb{P}}\left(1\right)$.
Let $\widetilde{J}_{T}^{*}\left(\beta\right)$ denote the estimator
that uses $\{V_{t}^{*}\left(\beta\right)\}$ where $\beta$ is elongated
to include $A_{\cdot,j}$ $(j=1,\ldots,\,p_{A})$. A mean-value expansion
of $\widetilde{J}_{T}^{*}(\widehat{\beta})(=\widehat{J}_{T}^{*})$
about $\beta_{0}$ (elongated to include $A_{\cdot,j}$ $(j=1,\ldots,\,p_{A})$)
yields 
\begin{align}
\sqrt{n_{T}}b_{\theta_{1},T}(\widehat{J}_{T}^{*}-\widetilde{J}_{T}^{*}) & =b_{\theta_{1},T}\frac{\partial}{\partial\beta'}\widetilde{J}_{T}^{*}(\bar{\beta})\sqrt{n_{T}}(\widehat{\beta}-\beta_{0})\nonumber \\
 & =b_{\theta_{1},T}\sum_{k=-T+1}^{T-1}K_{1}\left(b_{\theta_{1},T}k\right)\frac{\partial}{\partial\beta'}\widehat{\Gamma}^{*}\left(k\right)|_{\beta=\bar{\beta}}\sqrt{n_{T}}(\widehat{\beta}-\beta_{0}),\label{eq (A.9) Andrews-1}
\end{align}
for some $\bar{\beta}$ on the line segment joining $\widehat{\beta}$
and $\beta_{0}$. Note also that $\widehat{c}^{*}(rn_{T}/T,\,k)$
depends on $\beta$ although we have omitted it. We have for $k\geq0$
(the case $k<0$ is similar and omitted), 

\begin{align}
\sup_{0\leq k\leq b_{\theta_{1},T}^{-1}}\Bigl\Vert\frac{n_{T}}{T}\sum_{r=0}^{T/n_{T}} & \frac{\partial}{\partial\beta'}\widehat{c}^{*}\left(rn_{T}/T,\,k\right)\Bigr\Vert|_{\beta=\bar{\beta}}\label{Eq. A.10 Andrews 91-1}\\
 & =\sup_{0\leq k\leq b_{\theta_{1},T}^{-1}}\Biggl\Vert\frac{n_{T}}{T}\sum_{r=0}^{T/n_{T}}\left(Tb_{\theta_{2},T}\right)^{-1}\sum_{s=k+1}^{T}K_{2}^{*}\left(\frac{\left(r+1\right)n_{T}-\left(s+k/2\right)}{Tb_{\theta_{2},T}}\right)\nonumber \\
 & \quad\times\left(V_{s}^{*}\left(\beta\right)\frac{\partial}{\partial\beta'}V^{*}{}_{s-k}\left(\beta\right)+\frac{\partial}{\partial\beta'}V_{s}^{*}\left(\beta\right)V^{*}{}_{s-k}\left(\beta\right)\right)\Biggr\Vert|_{\beta=\bar{\beta}}\nonumber \\
 & \leq2\left(\frac{n_{T}}{T}\sum_{r=0}^{T/n_{T}}\left(Tb_{\theta_{2},T}\right)^{-1}\sum_{s=1}^{T}K_{2}^{*}\left(\frac{\left(r+1\right)n_{T}-s}{Tb_{\theta_{2},T}}\right)^{2}\sup_{\beta}\left(V_{s}^{*}\left(\beta\right)\right)^{2}\right)^{1/2}\nonumber \\
 & \quad\times\left(\left(Tb_{\theta_{2},T}\right)^{-1}\sum_{s=1}^{T}K_{2}^{*}\left(\frac{\left(r+1\right)n_{T}-s}{Tb_{\theta_{2},T}}\right)^{2}\sup_{\beta}\left\Vert \frac{\partial}{\partial\beta'}V_{s}^{*}\left(\beta\right)\right\Vert ^{2}\right)^{1/2}+o_{\mathbb{P}}\left(1\right)\nonumber \\
 & =O_{\mathbb{P}}\left(1\right),\nonumber 
\end{align}
where $O_{\mathbb{P}}\left(1\right)$ does not depend on $k$ and
we have used the boundedness of the kernel $K_{2}$, the uniform
Lipschitz continuity of $K_{2}$, the fact that $b_{\theta_{1},T}^{-1}/\left(Tb_{\theta_{2,T}}\right)\rightarrow0$,
Assumption \ref{Assumption B}-(ii,iii) and Markov's inequality to
each term in parentheses; also $\sup_{s\geq1}\mathbb{E}\sup_{\beta}||V_{s}^{*}\left(\beta\right)||^{2}<\infty$
under Assumption \ref{Assumption B}-(ii,iii) by a mean-value expansion
and, 
\begin{align}
\left(Tb_{\theta_{2},T}\right)^{-1}\sum_{s=1}^{T}K_{2}^{*}\left(\left(\left(r+1\right)n_{T}-s\right)/Tb_{\theta_{2},T}\right)^{2} & \rightarrow\int_{0}^{1}K_{2}^{2}\left(x\right)dx<\infty.\label{Eq. (A.10) Andrews 91 for K2}
\end{align}
 Then, \eqref{eq (A.9) Andrews-1} is such that
\begin{align*}
b_{\theta_{1},T} & \sum_{k=T+1}^{T-1}K_{1}\left(b_{\theta_{1},T},k\right)\frac{\partial}{\partial\beta'}\widehat{\Gamma}^{*}\left(k\right)|_{\beta=\bar{\beta}}\sqrt{n_{T}}\left(\widehat{\beta}-\beta_{0}\right)\\
 & =b_{\theta_{1},T}\sum_{k=-T+1}^{T-1}K_{1}\left(b_{\theta_{1},T}k\right)O_{\mathbb{P}}\left(1\right)O_{\mathbb{P}}\left(1\right)\\
 & =O_{\mathbb{P}}\left(1\right),
\end{align*}
where the last equality uses $b_{\theta_{1},T}\sum_{k=-T+1}^{T-1}|K_{1}(b_{\theta_{1},T}k)|\rightarrow\int|K_{1}\left(x\right)|dx<\infty.$
This concludes the proof of the lemma because $\sqrt{n_{T}}b_{\theta_{1},T}\rightarrow\infty$
by assumption. $\square$
\begin{lem}
\label{Lemma J*_That(opt_b1, 2) - J*_hat(b1,2 hat)=00003Do(1)}Under
the assumptions of Theorem \ref{Theorem 1 MSE Prew - Andrews 90}-(i),
we have
\begin{equation}
\widehat{J}_{T}^{*}\left(b_{\theta_{1},T},\,b_{\theta_{2},T}\right)-\widehat{J}_{T}^{*}\left(\widehat{b}_{1,T}^{*},\,\widehat{b}_{2,T}^{*}\right)=o_{\mathbb{P}}\left(1\right).\label{Eq. (J_hat* - J*) - 1st term =00003D op(1)}
\end{equation}
\end{lem}
\noindent\textit{Proof.} Let   $S_{T}=\left\lfloor b_{\theta_{1},T}^{-r}\right\rfloor $
and 
\begin{align*}
r\in( & \max\left\{ \left(12b-10q-5\right)/12\left(b-1\right),\,q/\left(l-1\right)\right\} \\
 & \min\left\{ \left(10q+17\right)/24,\,5q/6+5/12,\,1\right\} ).
\end{align*}
We will use the following decomposition, 
\begin{align}
\widehat{J}_{T}^{*}\left(\widehat{b}_{1,T}^{*},\,\widehat{b}_{2,T}^{*}\right)-\widehat{J}_{T}^{*}\left(b_{\theta_{1},T},\,b_{\theta_{2},T}\right) & =\left(\widehat{J}_{T}^{*}\left(\widehat{b}_{1,T}^{*},\,\widehat{b}_{2,T}^{*}\right)-\widehat{J}_{T}^{*}\left(b_{\theta_{1},T},\,\widehat{b}_{2,T}^{*}\right)\right)\label{eq (Decomposition J_T proof of Theorem 3 Andrews91)}\\
 & \quad+\left(\widehat{J}_{T}^{*}\left(b_{\theta_{1},T},\,\widehat{b}_{2,T}^{*}\right)-\widehat{J}_{T}^{*}\left(b_{\theta_{1},T},\,b_{\theta_{2},T}\right)\right).\nonumber 
\end{align}
Let $N_{1}\triangleq\left\{ -S_{T},\,-S_{T}+1,\ldots,\,-1,\,1,\ldots,\,S_{T}-1,\,S_{T}\right\} $,
and $N_{2}\triangleq\left\{ -T+1,\ldots,\,-S_{T}-1,\,S_{T}+1,\ldots,\,T-1\right\} $.
Let us consider the first term above,
\begin{align}
\widehat{J}_{T}^{*} & \left(\widehat{b}_{1,T}^{*},\,\widehat{b}_{2,T}^{*}\right)-\widehat{J}_{T}^{*}\left(b_{\theta_{1},T},\,\widehat{b}_{2,T}^{*}\right)\label{Eq. 23}\\
 & =\sum_{k\in N_{1}}\left(K_{1}\left(\widehat{b}_{1,T}^{*}k\right)-K_{1}\left(b_{\theta_{1},T}k\right)\right)\widehat{\Gamma}^{*}\left(k\right)\nonumber \\
 & \quad+\sum_{k\in N_{2}}K_{1}\left(\widehat{b}_{1,T}^{*}k\right)\widehat{\Gamma}^{*}\left(k\right)-\sum_{k\in N_{2}}K_{1}\left(b_{\theta_{1},T}k\right)\widehat{\Gamma}^{*}\left(k\right)\nonumber \\
 & \triangleq A_{1,T}+A_{2,T}-A_{3,T}.\nonumber 
\end{align}
We first show that $A_{1,T}\overset{\mathbb{P}}{\rightarrow}0$. Let
$A_{1,1,T}$ denote $A_{1,T}$ with the summation restricted over
positive integers $k$. Let $\widetilde{n}_{T}=\inf\left\{ T/n_{3,T},\,\sqrt{n_{2,T}}\right\} $.
We can use the Liptchitz condition on $K_{1}\left(\cdot\right)\in\boldsymbol{K}_{3}$
to yield, 
\begin{align}
\left|A_{1,1,T}\right| & \leq\sum_{k=1}^{S_{T}}C_{2}\left|\widehat{b}_{1,T}^{*}-b_{\theta_{1},T}\right|k\left|\widehat{\Gamma}^{*}\left(k\right)\right|\label{Eq. 24-1}\\
 & \leq C\left|\widehat{\phi}_{D}\left(q\right)^{1/\left(2q+1\right)}-\phi_{\theta^{*}}^{1/\left(2q+1\right)}\right|\left(\widehat{\phi}_{D}\left(q\right)\phi_{\theta^{*}}\right)^{-1/\left(2q+1\right)}\left(T\widehat{b}_{2,T}^{*}\right)^{-1/\left(2q+1\right)}\sum_{k=1}^{S_{T}}k\left|\widehat{\Gamma}^{*}\left(k\right)\right|,\nonumber 
\end{align}
for some $C<\infty$. By Assumption \ref{Assumption E-F-G}-(i),
\begin{align*}
\left|\widehat{\phi}_{D}\left(q\right)^{1/\left(2q+1\right)}-\phi_{\theta^{*}}^{1/\left(2q+1\right)}\right|\left(\widehat{\phi}_{D}\left(q\right)\phi_{\theta^{*}}\right)^{-1/\left(2q+1\right)} & =O_{\mathbb{P}}\left(1\right).
\end{align*}
Using the delta method, it suffices to show that $B_{1,T}+B_{2,T}+B_{3,T}\overset{\mathbb{P}}{\rightarrow}0$,
where 
\begin{align}
B_{1,T} & =\left(T\widehat{b}_{2,T}^{*}\right)^{-1/\left(2q+1\right)}\sum_{k=1}^{S_{T}}k\left|\widehat{\Gamma}^{*}\left(k\right)-\widetilde{\Gamma}^{*}\left(k\right)\right|\label{Eq. A.25 Andrews 91-1}\\
B_{2,T} & =\left(T\widehat{b}_{2,T}^{*}\right)^{-1/\left(2q+1\right)}\sum_{k=1}^{S_{T}}k\left|\widetilde{\Gamma}^{*}\left(k\right)-\Gamma_{T}^{*}\left(k\right)\right|\nonumber \\
B_{3,T} & =\left(T\widehat{b}_{2,T}^{*}\right)^{-1/\left(2q+1\right)}\sum_{k=1}^{S_{T}}k\left|\Gamma_{T}^{*}\left(k\right)\right|,\nonumber 
\end{align}
with $\Gamma_{T}^{*}\left(k\right)\triangleq\left(n_{T}/T\right)\sum_{r=0}^{\left\lfloor T/n_{T}\right\rfloor }c^{*}\left(rn_{T}/T,\,k\right)$
and $\widetilde{\Gamma}^{*}\left(k\right)$ is defined as $\widehat{\Gamma}^{*}\left(k\right)$
but with $\widehat{V}_{t}^{*}$ replaced by $V_{t}^{*}$.  By a mean-value
expansion, we have 
\begin{align}
B_{1,T} & \leq\left(T\widehat{b}_{2,T}^{*}\right)^{-1/\left(2q+1\right)}n_{T}^{-1/2}\sum_{k=1}^{S_{T}}k\left|\left(\frac{\partial}{\partial\beta'}\widehat{\Gamma}^{*}\left(k\right)|_{\beta=\overline{\beta}}\right)\sqrt{n_{T}}\left(\widehat{\beta}-\beta_{0}\right)\right|\label{Eq. A.26-1}\\
 & \leq C\left(T\widehat{b}_{2,T}^{*}\right)^{-1/\left(2q+1\right)}\left(Tb_{\theta_{2,T}}\right)^{2r/\left(2q+1\right)}n_{T}^{-1/2}\sup_{k\geq1}\left\Vert \frac{\partial}{\partial\beta}\widehat{\Gamma}^{*}\left(k\right)|_{\beta=\overline{\beta}}\right\Vert \sqrt{n_{T}}\left\Vert \widehat{\beta}-\beta_{0}\right\Vert ,\nonumber 
\end{align}
since $r<\left(10q+17\right)/24$, and $\sup_{k\geq1}||\left(\partial/\partial\beta\right)\widehat{\Gamma}^{*}\left(k\right)|_{\beta=\overline{\beta}}||=O_{\mathbb{P}}\left(1\right)$
using \eqref{Eq. A.10 Andrews 91-1} and Assumption \ref{Assumption B}-(ii,iii)
(the latter continues to hold for $\{V_{t}^{*}\}$). In addition,
\begin{align}
\mathbb{E}\left(B_{2,T}^{2}\right) & \leq\mathbb{E}\left(\left(T\widehat{b}_{2,T}^{*}\right)^{-2/\left(2q+1\right)}\sum_{k=1}^{S_{T}}\sum_{j=1}^{S_{T}}kj\left|\widetilde{\Gamma}^{*}\left(k\right)-\Gamma_{T}^{*}\left(k\right)\right|\left|\widetilde{\Gamma}^{*}\left(j\right)-\Gamma_{T}^{*}\left(j\right)\right|\right)\label{Eq. A.27-1}\\
 & \leq\left(T\widehat{b}_{2,T}^{*}\right)^{-2/\left(2q+1\right)-1}S_{T}^{4}\sup_{k\geq1}T\widehat{b}_{2,T}^{*}\mathrm{Var}\left(\widetilde{\Gamma}^{*}\left(k\right)\right)\nonumber \\
 & \leq\left(T\widehat{b}_{2,T}^{*}\right)^{-2/\left(2q+1\right)-1}\left(T\overline{b}_{\theta_{2,T}}\right)^{4r/\left(2q+1\right)}\sup_{k\geq1}T\widehat{b}_{2,T}^{*}\mathrm{Var}\left(\widetilde{\Gamma}^{*}\left(k\right)\right)\nonumber \\
 & \leq\left(\widehat{b}_{2,T}^{*}\right)^{-2/\left(2q+1\right)-1}T^{-1-2/\left(2q+1\right)}T^{16r/5\left(2q+1\right)}\sup_{k\geq1}T\widehat{b}_{2,T}^{*}\mathrm{Var}\left(\widetilde{\Gamma}^{*}\left(k\right)\right)\rightarrow0,\nonumber 
\end{align}
given that $r<\left(3+2q\right)/4$ and $\sup_{k\geq1}T\widehat{b}_{2,T}^{*}\mathrm{Var}(\widetilde{\Gamma}^{*}(k))=O\left(1\right)$
by Lemma S.A.5 in \citeReferencesSupp{casini_hac} that also holds
with $\widetilde{\Gamma}^{*}\left(k\right)$ in place of $\widetilde{\Gamma}\left(k\right)$.
Next,
\begin{align}
B_{3,T} & \leq\left(T\widehat{b}_{2,T}^{*}\right)^{-1/\left(2q+1\right)}S_{T}\sum_{k=1}^{\infty}\left|\Gamma_{T}^{*}\left(k\right)\right|\label{Eq. A.28-1}\\
 & \leq\left(T\widehat{b}_{2,T}^{*}\right)^{\left(r-1\right)/\left(2q+1\right)}O_{\mathbb{P}}\left(1\right)\rightarrow0,\nonumber 
\end{align}
using Assumption \ref{Assumption A - Dependence}-(i) since $r<1$.
 This gives $A_{1,T}\overset{\mathbb{P}}{\rightarrow}0$. Next,
we show that $A_{2,T}\overset{\mathbb{P}}{\rightarrow}0$. Let $A_{2,1,T}=L_{1,T}+L_{2,T}+L_{3,T}$,
where 
\begin{align}
L_{1,T} & =\sum_{k=S_{T}+1}^{T-1}K_{1}\left(\widehat{b}_{1,T}^{*}k\right)\left(\widehat{\Gamma}^{*}\left(k\right)-\widetilde{\Gamma}^{*}\left(k\right)\right),\label{Eq. A.29}\\
L_{2,T} & =\sum_{k=S_{T}+1}^{T-1}K_{1}\left(\widehat{b}_{1,T}^{*}k\right)\left(\widetilde{\Gamma}^{*}\left(k\right)-\Gamma_{T}^{*}\left(k\right)\right),\quad\mathrm{and}\nonumber \\
L_{3,T} & =\sum_{k=S_{T}+1}^{T-1}K_{1}\left(\widehat{b}_{1,T}^{*}k\right)\Gamma_{T}^{*}\left(k\right).\nonumber 
\end{align}
We apply a mean-value expansion and use $\sqrt{n_{T}}(\widehat{\beta}-\beta_{0})=O_{\mathbb{P}}\left(1\right)$
as well as \eqref{Eq. A.10 Andrews 91-1} to obtain
\begin{align}
\left|L_{1,T}\right| & =n_{T}^{-1/2}\sum_{k=S_{T}+1}^{T-1}C_{1}\left(\widehat{b}_{1,T}^{*}k\right)^{-b}\left|\left(\frac{\partial}{\partial\beta'}\widehat{\Gamma}^{*}\left(k\right)\right)|_{\beta=\overline{\beta}}\sqrt{n_{T}}\left(\widehat{\beta}-\beta_{0}\right)\right|\label{Eq. (30)-2}\\
 & =T^{-1/3+4b/5\left(2q+1\right)}\sum_{k=S_{T}+1}^{T-1}C_{1}k^{-b}\left|\left(\frac{\partial}{\partial\beta'}\widehat{\Gamma}^{*}\left(k\right)\right)|_{\beta=\overline{\beta}}\sqrt{n_{T}}\left(\widehat{\beta}-\beta_{0}\right)\right|\nonumber \\
 & =T^{-1/3+4b/5\left(2q+1\right)+4r\left(1-b\right)/5\left(2q+1\right)}\left|\left(\frac{\partial}{\partial\beta'}\widehat{\Gamma}^{*}\left(k\right)\right)|_{\beta=\overline{\beta}}\sqrt{n_{T}}\left(\widehat{\beta}-\beta_{0}\right)\right|\nonumber \\
 & =T^{-1/3+4b/5\left(2q+1\right)+4r\left(1-b\right)/5\left(2q+1\right)}O\left(1\right)O_{\mathbb{P}}\left(1\right),\nonumber 
\end{align}
 which converges to zero since $r>\left(12b-10q-5\right)/12\left(b-1\right)$.
Next, 
\begin{align}
\left|L_{2,T}\right| & =\sum_{k=S_{T}+1}^{T-1}C_{1}\left(\widehat{b}_{1,T}^{*}k\right)^{-b}\left|\widetilde{\Gamma}^{*}\left(k\right)-\Gamma_{T}^{*}\left(k\right)\right|\label{Eq. 31-1}\\
 & =C_{1}\left(qK_{1,q}^{2}\widehat{\phi}_{D}\left(q\right)\right)^{b/\left(2q+1\right)}T^{b/\left(2q+1\right)-1/2}\left(\widehat{b}_{2,T}^{*}\right){}^{b/\left(2q+1\right)-1/2}\left(\sum_{k=S_{T}+1}^{T-1}k^{-b}\right)\sqrt{T\widehat{b}_{2,T}^{*}}\left|\widetilde{\Gamma}^{*}\left(k\right)-\Gamma_{T}^{*}\left(k\right)\right|.\nonumber 
\end{align}
Note that,
\begin{align}
\mathbb{E} & \left(T^{b/\left(2q+1\right)-1/2}\left(\widehat{b}_{2,T}^{*}\right)^{b/\left(2q+1\right)-1/2}\sum_{k=S_{T}}^{T-1}k^{-b}\sqrt{T\widehat{b}_{2,T}^{*}}\left|\widetilde{\Gamma}^{*}\left(k\right)-\Gamma_{T}^{*}\left(k\right)\right|\right)^{2}\label{Eq. 32 Andrews 91-1}\\
 & \leq T^{2b/\left(2q+1\right)-1}\left(\widehat{b}_{2,T}^{*}\right){}^{2b/\left(2q+1\right)-1}\left(\sum_{k=S_{T}}^{T-1}k^{-b}\right)^{2}O\left(1\right)\nonumber \\
 & =T^{2b/\left(2q+1\right)-1}\left(\widehat{b}_{2,T}^{*}\right){}^{2b/\left(2q+1\right)-1}S_{T}^{2\left(1-b\right)}O\left(1\right)\rightarrow0,\nonumber 
\end{align}
 since $r>(b-1/2-q)/(b-1)$ and $T\widehat{b}_{2,T}^{*}\mathrm{Var}(\widetilde{\Gamma}^{*}(k))=O\left(1\right)$,
as above. Equations \eqref{Eq. 31-1}-\eqref{Eq. 32 Andrews 91-1}
combine to yield $L_{2,T}\overset{\mathbb{P}}{\rightarrow}0$, since
$\widehat{\phi}_{D}\left(q\right)=O_{\mathbb{P}}\left(1\right)$ by
Assumption \ref{Assumption E-F-G}-(i). Let us turn to $L_{3,T}$.
We have,
\begin{align}
\left|\sum_{k=S_{T}+1}^{T-1}K_{1}\left(\widehat{b}_{1,T}^{*}k\right)\Gamma_{T}^{*}\left(k\right)\right| & \leq\sum_{k=S_{T}+1}^{T-1}\frac{n_{T}}{T}\sum_{r=0}^{\left\lfloor T/n_{T}\right\rfloor }\left|c^{*}\left(rn_{T}/T,\,k\right)\right|\label{Eq. (33)-1}\\
 & \leq\sum_{k=S_{T}+1}^{T-1}\sup_{u\in\left[0,\,1\right]}\left|c^{*}\left(u,\,k\right)\right|\rightarrow0.\nonumber 
\end{align}
Equations \eqref{Eq. (30)-2}-\eqref{Eq. (33)-1} imply $A_{2,T}\overset{\mathbb{P}}{\rightarrow}0$.
An analogous argument yields $A_{3,T}\overset{\mathbb{P}}{\rightarrow}0$.
It remains to show that $(\widehat{J}_{T}(b_{\theta_{1},T},\,\widehat{b}_{2,T}^{*})-\widehat{J}_{T}(b_{\theta_{1},T},\,\overline{b}_{\theta_{2},T}))\overset{\mathbb{P}}{\rightarrow}0$.
Its proof is the same as in Theorem 5.1-(i) in \citeReferencesSupp{casini_hac}
which can be repeated given the conditions $n_{T}^{-1/2}/(\widehat{b}_{1,T}^{*})\rightarrow0,$
$r<5q/6+5/12,$ and $r>(b-1/2-q)/(b-1)$. $\square$
\begin{lem}
\label{Lemma 1 for Theorem 1 (ii)}Under the assumptions of Theorem
\ref{Theorem 1 MSE Prew - Andrews 90}-(ii), we have
\begin{align*}
\sqrt{Tb_{\theta_{1},T}b_{\theta_{2},T}}\left(\widehat{J}_{T}^{*}\left(b_{\theta_{1},T},\,b_{\theta_{2},T}\right)-J_{T}^{*}\right) & =O_{\mathbb{P}}\left(1\right).
\end{align*}
\end{lem}
\noindent\textit{Proof.} Write
\begin{align*}
\sqrt{Tb_{\theta_{1},T}b_{\theta_{2},T}}\left(\widehat{J}_{T}^{*}\left(b_{\theta_{1},T},\,b_{\theta_{2},T}\right)-J_{T}^{*}\right) & =\sqrt{Tb_{\theta_{1},T}b_{\theta_{2},T}}\left(\widehat{J}_{T}^{*}\left(b_{\theta_{1},T},\,b_{\theta_{2},T}\right)-\widetilde{J}_{T}^{*}+\widetilde{J}_{T}^{*}-J_{T}^{*}\right).
\end{align*}
Applying Theorem 3.1-(ii) in \citeReferencesSupp{casini_hac} with
$V_{s}^{*}$ in place of $V_{s}$, we have $\sqrt{Tb_{\theta_{1},T}b_{\theta_{2},T}}(\widetilde{J}_{T}^{*}-J_{T}^{*})=O_{\mathbb{P}}\left(1\right)$.
Thus, it is sufficient to show $\sqrt{Tb_{\theta_{1},T}b_{\theta_{2},T}}(\widehat{J}_{T}^{*}(b_{\theta_{1},T},\,b_{\theta_{2},T})-\widetilde{J}_{T}^{*})=o_{\mathbb{P}}\left(1\right).$
A second-order Taylor expansion gives
\begin{align*}
\sqrt{Tb_{\theta_{1},T}b_{\theta_{2},T}}\left(\widehat{J}_{T}^{*}-\widetilde{J}_{T}^{*}\right) & =\left[\frac{\sqrt{Tb_{\theta_{2},T}}}{\sqrt{n_{T}}}\sqrt{b_{\theta_{1},T}}\frac{\partial}{\partial\beta'}\widetilde{J}_{T}^{*}\left(\beta_{0}\right)\right]\sqrt{n_{T}}\left(\widehat{\beta}-\beta_{0}\right)\\
 & \quad+\frac{1}{2}\sqrt{n_{T}}\left(\widehat{\beta}-\beta_{0}\right)'\left[\frac{\sqrt{Tb_{\theta_{2},T}}}{n_{T}}\sqrt{b_{\theta_{1},T}}\frac{\partial^{2}}{\partial\beta\partial\beta'}\widetilde{J}_{T}^{*}\left(\overline{\beta}\right)\right]\sqrt{n_{T}}\left(\widehat{\beta}-\beta_{0}\right)\\
 & \triangleq G_{T}'\sqrt{n_{T}}\left(\widehat{\beta}-\beta_{0}\right)+\frac{1}{2}\sqrt{n_{T}}\left(\widehat{\beta}-\beta_{0}\right)'H_{T}\sqrt{n_{T}}\left(\widehat{\beta}-\beta_{0}\right).
\end{align*}
Using Assumption \ref{Assumption C Andrews 91}-(ii) and proceeding
as in the proof of Lemma \ref{Lemma J*(b1*, b2*) - J* =00003D o(1)},
\begin{align*}
\sup_{0\leq k\leq b_{\theta_{1},T}^{-1}}\biggl\Vert & \frac{\partial^{2}}{\partial\beta\partial\beta'}\widehat{c}^{*}\left(rn_{T}/T,\,k\right)\biggr\Vert\biggl|_{\beta=\bar{\beta}}\\
 & =\left\Vert \left(Tb_{\theta_{2},T}\right)^{-1}\sum_{s=k+1}^{T}K_{2}^{*}\left(\frac{\left(\left(r+1\right)n_{T}-\left(s+k/2\right)\right)/T}{b_{\theta_{2},T}}\right)\left(\frac{\partial^{2}}{\partial\beta\partial\beta'}V_{s}^{*}\left(\beta\right)V_{s-k}^{*}\left(\beta\right)\right)\right\Vert \biggl|_{\beta=\bar{\beta}}\\
 & =O_{\mathbb{P}}\left(1\right),
\end{align*}
 and thus, 
\begin{align*}
\left\Vert H_{T}\right\Vert  & \leq\left(\frac{Tb_{\theta_{2},T}b_{\theta_{1},T}}{n_{T}^{2}}\right)^{1/2}\sum_{k=-T+1}^{T-1}\left|K_{1}\left(b_{\theta_{1},T}k\right)\right|\sup_{\beta\in\Theta}\left\Vert \frac{\partial^{2}}{\partial\beta\partial\beta'}\widehat{\Gamma}^{*}\left(k\right)\right\Vert \\
 & \leq\left(\frac{Tb_{\theta_{2},T}b_{\theta_{1},T}}{n_{T}^{2}}\right)^{1/2}\sum_{k=-T+1}^{T-1}\left|K_{1}\left(b_{\theta_{1},T}k\right)\right|O_{\mathbb{P}}\left(1\right)\\
 & \leq\left(\frac{Tb_{\theta_{2},T}}{n_{T}^{2}b_{\theta_{1},T}}\right)^{1/2}b_{\theta_{1},T}\sum_{k=-T+1}^{T-1}\left|K_{1}\left(b_{\theta_{1},T}k\right)\right|O_{\mathbb{P}}\left(1\right)=o_{\mathbb{P}}\left(1\right),
\end{align*}
since $Tb_{\theta_{2},T}/(n_{T}^{2}b_{\theta_{1},T})\rightarrow0$.
Next, we want to show that $G_{T}=o_{\mathbb{P}}\left(1\right)$.
Following \citeReferencesSupp{andrews:91} (cf. the last paragraph
of p. 852), we apply the results of Theorem 3.1-(i,ii) in \citeReferencesSupp{casini_hac}
to $\widetilde{J}_{T}^{*}$ where the latter is constructed using
$(V_{t}^{*\prime},\,\partial V_{t}^{*}/\partial\beta'-\mathbb{E}(\partial V_{t}^{*}/\partial\beta'))'$
rather than just with $V_{t}^{*}$. The first row and column of the
off-diagonal elements of this $\widetilde{J}_{T}^{*}$ (written as
column vectors) are now 
\begin{align*}
A_{1} & \triangleq\sum_{k=-T+1}^{T-1}K_{1}\left(b_{\theta_{1},T}k\right)\frac{n_{T}}{T}\sum_{r=0}^{T/n_{T}}\frac{1}{Tb_{\theta_{2},T}}\\
 & \quad\times\sum_{s=k+1}^{T}K_{2}^{*}\left(\frac{\left(\left(r+1\right)n_{T}-\left(s+k/2\right)\right)/T}{b_{\theta_{2},T}}\right)V_{s}^{*}\left(\frac{\partial}{\partial\beta}V_{s-k}^{*}-\mathbb{E}\left(\frac{\partial}{\partial\beta}V_{s}^{*}\right)\right)\\
A_{2} & \triangleq\sum_{k=-T+1}^{T-1}K_{1}\left(b_{\theta_{1},T}k\right)\frac{n_{T}}{T}\sum_{r=0}^{T/n_{T}}\frac{1}{Tb_{2,T}}\\
 & \quad\times\sum_{s=k+1}^{T}K_{2}^{*}\left(\frac{\left(\left(r+1\right)n_{T}-\left(s+k/2\right)\right)/T}{b_{\theta_{2},T}}\right)\left(\frac{\partial}{\partial\beta}V_{s}^{*}-\mathbb{E}\left(\frac{\partial}{\partial\beta}V_{s}^{*}\right)\right)V_{s-k}^{*}.
\end{align*}
By Theorem 3.1-(i,ii) in \citeReferencesSupp{casini_hac} each expression
above is $O_{\mathbb{P}}\left(1\right)$. Given, 
\begin{align*}
G_{T} & \leq\frac{\sqrt{Tb_{\theta_{2},T}}}{\sqrt{n_{T}}}\sqrt{b_{\theta_{1},T}}\left(A_{1}+A_{2}\right)+\frac{\sqrt{Tb_{\theta_{2},T}}}{\sqrt{n_{T}}}\sqrt{b_{\theta_{1},T}}\sum_{k=-T+1}^{T-1}K_{1}\left(b_{\theta_{1},T}k\right)\frac{n_{T}}{T}\sum_{r=0}^{T/n_{T}}\frac{1}{Tb_{\theta_{2},T}}\\
 & \quad\times\sum_{s=k+1}^{T}K_{2}^{*}\left(\frac{\left(\left(r+1\right)n_{T}-\left(s+k/2\right)\right)/T}{b_{\theta_{2},T}}\right)\left|V_{s}^{*}+V_{s-k}^{*}\right|\left|\mathbb{E}\left(\frac{\partial}{\partial\beta}V_{s}^{*}\right)\right|\\
 & \triangleq\frac{\sqrt{Tb_{\theta_{2},T}}}{\sqrt{n_{T}}}\sqrt{b_{\theta_{1},T}}\left(A_{1}+A_{2}\right)+A_{3}\sup_{s}\left|\mathbb{E}\left(\frac{\partial}{\partial\beta}V_{s}^{*}\right)\right|,
\end{align*}
and the fact that $Tb_{\theta_{2},T}b_{\theta_{1},T}/n_{T}\rightarrow0$
it remains to show that $A_{3}$ is $o_{\mathbb{P}}\left(1\right).$
Note that 
\begin{align*}
\mathbb{E}\left(A_{3}^{2}\right) & \leq\frac{Tb_{\theta_{2},T}}{n_{T}}b_{1,T}\sum_{k=-T+1}^{T-1}\sum_{j=-T+1}^{T-1}\left|K_{1}\left(b_{\theta_{1},T}k\right)K_{1}\left(b_{\theta_{1},T}j\right)\right|4\left(\frac{n_{T}}{T}\right)^{2}\sum_{r=0}^{T/n_{T}}\sum_{b=0}^{T/n_{T}}\\
 & \quad\times\frac{1}{Tb_{\theta_{2},T}}\frac{1}{Tb_{\theta_{2},T}}\sum_{s=1}^{T}\sum_{l=1}^{T}K_{2}^{*}\left(\frac{\left(\left(r+1\right)n_{T}-\left(s+k/2\right)\right)/T}{b_{\theta_{2},T}}\right)\\
 & \quad\times K_{2}^{*}\left(\frac{\left(\left(b+1\right)n_{T}-\left(l+j/2\right)\right)/T}{b_{\theta_{2},T}}\right)\left|\mathbb{E}\left(V_{s}^{*}V_{l}^{*}\right)\right|,
\end{align*}
 and that $\mathbb{E}\left(V_{s}^{*}V_{l}^{*}\right)=c^{*}\left(u,\,h\right)+O\left(T^{-1}\right)$
uniformly in $h=s-l$ and $u=s/T$ by Lemma S.A.1 in \citeReferencesSupp{casini_hac}.
Since $\sum_{h=-\infty}^{\infty}\sup_{u\in\left[0,\,1\right]}\left|c^{*}\left(u,\,h\right)\right|<\infty$,
  
\begin{align*}
\mathbb{E}\left(A_{3}^{2}\right) & \leq\frac{1}{n_{T}b_{\theta_{1},T}}\left(b_{\theta_{1},T}\sum_{k=-T+1}^{T-1}\left|K_{1}\left(b_{\theta_{1},T}k\right)\right|\right)^{2}\int_{0}^{1}K_{2}^{2}\left(x\right)dx\int_{0}^{1}\sum_{h=-\infty}^{\infty}\left|c^{*}\left(u,\,h\right)\right|du=o\left(1\right).
\end{align*}
This implies $G_{T}=o_{\mathbb{P}}\left(1\right)$ which concludes
the proof. $\square$ 
\begin{lem}
\label{Lemma 2 for Theorem 1 (ii)}Under the assumptions of Theorem
\ref{Theorem 1 MSE Prew - Andrews 90}-(ii), we have
\begin{align*}
\sqrt{Tb_{\theta_{1},T}b_{\theta_{2},T}}\left(\widehat{J}_{T}^{*}\left(\widehat{b}_{1,T}^{*},\,\widehat{b}_{2,T}^{*}\right)-\widehat{J}_{T}^{*}\left(b_{\theta_{1},T},\,b_{\theta_{2},T}\right)\right) & =o_{\mathbb{P}}\left(1\right).
\end{align*}
\end{lem}
\noindent\textit{Proof.} Let  
\begin{align*}
r\in & (\max\{\{\{\left(-10+4q+24b\right)/(24\left(b-1\right))\},\,\left(8b-4\right)/\left(b-1\right)\left(10q+5\right)\},\\
 & \,\min\left\{ 2/3+q/3\right\} ),
\end{align*}
    and $S_{T}=\bigl\lfloor b_{\theta_{1},T}^{-r}\bigr\rfloor$.
We will use the following decomposition 
\begin{align}
\widehat{J}_{T}^{*}\left(\widehat{b}_{1,T}^{*},\,\widehat{b}_{2,T}^{*}\right)-\widehat{J}_{T}^{*}\left(b_{\theta_{1},T},\,b_{\theta_{2},T}\right) & =\left(\widehat{J}_{T}^{*}\left(\widehat{b}_{1,T}^{*},\,\widehat{b}_{2,T}^{*}\right)-\widehat{J}_{T}^{*}\left(b_{\theta_{1},T},\,\widehat{b}_{2,T}^{*}\right)\right)\label{eq (Decomposition J_T proof of Theorem 3 Andrews91)-1}\\
 & \quad+\left(\widehat{J}_{T}^{*}\left(b_{\theta_{1},T},\,\widehat{b}_{2,T}^{*}\right)-\widehat{J}_{T}^{*}\left(b_{\theta_{1},T},\,b_{\theta_{2},T}\right)\right).\nonumber 
\end{align}
Let
\begin{align*}
N_{1} & \triangleq\left\{ -S_{T},\,-S_{T}+1,\ldots,\,-1,\,1,\ldots,\,S_{T}-1,\,S_{T}\right\} \\
N_{2} & \triangleq\left\{ -T+1,\ldots,\,-S_{T}-1,\,S_{T}+1,\ldots,\,T-1\right\} .
\end{align*}
Let us consider the first term above, 
\begin{align}
T^{8q/10\left(2q+1\right)} & \left(\widehat{J}_{T}^{*}\left(\widehat{b}_{1,T}^{*},\,\widehat{b}_{2,T}^{*}\right)-\widehat{J}_{T}^{*}\left(b_{\theta_{1},T},\,\widehat{b}_{2,T}^{*}\right)\right)\label{Eq. 23-1}\\
 & =T^{8q/10\left(2q+1\right)}\sum_{k\in N_{1}}\left(K_{1}\left(\widehat{b}_{1,T}^{*}k\right)-K_{1}\left(b_{\theta_{1},T}k\right)\right)\widehat{\Gamma}^{*}\left(k\right)\nonumber \\
 & \quad+T^{8q/10\left(2q+1\right)}\sum_{k\in N_{2}}K_{1}\left(\widehat{b}_{1,T}^{*}k\right)\widehat{\Gamma}^{*}\left(k\right)\nonumber \\
 & \quad-T^{8q/10\left(2q+1\right)}\sum_{k\in N_{2}}K_{1}\left(b_{\theta_{1},T}k\right)\widehat{\Gamma}^{*}\left(k\right)\nonumber \\
 & \triangleq A_{1,T}+A_{2,T}-A_{3,T}.\nonumber 
\end{align}
 We first show that $A_{1,T}\overset{\mathbb{P}}{\rightarrow}0$.
Let $A_{1,1,T}$ denote $A_{1,T}$ with the summation restricted over
positive integers $k$. Let $\widetilde{n}_{T}=\inf\{T/n_{3,T},\,\sqrt{n_{2,T}}\}$.
We can use the Liptchitz condition on $K_{1}\left(\cdot\right)\in\boldsymbol{K}_{3}$
to yield, 
\begin{align}
\left|A_{1,1,T}\right| & \leq T^{8q/10\left(2q+1\right)}\sum_{k=1}^{S_{T}}C_{2}\left|\widehat{b}_{1,T}^{*}-b_{\theta_{1},T}\right|k\left|\widehat{\Gamma}^{*}\left(k\right)\right|\label{Eq. 24}\\
 & \leq C\widetilde{n}_{T}\left|\widehat{\phi}_{D}\left(q\right)^{1/\left(2q+1\right)}-\phi_{\theta^{*}}^{1/\left(2q+1\right)}\right|\left(\widehat{\phi}_{D}\left(q\right)\phi_{\theta^{*}}\right)^{-1/\left(2q+1\right)}\nonumber \\
 & \quad\left(\widehat{b}_{2,T}^{*}\right)^{-1/\left(2q+1\right)}T^{\left(8q-10\right)/10\left(2q+1\right)}\widetilde{n}_{T}^{-1}\sum_{k=1}^{S_{T}}k\left|\widehat{\Gamma}^{*}\left(k\right)\right|,\nonumber 
\end{align}
for some $C<\infty$. By Assumption \ref{Assumption E-F-G}-(ii),
($\widetilde{n}_{T}|\widehat{\phi}_{D}\left(q\right)-\phi_{\theta^{*}}|=O_{\mathbb{P}}\left(1\right)$)
and using the delta method, it suffices to show that $B_{1,T}+B_{2,T}+B_{3,T}\overset{\mathbb{P}}{\rightarrow}0$,
where 
\begin{align}
B_{1,T} & =\left(\widehat{b}_{2,T}^{*}\right)^{-1/\left(2q+1\right)}T^{\left(8q-10\right)/10\left(2q+1\right)}\widetilde{n}_{T}^{-1}\sum_{k=1}^{S_{T}}k\left|\widehat{\Gamma}^{*}\left(k\right)-\widetilde{\Gamma}^{*}\left(k\right)\right|,\label{Eq. A.25 Andrews 91}\\
B_{2,T} & =\left(\widehat{b}_{2,T}^{*}\right)^{-1/\left(2q+1\right)}T^{\left(8q-10\right)/10\left(2q+1\right)}\widetilde{n}_{T}^{-1}\sum_{k=1}^{S_{T}}k\left|\widetilde{\Gamma}^{*}\left(k\right)-\Gamma_{T}^{*}\left(k\right)\right|,\qquad\mathrm{and}\nonumber \\
B_{3,T} & =\left(\widehat{b}_{2,T}^{*}\right)^{-1/\left(2q+1\right)}T^{\left(8q-10\right)/10\left(2q+1\right)}\widetilde{n}_{T}^{-1}\sum_{k=1}^{S_{T}}k\left|\Gamma_{T}^{*}\left(k\right)\right|.\nonumber 
\end{align}
By a mean-value expansion, we have 
\begin{align}
B_{1,T} & \leq\left(\widehat{b}_{2,T}^{*}\right)^{-1/\left(2q+1\right)}T^{\left(8q-10\right)/10\left(2q+1\right)}\widetilde{n}_{T}^{-1}n_{T}^{-1/2}\sum_{k=1}^{S_{T}}k\left|\left(\frac{\partial}{\partial\beta'}\widehat{\Gamma}^{*}\left(k\right)|_{\beta=\overline{\beta}}\right)\sqrt{n_{T}}\left(\widehat{\beta}-\beta_{0}\right)\right|\label{Eq. A.26}\\
 & \leq C\left(\widehat{b}_{2,T}^{*}\right)^{-1/\left(2q+1\right)}T^{\left(8q-10\right)/10\left(2q+1\right)}\left(Tb_{\theta_{2,T}}\right)^{2r/\left(2q+1\right)}\widetilde{n}_{T}^{-1}n_{T}^{-1/2}\sup_{k\geq1}\left\Vert \frac{\partial}{\partial\beta}\widehat{\Gamma}^{*}\left(k\right)|_{\beta=\overline{\beta}}\right\Vert \sqrt{n_{T}}\left\Vert \widehat{\beta}-\beta_{0}\right\Vert \nonumber \\
 & \leq C\left(\widehat{b}_{2,T}^{*}\right)^{\left(-1+2r\right)/\left(2q+1\right)}T^{\left(8q-10\right)/10\left(2q+1\right)+2r/\left(2q+1\right)-1/3}\widetilde{n}_{T}^{-1}\sup_{k\geq1}\left\Vert \frac{\partial}{\partial\beta}\widehat{\Gamma}^{*}\left(k\right)|_{\beta=\overline{\beta}}\right\Vert \sqrt{n_{T}}\left\Vert \widehat{\beta}-\beta_{0}\right\Vert \overset{\mathbb{P}}{\rightarrow}0,\nonumber 
\end{align}
since $\widetilde{n}_{T}/T^{1/3}\rightarrow\infty$, $r<16q/48+44/48$,
$\sqrt{n_{T}}||\widehat{\beta}-\beta_{0}||=O_{\mathbb{P}}\left(1\right)$,
and $\sup_{k\geq1}||\left(\partial/\partial\beta\right)\widehat{\Gamma}^{*}\left(k\right)|_{\beta=\overline{\beta}}||=O_{\mathbb{P}}\left(1\right)$
using \eqref{Eq. A.10 Andrews 91-1} and Assumption \ref{Assumption B}-(ii,iii).
In addition, 
\begin{align}
\mathbb{E}\left(B_{2,T}^{2}\right) & \leq\mathbb{E}\left(\left(\widehat{b}_{2,T}^{*}\right)^{-2/\left(2q+1\right)}T^{\left(8q-10\right)/5\left(2q+1\right)}\widetilde{n}_{T}^{-2}\sum_{k=1}^{S_{T}}\sum_{j=1}^{S_{T}}kj\left|\widetilde{\Gamma}^{*}\left(k\right)-\Gamma_{T}^{*}\left(k\right)\right|\left|\widetilde{\Gamma}^{*}\left(j\right)-\Gamma_{T}^{*}\left(j\right)\right|\right)\label{Eq. A.27}\\
 & \leq\left(\widehat{b}_{2,T}^{*}\right)^{-2/\left(2q+1\right)-1}T^{\left(8q-10\right)/5\left(2q+1\right)-2/3-1}S_{T}^{4}\sup_{k\geq1}Tb_{2,T}\mathrm{Var}\left(\widetilde{\Gamma}^{*}\left(k\right)\right)\nonumber \\
 & \leq\left(\widehat{b}_{2,T}^{*}\right)^{-2/\left(2q+1\right)-1}T^{\left(8q-10\right)/5\left(2q+1\right)-2/3-1}\left(Tb_{2,T}\right)^{4r/\left(2q+1\right)}\sup_{k\geq1}Tb_{2,T}\mathrm{Var}\left(\widetilde{\Gamma}^{*}\left(k\right)\right)\nonumber \\
 & \leq T^{1/5}T^{2/5\left(2q+1\right)}T^{\left(8q-10\right)/5\left(2q+1\right)-2/3-1}T^{4r/\left(2q+1\right)}T^{-4r/5\left(2q+1\right)}\sup_{k\geq1}Tb_{2,T}\mathrm{Var}\left(\widetilde{\Gamma}^{*}\left(k\right)\right)\rightarrow0,\nonumber 
\end{align}
     given that $\sup_{k\geq1}Tb_{2,T}\mathrm{Var}(\widetilde{\Gamma}^{*}(k))=O\left(1\right)$
using Lemma S.A.5 in \citeReferencesSupp{casini_hac} and $r<46/48+20q/48$.
Assumption \ref{Assumption E-F-G}-(iii) and $\sum_{k=1}^{\infty}k^{1-l}<\infty$
for $l>2$ yield 
\begin{align}
B_{3,T} & \leq\widehat{b}_{2,T}^{-1/\left(2q+1\right)}T^{\left(8q-10\right)/10\left(2q+1\right)}\widetilde{n}_{T}^{-1}C_{3}\sum_{k=1}^{\infty}k^{1-l}\label{Eq. A.28}\\
 & \leq T^{\left(-21-14q\right)/10\left(2q+1\right)}C_{3}\sum_{k=1}^{\infty}k^{1-l}\rightarrow0,\nonumber 
\end{align}
 where we have used the fact that $\widetilde{n}_{T}/T^{1/3}\rightarrow\infty$.
Combining \eqref{Eq. 24}-\eqref{Eq. A.28} we deduce that $A_{1,1,T}\overset{\mathbb{P}}{\rightarrow}0$.
The same argument applied to $A_{1,T}$ where the summation now also
extends over negative integers $k$ gives $A_{1,T}\overset{\mathbb{P}}{\rightarrow}0$.
Next, we show that $A_{2,T}\overset{\mathbb{P}}{\rightarrow}0$. Again,
we use the notation $A_{2,1,T}$ (resp., $A_{2,2,T}$) to denote $A_{2,T}$
with the summation over positive (resp., negative) integers. Let $A_{2,1,T}=L_{1,T}+L_{2,T}+L_{3,T}$,
where 
\begin{align}
L_{1,T}=L_{1,T}^{A}+L_{1,T}^{B} & =T^{8q/10\left(2q+1\right)}\left(\sum_{k=S_{T}+1}^{\left\lfloor D_{T}T^{1/2}\right\rfloor }+\sum_{k=\left\lfloor D_{T}T^{1/2}\right\rfloor +1}^{T-1}\right)K_{1}\left(\widehat{b}_{1,T}^{*}k\right)\left(\widehat{\Gamma}^{*}\left(k\right)-\widetilde{\Gamma}^{*}\left(k\right)\right),\label{Eq. A.29-1}\\
L_{2,T}=L_{2,T}^{A}+L_{2,T}^{B} & =T^{8q/10\left(2q+1\right)}\left(\sum_{k=S_{T}+1}^{\left\lfloor D_{T}T^{1/2}\right\rfloor }+\sum_{k=\left\lfloor D_{T}T^{1/2}\right\rfloor +1}^{T-1}\right)K_{1}\left(\widehat{b}_{1,T}^{*}k\right)\left(\widetilde{\Gamma}^{*}\left(k\right)-\Gamma_{T}^{*}\left(k\right)\right),\quad\nonumber \\
\mathrm{and}\qquad\qquad L_{3,T} & =T^{8q/10\left(2q+1\right)}\sum_{k=S_{T}+1}^{T-1}K_{1}\left(\widehat{b}_{1,T}^{*}k\right)\Gamma_{T}^{*}\left(k\right).\nonumber 
\end{align}
We apply a mean-value expansion, use $\sqrt{n_{T}}(\widehat{\beta}-\beta_{0})=O_{\mathbb{P}}\left(1\right)$
as well as \eqref{Eq. A.10 Andrews 91-1} to obtain
\begin{align}
\left|L_{1,T}^{A}\right| & =T^{8q/10\left(2q+1\right)-1/3}\sum_{k=S_{T}+1}^{\left\lfloor D_{T}T^{1/2}\right\rfloor }C_{1}\left(\widehat{b}_{1,T}^{*}k\right)^{-b}\left|\left(\frac{\partial}{\partial\beta'}\widehat{\Gamma}^{*}\left(k\right)\right)|_{\beta=\overline{\beta}}\sqrt{n_{T}}\left(\widehat{\beta}-\beta_{0}\right)\right|\label{Eq. (30)}\\
 & =T^{8q/10\left(2q+1\right)-1/3+4b/5\left(2q+1\right)}\sum_{k=S_{T}+1}^{\left\lfloor D_{T}T^{1/2}\right\rfloor }C_{1}k^{-b}\left|\left(\frac{\partial}{\partial\beta'}\widehat{\Gamma}^{*}\left(k\right)\right)|_{\beta=\overline{\beta}}\sqrt{n_{T}}\left(\widehat{\beta}-\beta_{0}\right)\right|\nonumber \\
 & =T^{8q/10\left(2q+1\right)-1/3+4b/5\left(2q+1\right)+4r\left(1-b\right)/5\left(2q+1\right)}\left|\left(\frac{\partial}{\partial\beta'}\widehat{\Gamma}^{*}\left(k\right)\right)|_{\beta=\overline{\beta}}\sqrt{n_{T}}\left(\widehat{\beta}-\beta_{0}\right)\right|\nonumber \\
 & =T^{8q/10\left(2q+1\right)-1/3+4b/5\left(2q+1\right)+4r\left(1-b\right)/5\left(2q+1\right)}O_{\mathbb{P}}\left(1\right)O_{\mathbb{P}}\left(1\right),\nonumber 
\end{align}
   which goes to zero since $r>\left(-10+4q+24b\right)/24\left(b-1\right)$
with $b>\max\{1+1/q,\,4\}.$ We also have
\begin{align*}
\left|L_{1,T}^{B}\right| & =T^{8q/10\left(2q+1\right)-1/3}\sum_{k=\left\lfloor D_{T}T^{1/2}\right\rfloor +1}^{T-1}C_{1}\left(\widehat{b}_{1,T}^{*}k\right)^{-b}\left|\left(\frac{\partial}{\partial\beta'}\widehat{\Gamma}^{*}\left(k\right)\right)|_{\beta=\overline{\beta}}\sqrt{n_{T}}\left(\widehat{\beta}-\beta_{0}\right)\right|\\
 & =T^{8q/10\left(2q+1\right)-1/3+4b/5\left(2q+1\right)}\sum_{k=\left\lfloor D_{T}T^{1/2}\right\rfloor +1}^{T-1}C_{1}k^{-b}\left|\left(\frac{\partial}{\partial\beta'}\widehat{\Gamma}^{*}\left(k\right)\right)|_{\beta=\overline{\beta}}\sqrt{n_{T}}\left(\widehat{\beta}-\beta_{0}\right)\right|\\
 & =T^{8q/10\left(2q+1\right)-1/3+4b/5\left(2q+1\right)+\left(1-b\right)/2}\left|\left(\frac{\partial}{\partial\beta'}\widehat{\Gamma}^{*}\left(k\right)\right)|_{\beta=\overline{\beta}}\sqrt{n_{T}}\left(\widehat{\beta}-\beta_{0}\right)\right|\\
 & =T^{8q/10\left(2q+1\right)-1/3+4b/5\left(2q+1\right)+\left(1-b\right)/2}O_{\mathbb{P}}\left(1\right)\overset{\mathbb{P}}{\rightarrow}0,
\end{align*}
  given that $1-b<0$ and $b>1+1/q.$ Let us now consider $L_{2,T}$.
We have 
\begin{align}
\left|L_{2,T}^{A}\right| & =T^{\left(8q-1\right)/10\left(2q+1\right)}\sum_{k=S_{T}+1}^{\left\lfloor D_{T}T^{1/2}\right\rfloor }C_{1}\left(\widehat{b}_{1,T}^{*}k\right)^{-b}\left|\widetilde{\Gamma}^{*}\left(k\right)-\Gamma_{T}^{*}\left(k\right)\right|\label{Eq. 31}\\
 & =C_{1}\left(2qK_{1,q}^{2}\widehat{\phi}_{D}\left(q\right)\right)^{b/\left(2q+1\right)}T^{8q/10\left(2q+1\right)+b/\left(2q+1\right)-1/2}\left(\widehat{b}_{2,T}^{*}\right)^{b/\left(2q+1\right)-1/2}\sum_{k=S_{T}+1}^{\left\lfloor D_{T}T^{1/2}\right\rfloor }k^{-b}\nonumber \\
 & \quad\times\sqrt{T\widehat{b}_{2,T}^{*}}\left|\widetilde{\Gamma}^{*}\left(k\right)-\Gamma_{T}^{*}\left(k\right)\right|.\nonumber 
\end{align}
Note that 
\begin{align}
\mathbb{E} & \left(T^{8q/10\left(2q+1\right)+b/\left(2q+1\right)-1/2}\left(\widehat{b}_{2,T}^{*}\right)^{b/\left(2q+1\right)-1/2}\sum_{k=S_{T}+1}^{\left\lfloor D_{T}T^{1/2}\right\rfloor }k^{-b}\sqrt{T\widehat{b}_{2,T}^{*}}\left|\widetilde{\Gamma}^{*}\left(k\right)-\Gamma_{T}^{*}\left(k\right)\right|\right)^{2}\label{Eq. 32 Andrews 91}\\
 & \leq T^{8q/5\left(2q+1\right)+2b/\left(2q+1\right)-1}\left(\widehat{b}_{2,T}^{*}\right)^{b/\left(2q+1\right)-1/2}\left(\sum_{k=S_{T}+1}^{\left\lfloor D_{T}T^{1/2}\right\rfloor }k^{-b}\sqrt{T\widehat{b}_{2,T}^{*}}\left(\mathrm{Var}\left(\widetilde{\Gamma}^{*}\left(k\right)\right)\right)^{1/2}\right)^{2}\nonumber \\
 & =T^{8q/5\left(2q+1\right)+2b/\left(2q+1\right)-1}\left(\widehat{b}_{2,T}^{*}\right)^{2b/\left(2q+1\right)-1}\left(\sum_{k=S_{T}+1}^{\left\lfloor D_{T}T^{1/2}\right\rfloor }k^{-b}\right)^{2}O\left(1\right)\nonumber \\
 & =T^{8q/5\left(2q+1\right)+2b/\left(2q+1\right)-1}\widehat{b}_{2,T}^{2b/\left(2q+1\right)-1}D_{T}^{2\left(1-b\right)}S_{T}^{2\left(1-b\right)}O\left(1\right)\rightarrow0,\nonumber 
\end{align}
  since $r>\left(b-1/2\right)/\left(b-1\right)$ for $b>4$ and
$\sqrt{T\widehat{b}_{2,T}^{*}}\mathrm{Var}\left(\widetilde{\Gamma}^{*}\left(k\right)\right)=O\left(1\right)$
as above. Further, 

\begin{align}
\left|L_{2,T}^{B}\right| & =T^{\left(8q-1\right)/10\left(2q+1\right)}\sum_{k=\left\lfloor D_{T}T^{1/2}\right\rfloor +1}^{T-1}C_{1}\left(\widehat{b}_{1,T}^{*}k\right)^{-b}\left|\widetilde{\Gamma}^{*}\left(k\right)-\Gamma_{T}^{*}\left(k\right)\right|\label{Eq. 31-2}\\
 & =C_{1}\left(2qK_{1,q}^{2}\widehat{\phi}_{D}\left(q\right)\right)^{b/\left(2q+1\right)}T^{8q/10\left(2q+1\right)+b/\left(2q+1\right)-1/2}\left(\widehat{b}_{2,T}^{*}\right)^{b/\left(2q+1\right)-1/2}\nonumber \\
 & \quad\times\sum_{k=\left\lfloor D_{T}T^{1/2}\right\rfloor +1}^{T-1}k^{-b}\sqrt{T\widehat{b}_{2,T}^{*}}\left|\widetilde{\Gamma}^{*}\left(k\right)-\Gamma_{T}^{*}\left(k\right)\right|.\nonumber 
\end{align}
Note that 
\begin{align}
\mathbb{E} & \left(T^{8q/10\left(2q+1\right)+b/\left(2q+1\right)-1/2}\left(\widehat{b}_{2,T}^{*}\right)^{b/\left(2q+1\right)-1/2}\sum_{k=\left\lfloor D_{T}T^{1/2}\right\rfloor +1}^{T-1}k^{-b}\sqrt{T\widehat{b}_{2,T}^{*}}\left|\widetilde{\Gamma}^{*}\left(k\right)-\Gamma_{T}^{*}\left(k\right)\right|\right)^{2}\label{Eq. 32 Andrews 91-2}\\
 & \leq T^{8q/5\left(2q+1\right)+2b/\left(2q+1\right)-1}\left(\widehat{b}_{2,T}^{*}\right)^{2b/\left(2q+1\right)-1}\left(\sum_{k=\left\lfloor D_{T}T^{1/2}\right\rfloor +1}^{T-1}k^{-b}\sqrt{T\widehat{b}_{2,T}^{*}}\left(\mathrm{Var}\left(\widetilde{\Gamma}^{*}\left(k\right)\right)\right)^{1/2}\right)^{2}\nonumber \\
 & =T^{8q/5\left(2q+1\right)+2b/\left(2q+1\right)-1}\left(\widehat{b}_{2,T}^{*}\right)^{2b/\left(2q+1\right)-1}\left(\sum_{k=\left\lfloor D_{T}T^{1/2}\right\rfloor +1}^{T-1}k^{-b}\right)^{2}O\left(1\right)\nonumber \\
 & =T^{8q/5\left(2q+1\right)+2b/\left(2q+1\right)-1}\left(\widehat{b}_{2,T}^{*}\right)^{2b/\left(2q+1\right)-1}D_{T}^{2\left(1-b\right)}T^{\left(1-b\right)}O\left(1\right)\rightarrow0,\nonumber 
\end{align}
  since $r>\left(8b-4\right)/(\left(b-1\right)\left(10q+5\right))$
and $\sqrt{T\widehat{b}_{2,T}^{*}}\mathrm{Var}\left(\widetilde{\Gamma}^{*}\left(k\right)\right)=O\left(1\right)$
as above. Combining \eqref{Eq. 31}-\eqref{Eq. 32 Andrews 91} yields
$L_{2,T}\overset{\mathbb{P}}{\rightarrow}0$. Let us turn to $L_{3,T}$.
By Assumption \ref{Assumption E-F-G}-(iii) and $|K_{1}\left(\cdot\right)|\leq1$,
we have, 
\begin{align}
\left|L_{3,T}\right| & \leq T^{8q/10\left(2q+1\right)}\sum_{k=S_{T}}^{T-1}C_{3}k^{-l}\leq T^{8q/10\left(2q+1\right)}C_{3}S_{T}^{1-l}\label{Eq. (33)}\\
 & \leq C_{3}T^{8q/10\left(2q+1\right)}T^{-4r\left(l-1\right)/5\left(2q+1\right)}\rightarrow0,\nonumber 
\end{align}
since $r>q/\left(l-1\right)$.  In view of \eqref{Eq. A.29-1}-\eqref{Eq. (33)}
we deduce that $A_{2,1,T}\overset{\mathbb{P}}{\rightarrow}0$. Applying
the same argument to $A_{2,2,T}$, we have $A_{2,T}\overset{\mathbb{P}}{\rightarrow}0$.
Using similar arguments, one has $A_{3,T}\overset{\mathbb{P}}{\rightarrow}0$.
It remains to show that $T^{8q/10\left(2q+1\right)}(\widehat{J}_{T}^{*}(b_{\theta_{1},T},\,\widehat{b}_{2,T}^{*})-\widehat{J}_{T}^{*}(b_{\theta_{1},T},\,b_{\theta_{2},T}))\overset{\mathbb{P}}{\rightarrow}0$.
The proof of the latter result follows from the proof of the corresponding
result in Theorem 5.1-(ii) in \citeReferencesSupp{casini_hac} with
$r<2/3+q/3$ and $r>\left(b-2/3-q/3\right)/\left(b-1\right).$
$\square$

\bigskip{}

\noindent\textit{Proof of Theorem \ref{Theorem 1 MSE Prew - Andrews 90}.}
We begin with part (i). Note that 
\begin{align}
\widehat{J}_{T}^{*}\left(\widehat{b}_{1,T}^{*},\,\widehat{b}_{2,T}^{*}\right)-J_{T}^{*} & =\widehat{J}_{T}^{*}\left(\widehat{b}_{1,T}^{*},\,\widehat{b}_{2,T}^{*}\right)-\widehat{J}_{T}^{*}\left(b_{\theta_{1},T},\,b_{\theta_{2},T}\right)+\widehat{J}_{T}^{*}\left(b_{\theta_{1},T},\,b_{\theta_{2},T}\right)-J_{T}^{*}.\label{Eq. (J_hat* - J*)}
\end{align}
 By Lemma \ref{Lemma J*(b1*, b2*) - J* =00003D o(1)}-\ref{Lemma J*_That(opt_b1, 2) - J*_hat(b1,2 hat)=00003Do(1)}
the right-hand side is $o_{\mathbb{P}}\left(1\right).$ It follows
that the first term on the right-hand side of \eqref{Eq. (decomposition Theorem 1)}
is also $o_{\mathbb{P}}\left(1\right)$ because the presence of $\widehat{D}_{s}$
is irrelevant for the result to hold.  We have,
\begin{align}
J_{T,D}^{*} & =\frac{1}{T}\sum_{s=p_{A}+1}^{T}\sum_{t=p_{A}+1}^{T}D_{s}\mathbb{E}V_{s}^{*}\left(V_{t}^{*}D_{t}\right)'\nonumber \\
 & =\frac{1}{T}\sum_{s=p_{A}+1}^{T}\sum_{t=p_{A}+1}^{T}\left(I_{p}-\sum_{j=1}^{p_{A}}A_{D,s,j}\right)^{-1}\mathbb{E}\left(V_{s}-\sum_{j=1}^{p_{A}}A_{D,s,j}V_{s-j}\right)\left(V_{t}^{*}\left(I_{p}-\sum_{j=1}^{p_{A}}A_{D,t,j}\right)^{-1}\right)'\nonumber \\
 & =\frac{1}{T}\sum_{s=p_{A}+1}^{T}\sum_{t=p_{A}+1}^{T}\left(I_{p}-\sum_{j=1}^{p_{A}}A_{D,s,j}\right)^{-1}\nonumber \\
 & \quad\times\mathbb{E}\left(\left(V_{s}-\sum_{j=1}^{p_{A}}A_{D,s,j}V_{s}+\sum_{j=1}^{p_{A}}A_{D,s,j}V_{s}-\sum_{j=1}^{p_{A}}A_{D,s,j}V_{s-j}\right)\right)\left(V_{t}^{*}\left(I_{p}-\sum_{j=1}^{p_{A}}A_{D,t,j}\right)^{-1}\right)'\nonumber \\
 & =\frac{1}{T}\sum_{s=p_{A}+1}^{T}\sum_{t=p_{A}+1}^{T}\left(I_{p}-\sum_{j=1}^{p_{A}}A_{D,s,j}\right)^{-1}\mathbb{E}\left(\left(V_{s}-\sum_{j=1}^{p_{A}}A_{D,s,j}V_{s}+\sum_{j=1}^{p_{A}}A_{D,s,j}\left(V_{s}-V_{s-j}\right)\right)\right)\nonumber \\
 & \quad\times\left(V_{t}^{*}\left(I_{p}-\sum_{j=1}^{p_{A}}A_{D,t,j}\right)^{-1}\right)'\nonumber \\
 & =\frac{1}{T}\sum_{s=p_{A}+1}^{T}\sum_{t=p_{A}+1}^{T}\mathbb{E}\left(\left(V_{s}+\left(I_{p}-\sum_{j=1}^{p_{A}}A_{D,s,j}\right)^{-1}\sum_{j=1}^{p_{A}}A_{D,s,j}\left(V_{s}-V_{s-j}\right)\right)\right)\label{Eq. (J_T_D*)}\\
 & \quad\times\left(V_{t}^{*}\left(I_{p}-\sum_{j=1}^{p_{A}}A_{D,t,j}\right)^{-1}\right)'.\nonumber 
\end{align}
Now note that the sum involving $V_{s}-V_{s-j}$ has a telescopic
form to a sum. Using the smoothness of $A_{D,s,j},$ we have  that
the sum from any $s$ to $T$ is
\begin{align}
\left(I_{p}-\sum_{j=1}^{p_{A}}A_{D,s,j}\right)^{-1} & \sum_{j=1}^{p_{A}}A_{D,s,j}\left(V_{s}-V_{s-j}\right)\label{Eq.telescopic}\\
 & \quad+\left(I_{p}-\sum_{j=1}^{p_{A}}A_{D,s+1,j}\right)^{-1}\sum_{j=1}^{p_{A}}A_{D,s+1,j}\left(V_{s+1}-V_{s+1-j}\right)\nonumber \\
 & \quad\cdots\nonumber \\
 & \quad+\left(I_{p}-\sum_{j=1}^{p_{A}}A_{D,T,j}\right)^{-1}\sum_{j=1}^{p_{A}}A_{D,T,j}\left(V_{T}-V_{T-j}\right).\nonumber 
\end{align}
For $s\neq T_{r}^{0}$ $\left(r=1,\ldots,\,m_{0}\right)$ local stationarity
implies $A_{D,s+1,j}=A_{D,s,j}+O\left(1/T\right)$. There are only
a finite number of breaks $T_{r}^{0}\,\left(r=1,\ldots,\,m_{0}\right)$
so that \eqref{Eq.telescopic} is equal to
\begin{align*}
\left(I_{p}-\sum_{j=1}^{p_{A}}A_{D,p_{A}+1,j}\right)^{-1} & A_{D,p_{A}+1,p_{A}}V_{1}+\left(I_{p}-\sum_{j=1}^{p_{A}}A_{D,T,j}\right)A_{D,T,p_{A}}V_{T}\\
 & \quad+\sum_{r=1}^{m_{0}}\left(I_{p}-\sum_{j=1}^{p_{A}}A_{D,T_{r}^{0},j}\right)^{-1}\sum_{j=1}^{p_{A}}\left(A_{D,T_{r}^{0},j}-A_{D,T_{r}^{0}+1,j}\right)V_{T_{r}^{0}}\\
 & \triangleq C_{A,T}.
\end{align*}
It follows that
\begin{align*}
\frac{1}{T}\sum_{t=1}^{T}\mathbb{E}\left(C_{A,T}\right)\left(V_{t}^{*}\left(I_{p}-\sum_{j=1}^{p_{A}}A_{D,t,j}\right)^{-1}\right)' & \rightarrow0.
\end{align*}
Altogether, this implies $J_{T,D}^{*}\overset{\mathbb{P}}{\rightarrow}J_{T}$.
Using Assumption \ref{Assumption H Andrews 1990} and simple manipulations,
the second term on the right-hand side of \eqref{Eq. (decomposition Theorem 1)}
is $o_{\mathbb{P}}\left(1\right)$. Therefore, 
\begin{align}
\widehat{J}_{\mathrm{pw},T}-J_{T}=\left(\widehat{J}_{\mathrm{pw},T}-J_{T,\widehat{D}}^{*}\right)+\left(J_{T,\widehat{D}}^{*}-J_{T,D}^{*}\right)+\left(J_{T,D}^{*}-J_{T}\right) & =o_{\mathbb{P}}\left(1\right),\label{Eq. (A.1) Andrews 1990}
\end{align}
 which concludes the proof of part (i). 

 Next, we move to part (ii). Given the decomposition \eqref{Eq. (decomposition Theorem 1)},
we have to show
\begin{align}
\sqrt{Tb_{\theta_{1},T}b_{\theta_{2},T}}\left(\widehat{J}_{\mathrm{pw},T}-J_{T,\widehat{D}}^{*}\right) & =O_{\mathbb{P}}\left(1\right),\label{Eq. (1) Dec. Theorem 1-(ii)}\\
\sqrt{Tb_{\theta_{1},T}b_{\theta_{2},T}}\left(J_{T,\widehat{D}}^{*}-J_{T,D}^{*}\right) & =o_{\mathbb{P}}\left(1\right),\label{Eq. (2) Dec. Theorem 1-(ii)}\\
\sqrt{Tb_{\theta_{1},T}b_{\theta_{2},T}}\left(J_{T,D}^{*}-J_{T}\right) & =o_{\mathbb{P}}\left(1\right).\label{Eq. (3) Dec. Theorem 1-(ii)}
\end{align}
Equation \eqref{Eq. (1) Dec. Theorem 1-(ii)} follows from
\begin{align}
\sqrt{Tb_{\theta_{1},T}b_{\theta_{2},T}}\left(\widehat{J}_{T}^{*}\left(b_{\theta_{1},T},\,b_{\theta_{2},T}\right)-J_{T}^{*}\right) & =O_{\mathbb{P}}\left(1\right),\label{Eq. Tq(J_hat* - J*) - 2nd term =00003D op(1)}\\
\sqrt{Tb_{\theta_{1},T}b_{\theta_{2},T}}\left(\widehat{J}_{T}^{*}\left(\widehat{b}_{1,T}^{*},\,\widehat{b}_{2,T}^{*}\right)-\widehat{J}_{T}^{*}\left(b_{\theta_{1},T},\,b_{\theta_{2},T}\right)\right) & =o_{\mathbb{P}}\left(1\right),\label{Eq. Tq(J_hat* - J*) - 1st term =00003D op(1)}
\end{align}
since the presence of $\widehat{D}_{s}$ in $\widehat{V}_{D,s}^{*}$
is irrelevant. Thus, Lemma \ref{Lemma 1 for Theorem 1 (ii)}-\ref{Lemma 2 for Theorem 1 (ii)}
yield \eqref{Eq. (1) Dec. Theorem 1-(ii)}. Given that $\sqrt{Tb_{\theta_{1},T}b_{\theta_{2},T}/n_{T}}\rightarrow0$,
Assumption \ref{Assumption H Andrews 1990} and simple algebra yield
\eqref{Eq. (2) Dec. Theorem 1-(ii)}. From the proof of part (i),
it is easy to see that the multiplication by the factor $\sqrt{Tb_{\theta_{1},T}b_{\theta_{2},T}}$
in \eqref{Eq. (3) Dec. Theorem 1-(ii)} does not change the fact that
this term is $o_{\mathbb{P}}\left(1\right)$. Therefore, we conclude
that $T^{8q/10\left(2q+1\right)}(\widehat{J}_{\mathrm{pw},T}-J_{T})=O_{\mathbb{P}}\left(1\right)$.

We now move to part (iii). The estimator $\widehat{J}_{T,\mathrm{pw}}$
is actually a double kernel HAC estimator constructed using observations
$\{\widehat{V}_{D,s}\},$ where the latter is SLS. Thus, using Theorem
3.2 and 5.1 in \citeReferencesSupp{casini_hac} and Assumption \ref{Assumption H Andrews 1990},
we deduce that 
\begin{align}
\lim_{T\rightarrow\infty}\mathrm{MSE}\left(Tb_{\theta_{1},T}b_{\theta_{2},T},\,\widehat{J}_{\mathrm{pw},T},\,J_{T},\,W_{T}\right) & =\lim_{T\rightarrow\infty}\mathrm{MSE}\left(Tb_{\theta_{1},T}b_{\theta_{2},T},\,J_{T,D}^{*},\,J_{T},\,W_{T}\right).\label{Eq. MSE(Jpw) =00003D MSE(JTD)}
\end{align}
This implies that it is sufficient to determine the asymptotic MSE
of $J_{T,D}^{*}.$ Note that $J_{T,D}^{*}$ is simply a double kernel
HAC estimator constructed using observations $\{V_{D,t}^{*}\}.$ It
follows that $\{V_{D,t}^{*}\}$ is SLS and thus it satisfies the conditions
of Theorem 3.2 and 5.1 in \citeReferencesSupp{casini_hac}.  The
same argument in \citeReferencesSupp{casini_hac} now with reference
to Theorem \ref{Theorem 1 MSE Prew - Andrews 90}-(i,ii) yields 
\begin{align*}
\lim_{T\rightarrow\infty} & \mathrm{MSE}\left(Tb_{\theta_{1},T}b_{\theta_{2},T},\,J_{T,D}^{*},\,J_{T},\,W_{T}\right)\\
 & =4\pi^{2}\left[\gamma_{\theta}K_{1,q}^{2}\mathrm{vec}\left(\int_{0}^{1}f_{D}^{*\left(q\right)}\left(u,\,0\right)du\right)'W\mathrm{vec}\left(\int_{0}^{1}f_{D}^{*\left(q\right)}\left(u,\,0\right)du\right)\right]\\
 & \quad+\int K_{1}^{2}\left(y\right)dy\int K_{2}^{2}\left(x\right)dx\,\mathrm{tr}\left[W\left(I_{p_{\beta}^{2}}-C_{pp}\right)\left(\int_{0}^{1}f_{D}^{*}\left(u,\,0\right)du\right)\otimes\left(\int_{0}^{1}f_{D}^{*}\left(v,\,0\right)dv\right)\right].
\end{align*}
The latter relation and \eqref{Eq. MSE(Jpw) =00003D MSE(JTD)} conclude
the proof. $\square$ 

\section{\label{Section Proofs of the Results General Nonstationarity}Proofs
of the Results in Section \ref{Section: Extension-to-Unrestrcited Nonstationary}}

In the proofs below involving $\widehat{c}_{T}\left(u,\,k\right),\,\widetilde{c}_{T}\left(u,\,k\right)$
and $c\left(u,\,k\right)$, we assume $k\geq0$ unless otherwise stated.
The proofs for the case $k<0$ are similar and omitted. The novelty
of the proofs of the results of Section \ref{Section: Extension-to-Unrestrcited Nonstationary}
is twofold. First, they are provided for the double-kernel HAC estimator
and so there are two smoothing directions that are considered. Second,
the theoretical bounds are derived in terms of two distributions under
which $\left\{ V_{t}\right\} $ is a segmented locally stationary
process. This differs from early proofs in the literature that rely
on stationarity. Lemma \ref{Lemma 1 Andrews 88} and \ref{Lemma 2 Andrews 88}
are building blocks for our proofs as they establish upper and lower
bounds on the asymptotic variance and asymptotic bias, respectively,
of $\widetilde{J}_{T}$ under segmented local stationarity. The derivation
of the minimax MSE bounds is the key step for determining the optimal
kernels and bandwidths and for showing the MSE-optimality of the DK-HAC
estimator based on the proposed data-dependent bandwidths. 

\subsection{Proof of Theorem \ref{Theorem MSE J DKHAC Nonstationary}}

We first present upper and lower bounds on the asymptotic variance
of $\widetilde{J}_{T}$. Let $\mathrm{Var}_{\mathscr{P}}\left(\cdot\right)$
denote the variance of $\cdot$ under $\mathscr{P}$.
\begin{lem}
\label{Lemma 1 Andrews 88}Suppose that Assumption \ref{Assumption A*}
holds, $K_{2}\left(\cdot\right)\in\boldsymbol{K}_{2}$, $b_{1,T},\,b_{2,T}\rightarrow0$,
$n_{T}\rightarrow\infty,\,n_{T}/T\rightarrow0$ and $1/Tb_{1,T}b_{2,T}\rightarrow0$.
We have for all $a\in\mathbb{R}^{p_{\beta}}$:

(i) for any $K_{1}\left(\cdot\right)\in\boldsymbol{K}_{1}$, 
\begin{align*}
\underset{T\rightarrow\infty}{\lim} & \underset{\mathscr{P}\in\boldsymbol{P}_{U}}{\sup}Tb_{1,T}b_{2,T}\mathrm{Var}_{\mathscr{P}}\left(a'\widetilde{J}_{T}a\right)=\underset{T\rightarrow\infty}{\lim}Tb_{1,T}b_{2,T}\mathrm{Var}_{\mathscr{P}_{U}}\left(a'\widetilde{J}_{T}a\right)\\
 & =8\pi^{2}\int K_{1}^{2}\left(y\right)dy\int_{0}^{1}K_{2}^{2}\left(x\right)dx\left(\int_{0}^{1}f_{\mathscr{P}_{U},a}\left(u,\,0\right)du\right)^{2};
\end{align*}

(ii) for any $K_{1}\left(\cdot\right)\in\boldsymbol{K}_{1,+},$ 
\begin{align*}
\underset{T\rightarrow\infty}{\lim} & \underset{\mathscr{P}\in\boldsymbol{P}_{L}}{\inf}Tb_{1,T}b_{2,T}\mathrm{Var}_{\mathscr{P}}\left(a'\widetilde{J}_{T}a\right)=\underset{T\rightarrow\infty}{\lim}Tb_{1,T}b_{2,T}\mathrm{Var}_{\mathscr{P}_{L}}\left(a'\widetilde{J}_{T}a\right)\\
 & =8\pi^{2}\int K_{1}^{2}\left(y\right)dy\int_{0}^{1}K_{2}^{2}\left(x\right)dx\left(\int_{0}^{1}f_{\mathscr{P}_{L},a}\left(u,\,0\right)du\right)^{2}.
\end{align*}
\end{lem}
\noindent\textit{Proof of Lemma} \ref{Lemma 1 Andrews 88}. Let $Z_{t}=a'V_{t}$
and $c_{\mathscr{P},T}\left(rn_{T}/T,\,k\right)=\mathbb{E}_{\mathscr{P}}\widetilde{c}_{T}\left(rn_{T}/T,\,k\right)$.
For any $k\geq0$ and any $r=0,\ldots,\,\left\lfloor T/n_{T}\right\rfloor $,
\begin{align*}
a' & \left(\widetilde{c}_{T}\left(rn_{T}/T,\,k\right)-c_{\mathscr{P},T}\left(rn_{T}/T,\,k\right)\right)a\\
 & =\left(\left(Tb_{2,T}\right)^{-1}\sum_{s=k+1}^{T}K_{2}^{*}\left(\frac{\left(\left(r+1\right)n_{T}-\left(s-k/2\right)\right)/T}{b_{2,T}}\right)\left(Z_{s}Z{}_{s-k}-\mathbb{E}_{\mathscr{P}}\left(Z_{s}Z{}_{s-k}\right)\right)\right).
\end{align*}
 For any $k,\,j\geq0$ and any $r,\,b=0,\ldots,\,\left\lfloor T/n_{T}\right\rfloor $,
\begin{align*}
\underset{\mathscr{P}\in\boldsymbol{P}_{U}}{\sup} & \left|\mathbb{E}_{\mathscr{P}}\left(a'\left(\widetilde{c}_{T}\left(rn_{T}/T,\,k\right)-c_{\mathscr{P},T}\left(rn_{T}/T,\,k\right)\right)aa'\left(\widetilde{c}_{T}\left(bn_{T}/T,\,j\right)-c_{\mathscr{P},T}\left(bn_{T}/T,\,j\right)\right)a\right)\right|\\
 & =\Biggl|\left(Tb_{2,T}\right)^{-2}\sum_{s=k+1}^{T}\sum_{l=j+1}^{T}K_{2}^{*}\left(\frac{\left(\left(r+1\right)n_{T}-\left(s-k/2\right)\right)/T}{b_{2,T}}\right)K_{2}^{*}\left(\frac{\left(\left(b+1\right)n_{T}-\left(l-j/2\right)\right)/T}{b_{2,T}}\right)\\
 & \qquad\times\left(\mathbb{E}_{\mathscr{P}}\left(Z_{s}Z{}_{s-k}Z_{l}Z{}_{l-j}\right)-\mathbb{E}_{\mathscr{P}}\left(Z_{s}Z{}_{s-k}\right)\mathbb{E}_{\mathscr{P}}\left(Z_{l}Z{}_{l-j}\right)\right)\Biggr|.
\end{align*}
By definition of the fourth-order cumulant and by definition of $\boldsymbol{P}_{U}$,
\begin{align}
\sup_{\mathscr{P}\in\boldsymbol{P}_{U}} & \left|\mathbb{E}_{\mathscr{P}}\left(a'\left(\widetilde{c}_{T}\left(rn_{T}/T,\,k\right)-c_{\mathscr{P},T}\left(rn_{T}/T,\,k\right)\right)aa'\left(\widetilde{c}_{T}\left(bn_{T}/T,\,j\right)-c_{\mathscr{P},T}\left(bn_{T}/T,\,j\right)\right)a\right)\right|\nonumber \\
 & =\Biggl|\left(Tb_{2,T}\right)^{-2}\sum_{s=k+1}^{T}\sum_{l=j+1}^{T}K_{2}^{*}\left(\frac{\left(\left(r+1\right)n_{T}-\left(s-k/2\right)\right)/T}{b_{2,T}}\right)K_{2}^{*}\left(\frac{\left(\left(b+1\right)n_{T}-\left(l-j/2\right)\right)/T}{b_{2,T}}\right)\nonumber \\
 & \quad\times\biggl(\mathbb{E}_{\mathscr{P}}\left(Z_{s}Z{}_{s-k}\right)\mathbb{E}_{\mathscr{P}}\left(Z_{l}Z{}_{l-j}\right)+\mathbb{E}_{\mathscr{P}}\left(Z_{s}Z{}_{l}\right)\mathbb{E}_{\mathscr{P}}\left(Z_{s-k}Z{}_{l-j}\right)+\mathbb{E}_{\mathscr{P}}\left(Z_{s}Z{}_{l-j}\right)\mathbb{E}\left(Z_{s-k}Z{}_{l}\right)\nonumber \\
 & \quad+\kappa_{\mathscr{P},aV,s}\left(-k,\,l-s,\,l-j-s\right)-\mathbb{E}_{\mathscr{P}}\left(Z_{s}Z{}_{s-k}\right)\mathbb{E}\left(Z_{l}Z{}_{l-j}\right)\Biggr|\nonumber \\
 & \leq\left(Tb_{2,T}\right)^{-2}\sum_{s=k+1}^{T}\sum_{l=j+1}^{T}K_{2}^{*}\left(\frac{\left(\left(r+1\right)n_{T}-\left(s-k/2\right)\right)/T}{b_{2,T}}\right)K_{2}^{*}\left(\frac{\left(\left(b+1\right)n_{T}-\left(l-j/2\right)\right)/T}{b_{2,T}}\right)\nonumber \\
 & \quad\times\biggl(a'\Gamma_{\mathscr{P}_{U},s/T}\left(s-l\right)aa'\Gamma_{\mathscr{P}_{U},s-k}\left(s-k-l+j\right)a+a'\Gamma_{\mathscr{P}_{U},s/T}\left(s-l+j\right)aa'\Gamma_{\mathscr{P}_{U},s-k}\left(s-k-l\right)a\nonumber \\
 & \quad+\kappa_{s}^{*}\left(-k,\,l-s,\,l-j-s\right)\biggr)\nonumber \\
 & \leq\mathbb{E}_{\mathscr{P}_{U}}\left(a'\left(\widetilde{c}_{T}\left(rn_{T}/T,\,k\right)-c_{\mathscr{P}_{U},T}\left(rn_{T}/T,\,k\right)\right)aa'\left(\widetilde{c}_{T}\left(bn_{T}/T,\,j\right)-c_{\mathscr{P}_{U},T}\left(bn_{T}/T,\,j\right)\right)a\right)\label{eq A.1 Andrews 88}\\
 & \quad+2\left(\frac{1}{Tb_{2,T}}\right)^{2}\sum_{s=k+1}^{T}\sum_{l=j+1}^{T}K_{2}^{*}\left(\frac{\left(\left(r+1\right)n_{T}-\left(s-k/2\right)\right)/T}{b_{2,T}}\right)K_{2}^{*}\left(\frac{\left(\left(b+1\right)n_{T}-\left(l-j/2\right)\right)/T}{b_{2,T}}\right)\nonumber \\
 & \quad\times\kappa_{s}^{*}\left(-k,\,l-s,\,l-j-s\right),\nonumber 
\end{align}
 where the last inequality holds by reversing the argument of the
equality and the first inequality. 

By a similar argument, 
\begin{align}
\underset{\mathscr{P}\in\boldsymbol{P}_{L}}{\inf} & \left|\mathbb{E}_{\mathscr{P}}\left(a'\left(\widetilde{c}_{T}\left(rn_{T}/T,\,k\right)-c_{\mathscr{P},T}\left(rn_{T}/T,\,k\right)\right)aa'\left(\widetilde{c}_{T}\left(bn_{T}/T,\,j\right)-c_{\mathscr{P},T}\left(bn_{T}/T,\,j\right)\right)a\right)\right|\nonumber \\
 & \geq\mathbb{E}_{\mathscr{P}_{L}}\left(a'\left(\widetilde{c}_{T}\left(rn_{T}/T,\,k\right)-c_{\mathscr{P}_{L},T}\left(rn_{T}/T,\,k\right)\right)aa'\left(\widetilde{c}_{T}\left(bn_{T}/T,\,j\right)-c_{\mathscr{P}_{L},T}\left(bn_{T}/T,\,j\right)\right)a\right)\label{Eq. (A.2) Andrews 88}\\
 & \quad+2\left(\frac{1}{Tb_{2,T}}\right)^{2}\sum_{s=k+1}^{T}\sum_{l=j+1}^{T}\kappa_{s}^{*}\left(-k,\,l-s,\,l-j-s\right).\nonumber 
\end{align}
Let $\widetilde{J}_{T,K}$ be the same as $\widetilde{J}_{T}$ but
with $\left|K_{1}\left(\cdot\right)\right|$ and $\left|K_{2}\left(\cdot\right)\right|$
in place of $K_{1}\left(\cdot\right)$ and $K_{2}\left(\cdot\right)$,
respectively. Note that $K_{1}\left(\cdot\right)\in\boldsymbol{K}_{1}$
$\left(K_{2}\left(\cdot\right)\in\boldsymbol{K}_{2}\right)$ implies
$\left|K_{1}\left(\cdot\right)\right|\in\boldsymbol{K}_{1}$ $\left(\left|K_{2}\left(\cdot\right)\right|\in\boldsymbol{K}_{2}\right)$.
We have
\begin{align}
\underset{T\rightarrow\infty}{\lim} & Tb_{1,T}b_{2,T}\mathrm{Var}_{\mathscr{P}_{U}}\left(a'\widetilde{J}_{T}a\right)\nonumber \\
 & \leq\underset{T\rightarrow\infty}{\lim}\underset{\mathscr{P}\in\boldsymbol{P}_{U}}{\sup}Tb_{1,T}b_{2,T}\mathrm{Var}_{\mathscr{P}}\left(a'\widetilde{J}_{T}a\right)\nonumber \\
 & =\underset{T\rightarrow\infty}{\lim}\underset{\mathscr{P}\in\boldsymbol{P}_{U}}{\sup}Tb_{1,T}b_{2,T}\sum_{k=-T+1}^{T-1}\sum_{j=-T+1}^{T-1}K_{1}\left(b_{1,T}k\right)K_{1}\left(b_{1,T}j\right)\nonumber \\
 & \quad\times\left(\frac{n_{T}}{T}\right)^{2}\sum_{r=0}^{T/n_{T}}\sum_{b=0}^{T/n_{T}}\left(\frac{1}{Tb_{2,T}}\right)^{2}\sum_{s=k+1}^{T}\sum_{l=j+1}^{T}K_{2}^{*}\left(\frac{\left(rn_{T}+1\right)-\left(s+k/2\right)}{Tb_{2,T}}\right)K_{2}^{*}\left(\frac{\left(bn_{T}+1\right)-\left(l+j/2\right)}{Tb_{2,T}}\right)\nonumber \\
 & \quad\times\mathbb{E}_{\mathscr{P}}\left(a'\left(\Gamma_{s/T}\left(k\right)-\mathbb{E}_{\mathscr{P}}\left(\Gamma_{s/T}\left(k\right)\right)\right)aa'\left(\Gamma_{l/T}\left(k\right)-\mathbb{E}_{\mathscr{P}}\left(\Gamma_{l/T}\left(k\right)\right)\right)a\right)\nonumber \\
 & \leq\underset{T\rightarrow\infty}{\lim}Tb_{1,T}b_{2,T}\sum_{k=-T+1}^{T-1}\sum_{j=-T+1}^{T-1}\left|K_{1}\left(b_{1,T}k\right)K_{1}\left(b_{1,T}j\right)\right|\nonumber \\
 & \quad\times\left(\frac{n_{T}}{T}\right)^{2}\sum_{r=0}^{T/n_{T}}\sum_{b=0}^{T/n_{T}}\left(\frac{1}{Tb_{2,T}}\right)^{2}\sum_{s=k+1}^{T}\sum_{l=j+1}^{T}\left|K_{2}^{*}\left(\frac{\left(rn_{T}+1\right)-\left(s+k/2\right)}{Tb_{2,T}}\right)K_{2}^{*}\left(\frac{\left(bn_{T}+1\right)-\left(l+j/2\right)}{Tb_{2,T}}\right)\right|\nonumber \\
 & \quad\times\mathbb{E}_{\mathscr{P}_{U}}\left(a'\left(\Gamma_{s/T}\left(k\right)-\mathbb{E}_{\mathscr{P}_{U}}\left(\Gamma_{s/T}\left(k\right)\right)\right)aa'\left(\Gamma_{s/T}\left(k\right)-\mathbb{E}_{\mathscr{P}_{U}}\left(\Gamma_{s/T}\left(k\right)\right)\right)a\right)\nonumber \\
 & \quad+2\underset{T\rightarrow\infty}{\lim}Tb_{1,T}b_{2,T}\sum_{k=-T+1}^{T-1}\sum_{j=-T+1}^{T-1}\left|K_{1}\left(b_{1,T}k\right)K_{1}\left(b_{1,T}j\right)\right|\left(\frac{n_{T}}{T}\right)^{2}\sum_{r=0}^{T/n_{T}}\sum_{b=0}^{T/n_{T}}\left(\frac{1}{Tb_{2,T}}\right)^{2}\nonumber \\
 & \quad\left(\frac{1}{Tb_{2,T}}\right)^{2}\sum_{s=k+1}^{T}\sum_{l=j+1}^{T}\left|K_{2}^{*}\left(\frac{\left(\left(r+1\right)n_{T}-\left(s-k/2\right)\right)/T}{b_{2,T}}\right)K_{2}^{*}\left(\frac{\left(\left(b+1\right)n_{T}-\left(l-j/2\right)\right)/T}{b_{2,T}}\right)\right|\nonumber \\
 & \quad\times\kappa_{s}^{*}\left(-k,\,l-s,\,l-j-s\right)\nonumber \\
 & =\underset{T\rightarrow\infty}{\lim}Tb_{1,T}b_{2,T}\mathrm{Var}_{\mathscr{P}_{U}}\left(a'\widetilde{J}_{T,K}a\right),\label{Eq. (A.3) Andrews 88}
\end{align}
where $\Gamma_{s/T}\left(k\right)=\Gamma_{u}\left(k\right)$ with
$u=s/T$ and $\Gamma_{u}\left(k\right)$ was defined before eq. \eqref{Eq. c(u,k) =00003D f(u,w)},
and the last inequality uses \eqref{eq A.1 Andrews 88}. For $K_{1}\left(\cdot\right)\in\boldsymbol{K}_{1,+}$,
we can rely on an argument analogous to that of \eqref{Eq. (A.3) Andrews 88}
using \eqref{Eq. (A.2) Andrews 88} in place of \eqref{eq A.1 Andrews 88}
to yield, 
\begin{align}
\underset{T\rightarrow\infty}{\lim}Tb_{1,T}b_{2,T}\mathrm{Var}_{\mathscr{P}_{L}}\left(a'\widetilde{J}_{T}a\right) & \geq\underset{T\rightarrow\infty}{\lim}\underset{\mathscr{P}\in\boldsymbol{P}_{L}}{\inf}Tb_{1,T}b_{2,T}\mathrm{Var}_{\mathscr{P}}\left(a'\widetilde{J}_{T}a\right)\nonumber \\
 & \geq\underset{T\rightarrow\infty}{\lim}Tb_{1,T}b_{2,T}\mathrm{Var}_{\mathscr{P}_{L}}\left(a'\widetilde{J}_{T,K}a\right).\label{Eq. (A.4) Andrews 88}
\end{align}
By Theorem 3.1 in \citeReferencesSupp{casini_hac}, 
\begin{align}
\underset{T\rightarrow\infty}{\lim}Tb_{1,T}b_{2,T}\mathrm{Var}_{\mathscr{P}_{w}}\left(a'\widetilde{J}_{T}a\right) & =8\pi^{2}\int K_{1}^{2}\left(y\right)dy\int_{0}^{1}K_{2}^{2}\left(x\right)dx\left(\int_{0}^{1}f_{\mathscr{P}_{w},a}\left(u,\,0\right)du\right)^{2},\qquad\mathrm{and}\label{Eq. (A.5) Andrews 88}\\
\underset{T\rightarrow\infty}{\lim}Tb_{1,T}b_{2,T}\mathrm{Var}_{\mathscr{P}_{w}}\left(a'\widetilde{J}_{K,T}a\right) & =8\pi^{2}\int\left|K_{1}\left(y\right)\right|^{2}dy\int_{0}^{1}\left|K_{2}\left(x\right)\right|^{2}dx\left(\int_{0}^{1}f_{\mathscr{P}_{w},a}\left(u,\,0\right)du\right)^{2},\label{Eq. (A.6) Andrews 88}
\end{align}
for $w=L,\,U$. Equations \eqref{Eq. (A.3) Andrews 88}, \eqref{Eq. (A.5) Andrews 88}
and \eqref{Eq. (A.6) Andrews 88} combine to establish part (i) of
the lemma:
\begin{align*}
8\pi^{2} & \int K_{1}^{2}\left(y\right)dy\int_{0}^{1}K_{2}^{2}\left(x\right)dx\left(\int_{0}^{1}f_{\mathscr{P}_{U},a}\left(u,\,0\right)du\right)^{2}\\
 & =\underset{T\rightarrow\infty}{\lim}Tb_{1,T}b_{2,T}\mathrm{Var}_{\mathscr{P}_{U}}\left(a'\widetilde{J}_{T}a\right)\\
 & \leq\underset{T\rightarrow\infty}{\lim}\underset{\mathscr{P}\in\boldsymbol{P}_{U}}{\sup}Tb_{1,T}b_{2,T}\mathrm{Var}_{\mathscr{P}}\left(a'\widetilde{J}_{T}a\right)\\
 & \leq\underset{T\rightarrow\infty}{\lim}Tb_{1,T}b_{2,T}\mathrm{Var}_{\mathscr{P}_{U}}\left(a'\widetilde{J}_{T,K}a\right)\\
 & =8\pi^{2}\int K_{1}^{2}\left(y\right)dy\int_{0}^{1}K_{2}^{2}\left(x\right)dx\left(\int_{0}^{1}f_{\mathscr{P}_{U},a}\left(u,\,0\right)du\right)^{2}.
\end{align*}
 By a similar reasoning, equations \eqref{Eq. (A.4) Andrews 88} and
\eqref{Eq. (A.5) Andrews 88} yield part (ii). $\square$

\medskip{}

Upper and lower bounds on the asymptotic bias of $\widetilde{J}_{T}$
are given in the following lemma. Let $J_{\mathscr{P}_{w},T}$ be
equal to $J_{\mathscr{P},T}$ but with the expectation $\mathbb{E_{\mathscr{P}}}$
replaced by $\mathbb{E}_{\mathscr{P}_{w}}$, $w=U,\,L.$ 
\begin{lem}
\label{Lemma 2 Andrews 88}Let Assumption \ref{Assumption A*} hold,
$K_{1}\left(\cdot\right)\in\boldsymbol{K}_{1}$, $K_{2}\left(\cdot\right)\in\boldsymbol{K}_{2}$,
$b_{1,T},\,b_{2,T}\rightarrow0$, $n_{T}\rightarrow\infty,\,n_{T}/T\rightarrow0$,
$1/Tb_{1,T}b_{2,T}\rightarrow0$, $1/Tb_{1,T}^{q}b_{2,T}\rightarrow0$,
$n_{T}/Tb_{1,T}^{q}\rightarrow0$ and $b_{2,T}^{2}/b_{1,T}^{q}\rightarrow0$
for some $q\in[0,\,\infty)$ for which $K_{1,q},\,|\int_{0}^{1}f_{\mathscr{P}_{w},a}^{\left(q\right)}\left(u,\,0\right)du|$
$\in[0,\,\infty)$, $w=U,\,L$. We have for all $a\in\mathbb{R}^{p_{\beta}}$:

(i) $\underset{T\rightarrow\infty}{\lim}\underset{\mathscr{P}\in\boldsymbol{P}_{U}}{\sup}b_{1,T}^{-q}\left|\mathbb{E_{\mathscr{P}}}a'\widetilde{J}_{T}a-a'J_{\mathscr{P},T}a\right|=\underset{T\rightarrow\infty}{\lim}b_{1,T}^{-q}\left|\mathbb{E}_{\mathscr{P}_{U}}a'\widetilde{J}_{T}a-a'J_{\mathscr{P}_{U},T}a\right|=2\pi K_{1,q}f_{\mathscr{P}_{U},a}^{\left(q\right)}$
and 

(ii) $\underset{T\rightarrow\infty}{\lim}\underset{\mathscr{P}\in\boldsymbol{P}_{L}}{\inf}b_{1,T}^{-q}\left|\mathbb{E}_{\mathscr{P}}a'\widetilde{J}_{T}a-a'J_{\mathscr{P},T}a\right|=\underset{T\rightarrow\infty}{\lim}b_{1,T}^{-q}\left|\mathbb{E}_{\mathscr{P}_{L}}a'\widetilde{J}_{T}a-a'J_{\mathscr{P}_{L},T}a\right|=2\pi K_{1,q}f_{\mathscr{P}_{L},a}^{\left(q\right)}$.
\end{lem}
\noindent\textit{Proof of Lemma} \ref{Lemma 2 Andrews 88}. We begin
with part (i). We have, 
\begin{align*}
\underset{T\rightarrow\infty}{\lim} & \underset{\mathscr{P}\in\boldsymbol{P}_{U}}{\sup}b_{1,T}^{-q}\left|\mathbb{E}_{\mathscr{P}}a'\widetilde{J}_{T}a-a'J_{\mathscr{P},T}a\right|\\
 & =\underset{T\rightarrow\infty}{\lim}\underset{\mathscr{P}\in\boldsymbol{P}_{U}}{\sup}b_{1,T}^{-q}\left|\sum_{k=-T+1}^{T-1}K_{1}\left(b_{1,T}k\right)a'\mathbb{E}_{\mathscr{P}}\left(\widetilde{\Gamma}\left(k\right)\right)a-\sum_{k=-T+1}^{T-1}a'\Gamma_{\mathscr{P},T}\left(k\right)a\right|\\
 & =\underset{T\rightarrow\infty}{\lim}\underset{\mathscr{P}\in\boldsymbol{P}_{U}}{\sup}b_{1,T}^{-q}\left|\sum_{k=-T+1}^{T-1}\right.K_{1}\left(b_{1,T}k\right)a'\mathbb{E}_{\mathscr{P}}\left(\widetilde{\Gamma}\left(k\right)\right)a-\sum_{k=-T+1}^{T-1}K_{1}\left(b_{1,T}k\right)a'\Gamma_{\mathscr{P},T}\left(k\right)a\\
 & \quad+\sum_{k=-T+1}^{T-1}K_{1}\left(b_{1,T}k\right)a'\Gamma_{\mathscr{P},T}\left(k\right)a-\left.\sum_{k=-T+1}^{T-1}a'\Gamma_{\mathscr{P},T}\left(k\right)a\right|\\
 & =\underset{T\rightarrow\infty}{\lim}\underset{\mathscr{P}\in\boldsymbol{P}_{U}}{\sup}b_{1,T}^{-q}\left|G_{1,\mathscr{P},T}+G_{2,\mathscr{P},T}\right|.
\end{align*}
 Let us first consider $G_{1,\mathscr{P},T}$. Note that for $k\geq0$,
 
\begin{align*}
a' & \left(\mathbb{E}_{\mathscr{P}}\left(\widetilde{\Gamma}\left(k\right)\right)-\Gamma_{\mathscr{P},T}\left(k\right)\right)a\\
 & =\left(\frac{n_{T}}{T}\sum_{r=0}^{\left\lfloor T/n_{T}\right\rfloor }\sum_{s=k+1}^{T}T^{-1}\left(b_{2,T}^{-1}K_{2}\left(\frac{\left(\left(r+1\right)n_{T}-\left(s+k/2\right)\right)/T}{b_{2,T}}\right)-1\right)a'\mathbb{E}_{\mathscr{P}}\left(V_{s}V'_{s-k}\right)a\right).
\end{align*}
Thus,
\begin{align*}
\underset{\mathscr{P}\in\boldsymbol{P}_{U}}{\sup} & \left|a'\left(\mathbb{E}_{\mathscr{P}}\left(\widetilde{\Gamma}\left(k\right)\right)-\Gamma_{\mathscr{P},T}\left(k\right)\right)a\right|\\
 & \leq|\frac{n_{T}}{T}\sum_{r=0}^{\left\lfloor T/n_{T}\right\rfloor }\sum_{s=k+1}^{T}T^{-1}\left(b_{2,T}^{-1}K_{2}\left(\frac{\left(\left(r+1\right)n_{T}-\left(s+k/2\right)\right)/T}{b_{2,T}}\right)-1\right)a'\mathbb{E}_{\mathscr{P}_{U}}\left(V_{s}V'_{s-k}\right)a|.
\end{align*}
By Lemma S.A.1 in \citeReferencesSupp{casini_hac}, $\mathbb{E}_{\mathscr{P}_{U}}(V_{s}V'_{s-k})=c\left(s/T,\,k\right)+O\left(T^{-1}\right)$
uniformly in $s$ and $k$. By the proof of Lemma S.A.8 in \citeReferencesSupp{casini_hac},
\begin{align*}
\underset{\mathscr{P}\in\boldsymbol{P}_{U}}{\sup} & \left|a'\left(\mathbb{E}_{\mathscr{P}}\left(\widetilde{\Gamma}\left(k\right)\right)-\Gamma_{\mathscr{P},T}\left(k\right)\right)a\right|\\
 & \leq\biggl|\frac{n_{T}}{T}\sum_{r=0}^{\left\lfloor T/n_{T}\right\rfloor }\sum_{s=k+1}^{T}T^{-1}\left(\left(b_{2,T}\right)^{-1}K_{2}\left(\frac{\left(\left(r+1\right)n_{T}-\left(s+k/2\right)\right)/T}{b_{2,T}}\right)-1\right)a'\mathbb{E}_{\mathscr{P}_{U}}\left(V_{s}V'_{s-k}\right)a\biggl|\\
 & =O\left(\frac{n_{T}}{T}\right)+\left|\frac{1}{2}b_{2,T}^{2}\int_{0}^{1}x^{2}K_{2}\left(x\right)dx\int_{0}^{1}a'\left(\frac{\partial^{2}}{\partial^{2}u}c\left(u,\,k\right)\right)adu\right|+\Delta_{f}\left(0\right)O\left(b_{2,T}^{2}\right)+O\left(\frac{1}{Tb_{2,T}}\right).
\end{align*}
 It then follows that $\underset{T\rightarrow\infty}{\lim}\underset{\mathscr{P}\in\boldsymbol{P}_{U}}{\sup}b_{1,T}^{-q}|G_{1,\mathscr{P},T}|=0$
given the conditions $n_{T}/Tb_{1,T}^{q}\rightarrow0$ and $b_{2,T}^{2}/b_{1,T}^{q}\rightarrow0$.
Next, given that $1-K_{1}\left(b_{1,T}k\right)\geq0$, 
\begin{align*}
\underset{T\rightarrow\infty}{\lim} & \underset{\mathscr{P}\in\boldsymbol{P}_{U}}{\sup}b_{1,T}^{-q}\left|G_{2,\mathscr{P},T}\right|\\
 & =\underset{T\rightarrow\infty}{\lim}\underset{\mathscr{P}\in\boldsymbol{P}_{U}}{\sup}b_{1,T}^{-q}\left|\sum_{k=-T+1}^{T-1}\left(K_{1}\left(b_{1,T}k\right)-1\right)a'\Gamma_{\mathscr{P},T}\left(k\right)a\right|\\
 & =\underset{T\rightarrow\infty}{\lim}b_{1,T}^{-q}\sum_{k=-T+1}^{T-1}\left(1-K_{1}\left(b_{1,T}k\right)\right)a'\mathbb{E}_{\mathscr{P}_{U}}\left(\widetilde{\Gamma}\left(k\right)\right)a.
\end{align*}
Write the right-hand side above as, 
\begin{align}
\underset{T\rightarrow\infty}{\lim} & b_{1,T}^{-q}\sum_{k=-T+1}^{T-1}\left(1-K_{1}\left(b_{1,T}k\right)\right)a'\left(\mathbb{E}_{\mathscr{P}_{U}}\left(\widetilde{\Gamma}\left(k\right)\right)-\int_{0}^{1}c_{\mathscr{P}_{U}}\left(u,\,k\right)du\right)a\nonumber \\
 & +\underset{T\rightarrow\infty}{\lim}b_{1,T}^{-q}\sum_{k=-T+1}^{T-1}\left(1-K_{1}\left(b_{1,T}k\right)\right)a'\left(\int_{0}^{1}c_{\mathscr{P}_{U}}\left(u,\,k\right)du\right)a.\label{Eq. (1-K1)*Int_c}
\end{align}
By Lemma S.A.1 in \citeReferencesSupp{casini_hac}, the first term
above is less than, 
\begin{align}
\underset{T\rightarrow\infty}{\lim}b_{1,T}^{-q}\sum_{k=-T+1}^{T-1}\left(1-K_{1}\left(b_{1,T}k\right)\right)O\left(T^{-1}\right) & =0.\label{Eq. (A.11b)}
\end{align}
Thus, it remains to consider the second term of \eqref{Eq. (1-K1)*Int_c}.
Let $w\left(x\right)=\left(1-K_{1}\left(x\right)\right)/\left|x\right|^{q}$
for $x\neq0$ and $w\left(x\right)=K_{1,q}$ for $x=0.$ The following
properties hold: $w\left(x\right)\rightarrow K_{1,q}$ as $x\rightarrow0$;
$w\left(\cdot\right)$ is non-negative and bounded. The latter property
implies that there exists some constant $C<\infty$ such that $w\left(x\right)\leq C$
for all $x\in\mathbb{R}.$ Recall that $|\int_{0}^{1}f_{\mathscr{P}_{w},a}^{\left(q\right)}\left(u,\,0\right)du|\in[0,\,\infty)$,
$w=U,\,L$. Hence, given any $\varepsilon>0,$ we can choose a $\mathring{T}<\infty$
such that $\int_{0}^{1}\sum_{k=\mathring{T}+1}^{\infty}\left|k\right|^{q}(a'\Gamma_{\mathscr{P}_{U},u}\left(k\right)a)du<\varepsilon/\left(4C\right)$.
Then, using \eqref{Eq. (A.11b)}, we have 
\begin{align*}
\underset{T\rightarrow\infty}{\lim} & \underset{\mathscr{P}\in\boldsymbol{P}_{U}}{\sup}b_{1,T}^{-q}\left|G_{2,T}-2\pi K_{1,q}f_{U,a}^{\left(q\right)}\right|\\
 & \leq\limsup_{T\rightarrow\infty}\sum_{k=-\mathring{T}}^{\mathring{T}}\left|w\left(b_{1,T}k\right)-K_{1,q}\right|\left|k\right|^{q}a'\left(\int_{0}^{1}c\left(u,\,k\right)du\right)a\\
 & \quad+2\limsup_{T\rightarrow\infty}\sum_{k=-\mathring{T}+1}^{T}\left|w\left(b_{1,T}k\right)-K_{1,q}\right|\left|k\right|^{q}a'\left(\int_{0}^{1}c\left(u,\,k\right)du\right)a\\
 & \leq\varepsilon.
\end{align*}
This concludes the proof of part (i). The proof of part (ii) is identical
to that of part (i) except that $\underset{\mathscr{P}\in\boldsymbol{P}_{U}}{\sup},\,\Gamma_{\mathscr{P}_{U},u}$
and $f_{\mathscr{P}_{U},a}^{\left(q\right)}$ are replaced by $\underset{\mathscr{P}\in\boldsymbol{P}_{L}}{\inf},$
$\Gamma_{\mathscr{P}_{L},u}$ and $f_{\mathscr{P}_{L},a}^{\left(q\right)}$.
$\square$ 

\medskip{}

\noindent\textit{Proof of Theorem} \ref{Theorem MSE J DKHAC Nonstationary}.
Parts (i) and (ii) of the theorem follow from Lemma \ref{Lemma 1 Andrews 88}-(i)
and Lemma \ref{Lemma 2 Andrews 88}-(i), and Lemma \ref{Lemma 1 Andrews 88}-(ii)
and Lemma \ref{Lemma 2 Andrews 88}-(ii), respectively. $\square$

\subsection{Proof of Theorem \ref{Theorem 1 Andrews 91 - Consistency and Rate - Nonstationary}}

Lemma \ref{Lemma 1 Andrews 88}-\ref{Lemma 2 Andrews 88} {[}with
$q=0$ in part (ii){]} implies $\widetilde{J}_{T}-J_{\mathscr{P},T}=o_{\mathscr{P}}\left(1\right)$.
Noting that $\widehat{J}_{T}-\widetilde{J}_{T}=o_{\mathscr{P}}\left(1\right)$
if and only if $a'\widehat{J}_{T}a-a'\widetilde{J}_{T}a=o_{\mathscr{P}}\left(1\right)$
for arbitrary $a\in\mathbb{R}^{p}$ we shall provide the proof only
for the scalar case. We first show that $\sqrt{T}b_{1,T}(\widehat{J}_{T}-\widetilde{J}_{T})=O_{\mathscr{P}}\left(1\right)$
under Assumption \ref{Assumption B}. Let $\widetilde{J}_{T}(\beta)$
denote the estimator that uses $\{V_{t}\left(\beta\right)\}$. A
mean-value expansion of $\widetilde{J}_{T}(\widehat{\beta})\,(=\widehat{J}_{T})$
about $\beta_{0}$ yields, 
\begin{align}
\sqrt{T}b_{1,T}\left(\widehat{J}_{T}-\widetilde{J}_{T}\right) & =b_{1,T}\sum_{k=-T+1}^{T-1}K_{1}\left(b_{1,T}k\right)\frac{\partial}{\partial\beta'}\widehat{\Gamma}\left(k\right)|_{\beta=\bar{\beta}}\sqrt{T}\left(\widehat{\beta}-\beta_{0}\right),\label{eq (A.9) Andrews}
\end{align}
for some $\bar{\beta}$ on the line segment joining $\widehat{\beta}$
and $\beta_{0}$.  We have for $k\geq0$ (the case $k<0$ is similar
and omitted)   \eqref{Eq. A.10 Andrews 91-1}-\eqref{Eq. (A.10) Andrews 91 for K2}.
Proceeding as in the proof of Lemma \ref{Lemma J*(b1*, b2*) - J* =00003D o(1)},
it follows that \eqref{eq (A.9) Andrews} is 
\begin{align*}
b_{1,T} & \sum_{k=-T+1}^{T-1}K_{1}\left(b_{1,T}k\right)\frac{\partial}{\partial\beta'}\widehat{\Gamma}\left(k\right)|_{\beta=\bar{\beta}}\sqrt{T}\left(\widehat{\beta}-\beta_{0}\right)\\
 & \leq b_{1,T}\sum_{k=-T+1}^{T-1}K_{1}\left(b_{1,T}k\right)\frac{n_{T}}{T}\sum_{r=0}^{T/n_{T}}O_{\mathscr{P}}\left(1\right)O_{\mathscr{P}}\left(1\right)\\
 & =O_{\mathscr{P}}\left(1\right),
\end{align*}
where we have used $b_{1,T}\sum_{k=-T+1}^{T-1}|K_{1}(b_{1,T}k)|\rightarrow\int|K_{1}\left(x\right)|dx<\infty.$
Given $\sqrt{T}b_{1,T}\rightarrow\infty$, this concludes the proof
of Theorem \ref{Theorem 1 Andrews 91 - Consistency and Rate - Nonstationary}-(i).

Next, we show that $\sqrt{Tb_{1,T}}(\widehat{J}_{T}-\widetilde{J}_{T})=o_{\mathscr{P}}\left(1\right)$
under the assumptions of Theorem \ref{Theorem 1 Andrews 91 - Consistency and Rate - Nonstationary}-(ii).
A second-order Taylor expansion yields
\begin{align*}
\sqrt{Tb_{1,T}}\left(\widehat{J}_{T}-\widetilde{J}_{T}\right) & =\left[\sqrt{b_{1,T}}\frac{\partial}{\partial\beta'}\widetilde{J}_{T}\left(\beta_{0}\right)\right]\sqrt{T}\left(\widehat{\beta}-\beta_{0}\right)\\
 & \quad+\frac{1}{2}\sqrt{T}\left(\widehat{\beta}-\beta_{0}\right)'\left[\sqrt{b_{1,T}}\frac{\partial^{2}}{\partial\beta\partial\beta'}\widetilde{J}_{T}\left(\overline{\beta}\right)/\sqrt{T}\right]\sqrt{T}\left(\widehat{\beta}-\beta_{0}\right)\\
 & \triangleq G_{T}'\sqrt{T}\left(\widehat{\beta}-\beta_{0}\right)+\frac{1}{2}\sqrt{T}\left(\widehat{\beta}-\beta_{0}\right)'H_{T}\sqrt{T}\left(\widehat{\beta}-\beta_{0}\right).
\end{align*}
We can use the same argument as in \eqref{Eq. A.10 Andrews 91-1}
but now using Assumption \ref{Assumption C*}-(ii),  so that
\begin{align*}
\biggl\Vert & \frac{\partial^{2}}{\partial\beta\partial\beta'}\widehat{c}\left(rn_{T}/T,\,k\right)\biggr\Vert\biggl|_{\beta=\bar{\beta}}\\
 & =\left\Vert \left(Tb_{2,T}\right)^{-1}\sum_{s=k+1}^{T}K_{2}^{*}\left(\frac{\left(\left(r+1\right)n_{T}-\left(s+k/2\right)\right)/T}{b_{2,T}}\right)\left(\frac{\partial^{2}}{\partial\beta\partial\beta'}V_{s}\left(\beta\right)V{}_{s-k}\left(\beta\right)\right)\right\Vert \biggl|_{\beta=\bar{\beta}}\\
 & =O_{\mathscr{P}}\left(1\right),
\end{align*}
 and thus, 
\begin{align*}
\left\Vert H_{T}\right\Vert  & \leq\left(\frac{b_{1,T}}{T}\right)^{1/2}\sum_{k=-T+1}^{T-1}\left|K_{1}\left(b_{1,T}k\right)\right|\sup_{\beta\in\Theta}\left\Vert \frac{\partial^{2}}{\partial\beta\partial\beta'}\widehat{\Gamma}\left(k\right)\right\Vert \\
 & \leq\left(\frac{b_{1,T}}{T}\right)^{1/2}\sum_{k=-T+1}^{T-1}\left|K_{1}\left(b_{1,T}k\right)\right|O_{\mathscr{P}}\left(1\right)\\
 & \leq\left(\frac{1}{Tb_{1,T}}\right)^{1/2}b_{1,T}\sum_{k=-T+1}^{T-1}\left|K_{1}\left(b_{1,T}k\right)\right|O_{\mathscr{P}}\left(1\right)=o_{\mathscr{P}}\left(1\right),
\end{align*}
since $Tb_{1,T}\rightarrow\infty$. Next, we show that $G_{T}=o_{\mathscr{P}}\left(1\right)$.
We follow the argument in the last paragraph of p. 852 of \citeReferencesSupp{andrews:91}.
We apply Theorem \ref{Theorem 1 Andrews 91 - Consistency and Rate - Nonstationary}-(i,ii)
to $\widetilde{J}_{T}$ where the latter is constructed using $\left(V'_{t},\,\partial V_{t}/\partial\beta'-\mathbb{E}_{\mathscr{P}}\left(\partial V_{t}/\partial\beta'\right)\right)'$
rather than just with $V_{t}$. The first row and column of the off-diagonal
elements of this $\widetilde{J}_{T}$ are now 
\begin{align*}
A_{1} & \triangleq\sum_{k=-T+1}^{T-1}K_{1}\left(b_{1,T}k\right)\frac{n_{T}}{T}\sum_{r=0}^{T/n_{T}}\frac{1}{Tb_{2,T}}\\
 & \quad\times\sum_{s=k+1}^{T}K_{2}^{*}\left(\frac{\left(\left(r+1\right)n_{T}-\left(s+k/2\right)\right)/T}{b_{2,T}}\right)V_{s}\left(\frac{\partial}{\partial\beta}V{}_{s-k}-\mathbb{E}_{\mathscr{P}}\left(\frac{\partial}{\partial\beta}V{}_{s}\right)\right)\\
A_{2} & \triangleq\sum_{k=-T+1}^{T-1}K_{1}\left(b_{1,T}k\right)\frac{n_{T}}{T}\sum_{r=0}^{T/n_{T}}\frac{1}{Tb_{2,T}}\\
 & \quad\times\sum_{s=k+1}^{T}K_{2}^{*}\left(\frac{\left(\left(r+1\right)n_{T}-\left(s+k/2\right)\right)/T}{b_{2,T}}\right)\left(\frac{\partial}{\partial\beta}V{}_{s}-\mathbb{E}_{\mathscr{P}}\left(\frac{\partial}{\partial\beta}V{}_{s}\right)\right)V_{s-k},
\end{align*}
which are both $O_{\mathscr{P}}\left(1\right)$ by Theorem \ref{Theorem MSE J DKHAC Nonstationary}.
Note that 
\begin{align*}
G_{T} & \leq\sqrt{b_{1,T}}\left(A_{1}+A_{2}\right)+\sqrt{b_{1,T}}\sum_{k=-T+1}^{T-1}\left|K_{1}\left(b_{1,T}k\right)\right|\frac{n_{T}}{T}\sum_{r=0}^{T/n_{T}}\frac{1}{Tb_{2,T}}\\
 & \quad\times\sum_{s=k+1}^{T}K_{2}^{*}\left(\frac{\left(\left(r+1\right)n_{T}-\left(s+k/2\right)\right)/T}{b_{2,T}}\right)\left|\left(V_{s}+V_{s-k}\right)\right|\left|\mathbb{E}_{\mathscr{P}}\left(\frac{\partial}{\partial\beta}V{}_{s}\right)\right|\\
 & \triangleq\sqrt{b_{1,T}}\left(A_{1}+A_{2}\right)+A_{3}\sup_{1\leq s\leq T}\left|\mathbb{E}_{\mathscr{P}}\left(\frac{\partial}{\partial\beta}V{}_{s}\right)\right|.
\end{align*}
It remains to show that $A_{3}$ is $o_{\mathscr{P}}\left(1\right).$
We have,
\begin{align*}
\mathbb{E}_{\mathscr{P}}\left(A_{3}^{2}\right) & \leq b_{1,T}\sum_{k=-T+1}^{T-1}\sum_{j=-T+1}^{T-1}\left|K_{1}\left(b_{1,T}k\right)K_{1}\left(b_{1,T}j\right)\right|4\left(\frac{n_{T}}{T}\right)^{2}\sum_{r=0}^{T/n_{T}}\sum_{b=0}^{T/n_{T}}\\
 & \quad\times\frac{1}{Tb_{2,T}}\frac{1}{Tb_{2,T}}\sum_{s=1}^{T}\sum_{l=1}^{T}K_{2}^{*}\left(\frac{\left(\left(r+1\right)n_{T}-\left(s+k/2\right)\right)/T}{b_{2,T}}\right)\\
 & \quad\times K_{2}^{*}\left(\frac{\left(\left(b+1\right)n_{T}-\left(l+j/2\right)\right)/T}{b_{2,T}}\right)\left|\mathbb{E}_{\mathscr{P}}\left(V_{s}V_{l}\right)\right|.
\end{align*}
Since $\mathscr{P}\in\mathscr{P}_{U}$, $|\mathbb{E}_{\mathscr{P}}(V_{s}V_{l})|\leq|\Gamma_{\mathscr{P}_{U},s/T}(l-s)|.$
Given $\sum_{h=-\infty}^{\infty}\sup_{u\in\left[0,\,1\right]}|c_{\mathscr{P}_{U}}\left(u,\,h\right)|<\infty$,
we have   
\begin{align}
\mathbb{E}_{\mathscr{P}}\left(A_{3}^{2}\right) & \leq\frac{1}{Tb_{1,T}b_{2,T}}\left(b_{1,T}\sum_{k=-T+1}^{T-1}\left|K_{1}\left(b_{1,T}k\right)\right|\right)^{2}\int_{0}^{1}K_{2}^{2}\left(x\right)dx\int_{0}^{1}\sum_{h=-\infty}^{\infty}\left|c_{\mathscr{P}_{U}}\left(u,\,h\right)\right|du=o\left(1\right),\label{Eq. (A.15) Andrews 91, A.3 ->0}
\end{align}
from which it follows that $G_{T}=o_{\mathscr{P}}\left(1\right)$
and so $\sqrt{Tb_{1,T}}(\widehat{J}_{T}-\widetilde{J}_{T})=o_{\mathscr{P}}\left(1\right)$.
The latter concludes the proof of part (ii) because $\sqrt{Tb_{1,T}b_{2,T}}(\widetilde{J}_{T}-J_{T})=O_{\mathscr{P}}\left(1\right)$
by Theorem \ref{Theorem MSE J DKHAC Nonstationary}.

Let us consider part (iii). Let $\overline{G}_{T}=a'\widehat{J}_{T}a-a'\widetilde{J}_{T}a$.
We have, 
\begin{align}
\lim_{T\rightarrow\infty} & \sup_{\mathscr{P}\in\boldsymbol{P}_{U}}Tb_{1,T}b_{2,T}\left|\mathrm{MSE}_{\mathscr{P}}\left(a'\widehat{J}_{T}a\right)-\mathrm{MSE}_{\mathscr{P}}\left(a'\widetilde{J}_{T}a\right)\right|\label{Eq. (A.14) Andrews 88}\\
 & =\lim_{T\rightarrow\infty}\sup_{\mathscr{P}\in\boldsymbol{P}_{U}}Tb_{1,T}b_{2,T}\left|2\mathbb{E}_{\mathscr{P}}\left(a'\widetilde{J}_{T}a-a'J_{\mathscr{P},T}a\right)\overline{G}_{T}+\mathbb{E}_{\mathscr{P}}\left(\overline{G}_{T}^{2}\right)\right|\nonumber \\
 & \leq2\lim_{T\rightarrow\infty}\left(\sup_{\mathscr{P}\in\boldsymbol{P}_{U}}Tb_{1,T}b_{2,T}\mathrm{MSE}_{\mathscr{P}}\left(a'\widetilde{J}_{T}a\right)\right)^{1/2}\left(\sup_{\mathscr{P}\in\boldsymbol{P}_{U}}Tb_{1,T}b_{2,T}\mathbb{E}_{\mathscr{P}}\left(\overline{G}_{T}^{2}\right)\right)^{1/2}\nonumber \\
 & \quad+\lim_{T\rightarrow\infty}\sup_{\mathscr{P}\in\boldsymbol{P}_{U}}Tb_{1,T}b_{2,T}\mathbb{E}_{\mathscr{P}}\left(\overline{G}_{T}^{2}\right).\nonumber 
\end{align}
The right-hand side above equals zero if (a) $\lim_{T\rightarrow\infty}\sup_{\mathscr{P}\in\boldsymbol{P}_{U}}Tb_{1,T}b_{2,T}\mathbb{E}_{\mathscr{P}}(\overline{G}_{T}^{2})=0$
and (b) $\limsup_{T\rightarrow\infty}$ $\sup_{\mathscr{P}\in\boldsymbol{P}_{U}}Tb_{1,T}b_{2,T}\mathrm{MSE}_{\mathscr{P}}(a'\widetilde{J}_{T}a)<\infty$.
Result (b) follows by Lemma \ref{Lemma 1 Andrews 88}-(i).  A second-order
expansion yields, 
\begin{align}
\overline{G}_{T} & =\left[\frac{\partial}{\partial\beta}a'\widetilde{J}_{T}\left(\beta_{0}\right)a\right]\left(\widehat{\beta}-\beta_{0}\right)+\frac{1}{2}\left(\widehat{\beta}-\beta_{0}\right)'\left[\frac{\partial^{2}}{\partial\beta\partial\beta'}a'\widetilde{J}_{T}\left(\overline{\beta}\right)a\right]\left(\widehat{\beta}-\beta_{0}\right)=\overline{G}_{1,T}+\overline{G}_{2,T},\label{Eq. (A.30) Andrews 88}
\end{align}
 where $\overline{\beta}$ lies on the line segment joining $\widehat{\beta}$
and $\beta_{0}$. Note that $\mathbb{E}_{\mathscr{P}}(\overline{G}_{T}^{2})=\mathbb{E}_{\mathscr{P}}(\overline{G}_{1,T}^{2})+\mathbb{E}_{\mathscr{P}}(\overline{G}_{2,T}^{2})+2\mathbb{E}_{\mathscr{P}}(\overline{G}_{1,T}\overline{G}_{2,T})$.
Thus, using Assumption \ref{Assumption B Andrews 88},
\begin{align}
\sup_{\mathscr{P}\in\boldsymbol{P}_{U}} & Tb_{1,T}b_{2,T}\mathbb{E}_{\mathscr{P}}\left(\overline{G}_{1,T}^{2}\right)\label{Eq. (A.32) Andrews 88}\\
 & \leq Tb_{1,T}b_{2,T}p^{2}\max_{r\leq p}\sup_{\mathscr{P}\in\boldsymbol{P}_{U}}\mathbb{E}_{\mathscr{P}}\left(\frac{\partial}{\partial\beta^{\left(r\right)}}a'\widetilde{J}_{T}\left(\beta_{0}\right)a\left(\widehat{\beta}^{\left(r\right)}-\widehat{\beta}_{0}^{\left(r\right)}\right)\right)^{2}\nonumber \\
 & \leq\frac{1}{Tb_{1,T}}p^{2}\max_{r\leq p}\sup_{\mathscr{P}\in\boldsymbol{P}_{U}}\mathbb{E}_{\mathscr{P}}\left(H_{1,T}^{\left(r\right)}\sqrt{T}\left(\widehat{\beta}^{\left(r\right)}-\widehat{\beta}_{0}^{\left(r\right)}\right)\right)^{2}\nonumber \\
 & \rightarrow0,\nonumber 
\end{align}
 and 
\begin{align}
\sup_{\mathscr{P}\in\boldsymbol{P}_{U}} & Tb_{1,T}b_{2,T}\mathbb{E}_{\mathscr{P}}\left(\overline{G}_{2,T}^{2}\right)\label{Eq. (A.33) Andrews 88}\\
 & \leq\frac{1}{4}Tb_{1,T}b_{2,T}\sup_{\mathscr{P}\in\boldsymbol{P}_{U}}\mathbb{E}_{\mathscr{P}}\left(\biggl|\widehat{\beta}-\beta_{0}\biggr|\left|\frac{\partial^{2}}{\partial\beta\partial\beta^{\prime}}a'\widetilde{J}_{T}\left(\overline{\beta}\right)a\right|\biggl|\widehat{\beta}-\beta_{0}\biggr|\right)^{2}\nonumber \\
 & \leq\frac{b_{2,T}}{Tb_{1,T}}\sup_{\mathscr{P}\in\boldsymbol{P}_{U}}\mathbb{E}_{\mathscr{P}}\left(\sqrt{T}\biggl|\widehat{\beta}-\beta_{0}\biggr|H_{2,T}\sqrt{T}\biggl|\widehat{\beta}-\beta_{0}\biggr|\right)^{2}\nonumber \\
 & \rightarrow0.\nonumber 
\end{align}
Equations \eqref{Eq. (A.30) Andrews 88} to \eqref{Eq. (A.33) Andrews 88}
and the Cauchy-Schwartz inequality yield result (a) and thus the desired
result of the theorem. $\square$

\subsection{Proof of Proposition \ref{Proposition: Optimal Local Covariance - Nonstationary}}

For $K_{2}\left(\cdot\right)\in\boldsymbol{K}_{2}$, using the definition
of $\mathscr{P}_{U}$ and the arguments in \eqref{eq A.1 Andrews 88},
\begin{align}
\mathrm{Var}_{\mathscr{P}_{U}} & \left(a'\widetilde{c}_{T}\left(u_{0},\,k\right)a\right)\nonumber \\
 & \leq\sup_{\mathscr{P}\in\boldsymbol{P}_{U}}\mathrm{Var}_{\mathscr{P}}\left(a'\widetilde{c}_{T}\left(u_{0},\,k\right)a\right)\nonumber \\
 & =\sup_{\mathscr{P}\in\boldsymbol{P}_{U}}\mathbb{E}_{\mathscr{P}}\left(\left[\left(Tb_{2,T}\right)^{-1}\sum_{s=k+1}^{T}K_{2}^{*}\left(\frac{u_{0}-\left(s+k/2\right)/T}{b_{2,T}}\right)a'\left(\widetilde{V}_{s}\widetilde{V}'_{s-k}-\mathbb{E}_{\mathscr{P}}\left(\widetilde{V}_{s}\widetilde{V}'_{s-k}\right)\right)a\right]^{2}\right)\nonumber \\
 & =\sup_{\mathscr{P}\in\boldsymbol{P}_{U}}\mathbb{E}_{\mathscr{P}}\left(Tb_{2,T}\right)^{-2}\sum_{s=k+1}^{T}\sum_{l=j+1}^{T}K_{2}^{*}\left(\frac{u_{0}-\left(s+k/2\right)/T}{b_{2,T}}\right)K_{2}^{*}\left(\frac{u_{0}-\left(l+j/2\right)/T}{b_{2,T}}\right)\nonumber \\
 & \quad\times a'\left(\widetilde{V}_{s}\widetilde{V}'_{s-k}-\mathbb{E}_{\mathscr{P}}\left(\widetilde{V}_{s}\widetilde{V}'_{s-k}\right)\right)aa'\left(\widetilde{V}_{l}\widetilde{V}'_{l-j}-\mathbb{E}_{\mathscr{P}}\left(\widetilde{V}_{l}\widetilde{V}'_{l-j}\right)\right)a\nonumber \\
 & \leq\left(Tb_{2,T}\right)^{-2}\sum_{s=k+1}^{T}\sum_{l=j+1}^{T}\left|K_{2}^{*}\left(\frac{u_{0}-\left(s-k/2\right)/T}{b_{2,T}}\right)K_{2}^{*}\left(\frac{u_{0}-\left(l-j/2\right)/T}{b_{2,T}}\right)\right|\nonumber \\
 & \quad\times(a'\Gamma_{U,s/T}\left(s-l\right)aa'\Gamma_{U,s-k}\left(s-k-l+j\right)a\nonumber \\
 & \quad+a'\Gamma_{U,s/T}\left(s-l+j\right)aa'\Gamma_{U,s-k}\left(s-k-l\right)a+\kappa_{\mathscr{P}_{U},aV,s}\left(j,\,l-s,\,l-j-s\right))\nonumber \\
 & \leq\mathbb{E}_{\mathscr{P}_{U}}\left(a'\left(\overline{c}_{T}\left(u_{0},\,k\right)-\overline{c}_{\mathscr{P}_{U},T}\left(u_{0},\,k\right)\right)aa'\left(\overline{c}_{T}\left(u_{0},\,j\right)-\overline{c}_{\mathscr{P}_{U}T}\left(u_{0},\,j\right)\right)a\right)\nonumber \\
 & \quad+2\left(\frac{1}{Tb_{2,T}}\right)^{2}\sum_{s=k+1}^{T}\sum_{l=j+1}^{T}\left|K_{2}^{*}\left(\frac{\left(\left(r+1\right)n_{T}-\left(s-k/2\right)\right)/T}{b_{2,T}}\right)K_{2}^{*}\left(\frac{\left(\left(b+1\right)n_{T}-\left(l-j/2\right)\right)/T}{b_{2,T}}\right)\right|\nonumber \\
 & \quad\times\kappa_{\mathscr{P}_{U},aV,s}\left(j,\,l-s,\,l-j-s\right)\nonumber \\
 & =\mathrm{Var}_{\mathscr{P}_{U}}\left(a'\overline{c}_{T}\left(u_{0},\,k\right)a\right),\label{Eq. (A.1) Andrews 88 Local Autocovariance}
\end{align}
 where $\overline{c}_{T}(u_{0},\,k)$ (resp. $\overline{c}_{\mathscr{P}_{U},T}(u_{0},\,k)$)
is equal to $\widetilde{c}_{T}(u_{0},\,k)$ (resp. $c_{\mathscr{P}_{U},T}(u_{0},\,k)$)
but with $|K_{2}\left(\cdot\right)|$ in place of $K_{2}\left(\cdot\right)$.
Since $K_{2}\left(\cdot\right)\geq0$ by definition, Proposition 3.1
in \citeReferencesSupp{casini_hac} implies
\begin{align}
\mathrm{Var}_{\mathscr{P}_{U}} & \left(a'\widetilde{c}_{T}\left(u_{0},\,k\right)a\right)\nonumber \\
 & =\frac{1}{Tb_{2,T}}\int_{0}^{1}K_{2}^{2}\left(x\right)dx\sum_{l=-\infty}^{\infty}a'\left(c_{\mathscr{P}_{U}}\left(u_{0},\,l\right)\left[c_{\mathscr{P}_{U}}\left(u_{0},\,l\right)+c_{\mathscr{P}_{U}}\left(u_{0},\,l+2k\right)\right]'\right)a\nonumber \\
 & \quad+\frac{1}{Tb_{2,T}}\int_{0}^{1}K_{2}^{2}\left(x\right)dx\sum_{h_{1}=-\infty}^{\infty}\sum_{h_{2}=-\infty}^{\infty}\kappa_{\mathscr{P}_{U},aV,Tu_{0}}\left(h_{1},\,0,\,h_{2}\right)\nonumber \\
 & \quad+o\left(b_{2,T}^{4}\right)+O\left(1/\left(b_{2,T}T\right)\right)\nonumber \\
 & =\mathrm{Var}_{\mathscr{P}_{U}}\left(a'\overline{c}_{T}\left(u_{0},\,k\right)a\right).\label{Eq. (A.6) Andrews 88 Local Autocvariance Nonstationary}
\end{align}
 Next, we discuss the bias. We have, 
\begin{align}
\underset{\mathscr{P}\in\boldsymbol{P}_{U,2}}{\sup} & \left|\mathbb{E}_{\mathscr{P}}\left(a'\widetilde{c}_{T}\left(u_{0},\,k\right)a-a'c_{\mathscr{P}}\left(u_{0},\,k\right)a\right)\right|\nonumber \\
 & =\underset{T\rightarrow\infty}{\lim}\underset{\mathscr{P}\in\boldsymbol{P}_{U,2}}{\sup}\left|\left(Tb_{2,T}\right)^{-1}\sum_{s=k+1}^{T}K_{2}\left(\frac{\left(\left(r+1\right)n_{T}-\left(s+k/2\right)\right)/T}{b_{2,T}}\right)a'\mathbb{E}_{\mathscr{P}}\left(V_{s}V'_{s-k}\right)a-a'c_{\mathscr{P}}\left(u_{0},\,k\right)a\right|\nonumber \\
 & \leq\frac{1}{2}b_{2,T}^{2}\int_{0}^{1}x^{2}K_{2}\left(x\right)dx\int_{0}^{1}\left|a'\frac{\partial^{2}}{\partial^{2}u}c_{\mathscr{P}_{U}}\left(u_{0},\,k\right)a\right|du+o\left(b_{2,T}^{2}\right)+O\left(\frac{1}{Tb_{2,T}}\right),\label{Eq. (A.12) Andrews 88 Bias Local Autocovariance- Nonstationary}
\end{align}
where the inequality above follows from \eqref{Eq. Inequality for Pu2}.
Combining \eqref{Eq. (A.6) Andrews 88 Local Autocvariance Nonstationary}-\eqref{Eq. (A.12) Andrews 88 Bias Local Autocovariance- Nonstationary},
we have that $\sup_{\mathscr{P}\in\boldsymbol{P}_{U,2}}\mathrm{MSE}(a'\widetilde{c}_{T}$
$(u_{0},\,k)a)$ is equal to the right-hand side of \eqref{Eq. MSE Local Autocov in Prop - Nonstationary}.
The same result holds for $\widehat{c}_{T}(u_{0},\,k)$ since the
proof of Theorem \ref{Theorem 1 Andrews 91 - Consistency and Rate - Nonstationary}
and $\boldsymbol{P}_{U,2}\subseteq\boldsymbol{P}_{U}$ imply that
$\sup_{\mathscr{P}\in\boldsymbol{P}_{U,2}}\mathrm{MSE}_{\mathscr{P}}(a'\widehat{c}_{T}(u_{0},\,k)a)$
is asymptotically equivalent to $\sup_{\mathscr{P}\in\boldsymbol{P}_{U,2}}\mathrm{MSE}_{\mathscr{P}}(a'\widetilde{c}_{T}(u_{0},\,k)a)$.
This gives \eqref{Eq. MSE Local Autocov in Prop - Nonstationary}.
The form for the optimal $b_{2,T}\left(\cdot\right)$ and $K_{2}\left(\cdot\right)$
follow from the same argument as in Proposition 4.1 in \citeReferencesSupp{casini_hac}.
$\square$ 

\subsection{Proof of Theorem \ref{Theorem Optimal Kernels Nonstationarity}}

If $Tb_{1,T}^{2q+1}b_{2,T}\rightarrow\gamma\in\left(0,\,\infty\right)$
for some $q\in[0,\,\infty)$ for which $K_{1,q},\,|\int_{0}^{1}f_{U,a}^{\left(q\right)}\left(u,\,0\right)du|\in[0,\,\infty)$,
then by Lemma \ref{Lemma 1 Andrews 88}-(i) and Lemma \ref{Lemma 2 Andrews 88}-(i),
\begin{align*}
\lim_{T\rightarrow\infty} & Tb_{1,T}b_{2,T}\sup_{\mathscr{P}\in\boldsymbol{P}_{U}}\mathrm{MSE}_{\mathscr{P}}\left(a'\widehat{J}_{T}^{\mathrm{}}\left(b_{1,T,K_{1}}\right)a\right)\\
 & =4\pi^{2}\left[\gamma K_{1,q}^{2}\left(\int_{0}^{1}f_{U,a}^{\left(q\right)}\left(u,\,0\right)du\right)^{2}+\int K_{1}^{2}\left(y\right)dy\int_{0}^{1}\left(K_{2,0}\left(x\right)\right)^{2}dx\,\left(\int_{0}^{1}f_{U,a}\left(u,\,0\right)du\right)^{2}\right].
\end{align*}
 Assume $q=2$ so that $Tb_{1,T}^{5}b_{2,T}\rightarrow\gamma$. Then,
$Tb_{1,T,K_{1}}^{5}b_{2,T}\rightarrow\gamma/(\int K_{1}^{2}\left(y\right)dy)^{5}$
and 
\begin{align*}
Tb_{1,T}b_{2,T}=Tb_{1,T,K_{1}}b_{2,T}\int K_{1}^{2}\left(y\right)dy & .
\end{align*}
 Therefore, given $K_{1,2}<\infty$, 

\begin{align*}
\liminf_{T\rightarrow\infty}Tb_{1,T}b_{2,T} & \left(\sup_{\mathscr{P}\in\boldsymbol{P}_{U}}\mathrm{MSE}\left(a'\widehat{J}_{T}^{\mathrm{}}\left(b_{1,T,K_{1}}\right)a\right)-\sup_{\mathscr{P}\in\boldsymbol{P}_{U}}\mathrm{MSE}\left(a'\widehat{J}_{T}^{\mathrm{QS}}\left(b_{1,T}\right)a\right)\right)\\
 & =4\gamma\pi^{2}\left(\int_{0}^{1}f_{U,a}^{\left(q\right)}\left(u,\,0\right)du\right)^{2}\int_{0}^{1}\left(K_{2}^{\mathrm{}}\left(x\right)\right)^{2}dx\left[K_{1,2}^{2}\left(\int K_{1}^{2}\left(y\right)dy\right)^{4}-\left(K_{1,2}^{\mathrm{QS}}\right)^{2}\right].
\end{align*}
 The optimality of $K_{1}^{\mathrm{QS}}$ then follows from the
same argument as in the proof of Theorem 4.1 in \citeReferencesSupp{casini_hac}.
$\square$

\subsection{Proof of Theorem \ref{Theorem 3 Andrews 88}}

Suppose $\gamma\in\left(0,\,\infty\right)$. Under the conditions
of the theorem,
\[
(Tb_{2,T}^{\mathrm{}})^{2q/\left(2q+1\right)}=(\gamma^{-1/\left(2q+1\right)}+o\left(1\right))Tb_{1,T}b_{2,T}.
\]
 By Theorem \ref{Theorem MSE J DKHAC Nonstationary}-(i), 
\begin{align}
\liminf_{T\rightarrow\infty} & \left(Tb_{2,T}^{\mathrm{}}\right)^{2q/\left(2q+1\right)}\sup_{\mathscr{P}\in\boldsymbol{P}_{U}\left(\phi\left(q\right)\right)}\mathbb{E}_{\mathscr{P}}\mathrm{L}\left(\widetilde{J}_{T}\left(b_{1,T}\right),\,J_{\mathscr{P},T}\right)\label{Eq. (Jtilde) (A.14) Andrews 88}\\
 & =\liminf_{T\rightarrow\infty}\left(\gamma^{-1/\left(2q+1\right)}+o\left(1\right)\right)Tb_{1,T}b_{2,T}\sup_{\mathscr{P}\in\boldsymbol{P}_{U}\left(\phi\right)}\sum_{r=1}^{p}w_{r}\mathrm{MSE}_{\mathscr{P}}\left(a^{\left(r\right)\prime}\widetilde{J}_{T}\left(b_{1,T}\right)a^{\left(r\right)}\right)\nonumber \\
 & =\gamma^{-1/\left(2q+1\right)}4\pi^{2}\biggl[\sum_{r=1}^{p}w_{r}\biggl(\gamma K_{1,q}^{2}\left(\int_{0}^{1}f_{\mathscr{P}_{U},a^{\left(r\right)}}^{\left(q\right)}\left(u,\,0\right)du\right)^{2}\nonumber \\
 & \quad+2\int K_{1}^{2}\left(x\right)dx\int_{0}^{1}K_{2}^{2}\left(y\right)dy\left(\int_{0}^{1}f_{\mathscr{P}_{U},a^{\left(r\right)}}\left(u,\,0\right)du\right)^{2}\biggr)\biggr].\nonumber 
\end{align}
The right-hand side above is minimized at $\gamma^{\mathrm{opt}}=(2qK_{1,q}^{2}\phi\left(q\right))^{-1}(\int K_{1}^{2}\left(y\right)dy\int_{0}^{1}K_{2}^{2}\left(x\right)dx).$
Note that $\gamma^{\mathrm{opt}}>0$ provided that $f_{\mathscr{P}_{U},a^{\left(r\right)}}\left(u,\,0\right)>0$
and $f_{\mathscr{P}_{U},a^{\left(r\right)}}^{\left(q\right)}\left(u,\,0\right)>0$
for some $u\in\left[0,\,1\right]$ and some $r$ for which $w_{r}>0$.
Hence, $\{b_{1,T}\}$ is optimal in the sense that $Tb_{1,T}^{2q+1}b_{2,T}\rightarrow\gamma^{\mathrm{opt}}$
if and only if $b_{1,T}=b_{1,T}^{\mathrm{opt}}+o((Tb_{2,T})^{-1/\left(2q+1\right)})$.
In virtue of Theorem \ref{Theorem 1 Andrews 91 - Consistency and Rate - Nonstationary}-(iii),
eq. \eqref{Eq. (Jtilde) (A.14) Andrews 88} holds also when $\widetilde{J}_{T}(b_{1,T})$
is replaced by $\widehat{J}_{T}(b_{1,T})$. Thus, the final assertion
of the theorem follows. $\square$ 

\subsection{Proof of Theorem \ref{Theorem 6 Andrews 88}}

The proof of the theorem uses the following lemmas. 
\begin{lem}
\label{Lemma A.1 Andrews 88}Let $K_{1}\left(\cdot\right),\,K_{2}\left(\cdot\right)$,
$\{b_{1,\theta_{\mathscr{P}},T}\}$, $\{S_{\mathscr{P},T}\}$, $\widehat{\phi}\left(\cdot\right)$
and $q$ be as in Theorem \ref{Theorem 6 Andrews 88}. Then, for all
$a\in\mathbb{R}^{p}$, (i)
\begin{align*}
T^{8q/5\left(2q+1\right)}\sup_{\mathscr{P}\in\boldsymbol{P}_{U,3}}\mathbb{E}_{\mathscr{P}}\left(\sum_{k=S_{\mathscr{P},T}+1}^{T-1}K_{1}\left(\widehat{b}_{1,T}k\right)a'\widehat{\Gamma}\left(k\right)a\right)^{2} & \rightarrow0;
\end{align*}
(ii)
\begin{align*}
T^{8q/5\left(2q+1\right)}\sup_{\mathscr{P}\in\boldsymbol{P}_{U,3}}\mathbb{E}_{\mathscr{P}}\left(\sum_{k=1}^{S_{\mathscr{P},T}}\left(K_{1}\left(\widehat{b}_{1,T}k\right)-K_{1}\left(b_{1,\theta_{\mathscr{P}},T}k\right)\right)a'\widehat{\Gamma}\left(k\right)a\right)^{2} & \rightarrow0.
\end{align*}
\end{lem}
\noindent\textit{Proof of Lemma }\ref{Lemma A.1 Andrews 88}. First
we prove part (i). We have, 
\begin{align}
\Biggl( & T^{8q/5\left(2q+1\right)}\sup_{\mathscr{P}\in\boldsymbol{P}_{U,3}}\mathbb{E}_{\mathscr{P}}\left(\sum_{k=S_{\mathscr{P},T}+1}^{T-1}K_{1}\left(\widehat{b}_{1,T}k\right)a'\widehat{\Gamma}\left(k\right)a\right)^{2}\Biggr)^{1/2}\label{Eq. (A.37) Andrews 88}\\
 & \leq\left(T^{8q/5\left(2q+1\right)}\sup_{\mathscr{P}\in\boldsymbol{P}_{U,3}}\mathbb{E}_{\mathscr{P}}\left(\sum_{k=S_{\mathscr{P},T}+1}^{T-1}K_{1}\left(\widehat{b}_{1,T}k\right)\left(a'\widehat{\Gamma}\left(k\right)a-a'\Gamma_{\mathscr{P},T}a\right)\right)^{2}\right)^{1/2}\nonumber \\
 & \quad+\left(T^{8q/5\left(2q+1\right)}\sup_{\mathscr{P}\in\boldsymbol{P}_{U,3}}\mathbb{E}_{\mathscr{P}}\left(\sum_{k=S_{\mathscr{P},T}+1}^{T-1}K_{1}\left(\widehat{b}_{1,T}k\right)a'\Gamma_{\mathscr{P},T}a\right)^{2}\right)^{1/2}\nonumber \\
 & \triangleq B_{1,T}+B_{2,T}.\nonumber 
\end{align}
Since $|K_{1}\left(\cdot\right)|\leq1$ and $|a'\Gamma_{\mathscr{P},T}\left(k\right)a|\leq a'(\int_{0}^{1}\Gamma_{\mathscr{P}_{U},u}\left(k\right)du)a$,
we obtain
\begin{align}
B_{2,T} & \leq\left(T^{8q/5\left(2q+1\right)}\sup_{\mathscr{P}\in\boldsymbol{P}_{U,3}}\mathbb{E}_{\mathscr{P}}\left(\sum_{k=S_{\mathscr{P},T}+1}^{T-1}\left|K_{1}\left(\widehat{b}_{1,T}k\right)\right|a'\left(\int_{0}^{1}\Gamma_{\mathscr{P}_{U},u}\left(k\right)du\right)a\right)^{2}\right)^{1/2}\label{Eq. (A.38) Andrews 88}\\
 & \leq T^{8q/10\left(2q+1\right)}\sup_{\mathscr{P}\in\boldsymbol{P}_{U,3}}\sum_{k=S_{\mathscr{P},T}+1}^{T-1}\sup_{u\in\left[0,\,1\right]}a'\left(\int_{0}^{1}\Gamma_{\mathscr{P}_{U},u}\left(k\right)du\right)a\nonumber \\
 & \leq T^{8q/10\left(2q+1\right)}\sup_{\mathscr{P}\in\boldsymbol{P}_{U,3}}\sum_{k=S_{\mathscr{P},T}+1}^{T-1}C_{3}k^{-l}\nonumber \\
 & \leq C_{3,1}T^{8q/10\left(2q+1\right)}\sup_{\mathscr{P}\in\boldsymbol{P}_{U,3}}\int_{S_{\mathscr{P},T}}^{\infty}k^{-l}dk\nonumber \\
 & \leq C_{3,1}T^{8q/10\left(2q+1\right)}S_{T,\mathscr{P}}^{1-l}\nonumber \\
 & =T^{8q/10\left(2q+1\right)+4r\left(1-l\right)/5\left(2q+1\right)}\rightarrow0,\nonumber 
\end{align}
 for some constant $C_{3,1}\in\left(0,\,\infty\right)$, using the
fact that $\inf_{\mathscr{P}\in\boldsymbol{P}_{U,3}}\phi_{\mathscr{P}}\left(\cdot\right)\geq\underline{\phi}>0$
and $q/\left(l-1\right)<r$. Let
\begin{align*}
B_{1,1,T} & =\left(T^{8q/5\left(2q+1\right)}\sup_{\mathscr{P}\in\boldsymbol{P}_{U,3}}\mathbb{E}_{\mathscr{P}}\left(\sum_{k=S_{\mathscr{P},T}+1}^{\left\lfloor D_{T}T^{1/2}\right\rfloor }K_{1}\left(\widehat{b}_{1,T}k\right)a'\Gamma_{\mathscr{P},T}a\right)^{2}\right)^{1/2}\\
B_{1,2,T} & =\left(T^{8q/5\left(2q+1\right)}\sup_{\mathscr{P}\in\boldsymbol{P}_{U,3}}\mathbb{E}_{\mathscr{P}}\left(\sum_{k=\left\lfloor D_{T}T^{1/2}\right\rfloor +1}^{T}K_{1}\left(\widehat{b}_{1,T}k\right)a'\Gamma_{\mathscr{P},T}a\right)^{2}\right)^{1/2}.
\end{align*}
 We have 
\begin{align}
B_{1,1,T}^{2} & \leq T^{8q/5\left(2q+1\right)-4/5}\sup_{\mathscr{P}\in\boldsymbol{P}_{U,3}}\mathbb{E}_{\mathscr{P}}\left(\sum_{k=S_{\mathscr{P},T}+1}^{\left\lfloor D_{T}T^{1/2}\right\rfloor }C_{1}\left(\widehat{b}_{1,T}k\right)^{-b}\sqrt{T\overline{b}_{2,T}^{\mathrm{opt}}}\left|a'\widehat{\Gamma}\left(k\right)a-a'\Gamma_{\mathscr{P},T}\left(k\right)a\right|\right)^{2}\label{Eq. (A.39) Andrews 88}\\
 & \leq T^{8q/5\left(2q+1\right)-4/5+8b/5\left(2q+1\right)}\sup_{\mathscr{P}\in\boldsymbol{P}_{U,3}}\mathbb{E}_{\mathscr{P}}\left(\sum_{k=S_{\mathscr{P},T}+1}^{\left\lfloor D_{T}T^{1/2}\right\rfloor }C_{1}k^{-b}\sqrt{T\overline{b}_{2,T}^{\mathrm{opt}}}\left|a'\widehat{\Gamma}\left(k\right)a-a'\Gamma_{\mathscr{P},T}\left(k\right)a\right|\right)^{2}\nonumber \\
 & \quad\times\left(2qK_{1,q}^{2}\widehat{\phi}\left(q\right)/(\int K_{1}^{2}\left(y\right)dy\int_{0}^{1}K_{2}^{2}\left(x\right)dx)\right)^{2b/\left(2q+1\right)}\nonumber \\
 & \leq C_{1,2}T^{8q/5\left(2q+1\right)-4/5+8b/5\left(2q+1\right)}\nonumber \\
 & \quad\times\sup_{\mathscr{P}\in\boldsymbol{\mathscr{P}}_{U,3}}\left(\sum_{k=S_{\mathscr{P},T}+1}^{\left\lfloor D_{T}T^{1/2}\right\rfloor }\sum_{j=S_{\mathscr{P},T}+1}^{\left\lfloor D_{T}T^{1/2}\right\rfloor }k^{-b}j^{-b}T\overline{b}_{\theta_{2},T}\left(\mathrm{Var}_{\mathscr{P}}\left(a'\widehat{\Gamma}\left(k\right)a\right)\mathrm{Var}_{\mathscr{P}}\left(a'\widehat{\Gamma}\left(j\right)a\right)\right)^{1/2}\right)\nonumber \\
 & \leq C_{1,2}T^{8q/5\left(2q+1\right)-4/5+8b/5\left(2q+1\right)}\sup_{\mathscr{P}\in\boldsymbol{P}_{U,3}}\left(\left(\sum_{k=S_{\mathscr{P},T}+1}^{\left\lfloor D_{T}T^{1/2}\right\rfloor }k^{-b}\right)^{2}T\overline{b}_{2,T}^{\mathrm{opt}}\left(\sup_{k\geq1}\mathrm{Var}_{\mathscr{P}_{U}}\left(a'\widehat{\Gamma}\left(k\right)a\right)\right)\right)\nonumber \\
 & \leq C_{1,2}T^{8q/5\left(2q+1\right)-4/5+8b/5\left(2q+1\right)}\sup_{\mathscr{P}\in\boldsymbol{P}_{U,3}}\left(\left(\sum_{k=S_{\mathscr{P},T}+1}^{\left\lfloor D_{T}T^{1/2}\right\rfloor }k^{-b}\right)^{2}\right)O\left(1\right)\nonumber \\
 & \leq C_{1,3}T^{8q/5\left(2q+1\right)-4/5+8b/5\left(2q+1\right)-8\left(b-1\right)r/5\left(2q+1\right)}\rightarrow0,\nonumber 
\end{align}
for some constants $0<C_{1,2},\,C_{1,3}<\infty$, using the fact
that $\widehat{\phi}\left(q\right)\leq\overline{\phi}<\infty$, $\inf_{\mathscr{P}\in\boldsymbol{P}_{U,3}}\phi_{\mathscr{P}}\geq\underline{\phi}>0$
and $r>1.25.$ Using similar manipulations,
\begin{align}
B_{1,2,T}^{2} & \leq T^{8q/5\left(2q+1\right)-4/5}\sup_{\mathscr{P}\in\boldsymbol{P}_{U,3}}\mathbb{E}_{\mathscr{P}}\left(\sum_{k=\left\lfloor D_{T}T^{1/2}\right\rfloor +1}^{T}C_{1}\left(\widehat{b}_{1,T}k\right)^{-b}\sqrt{T\overline{b}_{2,T}^{\mathrm{opt}}}\left|a'\widehat{\Gamma}\left(k\right)a-a'\Gamma_{\mathscr{P},T}\left(k\right)a\right|\right)^{2}\label{Eq. (A.39) Andrews 88-1}\\
 & \leq C_{1,2}T^{8q/5\left(2q+1\right)-4/5+8b/5\left(2q+1\right)}\sup_{\mathscr{P}\in\boldsymbol{P}_{U,3}}\left(\left(\sum_{k=\left\lfloor D_{T}T^{1/2}\right\rfloor +1}^{T}k^{-b}\right)^{2}\right)O\left(1\right)\nonumber \\
 & \leq C_{1,3}T^{8q/5\left(2q+1\right)-4/5+8b/5\left(2q+1\right)-\left(b-1\right)}\rightarrow0,\nonumber 
\end{align}
for some constants $0<C_{1,2},\,C_{1,3}<\infty$ and with $q$ satisfying
$8/q-20q<6$.  Equations \eqref{Eq. (A.37) Andrews 88}-\eqref{Eq. (A.39) Andrews 88-1}
combine to establish part (i). We now prove part (ii). Using the
Lipschitz condition on $K_{1}\left(\cdot\right)$, we get 
\begin{align}
A_{1,T} & =T^{8q/5\left(2q+1\right)}\sup_{\mathscr{P}\in\boldsymbol{P}_{U,3}}\mathbb{E}_{\mathscr{P}}\left(\sum_{k=1}^{S_{\mathscr{P},T}}\left(K_{1}\left(\widehat{b}_{1,T}k\right)-K_{1}\left(b_{1,\theta_{\mathscr{P}},T}k\right)\right)a'\widehat{\Gamma}\left(k\right)a\right)^{2}\label{Eq. (A.40) Andrews 88}\\
 & \leq T^{8q/5\left(2q+1\right)}\sup_{\mathscr{P}\in\boldsymbol{P}_{U,3}}\mathbb{E}_{\mathscr{P}}\left(\sum_{k=1}^{S_{\mathscr{P},T}}C_{2}\left(\widehat{b}_{1,T}-b_{1,\theta_{\mathscr{P}},T}\right)ka'\widehat{\Gamma}\left(k\right)a\right)^{2}\nonumber \\
 & \leq C_{2,1}T^{8q/5\left(2q+1\right)-8/5\left(2q+1\right)}\widetilde{n}_{T}^{-1}\sup_{\mathscr{P}\in\boldsymbol{P}_{U,3}}\mathbb{E}_{\mathscr{P}}\left(\sum_{k=1}^{S_{\mathscr{P},T}}\left(\frac{\sqrt{\widetilde{n}_{T}}\left(\widehat{\phi}\left(q\right)^{1/\left(2q+1\right)}-\phi_{\theta_{\mathscr{P}}^{*}}\left(q\right)^{1/\left(2q+1\right)}\right)}{\left(\widehat{\phi}\left(q\right)\phi_{\theta_{\mathscr{P}}^{*}}\left(q\right)\right)^{1/\left(2q+1\right)}}\right)ka'\widehat{\Gamma}\left(k\right)a\right)^{2}\nonumber \\
 & \leq C_{2,1}T^{8q/5\left(2q+1\right)-8/5\left(2q+1\right)-6/10}\sup_{\mathscr{P}\in\boldsymbol{P}_{U,3}}\mathbb{E}_{\mathscr{P}}\left(\sum_{k=1}^{S_{\mathscr{P},T}}\left(\frac{\sqrt{\widetilde{n}_{T}}\left(\widehat{\phi}\left(q\right)^{1/\left(2q+1\right)}-\phi_{\theta_{\mathscr{P}}^{*}}\left(q\right)^{1/\left(2q+1\right)}\right)}{\left(\widehat{\phi}\left(q\right)\phi_{\theta_{\mathscr{P}}^{*}}\left(q\right)\right)^{1/\left(2q+1\right)}}\right)ka'\widehat{\Gamma}\left(k\right)a\right)^{2}\nonumber 
\end{align}
 for some constant $C_{2,1}\in\left(0,\,\infty\right)$, where $\widetilde{n}_{T}=(\inf\left\{ n_{3,T}/T,\,\sqrt{n_{2,T}}\right\} )^{2}$.
Now decompose the right-hand side above as follows, 
\begin{align}
A_{1,T}^{1/2} & \leq\Biggl(C_{2,1}T^{8q/5\left(2q+1\right)-8/5\left(2q+1\right)-6/10}\sup_{\mathscr{P}\in\boldsymbol{P}_{U,3}}\mathbb{E}_{\mathscr{P}}\left(\frac{\sqrt{\widetilde{n}_{T}}\left(\widehat{\phi}\left(q\right)^{1/\left(2q+1\right)}-\phi_{\theta_{\mathscr{P}}^{*}}\left(q\right)^{1/\left(2q+1\right)}\right)}{\left(\widehat{\phi}\left(q\right)\phi_{\theta_{\mathscr{P}}^{*}}\left(q\right)\right)^{1/\left(2q+1\right)}}\right)^{2}\label{Eq. (A.41) Andrews 88}\\
 & \quad\times\left(\sum_{k=1}^{S_{\mathscr{P},T}}k\left(a'\widehat{\Gamma}\left(k\right)a-a'\Gamma_{\mathscr{P},T}\left(k\right)a\right)\right)^{2}\Biggr)^{1/2}\nonumber \\
 & \quad+\Biggl(C_{2,1}T^{8q/5\left(2q+1\right)-8/5\left(2q+1\right)-6/10}\sup_{\mathscr{P}\in\boldsymbol{P}_{U,3}}\mathbb{E}_{\mathscr{P}}\left(\frac{\sqrt{\widetilde{n}_{T}}\left(\widehat{\phi}\left(q\right)^{1/\left(2q+1\right)}-\phi_{\theta_{\mathscr{P}}^{*}}\left(q\right)^{1/\left(2q+1\right)}\right)}{\left(\widehat{\phi}\left(q\right)\phi_{\theta_{\mathscr{P}}^{*}}\left(q\right)\right)^{1/\left(2q+1\right)}}\right)^{2}\nonumber \\
 & \quad\times\left(\sum_{k=1}^{S_{\mathscr{P},T}}ka'\Gamma_{\mathscr{P},T}\left(k\right)a\right)^{2}\Biggr)^{1/2}\nonumber \\
 & =A_{1,1,T}+A_{1,2,T}.\nonumber 
\end{align}
where we have used the fact that $n_{2,T}^{10/6}/T\rightarrow[c_{2},\,\infty),$
$n_{3,T}^{10/6}/T\rightarrow[c_{3},\,\infty)$ with $0<c_{2},\,c_{3}<\infty$.
Note that,
\begin{align}
A_{1,1,T}^{2} & \leq C_{2,1}T^{8q/5\left(2q+1\right)-8/5\left(2q+1\right)-3/5}S_{\mathscr{P},T}^{4}\sup_{\mathscr{P}\in\boldsymbol{P}_{U,3}}\mathbb{E}_{\mathscr{P}}\left(\frac{\sqrt{\widetilde{n}_{T}}\left(\widehat{\phi}\left(q\right)^{1/\left(2q+1\right)}-\phi_{\mathscr{\theta_{\mathscr{P}}^{*}}}\left(q\right)^{1/\left(2q+1\right)}\right)}{\left(\widehat{\phi}\left(q\right)\phi_{\theta_{\mathscr{P}}^{*}}\left(q\right)\right)^{1/\left(2q+1\right)}}\right)^{2}\label{Eq. (A.42) Andrews 88}\\
 & \quad\times\left(\frac{1}{S_{\mathscr{P},T}}\sum_{k=1}^{S_{\mathscr{P},T}}\frac{k}{S_{\mathscr{P},T}}\left(a'\widehat{\Gamma}\left(k\right)a-a'\Gamma_{\mathscr{P},T}\left(k\right)a\right)\right)^{2}\nonumber \\
 & \leq C_{2,1}T^{8q/5\left(2q+1\right)-8/5\left(2q+1\right)-3/5+16r/5\left(2q+1\right)-4/5}\nonumber \\
 & \quad\times\left(\sup_{\mathscr{P}\in\boldsymbol{P}_{U,3}}\mathbb{E}_{\mathscr{P}}\left(\frac{\sqrt{\widetilde{n}_{T}}\left(\widehat{\phi}\left(q\right)^{1/\left(2q+1\right)}-\phi_{\theta_{\mathscr{P}}^{*}}\left(q\right)^{1/\left(2q+1\right)}\right)}{\left(\widehat{\phi}\left(q\right)\phi_{\theta_{\mathscr{P}}^{*}}\left(q\right)\right)^{1/\left(2q+1\right)}}\right)^{4}\right)^{1/2}\nonumber \\
 & \quad\times\left(\sup_{\mathscr{P}\in\boldsymbol{P}_{U,3}}\mathbb{E}_{\mathscr{P}}\left(\frac{1}{S_{\mathscr{P},T}}\sum_{k=1}^{S_{\mathscr{P},T}}\sqrt{T\overline{b}_{2,T}^{\mathrm{opt}}}\left(a'\widehat{\Gamma}\left(k\right)a-a'\Gamma_{\mathscr{P},T}\left(k\right)a\right)\right)^{4}\right)^{1/2}\nonumber \\
 & \quad\times\left(2qK_{1,q}^{2}\phi_{\theta_{\mathscr{P}}^{*}}\left(q\right)/\int K_{1}^{2}\left(y\right)dy\int_{0}^{1}K_{2}^{2}\left(x\right)dx\right)^{4r/\left(2q+1\right)}\rightarrow0,\nonumber 
\end{align}
 for some constant $C_{2,1}\in\left(0,\,\infty\right)$, since
$\sup_{\mathscr{P}\in\boldsymbol{P}_{U,3}}\phi_{\theta_{\mathscr{P}}^{*}}<\infty$
and $r<15/16+3q/8$. In addition, we have 
\begin{align}
A_{1,2,T}^{2} & \leq C_{2,1}T^{8q/5\left(2q+1\right)-8/5\left(2q+1\right)-3/5}\sup_{\mathscr{P}\in\boldsymbol{P}_{U,3}}\mathbb{E}_{\mathscr{P}}\left(\frac{\sqrt{\widetilde{n}_{T}}\left(\widehat{\phi}\left(q\right)^{1/\left(2q+1\right)}-\phi_{\theta_{\mathscr{P}}^{*}}\left(q\right)^{1/\left(2q+1\right)}\right)}{\left(\widehat{\phi}\left(q\right)\phi_{\theta_{\mathscr{P}}^{*}}\left(q\right)\right)^{1/\left(2q+1\right)}}\right)^{2}\label{Eq. (A.43) Andrews 88}\\
 & \quad\times\sup_{\mathscr{P}\in\boldsymbol{P}_{U,3}}\left(\sum_{k=1}^{S_{\mathscr{P},T}}ka'\Gamma_{\mathscr{P},T}\left(k\right)a\right)^{2}\nonumber \\
 & \leq C_{2,1}T^{8q/5\left(2q+1\right)-8/5\left(2q+1\right)-3/5}\sup_{\mathscr{P}\in\boldsymbol{P}_{U,3}}\mathbb{E}_{\mathscr{P}}\left(\frac{\sqrt{\widetilde{n}_{T}}\left(\widehat{\phi}\left(q\right)^{1/\left(2q+1\right)}-\phi_{\theta_{\mathscr{P}}^{*}}\left(q\right)^{1/\left(2q+1\right)}\right)}{\left(\widehat{\phi}\left(q\right)\phi_{\theta_{\mathscr{P}}^{*}}\left(q\right)\right)^{1/\left(2q+1\right)}}\right)^{2}\nonumber \\
 & \quad\times\sup_{\mathscr{P}\in\boldsymbol{P}_{U,3}}\left(\sum_{k=1}^{S_{\mathscr{P},T}}k^{1-l}\right)^{2}\rightarrow0,\nonumber 
\end{align}
where we have used the definition of $\boldsymbol{P}_{U,3}$-(ii),
$q<11/2$ and $l>2$ which implies that $\sum_{k=1}^{\infty}k^{1-l}<\infty.$
Equations \eqref{Eq. (A.41) Andrews 88}-\eqref{Eq. (A.43) Andrews 88}
combine to establish part (ii) of the lemma. $\square$

\medskip{}

\noindent\textit{Proof of Theorem }\ref{Theorem 6 Andrews 88}. Let
$||\cdot||_{\mathscr{P}}=(\mathbb{E}_{\mathscr{P}}\left(\cdot\right)^{2})^{1/2}$.
For any constant $J$ and any random variables $\widehat{J}_{1}$
and $\widehat{J}_{2}$, the triangle inequality gives 
\begin{align}
\left\Vert \widehat{J}_{1}-\widehat{J}_{2}\right\Vert _{\mathscr{P}} & \geq\left|\left\Vert \widehat{J}_{1}-J\right\Vert _{\mathscr{P}}-\left\Vert J-\widehat{J}_{2}\right\Vert _{\mathscr{P}}\right|.\label{Eq. (A.34) Andrews 88}
\end{align}
Hence, it suffices to show that 
\begin{align}
T^{8q/5\left(2q+1\right)}\sup_{\mathscr{P}\in\boldsymbol{P}_{U,3}}\left\Vert a'\widehat{J}_{T}\left(\widehat{b}_{1,T},\,\widehat{\overline{b}}_{2,T}\right)a-a'\widehat{J}_{T}\left(b_{1,\theta_{\mathscr{P}},T},\,\overline{b}_{2,T}^{\mathrm{opt}}\right)a\right\Vert _{\mathscr{P}}^{2} & \rightarrow0.\label{Eq. (A.35) Andrews 88}
\end{align}
The latter follows from 
\begin{align}
T^{8q/5\left(2q+1\right)} & \sup_{\mathscr{P}\in\boldsymbol{P}_{U,3}}\left\Vert a'\widehat{J}_{T}\left(\widehat{b}_{1,T},\,\widehat{\overline{b}}_{2,T}\right)a-a'\widehat{J}_{T}\left(b_{1,\theta_{\mathscr{P}},T},\,\widehat{\overline{b}}_{2,T}\right)a\right\Vert _{\mathscr{P}}^{2}\label{Eq. (A.35) Andrews 88-1}\\
 & +T^{8q/5\left(2q+1\right)}\sup_{\mathscr{P}\in\boldsymbol{P}_{U,3}}\left\Vert a'\widehat{J}_{T}\left(b_{1,\theta_{\mathscr{P}},T},\,\widehat{\overline{b}}_{2,T}\right)a-a'\widehat{J}_{T}\left(b_{1,\theta_{\mathscr{P}},T},\,\overline{b}_{2,T}^{\mathrm{opt}}\right)a\right\Vert _{\mathscr{P}}^{2}\rightarrow0.\nonumber 
\end{align}
Note that 
\begin{align}
a' & \widehat{J}_{T}\left(\widehat{b}_{1,T},\,\widehat{\overline{b}}_{2,T}\right)a-a'\widehat{J}_{T}\left(b_{1,\theta_{\mathscr{P}},T},\,\widehat{\overline{b}}_{2,T}\right)a\label{Eq. (A.36) Andrews 88}\\
 & =2\sum_{k=S_{\mathscr{P},T}+1}^{T-1}\left(K_{1}\left(\widehat{b}_{1,T}k\right)-K_{1}\left(b_{1,\theta_{\mathscr{P}},T}k\right)\right)a'\widehat{\Gamma}\left(k\right)a\nonumber \\
 & \quad+2\sum_{k=1}^{S_{\mathscr{P},T}}K_{1}\left(\widehat{b}_{1,T}k\right)a'\widehat{\Gamma}\left(k\right)a-2\sum_{k=1}^{S_{\mathscr{P},T}}K_{1}\left(b_{1,\theta_{\mathscr{P}},T}k\right)a'\widehat{\Gamma}\left(k\right)a.\nonumber 
\end{align}
 We can apply Lemma \ref{Lemma A.1 Andrews 88}-(ii) to the first
term of \eqref{Eq. (A.36) Andrews 88} and Lemma \ref{Lemma A.1 Andrews 88}-(i)
to second and third terms (with $\{b_{1,\theta_{\mathscr{P}},T}\}$
in place of $\{\widehat{b}_{1,T}\}$ for the third term). It remains
to show that the second summand of \eqref{Eq. (A.35) Andrews 88-1}
converges to zero. Let $\widehat{c}_{\theta_{2},T}\left(rn_{T}/T,\,k\right)$
denote the estimator that uses $b_{2,T}^{\mathrm{opt}}\left(u\right)$
in place of $\widehat{b}_{2,T}\left(u\right).$ We have for $k\geq0,$
\begin{align}
\widehat{c}_{T} & \left(rn_{T}/T,\,k\right)-\widehat{c}_{\theta_{2},T}\left(rn_{T}/T,\,k\right)\nonumber \\
 & =\left(T\overline{b}_{2,T}^{\mathrm{opt}}\right)^{-1}\sum_{s=k+1}^{T}\left(K_{2}^{*}\left(\frac{\left(\left(r+1\right)n_{T}-\left(s-k/2\right)\right)/T}{\widehat{b}_{2,T}\left(\left(r+1\right)n_{T}/T\right)}\right)-K_{2}^{*}\left(\frac{\left(\left(r+1\right)n_{T}-\left(s-k/2\right)\right)/T}{b_{2,T}^{\mathrm{opt}}\left(\left(r+1\right)n_{T}/T\right)}\right)\right)\widehat{V}_{s}\widehat{V}{}_{s-k}\nonumber \\
 & \quad+O_{\mathbb{P}}\left(1/T\overline{b}_{2,T}^{\mathrm{opt}}\right).\label{eq (c_hat - c_theta2)}
\end{align}
Given Assumption \ref{Assumption E-F-G}-(v) \ref{Assumption C Andrews 88}-(ii,iii)
and using the delta method, we have for $s\in\{Tu-\bigl\lfloor T\overline{b}_{2,T}^{\mathrm{opt}}\bigr\rfloor,\ldots,\,Tu+\bigl\lfloor T\overline{b}_{2,T}^{\mathrm{opt}}\bigr\rfloor\}$:
\begin{align}
K_{2} & \left(\frac{\left(Tu-\left(s-k/2\right)\right)/T}{\widehat{b}_{2,T}\left(u\right)}\right)-K_{2}\left(\frac{\left(Tu-\left(s-k/2\right)\right)/T}{b_{2,T}^{\mathrm{opt}}\left(u\right)}\right)\label{Eq. K2-K2 fort part (ii)}\\
 & \leq C_{4}\left|\frac{Tu-\left(s-k/2\right)}{T\widehat{b}_{2,T}\left(u\right)}-\frac{Tu-\left(s-k/2\right)}{Tb_{2,T}^{\mathrm{opt}}\left(u\right)}\right|\nonumber \\
 & \leq CT^{-4/5-2/5}T^{2/5}\left|\left(\frac{\widehat{D}_{2}\left(u\right)}{\widehat{D}_{1}\left(u\right)}\right)^{-1/5}-\left(\frac{D_{2}\left(u\right)}{D_{1,\theta}\left(u\right)}\right)^{-1/5}\right|\left|Tu-\left(s-k/2\right)\right|\nonumber \\
 & \leq CT^{-4/5-2/5}O_{\mathbb{P}}\left(1\right)\left|Tu-\left(s-k/2\right)\right|.\nonumber 
\end{align}
  Therefore, 
\begin{align}
T^{8q/10\left(2q+1\right)} & \left(a'\widehat{J}_{T}\left(b_{1,\theta_{\mathscr{P}},T},\,\widehat{\overline{b}}_{2,T}\right)a\right.-\left.a'\widehat{J}_{T}\left(b_{1,\theta_{\mathscr{P}},T},\,\overline{b}_{2,T}^{\mathrm{opt}}\right)a\right)\label{Eq. (H1+H2+H3)}\\
 & =T^{8q/10\left(2q+1\right)}\sum_{k=-T+1}^{T-1}K_{1}\left(b_{1,\theta_{\mathscr{P}},T}k\right)\frac{n_{T}}{T}\sum_{r=0}^{\left\lfloor T/n_{T}\right\rfloor }\left(a'\widehat{c}\left(rn_{T}/T,\,k\right)a-a'\widehat{c}_{\theta_{2},T}\left(rn_{T}/T,\,k\right)a\right)\nonumber \\
 & \leq T^{8q/10\left(2q+1\right)}C\sum_{k=-T+1}^{T-1}\bigl|K_{1}\left(b_{1,\theta_{\mathscr{P}},T}k\right)\bigr|\frac{n_{T}}{T}\sum_{r=0}^{\left\lfloor T/n_{T}\right\rfloor }\frac{1}{T\overline{b}_{2,T}^{\mathrm{opt}}}\nonumber \\
 & \quad\times\sum_{s=k+1}^{T}\left|K_{2}^{*}\left(\frac{\left(\left(r+1\right)n_{T}-\left(s-k/2\right)\right)/T}{\widehat{b}_{2,T}\left(\left(r+1\right)n_{T}/T\right)}\right)-K_{2}^{*}\left(\frac{\left(\left(r+1\right)n_{T}-\left(s-k/2\right)\right)/T}{b_{2,T}^{\mathrm{opt}}\left(\left(r+1\right)n_{T}/T\right)}\right)\right|\nonumber \\
 & \quad\times\left|\left(a'\widehat{V}_{s}\widehat{V}'_{s-k}a-\mathbb{E}_{\mathscr{P}}\left(a'V_{s}V'_{s-k}a\right)\right)+\mathbb{\mathbb{E}_{\mathscr{P}}}\left(a'V_{s}V'_{s-k}a\right)\right|\nonumber \\
 & \triangleq H_{1,T}+H_{2,T}.\nonumber 
\end{align}
We have to show that $H_{1,T}+H_{2,T}\overset{\mathbb{P}}{\rightarrow}0$.
Let $H_{1,1,T}$ (resp. $H_{1,2,T}$) be defined as $H_{1,T}$ but
with the sum over $k$ restricted to $k=1,\ldots,\,S_{T}$ (resp.
$k=S_{T}+1,\ldots,\,T$). Let $H_{2,1,T}$ (resp. $H_{2,2,T}$) be
defined as $H_{2,T}$ but with the sum over $k$ be restricted to
$k=1,\ldots,\,S_{T}$ (resp. $k=S_{T}+1,\ldots,\,T$). Using the definition
of $\boldsymbol{P}_{U,3}$,
\begin{align}
\mathbb{E}\left(H_{1,1,T}^{2}\right) & \leq T^{8q/5\left(2q+1\right)}\sum_{k=1}^{S_{T}}\sum_{j=1}^{S_{T}}K_{1}\left(b_{1,\theta_{\mathscr{P}},T}k\right)K_{1}\left(b_{1,\theta_{\mathscr{P}},T}j\right)\left(\frac{n_{T}}{T}\right)^{2}\sum_{r_{1}=0}^{\left\lfloor T/n_{T}\right\rfloor }\sum_{r_{2}=0}^{\left\lfloor T/n_{T}\right\rfloor }\frac{1}{\left(T\overline{b}_{2,T}^{\mathrm{opt}}\right)^{2}}\label{Eq. (H21) =00003D 0 part (ii)}\\
 & \quad\times\sum_{s=k+1}^{T}\sum_{t=j+1}^{T}\left(K_{2}^{*}\left(\frac{\left(\left(r_{1}+1\right)n_{T}-\left(s-k/2\right)\right)/T}{\widehat{b}_{2,T}\left(\left(r_{1}+1\right)n_{T}/T\right)}\right)-K_{2}^{*}\left(\frac{\left(\left(r_{1}+1\right)n_{T}-\left(s-k/2\right)\right)/T}{b_{2,T}^{\mathrm{opt}}\left(\left(r_{1}+1\right)n_{T}/T\right)}\right)\right)\nonumber \\
 & \quad\times\left(K_{2}^{*}\left(\frac{\left(\left(r_{2}+1\right)n_{T}-\left(t-j/2\right)\right)/T}{\widehat{b}_{2,T}\left(\left(r_{2}+1\right)n_{T}/T\right)}\right)-K_{2}^{*}\left(\frac{\left(\left(r_{2}+1\right)n_{T}-\left(t-j/2\right)\right)/T}{b_{2,T}^{\mathrm{opt}}\left(\left(r_{2}+1\right)n_{T}/T\right)}\right)\right)\nonumber \\
 & \quad\times\mathbb{E}_{\mathscr{P}}\left(a'\widehat{V}_{s}\widehat{V}'_{s-k}a-\mathbb{E}_{\mathscr{P}}\left(V_{s}V_{s-k}\right)\right)\left(a'\widehat{V}_{t}\widehat{V}'_{t-j}a-\mathbb{E}_{\mathscr{P}}\left(V_{t}V_{t-j}\right)\right)\nonumber \\
 & \leq CT^{8q/5\left(2q+1\right)}S_{T}^{2}T^{-2/5}\left(T\overline{b}_{2,T}^{\mathrm{opt}}\right)^{-1}\sup_{k\geq1}T\overline{b}_{2,T}^{\mathrm{opt}}\mathrm{Var}_{\mathscr{P}}\left(\widehat{\Gamma}\left(k\right)\right)O_{\mathbb{P}}\left(1\right)\nonumber \\
 & \leq CT^{8q/5\left(2q+1\right)}S_{T}^{2}T^{-2/5}\left(T\overline{b}_{2,T}^{\mathrm{opt}}\right)^{-1}\sup_{k\geq1}T\overline{b}_{2,T}^{\mathrm{opt}}\mathrm{Var}_{\mathscr{P}_{U}}\left(\widehat{\Gamma}\left(k\right)\right)O_{\mathbb{P}}\left(1\right)\nonumber \\
 & \leq CT^{\left(8q+8r\right)/5\left(2q+1\right)-2/5-1}O_{\mathbb{P}}\left(\left(\overline{b}_{2,T}^{\mathrm{opt}}\right)^{-1}\right)\rightarrow0,\nonumber 
\end{align}
where we have used $r<\left(6+4q\right)/8$. Turning to $H_{1,2,T},$
\begin{align}
\mathbb{E}\left(H_{1,2,T}^{2}\right) & \leq T^{8q/5\left(2q+1\right)-2/5}\left(T\overline{b}_{2,T}^{\mathrm{opt}}\right)^{-1}b_{1,\theta_{\mathscr{P}},T}^{-2b}\left(\sum_{k=S_{T}+1}^{T-1}k^{-b}\sqrt{T\overline{b}_{2,T}^{\mathrm{opt}}}\left(\mathrm{Var}_{\mathscr{P}}\left(\widehat{\Gamma}\left(k\right)\right)\right)^{1/2}O\left(1\right)\right)^{2}\label{Eq. (H22) =00003D0 part (ii)}\\
 & \leq T^{8q/5\left(2q+1\right)}T^{-2/5-1}\left(\overline{b}_{2,T}^{\mathrm{opt}}\right)^{-1}b_{1,\theta_{\mathscr{P}},T}^{-2b}\left(\sum_{k=S_{T}+1}^{T-1}k^{-b}\sqrt{T\overline{b}_{2,T}^{\mathrm{opt}}}\left(\mathrm{Var}_{\mathscr{P}_{U}}\left(\widehat{\Gamma}\left(k\right)\right)\right)^{1/2}\right)^{2}\nonumber \\
 & \leq T^{8q/5\left(2q+1\right)}T^{-2/5-1}\left(\overline{b}_{2,T}^{\mathrm{opt}}\right)^{-1}b_{1,\theta_{\mathscr{P}},T}^{-2b}\left(\sum_{k=S_{T}+1}^{T-1}k^{-b}O\left(1\right)\right)^{2}\nonumber \\
 & \leq T^{8q/5\left(2q+1\right)}T^{-2/5-1}\left(\overline{b}_{2,T}^{\mathrm{opt}}\right)^{-1}b_{1,\theta_{\mathscr{P}},T}^{-2b}S_{T}^{2\left(1-b\right)}\rightarrow0,\nonumber 
\end{align}
since $r>(b-3/4-q/2)/\left(b-1\right).$ Eq. \eqref{Eq. (H21) =00003D 0 part (ii)}
and \eqref{Eq. (H22) =00003D0 part (ii)} yield $H_{1,T}\overset{\mathbb{P}}{\rightarrow}0.$
Given $\left|K_{1}\left(\cdot\right)\right|\leq1$ and \eqref{Eq. K2-K2 fort part (ii)},
we have 
\begin{align*}
\left|H_{2,1,T}\right| & \leq CT^{8q/10\left(2q+1\right)}T^{-2/5}\sum_{k=1}^{\infty}k^{-l}\rightarrow0,
\end{align*}
since $\sum_{k=1}^{\infty}k^{-l}<\infty$ for $l>1$ and $T^{8q/10\left(2q+1\right)}T^{-2/5}\rightarrow0.$
Finally, 
\begin{align*}
\left|H_{2,2,T}\right| & \leq CT^{8q/10\left(2q+1\right)}T^{-2/5}\sum_{k=S_{T}+1}^{T-1}\left|\Gamma_{\mathscr{P}_{U},T}\left(k\right)\right|\\
 & \leq CT^{8q/10\left(2q+1\right)}T^{-2/5}S_{T}^{1-l}\\
 & \leq CT^{8q/10\left(2q+1\right)}T^{-2/5}T^{4r\left(1-l\right)/5\left(2q+1\right)}\rightarrow0,
\end{align*}
which completes the proof. $\square$

\section{\label{Section Proofs of Section Power}Proof of the Results of Section
\ref{Section Power}}

\subsection{Proof of Theorem \ref{Theorem Power DM HAR Tests}}

Consider first the numerator of $t_{\mathrm{DM},i}$.  We have 
\begin{align*}
T_{n}^{1/2}\overline{d}_{L} & =\delta^{2}O_{\mathbb{P}}\left(T_{n}^{1/2}T_{n}^{-1}n_{\delta}\right)+O_{\mathbb{P}}\left(T_{n}^{1/2}T_{n}^{-1}\left(T_{n}-n_{\delta}\right)^{1/2}\right)\mathscr{N}\left(0,\,J_{\mathrm{DM}}\right)\\
 & =\delta^{2}O_{\mathbb{P}}\left(T_{n}^{-1/2}n_{\delta}\right)+O_{\mathbb{P}}\left(1\right),
\end{align*}
 for some $J_{\mathrm{DM}}\in\left(0,\,\infty\right)$ where the
factor $\delta^{2}$ follows from the quadratic loss.

Next, we focus on the expansion of the denominator of $t_{\mathrm{DM},i}$
which hinges on which LRV estimator is used. We begin with part (i).
Under $b_{T}\rightarrow0$ as $T\rightarrow\infty$, Theorem 3.1 in
\citeReferencesSupp{casini/perron_Low_Frequency_Contam_Nonstat:2020}
yields 
\begin{align*}
\widehat{J}_{d_{L},\mathrm{NW87},T} & =\sum_{k=-\left\lfloor b_{T}^{-1}\right\rfloor }^{\left\lfloor b_{T}^{-1}\right\rfloor }\left(1-\left|b_{T}k\right|\right)\widehat{\Gamma}\left(k\right)\\
 & =\sum_{k=-\left\lfloor b_{T}^{-1}\right\rfloor }^{\left\lfloor b_{T}^{-1}\right\rfloor }\left(1-\left|b_{T}k\right|\right)\int_{0}^{1}c\left(u,\,k\right)du\\
 & \quad+\sum_{k=-\left\lfloor b_{T}^{-1}\right\rfloor }^{\left\lfloor b_{T}^{-1}\right\rfloor }\left(1-\left|b_{T}k\right|\right)\left(2^{-1}\left(\frac{T_{b}-T_{m}-1}{T_{n}}\right)\left(\frac{T_{n}-T_{b}-2}{T_{n}}\right)\delta^{4}+o_{\mathbb{P}}\left(1\right)\right)\\
 & =CJ_{\mathrm{DM}}+\sum_{k=-\left\lfloor b_{T}^{-1}\right\rfloor }^{\left\lfloor b_{T}^{-1}\right\rfloor }\left(1-\left|b_{T}k\right|\right)\left(2^{-1}\left(\frac{T_{b}-T_{m}-1}{T_{n}}\right)\left(\frac{T_{n}-T_{b}-2}{T_{n}}\right)\delta^{4}+o_{\mathbb{P}}\left(1\right)\right),
\end{align*}
for some $C>0$ such that $C<\infty$. By Exercise 1.7.12 in \citeReferencesSupp{brillinger:75},
\begin{align*}
\sum_{k=-\left\lfloor b_{T}^{-1}\right\rfloor }^{\left\lfloor b_{T}^{-1}\right\rfloor }\left(1-\left|b_{T}k\right|\right)\exp\left(-i\omega k\right) & =b_{T}\left(\frac{\sin\frac{\left\lfloor b_{T}^{-1}\right\rfloor \omega}{2}}{\sin\frac{\omega}{2}}\right)^{2}.
\end{align*}
 Evaluating the expression above at $\omega=0$ and applying L'Hôpital's
rule we yield,
\begin{align*}
\sum_{k=-\left\lfloor b_{T}^{-1}\right\rfloor }^{\left\lfloor b_{T}^{-1}\right\rfloor }\left(1-\left|b_{T}k\right|\right) & =b_{T}\left(\frac{\frac{\left\lfloor b_{T}^{-1}\right\rfloor }{2}}{\frac{1}{2}}\right)^{2}=\left\lfloor b_{T}^{-1}\right\rfloor .
\end{align*}
 Therefore, $\widehat{J}_{d_{L},\mathrm{NW87},T}=CJ_{\mathrm{DM}}+\delta^{4}O_{\mathbb{P}}(b_{T}^{-1})$
and 

\begin{align}
\left|t_{\mathrm{DM},\mathrm{NW87}}\right| & \leq\frac{\delta^{2}O_{\mathbb{P}}\left(T_{n}^{-1/2}n_{\delta}\right)+O_{\mathbb{P}}\left(1\right)}{\left(\delta^{4}O\left(b_{T}^{-1}\right)\right)^{1/2}}\label{Eq. t-test NW87}\\
 & =\frac{\delta^{2}O\left(T_{n}^{\zeta}\right)}{\delta^{2}O\left(b_{T}^{-1/2}\right)}=O\left(T_{n}^{\zeta}b_{T}^{1/2}\right),\nonumber 
\end{align}
 which implies $\mathbb{P}_{\delta}(|t_{\mathrm{DM},\mathrm{NW87}}|>z_{\alpha})\rightarrow0$
since $T_{n}^{\zeta}b_{T}^{1/2}\rightarrow0$. 

If $b_{T}=O(T^{-1/3})$, similar derivations yield $|t_{\mathrm{DM},\mathrm{NW87}}|=O(T_{n}^{\zeta-1/6})$
and $\mathbb{P}_{\delta}(|t_{\mathrm{DM},\mathrm{NW87}}|>z_{\alpha})\rightarrow0$. 

We now consider part (ii). We have $b_{T}\rightarrow0$ as $T\rightarrow\infty$.
The whitening step for $\widehat{J}_{d_{L},\mathrm{pwNW87},T}$ involves
the following fitted least-squares regression, 
\begin{align*}
\widehat{V}_{t} & =\widehat{A}_{1}\widehat{V}_{t-1}+\widehat{V}_{t}^{*}\quad\mathrm{for\quad}t=2,\ldots,\,T_{n}-1,
\end{align*}
 where $\widehat{A}_{1}$ is the least-squares estimate and $\{\widehat{V}_{t}^{*}\}$
is the corresponding least-squares residual. Under $H_{1},$ $\widehat{V}_{t}^{*}$
exhibits a break in the mean of magnitude $\delta^{2}$ because $\widehat{V}_{t}$
has a break in the mean of the same magnitude after $T_{b}$. From
\citeReferencesSupp{casini/perron_Low_Frequency_Contam_Nonstat:2020}
it follows that $\widehat{A}_{1}\overset{\mathbb{P}}{\rightarrow}A_{1,B}$
with $|1-A_{1,B}|<|1-A_{1}|$ where $A_{1}$ is such that $\widehat{A}_{1}\overset{\mathbb{P}}{\rightarrow}A_{1}$
under the null hypothesis (i.e., under $\delta=0$). Let 
\begin{align*}
\widehat{J}_{T}^{*}=\sum_{k=-\left\lfloor b_{T}^{-1}\right\rfloor }^{\left\lfloor b_{T}^{-1}\right\rfloor }K_{1}\left(b_{T}k\right)\widehat{\Gamma}^{*}\left(k\right), & \qquad\widehat{\Gamma}^{*}\left(k\right)=\left(T_{n}-2\right)^{-1}\sum_{t=|k|+2}^{T_{n}-1}\widehat{V}_{t}^{*}\widehat{V}_{t-|k|}^{*},
\end{align*}
and $c^{*}\left(u,\,k\right)=\mathrm{Cov}(\widehat{V}_{\left\lfloor Tu\right\rfloor }^{*},\,\widehat{V}_{\left\lfloor Tu\right\rfloor -k}^{*})$.
Using Theorem 3.1 in \citeReferencesSupp{casini/perron_Low_Frequency_Contam_Nonstat:2020},
\begin{align*}
\widehat{J}_{d_{L},\mathrm{pwNW87},T} & =\left(1-\widehat{A}_{1}\right)^{-2}\widehat{J}_{T}^{*}\\
 & =\left(1-\widehat{A}_{1}\right)^{-2}\sum_{k=-\left\lfloor b_{T}^{-1}\right\rfloor }^{\left\lfloor b_{T}^{-1}\right\rfloor }\left(1-\left|b_{T}k\right|\right)\widehat{\Gamma}^{*}\left(k\right)\\
 & =\left(1-\widehat{A}_{1}\right)^{-2}\sum_{k=-\left\lfloor b_{T}^{-1}\right\rfloor }^{\left\lfloor b_{T}^{-1}\right\rfloor }\left(1-\left|b_{T}k\right|\right)\int_{0}^{1}c^{*}\left(u,\,k\right)du\\
 & \quad+\left(1-\widehat{A}_{1}\right)^{-2}\sum_{k=-\left\lfloor b_{T}^{-1}\right\rfloor }^{\left\lfloor b_{T}^{-1}\right\rfloor }\left(1-\left|b_{T}k\right|\right)\left(2^{-1}\left(\frac{T_{b}-T_{m}}{T_{n}}\right)\left(\frac{T_{n}-T_{b}}{T_{n}}\right)\delta^{4}+o_{\mathbb{P}}\left(1\right)\right)\\
 & =C\left(1-A_{1,B}\right)^{-2}J_{\mathrm{DM}}\\
 & \quad+\left(1-A_{1,B}\right)^{-2}\sum_{k=-\left\lfloor b_{T}^{-1}\right\rfloor }^{\left\lfloor b_{T}^{-1}\right\rfloor }\left(1-\left|b_{T}k\right|\right)\left(2^{-1}\left(\frac{T_{b}-T_{m}}{T_{n}}\right)\left(\frac{T_{n}-T_{b}}{T_{n}}\right)\delta^{4}+o_{\mathbb{P}}\left(1\right)\right),
\end{align*}
for some finite $C>0$. Thus, $\widehat{J}_{d_{L},\mathrm{pwNW87},T}=C\left(1-A_{1,B}\right)^{-2}J_{\mathrm{DM}}+\left(1-A_{1,B}\right)^{-2}\delta^{4}O_{\mathbb{P}}(b_{T}^{-1})$
and 

\begin{align}
\left|t_{\mathrm{DM},\mathrm{pwNW87}}\right| & \leq\frac{\delta^{2}O_{\mathbb{P}}\left(T_{n}^{-1/2}n_{\delta}\right)+O_{\mathbb{P}}\left(1\right)}{\left(\left(1-A_{1,B}\right)^{-2}\delta^{4}O\left(b_{T}^{-1}\right)\right)^{1/2}}\label{Eq. t-test NW87-1}\\
 & =\frac{\left(1-A_{1,B}\right)\delta^{2}O\left(T_{n}^{\zeta}\right)}{\delta^{2}O\left(b_{T}^{-1/2}\right)}=O\left(T_{n}^{\zeta}b_{T}^{1/2}\right),\nonumber 
\end{align}
 which implies $\mathbb{P}_{\delta}(|t_{\mathrm{pwDM},\mathrm{NW87}}|>z_{\alpha})\rightarrow0$
since $T_{n}^{\zeta}b_{1,T}^{1/2}\rightarrow0.$ 

In part (iii), $b_{T}=T^{-1}$. Proceeding as in \eqref{Eq. t-test NW87}
we have $|t_{\mathrm{DM},\mathrm{KVB}}|=O(T_{n}^{\zeta-1})$ and $\mathbb{P}_{\delta}(|t_{\mathrm{DM},\mathrm{KVB}}|>z_{\alpha})\rightarrow0$
since $T_{n}^{\zeta-1}\rightarrow0.$ 

We consider part (iv). Using Theorem 3.3 in \citeReferencesSupp{casini/perron_Low_Frequency_Contam_Nonstat:2020},
we have 
\begin{align*}
\widehat{J}_{d_{L},\mathrm{DK},T} & =\sum_{k=-T_{n}+2}^{T_{n}-2}K_{1}\left(\widehat{b}_{1,T}k\right)\frac{n_{T_{n}}}{T_{n}}\sum_{r=1}^{\left\lfloor T_{n}/n_{T_{n}}\right\rfloor }\widehat{c}_{T}\left(rn_{T_{n}}/T_{n},\,k\right)\\
 & =\sum_{k=-T_{n}+2}^{T_{n}-2}K_{1}\left(\widehat{b}_{1,T}k\right)\frac{n_{T_{n}}}{T_{n}}\sum_{r=1}^{\left\lfloor T_{n}/n_{T_{n}}\right\rfloor }\biggl(c\left(rn_{T_{n}}/T_{n},\,k\right)\\
 & \quad+\delta^{2}\mathbf{1}\left\{ \left(|rn_{T_{n}}+k/2+T_{n}\widehat{b}_{2,T}/2+1)-T_{b}|/(T_{n}\widehat{b}_{2,T})\right)\in\left(0,\,1\right)\right\} \biggr)+o_{\mathbb{P}}\left(1\right)\\
 & =J_{\mathrm{DM}}+\delta^{2}O_{\mathbb{P}}\left(\left(\widehat{b}_{1,T}\right)^{-1}\widehat{b}_{2,T}\right)+o_{\mathbb{P}}\left(1\right).
\end{align*}
Using $\widehat{b}_{2,T}/\widehat{b}_{1,T}\rightarrow0$ it follows
that 
\begin{align*}
\left|t_{\mathrm{DM},\mathrm{DK}}\right| & =\frac{\delta^{2}O_{\mathbb{P}}\left(T_{n}^{-1/2}n_{\delta}\right)+O_{\mathbb{P}}\left(1\right)}{\left(J_{\mathrm{DM}}+\delta^{2}O_{\mathbb{P}}\left(\left(\widehat{b}_{1,T}\right)^{-1}\widehat{b}_{2,T}\right)\right)^{1/2}}\\
 & =\delta^{2}O\left(T_{n}^{\zeta}\right).
\end{align*}
Since $T_{n}^{\zeta}\rightarrow\infty$, we have $\mathbb{P}_{\delta}(|t_{\mathrm{DM},\mathrm{DK}}|>z_{\alpha})\rightarrow1$. 

Finally, we consider part (v). The whitening step for $\widehat{J}_{d_{L},\mathrm{pwDK},T}$
involves the following fitted least-squares regression, 
\begin{align*}
\widehat{V}_{t} & =\widehat{A}_{r,1}\widehat{V}_{t-1}+\widehat{V}_{t}^{*}\quad\mathrm{for\quad}t=rn_{T_{n}}+1,\ldots,\,\left(r+1\right)n_{T_{n}},
\end{align*}
(for the last block, $t=\left\lfloor T_{n}/n_{T_{n}}\right\rfloor n_{T_{n}}+1,\ldots,\,T_{n}$)
where $\widehat{A}_{r,1}$ is the least-squares estimate and $\{\widehat{V}_{t}^{*}\}$
is the corresponding least-squares residual. Under $H_{1},$ $\widehat{V}_{t}^{*}$
exhibits a break in the mean of magnitude $\delta^{2}$ in the $r^{*}$th
block such that $T_{b}\in\{r^{*}n_{T_{n}}+1,\ldots,\,\left(r^{*}+1\right)n_{T_{n}}\}.$
This follows because $\widehat{V}_{t}$ has a break in the mean of
the same magnitude after $T_{b}$. Note that over the blocks $r\neq r^{*}$,
$\widehat{V}_{t}$ does not have a break in the mean. Using Theorem
3.3 in \citeReferencesSupp{casini/perron_Low_Frequency_Contam_Nonstat:2020},
we have  
\begin{align*}
\widehat{J}_{d_{L},\mathrm{pwDK},T} & =\sum_{k=-T_{n}+1}^{T_{n}-1}K_{1}\left(\widehat{b}_{1,T}^{*}k\right)\frac{n_{T_{n}}}{T_{n}}\sum_{r=1}^{\left\lfloor T_{n}/n_{T_{n}}\right\rfloor }\widehat{c}_{T,D}^{*}\left(rn_{T_{n}}/T_{n},\,k\right)\\
 & =\sum_{k=-T_{n}+1}^{T_{n}-1}K_{1}\left(\widehat{b}_{1,T}^{*}k\right)\frac{n_{T_{n}}}{T_{n}}\sum_{r=1}^{\left\lfloor T_{n}/n_{T_{n}}\right\rfloor }\biggl(c_{D}^{*}\left(rn_{T_{n}}/T_{n},\,k\right)\\
 & \quad+\delta^{2}\mathbf{1}\left\{ \left(|r^{*}n_{T_{n}}+k/2+T_{n}\widehat{b}_{2,T}^{*}/2+1)-T_{b}|/(T_{n}\widehat{b}_{2,T}^{*})\right)\in\left(0,\,1\right)\right\} \biggr)+o_{\mathbb{P}}\left(1\right)\\
 & =J_{\mathrm{DM}}+\delta^{2}O_{\mathbb{P}}\left(\left(\widehat{b}_{1,T}^{*}\right)^{-1}\widehat{b}_{2,T}^{*}\right)+o_{\mathbb{P}}\left(1\right).
\end{align*}
 It follows that 
\begin{align*}
\left|t_{\mathrm{DM},\mathrm{pwDK}}\right| & =\frac{\delta^{2}O_{\mathbb{P}}\left(T_{n}^{-1/2}n_{\delta}\right)+O_{\mathbb{P}}\left(1\right)}{\left(J_{\mathrm{DM}}+\delta^{2}O_{\mathbb{P}}\left(\left(\widehat{b}_{1,T}^{*}\right)^{-1}\widehat{b}_{2,T}^{*}\right)\right)^{1/2}}\\
 & =\delta^{2}O\left(T_{n}^{\zeta}\right).
\end{align*}
Since $T_{n}^{\zeta}\rightarrow\infty$ we have $\mathbb{P}_{\delta}(|t_{\mathrm{DM},\mathrm{pwDK}}|>z_{\alpha})\rightarrow1$.
$\square$ 

\bibliographystyleReferencesSupp{elsarticle-harv}  
\bibliographyReferencesSupp{References_Supp}

\clearpage{}

\end{singlespace}
\end{document}